\documentclass[showpacs,preprintnumbers,eqsecnum,floatfix]{revtex4}
\usepackage{amsmath}
\usepackage{color}
\usepackage{graphicx}
\usepackage{dcolumn}
\usepackage{bm}

\graphicspath{{c:/utsi/tex/mesondprd/}}
\input{tcilatex}

\begin{document}

\title{Tests of Two-Body Dirac Equation Wave Functions in the Decays of
Quarkonium and Positronium into Two Photons }
\author{Horace W. Crater}
\affiliation{The University of Tennessee Space Institute, Tullahoma, TN 37388\footnote{%
hcrater@utsi.edu}}
\author{Cheuk-Yin Wong}
\affiliation{Physics Division, Oak Ridge National Laboratory, Oak Ridge, TN 37831\footnote{%
wongc@ornl.gov}}
\affiliation{Department of Physics, University of Tennessee, Knoxville, TN 37996}
\author{ Peter Van Alstine}
\affiliation{Moorpark, Ca.}
\date{\today}

\begin{abstract}
Two-Body Dirac equations of constraint dynamics provide a covariant
framework to investigate the problem of highly relativistic quarks in meson
bound states. \ This formalism eliminates automatically the problems of
relative time and energy, leading to a covariant three dimensional formalism
with the same number of degrees of freedom as appears in the corresponding
nonrelativistic problem. It provides bound state wave equations with the
simplicity of the nonrelativistic Schr\"{o}dinger equation. \ Here we begin
important tests of the relativistic sixteen component wave function
solutions obtained in a recent work on meson spectroscopy, extending a
method developed previously for positronium decay into two photons. \
Preliminary to this we examine the positronium decay in the $^{3}P_{0,2}$
states as well as the $^{1}S_{0}$. The two-gamma quarkonium decays that we
investigate are for the $\eta _{c},\eta _{c}^{\prime },\chi _{c0},\chi
_{c2},\pi ^{0},\pi _{2},a_{2},$and $f_{2}^{\prime }$ mesons. Our results for
the four charmonium states compare well with those from other quark models
and show the particular importance of including all components of the wave
function as well as strong and CM energy dependent potential effects on the
norm and amplitude. \ The results for the $\pi ^{0}$, although off the
experimental rate by 15\%, is much closer than the usual expectations from a
potential model. We conclude that the Two-Body Dirac equations lead to wave
functions which provide good descriptions of the two-gamma decay amplitude
and can be used with some confidence for other purposes.
\end{abstract}

\pacs{ 12.39.Ki,03.65.Pm,12.39.Pn \ }
\eid{}
\startpage{1}
\endpage{1}
\maketitle

\section{Introduction}

\setcounter{section}{1}Relativistic treatments of the two-body problem arise
in many problems in particle and nuclear physics. Relativistic effects are
important for composite systems with light quarks, in systems with large
coupling strength, and in reactions of these composite objects. In recent
years, there is much interest in the dissociation and the recombinations of
the $J/\psi $ particle in hadron matter or in the quark-gluon plasma \cite%
{Won02}. Reactions of the form 
\begin{equation}
J/\psi +\pi \leftrightarrow D+\bar{D}^{\ast },
\end{equation}%
provide useful information on the suppression or the enhancement of $J/\psi $
in high-energy heavy-ion collisions and are relevant to the use of heavy
quarkonium as a diagnostic tool for the quark-gluon plasma \cite{Mat86}.

Previously, Wong, Barnes, and Swanson studied the above reactions using a
non-relativistic model of the reacting composite objects including pions 
\cite{Won00,Won02}. While the results have been calibrated with the $\pi \pi 
$ scattering phase shifts for the $I=2$ S-wave channel, the use of the
non-relativistic formalism for pions with light constituents may be subject
to question. One should examine the reaction process using a well tested
relativistic formalism. The Two-Body Dirac equations (TBDE) of constraint
dynamics has had successful applications to relativistic two body bound
states in QED \cite{exct,bckr}, QCD \cite{cra94,crater2}, and two body
nucleon-nucleon scattering \cite{liu,geramb}. But its relativistic extension 
\cite{seaps} of the nonrelativistic four-body scattering formalism of Barnes
and Swanson \cite{barnes,wong} involves untested assumptions beyond the
standard constraint formalism. The reaction process is sensitive to the
spatial distribution of the reacting objects. It is thus important to have a
sensitive test of the wave functions obtained in \cite{crater2}.

We perform this test in this paper by examining the application of the
relativistic constraint formalism in the description of decays of mesons
into two photons. \ In the next section we present a brief review of the
constraint formalism as it applies to quark-antiquark bound states. Part of
the purpose of this review section is to outline some of the numerous tests
made so far on the formalism. \ We give the Pauli forms of the Two-Body
Dirac equations of constraint dynamics that we used in \cite{crater2} to
describe the entire meson spectrum (exceptions being light quark isoscalars
such as the $\eta ,\omega ,\eta ^{\prime }$ and their orbital and radial
excitations). \ We review those aspects of the formalism which give one
confidence in the accurate accounts for all bound states from the excited
states of bottomonium to the pion. \ We also list those aspects particularly
related to the pion and its Goldstone boson behavior. \ The constraint
formalism, unlike most of the other ones that purport to account for the
entire meson spectrum, accounts well in standard perturbative approaches as
well as in nonperturbative and numerical approaches for the QED bound state
spectrum \cite{bckr}. \ We emphasize the importance of this by showing the
correlation between that agreement for the singlet and triplet positronium
systems and the $\pi $ and $\rho $ states, including the Goldstone behavior
of the former.

This connection to QED brings us in Section 3 to our treatment of the $%
2~\gamma $ decay of singlet positronium. \ In constraint dynamics, the Two
Body Dirac equations lead to an analytic solution of the singlet states of
positronium. \ For the singlet ground state the wave function is mildly
singular. \ Standard formalisms \cite{sak} will fail with wave functions
that are singular at the origin, 
\begin{equation}
\Gamma (e^{+}e^{-}\rightarrow 2\gamma )=\sigma _{tot}\upsilon
_{e^{+}}\left\vert \psi (0)\right\vert ^{2}.
\end{equation}%
\ 

Independent treatments by Crater \cite{posit} and Ackleh and Barnes \cite%
{ack} develop related (but distinctly different) approaches for folding in
the effect of the Yukawa fermion exchange mass, giving a smearing of the
singularity over the corresponding Compton wave length. \ We give a brief
review of the first of these approaches and how we extend it to include the
effects of the full sixteen component Two-Body Dirac wave function. This
extension does not have any significant effect on the $^{1}S_{0}$
positronium decay rate. \ However, the effects on the decays of the more
relativistic quark-antiquark systems is significant. \ 

We include in Section 3 technical aspects in which we establish in the
context of a $4\times 4$ matrix wave function, more natural for use in the
decay formalism of a particle-antiparticle system than the 16 component
form, the relation between the sector of the full wave function used in the
Pauli form of the bound state equations and the remaining sectors necessary
for a complete description of the decay. \ \ We review our $^{1}S_{0}$
positronium decay results as well as those of our constraint approach for $%
^{3}P_{0}$ and $^{3}P_{2}$ positronium decay. \ Finally we present the
results for the decay rates of the $\eta _{c},\eta _{c}^{\prime },\chi
_{c0},\chi _{c2},\pi ^{0},\pi _{2},a_{2},f_{2}^{\prime }$ mesons. \ We
conclude in Section 4 with a discussion of our results and a comparison with
other approaches. \ 

\section{Constraint Dynamics and Meson\ Bound States~}

\subsection{ Constraint Dynamics for Two Classical Spinless Particles}

Here we give a brief review of the highlights of the constraint approach
serving also to introduce notations. Although Sazdjian has shown that the
bound state equations of \ constraint dynamics are to be viewed as
\textquotedblleft quantum mechanical transforms\textquotedblright\ of the
Bethe-Salpeter equation \cite{saz85}-\cite{saz97} the constraint approach to
the two body problem has its origins in classical relativistic physics \cite%
{di64}-\cite{drz75}. \ Our review here is base on \cite{cra82} and \cite%
{tod76}. Two free spinless particles are described by the mass shell
constraints 
\begin{equation}
\mathcal{H}_{1}^{0}~\equiv p_{1}^{2}+m_{1}^{2} ~~ \mathcal{\approx }~~0~,\ 
\mathcal{H}_{2}^{0}~\equiv p_{2}^{2}+m_{2}^{2} ~~ \mathcal{\approx }~~0.
\end{equation}
We introduce Poincare' invariant world scalar interactions (to display most
simply the basic ideas) by 
\begin{eqnarray}
m_{1} &\rightarrow &m_{1}+S_{1}(x,p_{1},p_{2})\equiv M_{1}(x,p_{1},p_{2}), 
\notag \\
m_{2} &\rightarrow &m_{2}+S_{2}(x,p_{1},p_{2})\equiv M_{2}(x,p_{1},p_{2}), 
\notag \\
x &=&x_{1}-x_{2}.
\end{eqnarray}
Kinematical constraints then become dynamical mass shell constraints: 
\begin{equation}
\mathcal{H}_{i}^{0}=p_{i}^{2}+m_{i}^{2}\rightarrow p_{i}^{2}+M_{i}^{2}\equiv 
\mathcal{H}_{i}\equiv p_{i}^{2}+m_{i}^{2}+\Phi _{i}(x,p_{1},p_{2}).
\end{equation}

Each constraint must be conserved, implying that the two constraints must be
compatible 
\begin{eqnarray}
0 &\approx &\{\mathcal{H}_{1},\mathcal{H}_{2}\}  \notag \\
&=&-(p_{1}+p_{2})\cdot \frac{\partial }{\partial x}(\Phi _{2}+\Phi
_{1})-(p_{1}-p_{2})\cdot \frac{\partial }{\partial x}(\Phi _{2}-\Phi
_{1})+\{\Phi _{1},\Phi _{2}\}.
\end{eqnarray}
Its simplest solution is 
\begin{equation}
\Phi _{1}=\Phi _{2}=\Phi (x_{\perp },p_{1},p_{2})\equiv \Phi _{w},
\label{thrd}
\end{equation}
and requires abandoning $x=x_{1}-x_{2}$ in favor of 
\begin{eqnarray}
x_{12\perp }^{\mu } &=&(\eta ^{\mu \nu }+\hat{P}^{\mu }\hat{P}^{\nu
})(x_{1}-x_{2})_{\nu }\equiv \eta _{\perp }^{\mu \nu }(x_{1}-x_{2})_{\nu }, 
\notag \\
~\hat{P}^{\mu } &\equiv &\frac{P^{\mu }}{\sqrt{-P^{2}}}~~;~P^{\mu
}=p_{1}^{\mu }+p_{2}^{\mu }~~;\ ~x_{12\perp }\cdot ~\hat{P}=0.  \label{tp}
\end{eqnarray}
Thus we have a \textquotedblleft third law\textquotedblright\ condition (\ref%
{thrd}) of action and reaction plus a restriction on how the quasipotential $
\Phi _{w}$ may depend on relative separation. The invariant $r$ defined
below is the interparticle separation in the CM frame $\hat{P}=(1,\mathbf{0)}
$ 
\begin{equation}
r\equiv \sqrt{x_{\perp }^{2}}=\sqrt{\mathbf{r}^{2}}\text{ in CM frame }\hat{%
P }=(1,\mathbf{0)}\text{ },  \label{t}
\end{equation}
since $t_{1}-t_{2}=0$ in that frame. Relative time is thus controlled in a
covariant way. Assume the two invariants $M_{i},~i=1,2$ are simply functions
of $r$ and the CM energy 
\begin{equation}
w=\sqrt{-P^{2}}.
\end{equation}
The invariant potentials $M_{i}$ are not independent. The third law
condition implies they are related by 
\begin{equation}
M_{1}^{2}-M_{2}^{2}=m_{1}^{2}-m_{2}^{2}.
\end{equation}
Hence there is only one independent invariant function controlling the
scalar interaction which we designate by 
\begin{equation}
S(r),
\end{equation}
the underlying scalar interaction. Alternatively, the third law allows us to
recast the mass potentials into the hyperbolic function solutions depending
on a single invariant function $L$, 
\begin{eqnarray}
M_{1} &=&m_{1}\cosh L(S(r))+m_{2}\sinh L(S(r)),  \notag \\
M_{2} &=&m_{2}\cosh L(S(r))+m_{1}\sinh L(S(r)).  \label{hyper}
\end{eqnarray}

Subtracting the constraints gives us a complimentary covariant restriction
(to Eq. (\ref{t})) on the relative energy 
\begin{equation}
\mathcal{H}_{1}-\mathcal{H}%
_{2}=p_{1}^{2}+M_{1}^{2}-p_{2}^{2}-M_{2}^{2}=p_{1}^{2}+m_{1}^{2}-p_{2}^{2}-m_{2}^{2}=2P\cdot p~%
\mathcal{\approx }0,  \label{rle}
\end{equation}%
with relative momentum 
\begin{eqnarray}
\ p^{\mu } &=&\frac{(\varepsilon _{2}p_{1}^{\mu }-\varepsilon _{1}p_{2}^{\mu
})}{w}\ \ ,\ \varepsilon _{1}+\varepsilon _{2}=w\ \ ,\ \varepsilon
_{1}-\varepsilon _{2}=\frac{(m_{1}^{2}-m_{2}^{2})}{w},  \notag \\
\varepsilon _{i} &=&\text{CM energy of particle }i.  \label{rp}
\end{eqnarray}%
The relative momentum is canonically conjugate to $x_{\perp },$ 
\begin{equation}
\{x_{\perp }^{\mu },p^{\nu }\}=\eta _{\perp }^{\mu \nu }.
\end{equation}%
The other combination of our constraints is the primary dynamical equation \ 
\begin{equation}
\mathcal{H}\equiv \frac{(\varepsilon _{2}\mathcal{H}_{1}+\varepsilon _{1}%
\mathcal{H}_{2})}{w}=p_{\perp }^{2}+\Phi _{w}-b^{2}(w)~\mathcal{\approx }~0,
\end{equation}%
and incorporates exact two-body kinematics with 
\begin{equation}
b^{2}(w)=\frac{(w^{4}-2w^{2}(m_{1}^{2}+m_{2}^{2})+(m_{1}^{2}-m_{2}^{2})^{2})%
}{4w^{2}}=\varepsilon _{w}^{2}-m_{w}^{2},
\end{equation}%
and 
\begin{equation}
m_{w}=\frac{m_{1}m_{2}}{w}\ \ ,~~\varepsilon _{w}=\frac{%
(w^{2}-m_{1}^{2}-m_{2}^{2})}{2w},  \label{emw}
\end{equation}%
defined as the mass and energy of the fictitious particle of relative
motion. Under quantization all of the constraints become equations the wave
functions must satisfy.

\subsection{Two-Body Dirac Equations}

The constraint formalism embodies spin in a system of two coupled,
compatible Dirac equations on a single wave-function. For particles
interacting through world vector and scalar interactions the TBDE take this
general minimal-coupling form 
\begin{eqnarray}
\mathcal{S}_{1}\psi &\equiv &\gamma _{51}(\gamma _{1}\cdot (p_{1}-\tilde{A}%
_{1})+m_{1}+\tilde{S}_{1})\psi =0,  \notag \\
\mathcal{S}_{2}\psi &\equiv &\gamma _{52}(\gamma _{2}\cdot (p_{2}-\tilde{A}%
_{2})+m_{2}+\tilde{S}_{2})\psi =0.  \label{tbde}
\end{eqnarray}%
The wave function has sixteen components 
\begin{equation}
\psi =[\psi _{1},\psi _{2},\psi _{3},\psi _{4}],
\end{equation}%
in which each $\psi _{i}$ is a four component Pauli spinor for two
spin-one-half particles. \ The two equations are compatible: 
\begin{equation}
\lbrack \mathcal{S}_{1},\mathcal{S}_{2}]\psi =0.~~~~~
\end{equation}

This is a result of the presence of spin supersymmetries \cite{cra82},\cite%
{cra87}, in addition to the relativistic third law, and covariant
restrictions on the relative time and energy appearing in the spinless case.
There is automatic incorporation of correct spin-dependent recoil terms \cite%
{a}, 
\begin{equation}
\tilde{A}_{i}^{\mu }=\tilde{A}_{i}^{\mu }(A(r),p_{\perp },\hat{P},w,\gamma
_{1},\gamma _{2}),~\ \tilde{S}_{i}=\tilde{S}_{i}(S(r),A(r),p_{\perp },\hat{P}%
,w,\gamma _{1},\gamma _{2}).
\end{equation}%
This two-body formalism has many advantages over the traditional
Bethe-Salpeter equation and its numerous three dimensional truncations. One
is its simplicity. A Pauli reduction and scale transformation brings our
equations to this covariant Schr\"{o}dinger-like form 
\begin{equation}
{\biggl (}p^{2}+\Phi _{w}(\sigma _{1},\sigma _{2},p_{\perp },A(r),S(r)){%
\biggr )}\psi =b^{2}(w)\psi .
\end{equation}

\subsubsection{Schr\"{o}dinger-Like Form of the Two-Body Dirac Equations}

From separate classical \cite{fw} or quantum field theories \cite{saz97} one
can show that the quasipotential in the combination $\Phi _{w}-b^{2}(w)$
depends on the difference of squares of the invariant mass and energy
potentials ($M_{i}$ and $E_{i}$ respectively) \ 
\begin{equation}
M_{i}^{2}=m_{i}^{2}+2m_{w}S+S^{2};\ E_{i}^{2}=\varepsilon
_{i}^{2}-2\varepsilon _{w}A+A^{2},  \label{em}
\end{equation}
with $A$ playing the same role for vector interactions that $S$ does for
scalar ones. \textquotedblleft Squaring\textquotedblright\ \ the TBDE (\ref%
{tbde}) yields a Schr\"{o}dinger-like equation \cite{bckr} for the
upper-upper $\psi _{1}$ component

\begin{eqnarray}
&&\{p^{2}+2m_{w}S+S^{2}+2\mathcal{\varepsilon }_{w}A-A^{2}  \notag \\
&&+\Phi _{D}i\hat{r}\cdot p+\Phi _{D^{\prime }}+\Phi _{SO1}L\cdot \sigma
_{1}+\Phi _{SO2}L\cdot \sigma _{2}+\Phi _{SS}\sigma _{1}\cdot \sigma
_{2}+\Phi _{T}S_{T}\}\psi _{1}  \notag \\
&&+\{\Phi _{SS}^{\prime }\sigma _{1}\cdot \sigma _{2}+\Phi _{T}^{\prime
}S_{T}\}\psi _{4}  \notag \\
&=&b^{2}(w)\psi _{1},  \label{pl1}
\end{eqnarray}
coupled to a Schr\"{o}dinger-like wave equation for the lower-lower
component $\psi _{4}$ \cite{phi} 
\begin{eqnarray}
&&\{p^{2}+2m_{w}S+S^{2}+2\mathcal{\varepsilon }_{w}A-A^{2}  \notag \\
&&+\tilde{\Phi}_{D}i\hat{r}\cdot p+\tilde{\Phi}_{D^{\prime }}+\tilde{\Phi}
_{SO1}L\cdot \sigma _{1}+\tilde{\Phi}_{SO2}L\cdot \sigma _{2}+\tilde{\Phi}
_{SS}\sigma _{1}\cdot \sigma _{2}+\tilde{\Phi}_{T}S_{T}\}\psi _{4}  \notag \\
&&+\{\tilde{\Phi}_{SS}^{\prime }\sigma _{1}\cdot \sigma _{2}+\tilde{\Phi}
_{T}^{\prime }S_{T}\}\psi _{1}  \notag \\
&=&b^{2}(w)\psi _{4}.  \label{pl2}
\end{eqnarray}

These equations can be solved nonperturbatively for QED ($S=0$) or quark
model calculations since everyone of the quasipotentials terms $\Phi _{i}$
(including the Darwin pieces $\Phi _{D}$) is quantum mechanically well
defined (less singular than $-1/4r^{2}$).

\subsubsection{ Nonperturbative Solutions of the Two-Body Dirac Equations}

For Two-Body Dirac equations of constraint dynamics applied to QED we have 
\begin{equation}
A(r)=-\frac{\alpha }{r}.  \label{aal}
\end{equation}%
For singlet positronium system we can obtain an exact solution \cite{exct}
for the total CM energy $w$ 
\begin{eqnarray}
w &=&m\sqrt{2+2/\sqrt{1+{\alpha ^{2}}/{(}n+\sqrt{(l+\frac{1}{2})^{2}-\alpha
^{2}}-l-\frac{1}{2}{)^{2}}}}  \notag \\
&=&2m-m{\alpha ^{2}}/{4}n^{2}-m\alpha ^{4}/2n^{3}(2l+1)+11/64m\alpha
^{4}/n^{4}+O(\alpha ^{6}),  \label{exct}
\end{eqnarray}%
that agrees through order $\alpha ^{4}$ with standard spectrum found by
perturbative treatment of the Darwin and spin-dependent terms in our Pauli
form. Numerical triplet state calculations agree equally well with
perturbative QED\cite{bckr}.

Many of the standard approaches to QED bound states have been applied in QCD
in nonperturbative numerical calculations of the meson spectra without first
testing them nonperturbatively in QED. Sommerer $et~al$. \cite{vary} have
shown that the Blankenbecler-Sugar equation and the Gross equations fail
this test. This indicates danger in applying such three dimensional
truncations of the Bethe Salpeter equation to quark models, for if failure
occurs in their applications to QED how can similar non-perturbative (i.e.
numerical) approaches based on the same truncations (but with QCD kernels)
give results that are trustworthy representations of the physics for meson
spectroscopy?

\subsection{Two Body Dirac Equations for Meson Spectroscopy - The
Adler-Piran Potential}

We obtain a constraint version of the naive quark model for mesons by
employing a covariant adaptation of a static quark potential due to Adler
and Piran \cite{adler}. From an effective non-linear field theory derived
from QCD they obtain 
\begin{equation}
V_{AP}(r)=\Lambda (U(\Lambda r)+U_{0})\ (=A+S).
\end{equation}%
The original $V_{AP}$ is nonrelativistic, and appears in our equations in
that limit as the sum of world vector and scalar potentials with 
\begin{eqnarray}
\Lambda U(\Lambda r &<&<1)~\sim \frac{1}{r\ln \Lambda r}  \notag \\
V_{AP}(r) &=&\Lambda \lbrack c_{1}\Lambda r+c_{2}\log (\Lambda r)+\frac{c_{3}%
}{\sqrt{\Lambda r}}+\frac{c_{4}}{\Lambda r}+c_{5}],\ \ \Lambda r>2.
\end{eqnarray}%
The explicit form for $V_{AP}(r)$ at all distances is given in \cite{adler}
and \cite{cra88}.

\subsubsection{Relativistic Naive Quark Model}

We reinterpret the static $V_{AP}$ covariantly by replacing the
nonrelativistic $r~$by $\sqrt{x_{\perp }^{2}}\equiv r$, and parceling out
the static potential $V_{AP}$ into the invariant functions $A(r)$ and $S(r)$ 
\cite{crater2} as follows: 
\begin{equation}
A=\exp (-\beta r)[V_{AP}-\frac{c_{4}}{r}]+\frac{c_{4}}{r}+\frac{e_{1}e_{2}}{r%
},~~S=V_{AP}+\frac{e_{1}e_{2}}{r}-A.
\end{equation}%
(The constants $c_{1},c_{2},c_{3},c_{4}~$\ are fixed by the Adler-Piran
formalism while $e_{1},e_{2}$ are the quark and anti-quark electric
charges.) Thus at short distances the potential is strictly vector while at
long distances the vector portion is strictly Coulombic with the confining
portion at long distance (including subdominant portions) strictly scalar.
Once $A$ and $S$ have been determined, so are all the accompanying
spin-dependent interactions 
\begin{equation}
\Phi _{i}=\Phi _{i}(\sigma _{1},\sigma _{2},p_{\perp
},A(r),S(r));~i=D,D^{\prime },SO1,SO2,SS,T,..
\end{equation}

Our bound state results are quite accurate, from the heaviest bottomonium
states to the pion. They compare quite favorably with the results of Godfrey
and Isgur \cite{isg}, but with only two parametric functions ($A,S$) as
opposed to the six or so used in their approach. In the table below we
reproduce a portion of the entire spectrum given in \cite{crater2}. \ The
quark masses and potential parameters are given by $m_{u}\sim 55$ \textrm{%
MeV }, $m_{c}\sim 1.5$ \textrm{GeV}, $m_{d}\sim 58$ \textrm{MeV}, $\Lambda
=0.216\ \mathrm{GeV},$ and $\Lambda U_{0}=1.865\ \mathrm{GeV}$.

\bigskip 
\begin{tabular}{llll}
MESON & EXP(GeV) & \ ($\pm $MeV$)$\  & THEORY \\ 
$\eta _{c}:c\overline{c}\ 1^{1}S_{0}$ & 2.980 \ \ \ \ \ \ \ \ \  & ( 2.1) \
\ \ \ \ \  & 2.978 \\ 
$\psi :c\overline{c}\ 1^{3}S_{1}$ \ \ \ \  & 3.097 \ \ \ \ \ \ \ \ \ \  & 
(0.0) \ \ \ \ \ \ \ \ \  & 3.129 \\ 
$\chi _{0}:c\overline{c}\ 1^{1}P_{1}$ \ \ \ \ \ \ \ \ \ \ \  & 3.526 \  & 
(0.2) \ \ \ \ \ \ \  & 3.520 \\ 
$\chi _{0}:c\overline{c}\ 1^{3}P_{0}$ \ \ \ \ \  & 3.415 \ \ \ \ \ \ \ \ \ \
\ \  & (1.0) \ \ \ \ \ \ \ \ \  & 3.407 \\ 
$\chi _{1}:c\overline{c}\ 1^{3}P_{1}$ \ \ \ \ \ \ \  & 3.510 \ \ \ \ \ \ \ \
\ \ \ \  & (0.1) \ \ \ \ \ \ \ \ \ \  & 3.507 \\ 
$\chi _{2}:c\overline{c}\ 1^{3}P_{2}$ \ \ \ \ \ \ \  & 3.556 \ \ \ \ \ \ \ \
\ \ \ \  & (0.1) \ \ \ \ \ \ \ \ \ \  & 3.549 \\ 
$\eta _{c}:c\overline{c}\ 2^{1}S_{0}$ \ \ \ \ \ \ \  & 3.594 \ \ \ \ \ \ \ \
\ \ \ \ \  & (5.0) \ \ \ \ \ \ \ \ \ \  & 3.610 \\ 
$\psi :c\overline{c}\ 2^{3}S_{1}$ \ \ \ \ \ \ \  & 3.686 \ \ \ \ \ \ \ \ \ \
\ \  & (0.1) \ \ \ \ \ \ \ \ \ \  & 3.688 \\ 
$\psi :c\overline{c}\ 1^{3}D_{1}$ \ \ \ \ \ \ \  & 3.770 \ \ \ \ \ \ \ \ \ \
\ \ \  & (2.5) \ \ \ \ \ \ \ \ \  & 3.808 \\ 
$\psi :c\overline{c}\ 3^{3}S_{1}$ \ \ \ \ \ \ \  & 4.040 \ \ \ \ \ \ \ \ \ \
\ \ \  & (10.0) \ \ \ \ \ \ \ \ \ \  & 4.081 \\ 
$\psi :c\overline{c}\ 2^{3}D_{1}$ \ \ \ \ \ \ \  & 4.159 \ \ \ \ \ \ \ \ \ \
\ \ \  & (20.0) \ \ \ \ \ \ \ \ \ \  & 4.157 \\ 
$\psi :c\overline{c}\ 3^{3}D_{1}$ \ \ \ \ \ \ \  & 4.415 \ \ \ \ \ \ \ \ \ \
\ \ \  & (6.0) \ \ \ \ \ \ \ \ \ \  & 4.454 \\ 
$\pi :u\overline{d}\ 1^{1}S_{0}$\ \ \ \ \ \ \ \  & 0.140 \ \ \ \ \ \ \ \ \ \
\ \  & (0.0) \ \ \ \ \ \ \ \ \ \  & 0.144 \\ 
$\rho :u\overline{d}\ 1^{3}S_{1}$\ \ \ \ \ \ \ \ \  & 0.767 \ \ \ \ \ \ \ \
\ \ \ \  & ( 1.2) \ \ \ \ \ \ \ \ \ \  & 0.792 \\ 
$b_{1}:u\overline{d}\ 1^{1}P_{1}$\ \ \ \ \ \ \  & 1.231 \ \ \ \ \ \ \ \ \ \
\ \  & (10.0) \ \ \ \ \ \ \ \ \ \  & 1.392 \\ 
$a_{0}:u\overline{d}\ 1^{3}P_{0}$\ \ \ \ \ \ \ \ \ \  & 1.450 \ \ \ \ \ \ \
\ \ \ \  & (40.0) \ \ \ \ \ \ \ \ \ \  & 1.491 \\ 
$a_{1}:u\overline{d}\ 1^{3}P_{1}$\ \ \ \ \ \ \  & 1.230 \ \ \ \ \ \ \ \ \ \
\  & (40.0) \ \ \ \ \ \ \ \ \ \  & 1.568 \\ 
$a_{2}:u\overline{d}\ 1^{3}P_{2}$\ \ \ \ \ \ \  & 1.318 \ \ \ \ \ \ \ \ \ \
\  & (\ 0.7) \ \ \ \ \ \ \ \ \ \ \  & 1.310 \\ 
$\pi :u\overline{d}\ 2^{1}S_{0}$\ \ \ \ \ \ \ \ \  & 1.300 \ \ \ \ \ \ \ \ \
\ \ \ \  & (100.0) \ \ \ \ \ \ \ \ \  & 1.536 \\ 
$\rho :u\overline{d}\ 2^{3}S_{1}$\ \ \ \ \ \ \ \  & 1.465 \ \ \ \ \ \ \ \ \
\ \ \  & (25.0) \ \ \ \ \ \ \ \ \ \ \  & 1.775 \\ 
$\pi _{2}:u\overline{d}\ 1^{1}D_{2}$\ \ \ \ \ \  & 1.670 \ \ \ \ \ \ \ \ \ \
\ \  & (20.0) \ \ \ \ \ \ \ \ \ \ \  & 1.870 \\ 
$\rho :u\overline{d}\ 1^{3}D_{1}$\ \ \ \ \ \ \  & 1.700 \ \ \ \ \ \ \ \ \ \
\ \  & (20.0) \ \ \ \ \ \ \ \ \ \ \  & 1.986 \\ 
$\rho _{3}:u\overline{d}\ 1^{3}D_{3}$\ \ \ \ \ \ \ \  & 1.691 \ \ \ \ \ \ \
\ \ \ \ \ \ \  & (5.0) \ \ \ \ \ \ \ \ \ \ \ \ \ \  & 1.710%
\end{tabular}

\ 

\subsubsection{\ Positronium and the Pion}

Positronium numerical spectral predictions of the constraint approach for
hyperfine splittings are inadequate if we ignore coupling to the small
(including the lower-lower one $\psi _{4}$) components of the wave function 
\cite{bckr}. \ In the table below, $N_{c}$ refers to the number of coupled
equations, which for the fully coupled system is two for the singlet and
four for the triplet states \cite{bckr}. \ Units are in \textrm{eV}$.~$As
seen in the table, only the fully coupled system of equations (lower-lower
and upper-upper for the singlet, the same in addition to tensor coupling for
the triplet) produces accurate results to the require precision.

\begin{tabular}{llllllll}
$l~\ $ & $s~\ \ $ & $j~\ \ $ & $n~$ & $N_{c}~\ \ \ \ \ $ & Perturbative\
(eV)\  & \ Numerical (eV)\ \ \  & \ Diff/$\mu \alpha ^{4}$ \\ 
0 & 0 & 0 & 1 & 1 & \ -6.8033256279 \  & \ -6.8032861579 \ \  & ~5.45E-02 \\ 
0 & 0 & 0 & 1 & 2 & \ -6.8033256279 \  & \ -6.8033256719 \ \  & -6.08E-05 \\ 
0 & 1 & 1 & 1 & 1 & \ -6.8028426132 \  & \ -6.8028074990 \ \  & -0.84E-02 \\ 
0 & 1 & 1 & 1 & 2 & \ -6.8028426132 \  & \ -6.8028082195 \ \  & -4.75E-02 \\ 
0 & 1 & 1 & 1 & 2 & \ -6.8028426132 \  & \ -6.8028239499 \ \  & -2.58E-02 \\ 
0 & 1 & 1 & 1 & 4 & \ -6.8028426132 \  & \ -6.8028426636 \ \  & -6.97E-05%
\end{tabular}

The corresponding good $\pi -\rho $ splitting obtained in \cite{crater2}\ is
spoiled if the we ignore these couplings, leading to $\ \ m_{\pi } \sim 850~$%
\textrm{MeV}$, ~m_{\rho }\sim 1060\ $\textrm{MeV.} \ The same relativistic
structure in the constraint equations responsible for the success of the
Sommerfeld-Balmer formula for positronium spin singlet states appears to be
important for bringing the pion mass down to its observed value.

\subsubsection{Golstone Behavior of the Pion.}

As a bonus, we find \cite{cra88}, \cite{cra94} ,\cite{crater2} that the pion
is a Goldstone boson in the sense that

\begin{equation}
m_{\pi }(m_{q}\rightarrow 0)\rightarrow 0,
\end{equation}%
while the $\rho $ and excited $\pi $ have finite mass in this limit.
However, if the TBDE for the pion is truncated so that the coupling to the
lower-lower component is dropped, then the pion loses its Goldstone boson
behavior. Its mass no longer decreases toward zero with vanishing quark
mass. \ This and the $\pi -\rho $ result above support our contention that
the pion does not need to be treated in a special way insofar as the binding
mechanism is concerned. \ The light pion mass as well as its Goldstone
behavior is a natural outgrowth of the covariant Two-Body Dirac formalism. \
We now see how this model for the pion and other mesons holds up for a
different probe, that of $2\gamma $ decays.

\section{ Two Gamma Decay Amplitudes for Positronium and Quarkonium}

\bigskip Our treatment of decays in the sections below are for general
angular momentum states but for illustrative purposes we begin by
considering a treatment of\ singlet positronium or quarkonium systems. \
They can be viewed as bosons given by the state vector 
\begin{equation}
|^{1}S_{0}\rangle =\frac{1}{\sqrt{2}}\int d^{3}p\tilde{\psi}(\mathbf{p}){(b_{%
\mathbf{p}}^{\dagger 1}d_{-\mathbf{p}}^{\dagger 2}-b_{\mathbf{p}}^{\dagger
2}d_{-\mathbf{p}}^{\dagger 1})}|0\rangle .
\end{equation}%
Both the electron and positron (or quark and antiquark) are off shell but on
energy shell. The amplitudes for the annihilation of a quark-antiquark pair
into two photons are given by the Feynman diagrams in Fig.\ 1.

\vspace*{0.1cm} 
\begin{figure}[h]
\includegraphics[scale=0.50]{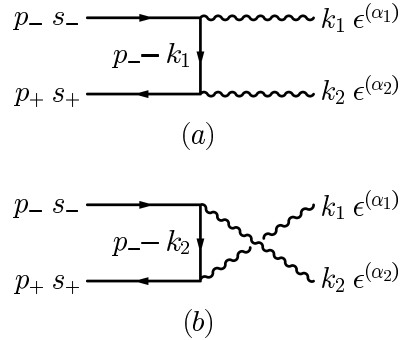} \vspace*{-8.6cm}
\caption{Feynman diagrams for the annihilation of a quark-antiquak pair into
two photons.}
\end{figure}

The singlet amplitude for annihilation of a free $e^{+}e^{-}$ pair with
momenta $p_{+}$ and $p_{-}$ into two photons with polarizations $\epsilon
^{\alpha _{1}},\epsilon ^{\alpha _{2}}$ and momenta $k_{1}=(w/2,\mathbf{k),~}%
k_{2}=(w/2,-\mathbf{k)~}$is

\begin{eqnarray}
M_{\alpha \beta } &=&\frac{e^{2}}{(2\pi )^{3}w\sqrt{2}}\{\bar{v}%
^{(s_{+})}(p_{+}){[}\gamma \cdot \epsilon ^{(\alpha _{1})}\frac{m-\gamma
\cdot (p_{-}-k_{1})}{(p_{-}-k_{1})^{2}+m^{2}-i0}\gamma \cdot \epsilon
^{(\alpha _{2})}  \notag \\
&&+\gamma \cdot \epsilon ^{(\alpha _{2})}\frac{m-\gamma \cdot (p_{-}-k_{2})}{%
(p_{-}-k_{2})^{2}+m^{2}-i0}\gamma \cdot \epsilon ^{(\alpha _{1})}{]}%
u^{(s_{-})}(p_{-})-(s_{+}\Leftrightarrow s_{-})\}.
\end{eqnarray}%
For positronium or quarkonium, we would replace this decay amplitude by 
\begin{equation}
M_{\alpha \beta }\rightarrow \int d^{3}p\tilde{\psi}_{^{1}S_{0}}(\mathbf{p)}%
M_{\alpha \beta }\equiv \frac{1}{(2\pi )^{3}w}\mathcal{M}_{^{1}S_{0}%
\rightarrow 2\gamma }.  \label{3}
\end{equation}%
Unlike free amplitudes, the fermion spinors and momenta in $\mathcal{M}%
_{^{1}S_{0}\rightarrow 2\gamma }$ \ are not on shell.

\subsection{Sixteen Component Two Gamma Decay Formalism}

\ The amplitude in Eq. (\ref{3}) above is\ of the form (in CM) 
\begin{eqnarray}
\mathcal{M}_{X\rightarrow 2\gamma } &=&\int d^{3}p\psi (\mathbf{p)}\frac{1}{%
\sqrt{2}}[\bar{v}^{(s_{+})}(-\mathbf{p)\Gamma (p,k)}u^{(s_{-})}(\mathbf{p})-%
\bar{v}^{(s_{-})}(-\mathbf{p)\Gamma (p,k)}u^{(s_{+})}(\mathbf{p})]  \notag \\
&=&\frac{1}{\sqrt{2}}\int d^{3}p\psi (\mathbf{p)}Tr\mathbf{\Gamma (p,k)[}%
u^{(s_{-})}(\mathbf{p})\bar{v}^{(s_{+})}(-\mathbf{p)}-u^{(s_{+})}(\mathbf{p})%
\bar{v}^{(s_{-})}(-\mathbf{p)}],  \label{anamp}
\end{eqnarray}%
in which 
\begin{equation}
\mathbf{\Gamma (p,k)=}e^{2}[\boldsymbol{\gamma}\mathbf{\cdot }%
\boldsymbol{\epsilon }^{(\alpha _{1})}\frac{m-\boldsymbol{\gamma}\mathbf{%
\cdot (p-k)}}{(\mathbf{p-k)}^{2}+m^{2}}\boldsymbol{\gamma}\mathbf{\cdot }%
\boldsymbol{\epsilon }^{(\alpha _{2})}+\boldsymbol{\gamma}\mathbf{\cdot }%
\boldsymbol{\epsilon }^{(\alpha _{2})}\frac{m-\boldsymbol{\gamma}\mathbf{%
\cdot (p+k)}}{(\mathbf{p+k)}^{2}+m^{2}}\boldsymbol{\gamma}\mathbf{\cdot }%
\boldsymbol{\epsilon }^{(\alpha _{1})}].  \label{gamp}
\end{equation}%
We replace this amplitude for general angular momentum states by 
\begin{equation}
\int d^{3}pTr\mathbf{\Gamma (p,k)}\psi \mathbf{(\mathbf{p),}}  \label{3m}
\end{equation}%
where $\psi (\mathbf{p)}$ is our bound state wave function in matrix form in
an arbitrary angular momentum state. \ Thus we are expanding our
investigation from $^{1}S_{0}$ states to general $^{1}L_{l}$ and $%
^{3}L_{l\pm 1}$ states. In the case of $^{1}S_{0}$ states\ what we are doing
amounts to replacing the matrix wave function $\psi (\mathbf{p)[}u^{(s_{-})}(%
\mathbf{p})\bar{v}^{(s_{+})}(-\mathbf{p)}-u^{(s_{+})}(\mathbf{p})\bar{v}%
^{(s_{-})}(-\mathbf{p)}]$ having a spin structure governed by free Dirac
spinors by the matrix wave function $\psi \mathbf{(\mathbf{p)}}$ which,
unlike the solution constructed from the free spinors, is a solution of the
full interacting set of Two-Body Dirac equations. Similar comments apply for
the other angular momentum states. In terms of the Fourier transformed
matrix wave function $\psi (\mathbf{r)}$ (defined below in Eqs. (\ref{psin})
and (\ref{spn})) 
\begin{equation}
\int d^{3}pTr\mathbf{\Gamma (p,k)}\psi (\mathbf{p)=}\int d^{3}rTr[\psi (%
\mathbf{r)}\int d^{3}p\frac{\exp (-i\mathbf{p\cdot r)}}{(2\pi )^{3/2}}%
\mathbf{\Gamma (p,k)].}
\end{equation}%
Now 
\begin{eqnarray}
&&\int d^{3}p\frac{\exp (-i\mathbf{p\cdot r)}}{(2\pi )^{3/2}}\mathbf{\Gamma
(p,k)}  \notag \\
&=&\exp (-i\mathbf{k\cdot r)}\int d^{3}q\frac{\exp (-i\mathbf{q\cdot r)}}{%
(2\pi )^{3/2}}\mathbf{\Gamma }_{D}\mathbf{(q)+}\exp (i\mathbf{k\cdot r)}\int
d^{3}p\frac{\exp (-i\mathbf{q\cdot r)}}{(2\pi )^{3/2}}\mathbf{\Gamma }_{C}%
\mathbf{(q),}
\end{eqnarray}%
where 
\begin{eqnarray}
\mathbf{\Gamma }_{D}\mathbf{(q)} &\mathbf{=}&e^{2}\frac{\boldsymbol{\gamma}%
\mathbf{\cdot }\boldsymbol{\epsilon }^{(\alpha _{1})}[m-\mathbf{q\cdot }%
\boldsymbol{\gamma}\mathbf{]}\boldsymbol{\gamma}\mathbf{\cdot }%
\boldsymbol{\epsilon }^{(\alpha _{2})}}{m^{2}+\mathbf{q}^{2}}, \\
\mathbf{\Gamma }_{C}\mathbf{(q)} &\mathbf{=}&e^{2}\frac{\boldsymbol{\gamma}%
\mathbf{\cdot }\boldsymbol{\epsilon }^{(\alpha _{2})}[m-\mathbf{q\cdot }%
\boldsymbol{\gamma}\mathbf{]}\boldsymbol{\gamma}\mathbf{\cdot }%
\boldsymbol{\epsilon }^{(\alpha _{1})}}{m^{2}+\mathbf{q}^{2}}.  \notag
\end{eqnarray}%
Performing the Fourier transforms gives 
\begin{eqnarray}
&&\int d^{3}q\frac{\exp (-i\mathbf{q\cdot r)}}{(2\pi )^{3/2}}\mathbf{\Gamma }%
_{D}\mathbf{(q)}  \notag \\
&=&e^{2}\sqrt{\frac{\pi }{2}}\boldsymbol{\gamma}\mathbf{\cdot }%
\boldsymbol{\epsilon }^{(\alpha _{1})}(m-i\boldsymbol{\gamma}\mathbf{\cdot
\nabla )}\frac{\exp (-mr)}{r}\boldsymbol{\gamma}\mathbf{\cdot }%
\boldsymbol{\epsilon }^{(\alpha _{2})},
\end{eqnarray}%
and 
\begin{eqnarray}
&&\int d^{3}q\frac{\exp (-i\mathbf{q\cdot r)}}{(2\pi )^{3/2}}\mathbf{\Gamma }%
_{C}\mathbf{(q)}  \notag \\
&=&e^{2}\sqrt{\frac{\pi }{2}}\boldsymbol{\gamma}\mathbf{\cdot }%
\boldsymbol{\epsilon }^{(\alpha _{2})}(m-i\boldsymbol{\gamma}\mathbf{\cdot
\nabla )}\frac{\exp (-mr)}{r}\boldsymbol{\gamma}\mathbf{\cdot }%
\boldsymbol{\epsilon }^{(\alpha _{1})}.
\end{eqnarray}

This generalizes the configuration space form given in \cite{posit} to the
amplitude below depending on the full $4\times 4$ matrix wave function 
\begin{eqnarray}
\mathcal{M}_{X\rightarrow 2\gamma } &\mathbf{=}&\int d^{3}r\exp (-i\mathbf{%
k\cdot r)}Tr[\psi (\mathbf{r)}\int d^{3}\mathbf{p}\frac{\exp (-i\mathbf{%
p\cdot r)}}{(2\pi )^{3/2}}\mathbf{\Gamma (p,k)]}  \notag \\
&=&e^{2}\sqrt{\frac{\pi }{2}}\int d^{3}rTr\{\psi (\mathbf{r)[}\exp (-i%
\mathbf{k\cdot r})\boldsymbol{\gamma}\mathbf{\cdot }\boldsymbol{\epsilon }%
^{(\alpha _{1})}(m-i\boldsymbol{\gamma}\mathbf{\cdot \nabla )}\frac{\exp
(-mr)}{r}\boldsymbol{\gamma}\mathbf{\cdot }\boldsymbol{\epsilon }^{(\alpha
_{2})}  \notag \\
&&+\exp (i\mathbf{k\cdot r)}\boldsymbol{\gamma}\mathbf{\cdot }%
\boldsymbol{\epsilon }^{(\alpha _{2})}(m-i\boldsymbol{\gamma}\mathbf{\cdot
\nabla )}\frac{\exp (-mr)}{r}\boldsymbol{\gamma}\mathbf{\cdot }%
\boldsymbol{\epsilon }^{(\alpha _{1})}]\}.  \label{dcmp}
\end{eqnarray}%
The wave function will often display mild singularities at the origin
typical for relativistic wave functions. For example the ground state
solution corresponding to Eq. (\ref{exct}) is 
\begin{equation}
\psi (r)=\frac{{(}m\alpha {)^{3/2}}}{\sqrt{4\pi \Gamma (2+2\sqrt{1/4-\alpha
^{2}})}}(rm\alpha )^{+(1/2+\sqrt{1/4-\alpha ^{2}})}\exp (-\alpha mr/2).
\label{grnd}
\end{equation}%
The mild singularity at the origin appearing in this equation is rendered
harmless by the smearing action of the Yukawa distribution that comes from
folding the effects of the decay amplitude with that of the wave function 
\cite{posit}. \ 

\subsubsection{ $4\times 4$ Matrix Form of Solutions of the Two-Body Dirac
Equations}

To accommodate the structure of the TBDE to the above decay amplitude we
explicitly construct the $4\times 4$ matrix wave function solution $\psi (%
\mathbf{r)~}$of the equation. \ First we observe that one can write Eqs. (%
\ref{tbde}) in terms of mass and energy potentials and their derivatives
analogous to what is done in the case of two-spinless particles \cite%
{bckr,cra94},\cite{long},\cite{liu}. \ In analogy to the solution (\ref%
{hyper}) we gave to the third law condition in the spinless case we define\ 

\begin{eqnarray}
M_{1} &=&m_{1}\ \cosh L(S,A)\ +m_{2}\sinh L(S,A),  \notag \\
M_{2} &=&m_{2}\ \cosh L(S,A)\ +m_{1}\ \sinh L(S,A),
\end{eqnarray}
\begin{eqnarray}
E_{1} &=&\varepsilon _{1}\ \cosh \mathcal{G}(A)\ -\varepsilon _{2}\sinh 
\mathcal{G}(A),  \notag \\
E_{2} &=&\varepsilon _{2}\ \cosh \mathcal{G}(A)-\varepsilon _{1}\ \sinh 
\mathcal{G}(A).
\end{eqnarray}

In terms of these functions the coupled Two-Body Dirac equations in an
arbitrary frame have the form $\mathcal{S}_{i}\psi =0$ in which 
\begin{align}
\mathcal{S}_{1}& =\exp \left( \mathcal{G}\right) \beta _{1}\Sigma _{1}\cdot 
\mathcal{P}_{1}+E_{1}\beta _{1}\gamma _{51}+M_{1}\gamma _{51}-\exp \left( 
\mathcal{G}\right) \frac{i}{2}\Sigma _{2}\cdot \partial (\mathcal{G}\beta
_{1}+L\beta _{2})\gamma _{51}\gamma _{52},  \notag \\
\mathcal{S}_{2}& =-\exp \left( \mathcal{G}\right) \beta _{2}\Sigma _{2}\cdot 
\mathcal{P}_{2}+E_{2}\beta _{2}\gamma _{52}+M_{2}\gamma _{52}+\exp \left( 
\mathcal{G}\right) \frac{i}{2}\Sigma _{1}\cdot \partial (\mathcal{G}\beta
_{2}+L\beta _{1})\gamma _{51}\gamma _{52},  \label{ctbde}
\end{align}
with 
\begin{equation}
\mathcal{P}_{i}\equiv p-\frac{i}{2}\Sigma _{i}\cdot \partial \mathcal{G}
\Sigma _{i}.
\end{equation}
The gamma matrices have block forms given in Appendix A.

If we use the combinations $\phi _{\pm }=\psi _{1}\pm \psi _{4}$ and $\chi
_{\pm }=\psi _{2}\pm \psi _{3}$, then unlike Eqs.\ (\ref{pl1}) and (\ref{pl2}%
), the corresponding Schr\"{o}dinger-like equations decouple \cite{cra94},%
\cite{long},\cite{crater2}. We obtain \cite{crater2} 
\begin{eqnarray}
&&[p^{2}+2m_{w}S+S^{2}+2\varepsilon _{w}A-A^{2}  \notag \\
&&-\frac{1}{2}\nabla ^{2}\mathcal{G}+\frac{3}{4}{\mathcal{G}^{\prime }}^{2}-(%
\mathcal{G}^{\prime }+L^{\prime })^{2}+\mathcal{G}^{\prime }F^{\prime } 
\notag \\
&&-\frac{L\cdot (\sigma _{1}+\sigma _{2})}{r}F^{\prime }+L\cdot (\sigma
_{1}-\sigma _{2})l^{\prime }+iq^{\prime }L\cdot (\sigma _{1}\times \sigma
_{2})  \notag \\
&&+2F^{\prime }i\hat{r}\cdot p+iK^{\prime }(\sigma _{1}\cdot \hat{r}\sigma
_{2}\cdot p+\sigma _{2}\cdot \hat{r}\sigma _{1}\cdot p)  \notag \\
&&+\sigma _{1}\cdot \sigma _{2}(\frac{1}{2}\nabla ^{2}\mathcal{G}+\frac{1}{2r%
}L^{\prime }-\frac{1}{2}\mathcal{G}^{\prime }{}^{2}-\mathcal{G}^{\prime
}F^{\prime })+\sigma _{1}\cdot \hat{r}\sigma _{2}\cdot \hat{r}(\frac{1}{2}%
\nabla ^{2}L-(\frac{3}{2r}+F^{\prime }))]\phi _{+}  \notag \\
&=&b^{2}(w)\phi _{+},  \label{szin}
\end{eqnarray}%
where the prime symbol stands for $d/dr$. \ We have used the abbreviations%
\begin{eqnarray}
F &=&\frac{1}{2}\log \mathcal{D}-\mathcal{G},  \notag \\
\mathcal{D} &\mathcal{=}&E_{2}M_{1}+E_{1}M_{2},  \notag \\
K &=&\frac{(\mathcal{G}+L)}{2},  \notag \\
l^{\prime }(r) &=&-\frac{1}{2r}\frac{E_{2}M_{2}-E_{1}M_{1}}{%
E_{2}M_{1}+E_{1}M_{2}}(L-\mathcal{G})^{\prime },  \notag \\
q^{\prime }(r) &=&\frac{1}{2r}\frac{E_{1}M_{2}-E_{2}M_{1}}{%
E_{2}M_{1}+E_{1}M_{2}}(L-\mathcal{G})^{\prime }.  \label{kf}
\end{eqnarray}%
We work in the CM frame in which $\hat{P}=(1,\mathbf{0)}$ and $\hat{r}=(0,%
\mathbf{\hat{r}).}$ Once we find the four component solutions $\phi _{+}~$to
this equation we can obtain the other twelve components $\phi _{-},\chi
_{\pm }$. \ In Appendix B we find from Eq. (\ref{ctbde}) 
\begin{eqnarray}
\chi _{+} &=&\frac{\exp (\mathcal{G)}}{\mathcal{D}}\{M_{2}[%
\boldsymbol{\sigma}_{1}\cdot \mathbf{p}-\frac{i}{2}\boldsymbol{\sigma}%
_{2}\cdot \boldsymbol{\nabla }(-\mathcal{G}-L+\mathcal{G}\boldsymbol{\sigma}%
_{1}\cdot \boldsymbol{\sigma}_{2})]  \notag \\
&&-M_{1}[\boldsymbol{\sigma}_{2}\cdot \mathbf{p}-\frac{i}{2}%
\boldsymbol{\sigma}_{1}\cdot \boldsymbol{\nabla }(-\mathcal{G}-L+\mathcal{G}%
\boldsymbol{\sigma}_{1}\cdot \boldsymbol{\sigma}_{2})]\}\phi _{+},
\label{chip}
\end{eqnarray}%
and similarly 
\begin{eqnarray}
\chi _{-} &=&-\frac{\exp (\mathcal{G)}}{\mathcal{D}}\{E_{2}[%
\boldsymbol{\sigma}_{1}\cdot \mathbf{p}-\frac{i}{2}\boldsymbol{\sigma}%
_{2}\cdot \boldsymbol{\nabla }(-\mathcal{G}-L+\mathcal{G}\boldsymbol{\sigma}%
_{1}\cdot \boldsymbol{\sigma}_{2})]  \notag \\
&&+E_{1}[\boldsymbol{\sigma}_{2}\cdot \mathbf{p}-\frac{i}{2}%
\boldsymbol{\sigma}_{1}\cdot \boldsymbol{\nabla }(-\mathcal{G}-L+\mathcal{G}%
\boldsymbol{\sigma}_{1}\cdot \boldsymbol{\sigma}_{2})]\}\phi _{+},
\label{chim}
\end{eqnarray}%
and 
\begin{eqnarray}
\phi _{-} &=&\frac{(E_{2}E_{1}+M_{2}M_{1})}{\mathcal{D}}\phi _{+}  \notag \\
&&-\frac{1}{2\mathcal{D}}[(E_{2}D_{1}^{-+}-E_{1}D_{2}^{-+}){\frac{1}{%
\mathcal{D}}}(M_{2}D_{1}^{++}-M_{1}D_{2}^{++})  \notag \\
&&-(M_{2}D_{1}^{--}+M_{1}D_{2}^{--}){\frac{1}{\mathcal{D}}}%
(E_{2}D_{1}^{++}+E_{1}D_{2}^{++})]\phi _{+},  \label{54}
\end{eqnarray}%
in which 
\begin{subequations}
\begin{eqnarray}
D_{1}^{++} &=&\exp (\mathcal{G)}[\boldsymbol{\sigma}_{1}\mathbf{\cdot p-}%
\frac{i}{2}\boldsymbol{\sigma}_{2}\mathbf{\cdot }{\boldsymbol{\nabla }}(-%
\mathcal{G}-L+{\mathcal{G}}\boldsymbol{\sigma}_{1}\mathbf{\cdot }%
\boldsymbol{\sigma}_{2})],  \notag \\
D_{1}^{-+} &=&\exp (\mathcal{G)}[\boldsymbol{\sigma}_{1}\mathbf{\cdot p+}%
\frac{i}{2}\boldsymbol{\sigma}_{2}\mathbf{\cdot }{\boldsymbol{\nabla }}(%
\mathcal{G}-L-{\mathcal{G}}\boldsymbol{\sigma}_{1}\mathbf{\cdot }%
\boldsymbol{\sigma}_{2})],  \notag \\
D_{1}^{--} &=&\exp (\mathcal{G)}[\boldsymbol{\sigma}_{1}\mathbf{\cdot p+}%
\frac{i}{2}\boldsymbol{\sigma}_{2}{\cdot \boldsymbol{\nabla }}(\mathcal{G}+L-%
{\mathcal{G}}\boldsymbol{\sigma}_{1}\mathbf{\cdot }\boldsymbol{\sigma}_{2})],
\label{d1}
\end{eqnarray}%
and 
\begin{eqnarray}
D_{2}^{++} &=&\exp (\mathcal{G)}[\boldsymbol{\sigma}_{2}\mathbf{\cdot p-}%
\frac{i}{2}\boldsymbol{\sigma}_{1}{\cdot \boldsymbol{\nabla }}(\mathcal{G}-L+%
{\mathcal{G}}\boldsymbol{\sigma}_{1}\mathbf{\cdot }\boldsymbol{\sigma}_{2})],
\notag \\
D_{2}^{-+} &=&\exp (\mathcal{G)}[\boldsymbol{\sigma}_{2}\cdot \mathbf{p}+%
\frac{i}{2}{\boldsymbol{\sigma}_{1}\cdot \boldsymbol{\nabla }}(\mathcal{G}-L-%
{\mathcal{G}\boldsymbol{\sigma}_{1}\cdot \boldsymbol{\sigma}_{2}})],  \notag
\\
D_{2}^{--} &=&\exp (\mathcal{G)}[\boldsymbol{\sigma}_{2}\cdot \mathbf{p}+%
\frac{i}{2}{\boldsymbol{\sigma}_{1}\cdot \boldsymbol{\nabla }}(\mathcal{G}+L-%
{\mathcal{G}\boldsymbol{\sigma}_{1}\cdot \boldsymbol{\sigma}_{2}})].
\label{d2}
\end{eqnarray}

We then further define four component wave functions $\psi _{\pm },\eta
_{\pm }$ related to the above by \cite{liu} 
\end{subequations}
\begin{eqnarray}
\phi _{\pm } &=&\exp (F+K\boldsymbol{\sigma}_{1}\mathbf{\cdot \hat{r}}%
\boldsymbol{\sigma}_{2}\mathbf{\cdot \hat{r}})\psi _{\pm }=\exp F(\cosh
K+\sinh K\boldsymbol{\sigma}_{1}\mathbf{\cdot \hat{r}}\boldsymbol{\sigma}_{2}%
\mathbf{\cdot \hat{r}})\psi _{\pm },  \notag \\
\chi _{\pm } &=&\exp (F+K\boldsymbol{\sigma}_{1}\mathbf{\cdot \hat{r}}%
\boldsymbol{\sigma}_{2}\mathbf{\cdot \hat{r}})\eta _{\pm }=\exp F(\cosh
K+\sinh K\boldsymbol{\sigma}_{1}\mathbf{\cdot \hat{r}}\boldsymbol{\sigma}_{2}%
\mathbf{\cdot \hat{r}})\eta _{\pm },  \label{fk}
\end{eqnarray}%
In this case the decoupled form of the Schr\"{o}dinger-like equation for $%
\psi _{+}$ has the convenient property that \ the coefficients of the first
order relative momentum terms $\mathbf{\hat{r}\cdot p}$ and $(%
\boldsymbol{\sigma}_{1}\mathbf{\cdot \hat{r}}\boldsymbol{\sigma}_{2}\mathbf{%
\cdot p+}\boldsymbol{\sigma}_{2}\mathbf{\cdot \hat{r}}\boldsymbol{\sigma}_{1}%
\mathbf{\cdot p})$ as appear in Eq. (\ref{szin}) vanish. \ We obtain \cite%
{liu} 
\begin{align}
& \{\mathbf{p}^{2}+2m_{w}S+S^{2}+2\varepsilon _{w}A-A^{2}-\frac{2F^{\prime
}(\cosh 2K-1)}{r}+2F^{\prime 2}+2K^{\prime 2}+\frac{2K^{\prime }\sinh 2K}{r}
\notag \\
& -\boldsymbol{\nabla }^{2}F-F^{\prime 2}-K^{\prime 2}-\frac{2(\cosh 2K-1)}{%
r^{2}}+m(r)  \notag \\
& +\mathbf{L\cdot (}\boldsymbol{\sigma}_{1}\mathbf{+}\boldsymbol{\sigma}_{2}%
\mathbf{)}[-\frac{F^{\prime }}{r}-\frac{F^{\prime }(\cosh 2K-1)}{r}-\frac{%
(\cosh 2K-1)}{r^{2}}+\frac{K^{\prime }\sinh 2K}{r}]  \notag \\
& +\mathbf{L\cdot (}\boldsymbol{\sigma}_{1}\mathbf{-}\boldsymbol{\sigma}_{2}%
\mathbf{)}(l^{\prime }\cosh 2K-q^{\prime }\sinh 2K)  \notag \\
& +i\mathbf{L\cdot }\boldsymbol{\sigma}_{1}\mathbf{\times }%
\boldsymbol{\sigma}_{2}(q^{\prime }\cosh 2K+l^{\prime }\sinh 2K)  \notag \\
& +\boldsymbol{\sigma}_{1}\mathbf{\cdot \hat{r}}\boldsymbol{\sigma}_{2}%
\mathbf{\cdot \hat{r}L\cdot (}\boldsymbol{\sigma}_{1}\mathbf{+}%
\boldsymbol{\sigma}_{2}\mathbf{)}(-\frac{K^{\prime }(\cosh 2K-1)}{r}+\frac{%
\sinh 2K}{r^{2}}-\frac{K^{\prime }}{r}+\frac{F^{\prime }\sinh 2K}{r})  \notag
\\
& +\boldsymbol{\sigma}_{1}\mathbf{\cdot }\boldsymbol{\sigma}_{2}[k(r)-\frac{%
F^{\prime }\sinh 2K}{2r}-\frac{F^{\prime }(\cosh 2K-1)}{r}+\frac{K^{\prime
}\sinh 2K}{r}  \notag \\
& +\frac{K^{\prime }(\cosh 2K-1)}{r}+\frac{\sinh 2K}{r^{2}}-\frac{(\cosh
2K-1)}{r^{2}}]  \notag \\
+& \boldsymbol{\sigma}_{1}\mathbf{\cdot \hat{r}}\boldsymbol{\sigma}_{2}%
\mathbf{\cdot \hat{r}}[n(r)+\frac{3F^{\prime }\sinh 2K}{r}+\frac{F^{\prime
}(\cosh 2K-1)}{r}+2F^{\prime }K^{\prime }-\frac{K^{\prime }\sinh 2K}{r} 
\notag \\
& -\frac{3K^{\prime }(\cosh 2K-1)}{r}-\boldsymbol{\nabla }^{2}K+\frac{3\sinh
2K}{r^{2}}+\frac{(\cosh 2K-1)}{r^{2}}]\}\psi _{+}  \notag \\
& =b^{2}\psi _{+},  \label{57}
\end{align}%
in which

\begin{eqnarray}
k(r) &=&\frac{1}{2}\nabla ^{2}\mathcal{G-}\frac{1}{2}\mathcal{G}^{\prime 2}- 
\frac{1}{2}\mathcal{G}^{\prime }K^{\prime }-\frac{1}{2}\frac{\mathcal{G}
^{\prime }}{r}+\frac{K^{\prime }}{r},  \notag \\
n(r) &=&\nabla ^{2}K-\frac{1}{2}\nabla ^{2}\mathcal{G}-2K^{\prime }F^{\prime
}+\mathcal{G}^{\prime }F^{\prime }-\frac{3}{2r}\mathcal{G}^{\prime },  \notag
\\
m(r) &=&-\frac{1}{2}\nabla ^{2}\mathcal{G+}\frac{3}{4}\mathcal{G}^{\prime
2}+ \mathcal{G}^{\prime }F^{\prime }-K^{\prime 2},
\end{eqnarray}

For equal mass singlet states, the hyperbolic terms cancel. \ The spin-orbit
\ difference terms in general produce spin mixing.

\subsubsection{Matrix Form of the Wave Functions}

We now construct the $4\times 4$ matrix forms of the wave functions
(appropriate for a spin-one-half particle-antiparticle system) from the
sixteen component forms (appropriate for system of two spin-one-half
particles). We begin by writing the 16 component spinor wave function as 
\begin{eqnarray}
\psi &=&\psi _{1}%
\begin{pmatrix}
1 \\ 
0 \\ 
0 \\ 
0%
\end{pmatrix}%
+\psi _{2}%
\begin{pmatrix}
0 \\ 
1 \\ 
0 \\ 
0%
\end{pmatrix}%
+\psi _{3}%
\begin{pmatrix}
0 \\ 
0 \\ 
1 \\ 
0%
\end{pmatrix}%
+\psi _{4}%
\begin{pmatrix}
0 \\ 
0 \\ 
0 \\ 
1%
\end{pmatrix}
\notag \\
&=&\frac{\phi _{+}}{2}[%
\begin{pmatrix}
1 \\ 
0%
\end{pmatrix}%
\otimes 
\begin{pmatrix}
1 \\ 
0%
\end{pmatrix}%
+%
\begin{pmatrix}
0 \\ 
1%
\end{pmatrix}%
\otimes 
\begin{pmatrix}
0 \\ 
1%
\end{pmatrix}%
]+\frac{\phi _{-}}{2}[%
\begin{pmatrix}
1 \\ 
0%
\end{pmatrix}%
\otimes 
\begin{pmatrix}
1 \\ 
0%
\end{pmatrix}%
-%
\begin{pmatrix}
0 \\ 
1%
\end{pmatrix}%
\otimes 
\begin{pmatrix}
0 \\ 
1%
\end{pmatrix}%
]  \notag \\
&&+\frac{\chi _{+}}{2}[%
\begin{pmatrix}
1 \\ 
0%
\end{pmatrix}%
\otimes 
\begin{pmatrix}
0 \\ 
1%
\end{pmatrix}%
+%
\begin{pmatrix}
0 \\ 
1%
\end{pmatrix}%
\otimes 
\begin{pmatrix}
1 \\ 
0%
\end{pmatrix}%
]+\frac{\chi _{-}}{2}[%
\begin{pmatrix}
1 \\ 
0%
\end{pmatrix}%
\otimes 
\begin{pmatrix}
0 \\ 
1%
\end{pmatrix}%
-%
\begin{pmatrix}
0 \\ 
1%
\end{pmatrix}%
\otimes 
\begin{pmatrix}
1 \\ 
0%
\end{pmatrix}%
].  \label{psin}
\end{eqnarray}%
The spinors $\psi _{i}$ as well as $\phi _{\pm }=\psi _{1}\pm \psi _{4},\ \
\chi _{\pm }=\psi _{2}\pm \psi _{3}$ are themselves four component Pauli
spinors (upon which $\sigma _{1i},\sigma _{2i}$ operate). The conversion \
from sixteen \ component spinor wave functions to four by four matrix wave
functions now can be carried out in a two-step process. First, as in \cite%
{long,cww}, the \textquotedblleft energy\textquotedblright\ or $q$ space
column vector direct products are converted to 4x4 matrices as follows
(recall the factor of $i\alpha _{y}$ plus the transpose operation changes
particle spinor into antiparticle spinor) 
\begin{equation}
\Psi _{(1)}\otimes \Psi _{(2)}\rightarrow \Psi _{(1)}\Psi _{(2)}^{T}i\alpha
_{2}=\Psi _{(1)}\Psi _{(2)}^{T}q_{1}\otimes i\sigma _{2},  \label{ww}
\end{equation}%
in which $\sigma _{0},\sigma _{i},q_{0},q_{i},;i=1,2,3$ are the $2\times 2~$
unit and three Pauli matrices in commuting spaces (spin and energy space) 
\begin{eqnarray}
\sigma _{i}\sigma _{j} &=&\delta _{ij}\sigma _{0}+i\varepsilon _{ijk}\sigma
_{k},  \notag \\
q_{i}q_{j} &=&\delta _{ij}q_{0}+i\varepsilon _{ijk}q_{k},
\end{eqnarray}%
and whose direct products form the Dirac matrices. Second, the $\phi _{\pm
},\chi _{\pm }$ four component Pauli spinors are converted to $2\times 2$
matrices in $\sigma $ and $q$ space by 
\begin{eqnarray}
\phi _{\pm } &\rightarrow &\mathcal{\phi }_{\pm }=\left( \mathcal{\phi }%
_{\pm 0}\sigma _{0}+\boldsymbol{\phi }_{\pm }\mathbf{\cdot }%
\boldsymbol{\sigma}\right) ,  \notag \\
\chi _{\pm } &\rightarrow &\mathcal{\chi }_{\pm }=\left( \mathcal{\chi }%
_{\pm 0}\sigma _{0}+\boldsymbol{\chi }_{\pm }\mathbf{\cdot }%
\boldsymbol{\sigma}\right) .  \label{fch}
\end{eqnarray}%
Together, the 4x4 matrix wave function in $\sigma ,q$ space is 
\begin{eqnarray}
&&(\frac{\phi _{+}}{2}\otimes q_{0}\mathbf{+}\frac{\phi _{-}}{2}\otimes
q_{3}+\frac{\chi _{+}}{2}\otimes q_{1}+\frac{\chi _{-}}{2}\otimes
iq_{2})q_{1}\otimes i\sigma _{2}  \notag \\
&=&(\frac{\phi _{+}}{2}\otimes q_{1}\mathbf{+}\frac{\phi _{-}}{2}\otimes
iq_{2}+\frac{\chi _{+}}{2}\otimes q_{0}+\frac{\chi _{-}}{2}\otimes
q_{3})i\sigma _{2}\otimes 1  \notag \\
&=&(\frac{\phi _{+}i\sigma _{2}}{2}\otimes q_{1}\mathbf{+}\frac{\phi
_{-}i\sigma _{2}}{2}\otimes iq_{2}+\frac{\chi _{+}i\sigma _{2}}{2}\otimes
q_{0}+\frac{\chi _{-}i\sigma _{2}}{2}\otimes q_{3})  \notag \\
&\rightarrow &(\frac{\mathcal{\phi }_{+}}{2}\otimes q_{1}\mathbf{+}\frac{%
\mathcal{\ \phi }_{-}}{2}\otimes iq_{2}+\frac{\mathcal{\chi }_{+}}{2}\otimes
q_{0}+\frac{\mathcal{\chi }_{-}}{2}\otimes q_{3}),  \label{spn}
\end{eqnarray}%
where for convenience we have absorbed the factor $i\sigma _{2}$ into the
wave functions as the wave function is arbitrary up to a constant
multiplicative matrix and we have used the same symbol for each of the
transformed wave functions to simplify notation. \ In our work below we drop
the direct product symbol $\otimes $, it being understood to apply whenever $%
\sigma $ and $q$ space matrices multiply one another. \ 

The four component spinors $\psi _{\pm }$ and $\eta _{\pm }$ are similarly
transformed into matrices which can be expanded in terms of $\sigma _{0}$
and $\mathbf{\sigma }$. 
\begin{eqnarray}
\psi _{\pm } &\rightarrow &\mathcal{\psi }_{\pm }=\left( \mathcal{\psi }%
_{\pm 0}\sigma _{0}+\boldsymbol{\psi }_{\pm }\mathbf{\cdot }%
\boldsymbol{\sigma}\right) ,  \notag \\
\eta _{\pm } &\rightarrow &\mathcal{\eta }_{\pm }=\left( \mathcal{\eta }%
_{\pm 0}\sigma _{0}+\boldsymbol{\eta }_{\pm }\mathbf{\cdot }%
\boldsymbol{\sigma}\right) .  \label{sie}
\end{eqnarray}

With $\mathbf{A}$ a generic matrix, Eq. (\ref{ww}) leads to 
\begin{eqnarray}
\boldsymbol{\sigma}_{1}\cdot \mathbf{A}\psi _{+} &\rightarrow &\mathbf{%
A\cdot }\boldsymbol{\sigma}\left( \mathcal{\psi }_{+0}\sigma _{0}+%
\boldsymbol{\psi }_{+}\mathbf{\cdot }\boldsymbol{\sigma}\right)  \notag \\
&=&\mathbf{A}\cdot \boldsymbol{\psi }_{+}\sigma _{0}+\mathbf{A\cdot }%
\boldsymbol{\sigma}\mathcal{\psi }_{+0}+i\mathbf{A}\times \boldsymbol{\psi }%
_{+}\mathbf{\cdot }\boldsymbol{\sigma},  \notag \\
\boldsymbol{\sigma}_{2}\cdot \mathbf{A}\psi _{+} &\rightarrow &-\mathbf{A}%
\cdot \left( \mathcal{\psi }_{+0}\sigma _{0}+\boldsymbol{\psi }_{+}\mathbf{%
\cdot }\boldsymbol{\sigma}\right) \sigma  \notag \\
&=&-\mathbf{A}\cdot \boldsymbol{\psi }_{+}\sigma _{0}-\mathbf{A\cdot }%
\boldsymbol{\sigma}\mathcal{\psi }_{+0}+i\mathbf{A}\times \boldsymbol{\psi }%
_{+}\mathbf{\cdot }\boldsymbol{\sigma},  \notag \\
\boldsymbol{\sigma}_{1}\mathbf{\cdot A}\boldsymbol{\sigma}_{2}\mathbf{\cdot A%
}\psi _{+} &\rightarrow &-\boldsymbol{\sigma}\mathbf{\cdot A}\left( \mathcal{%
\psi }_{+0}\sigma _{0}+\boldsymbol{\psi }_{+}\mathbf{\cdot }%
\boldsymbol{\sigma}\right) \boldsymbol{\sigma}\mathbf{\cdot A}  \notag \\
&=&-\mathbf{A}^{2}\mathcal{\phi }_{+0}\sigma _{0}+(\mathbf{A}^{2}%
\boldsymbol{ \psi }_{+}\mathbf{-}2\mathbf{A\cdot }{\boldsymbol{\psi}}_{+}%
\mathbf{A)\cdot }\boldsymbol{\sigma}\mathbf{,}  \label{smtrx}
\end{eqnarray}%
which are needed to convert Eqs. (\ref{chip}), (\ref{chim}), and (\ref{54})
into their matrix counterparts. \ In terms of matrix wave functions $%
\mathcal{\phi }_{\pm },\mathcal{\chi }_{\pm },\mathcal{\psi }_{\pm },$ and $%
\mathcal{\eta }_{\pm }$, we find that Eq. (\ref{fk} ) becomes 
\begin{eqnarray}
\mathcal{\phi }_{\pm 0} &=&\exp (F-K)\mathcal{\psi }_{\pm 0};~\ %
\boldsymbol{\phi }_{\pm }=\exp (F+K)(\mathbf{1}-(1-\exp (-2K))\mathbf{\hat{r}%
\hat{r}})\cdot \boldsymbol{\psi }_{\pm },  \notag \\
\boldsymbol{\chi }_{\pm 0} &=&\exp (F-K)\mathcal{\eta }_{\pm 0};\mathbf{~\
\chi }_{\pm }=\exp (F+K)(\mathbf{1}-(1-\exp (-2K))\mathbf{\hat{r}\hat{r}%
)\cdot }{\boldsymbol{\eta}}_{\pm }.  \label{mspn}
\end{eqnarray}

We write the $4\times 4$ matrix wave function form $\psi $ of the sixteen
component $\psi ~$to be used in our decay \ amplitude in terms of the matrix
form 
\begin{equation}
\Psi (\mathbf{r)}\mathcal{=}\frac{1}{2\sqrt{2}}(\mathcal{\psi }_{+}q_{1}+%
\mathcal{\psi }_{-}iq_{2}+\eta _{+}q_{0}\mathbf{+}\eta _{-}q_{3}),
\label{simp}
\end{equation}%
where $\mathcal{\psi }_{+}=\mathcal{\psi }_{+0}\sigma _{0}+\boldsymbol{\psi }%
_{+}\mathbf{\cdot }\boldsymbol{\sigma}~$is the $2\times 2$ matrix form of
the solution of the above Schr\"{o}dinger-like Pauli equation (\ref{57}). \ 

Using these four components, the remaining twelve components $\mathcal{\psi }%
_{-0},$ $\boldsymbol{\psi }_{-},\eta _{\pm 0},$ and $\boldsymbol{\eta }_{\pm
}$ are obtained from Eqs. (\ref{chip}), (\ref{chim}), (\ref{54}), and (\ref%
{mspn}). \ In all of \ our \ decays the particle antiparticle pairs have the
same mass: $m_{1}=m_{1}\equiv m$ and so $\varepsilon _{1}=\varepsilon
_{2}\equiv \varepsilon =w/2$. \ Using the definition 
\begin{eqnarray}
M_{1} &=&M_{2}\equiv M=m\exp (L),  \notag \\
E_{1} &=&E_{2}\equiv E=\varepsilon \exp (-\mathcal{G)},  \label{me}
\end{eqnarray}%
we show in Appendix B 
\begin{eqnarray}
\mathcal{\eta }_{+0} &=&\frac{\exp (\mathcal{G}+2K\mathcal{)}}{E}[\mathbf{p}-%
\frac{i}{2}\boldsymbol{\nabla }(L+2F+2K)]\cdot \lbrack \mathbf{1}+Q_{m}%
\mathbf{\hat{r}\hat{r}}]\cdot \boldsymbol{\psi }_{+},  \notag \\
\boldsymbol{\eta }_{+} &=&\frac{\exp (\mathcal{-}L)}{E}\{(\mathbf{p}-\frac{i%
}{2}\boldsymbol{\nabla }L)+Q_{p}(\mathbf{\hat{r}\hat{r}}\cdot \mathbf{p}-%
\frac{i}{2}\boldsymbol{\nabla }L)\}\mathcal{\psi }_{+0},  \label{etap}
\end{eqnarray}%
and 
\begin{eqnarray}
\mathcal{\eta }_{-0} &=&0,  \notag \\
\boldsymbol{\eta }_{-} &=&-\frac{\exp (\mathcal{G)}}{M}[\mathbf{1}+Q_{p}%
\mathbf{\hat{r}\hat{r}}]\cdot \lbrack i\mathbf{p}+\frac{1}{2}%
\boldsymbol{\nabla }(L-2\mathcal{G})]\times \lbrack \mathbf{1}+Q_{m}\mathbf{%
\hat{r}\hat{r}}]\cdot \boldsymbol{\psi }_{+}.  \label{etam}
\end{eqnarray}%
The final four components of the four by four matrix wave function found in
Appendix B are 
\begin{eqnarray}
\mathcal{\psi }_{-0} &=&\{\frac{(E^{2}+M^{2})}{2EM}-\frac{\exp (2\mathcal{G)}%
}{2ME}[\mathbf{p}+\frac{i}{2}\boldsymbol{\nabla }L]\cdot \lbrack \mathbf{p}-%
\frac{i}{2}\boldsymbol{\nabla }L]\}\mathcal{\psi }_{+0},  \notag \\
\boldsymbol{\psi }_{-} &=&\frac{(E^{2}+M^{2})}{2EM}\boldsymbol{\psi }_{+}-%
\frac{\exp (2\mathcal{G})}{2EM}[\mathbf{1}+Q_{p}\mathbf{\hat{r}\hat{r}}] 
\notag \\
&&\cdot {\biggl (}[\mathbf{p}-\frac{i}{2}\boldsymbol{\nabla }(L+6\mathcal{G}%
)][\mathbf{p}-\frac{i}{2}\boldsymbol{\nabla }(3L-2\mathcal{G})]\times
\lbrack \mathbf{1}+Q_{m}\mathbf{\hat{r}\hat{r}}]\cdot \boldsymbol{\psi }_{+}
\notag \\
&&+[\mathbf{p}-\frac{i}{2}\boldsymbol{\nabla }(L+2\mathcal{G})]\times \{[%
\mathbf{p}-\frac{i}{2}\boldsymbol{\nabla }(L-2\mathcal{G})]\times \lbrack 
\mathbf{1}+Q_{m}\mathbf{\hat{r}\hat{r}}]\cdot \boldsymbol{\psi }_{+}\}{%
\biggr ),}  \label{psim}
\end{eqnarray}%
where 
\begin{eqnarray}
Q_{p} &\equiv &\exp (2K)-1,  \notag \\
Q_{m} &\equiv &\exp (-2K)-1.
\end{eqnarray}%
We also show in Appendix B how for both singlet and triplet states these
solutions together with Eq. (\ref{simp}) are related to the solutions
governed by the free Dirac spinors in the absence of interactions (see also
Eq. (\ref{anamp}) and discussion below (\ref{3m})).

\subsubsection{\protect\bigskip Covariant Normalization Conditions for the
Matrix Wave Function}

In this section we discuss how the norm of our matrix \ wave function will
differ from the naive form of\ 
\begin{equation}
\frac{1}{8}\int d^{3}xTr_{q\sigma }\Psi ^{\dag }\Psi \mathcal{=}\frac{1}{4}%
\int d^{3}xTr_{\sigma }(\mathcal{\psi }_{+}^{\dag }\mathcal{\psi }_{+}+%
\mathcal{\psi }_{-}^{\dag }\mathcal{\psi }_{-}+\mathcal{\eta }_{+}^{\dag }%
\mathcal{\eta }_{+}+\mathcal{\eta }_{-}^{\dag }\mathcal{\eta }_{-})=1.
\label{nrm}
\end{equation}%
In a series of papers in the context of constraint dynamics, H. Sazdjian has
shown \cite{saznm} how this norm must be modified so that, like its
nonrelativistic counterpart, its constancy is connected to a conserved, in
this two-body case, tensor current. The norm he developed was not for the
solution of a quasipotential equation like Eq. (\ref{57}) but rather
developed from a set of two-body Dirac equations similar to those we use
here. \ It deviated from one like the above by terms that depend on the
interaction as well as the way in which the interaction depends on the CM
energy. Later work \cite{jmp} simplified the norm to one that is interaction
independent when the interaction is independent of the energy. \ In terms of
the 16 component spinor solutions $\psi $ of the Two-Body Dirac equations
given in Eqs. (\ref{tbde}) we found the norm condition of 
\begin{equation}
\int d^{3}x[\psi ^{\dag }(1+4w^{2}\beta _{1}\beta _{2}\frac{\partial \Delta 
}{\partial w^{2}})\psi ]\equiv \int d^{3}x\psi ^{\dag }\mathcal{L}\psi =1.
\label{nrm2}
\end{equation}%
If the matrix $\Delta $ is CM energy independent, then the norm is like that
of the (one-body) Dirac equation (with no energy dependence of the
interactions). We call the norm of Eq. (\ref{nrm}) the naive norm (NN) and
that of Eq. (\ref{nrm2}) the two-body Dirac norm (TBDN).\ The connection
between the matrix interaction function $\Delta $ and the core scalar and
vector interactions appearing in Eqs. (\ref{tbde}) were found in \cite{jmp}.
There we showed that \ref{tbde} has the hyperbolic structure%
\begin{align}
\mathcal{S}_{1}\psi & =(\cosh (\Delta )\mathbf{S}_{1}+\sinh (\Delta )\mathbf{%
S}_{2})\psi =0,  \label{cnhyp} \\
\mathcal{S}_{2}\psi & =(\cosh (\Delta )\mathbf{S}_{2}+\sinh (\Delta )\mathbf{%
S}_{1})\psi =0,  \notag
\end{align}%
in which 
\begin{align}
\mathbf{S}_{1}\psi & \equiv (\mathcal{S}_{10}\cosh (\Delta )+\mathcal{S}%
_{20}\sinh (\Delta ))\psi =0,  \notag \\
\mathbf{S}_{2}\psi & \equiv (\mathcal{S}_{20}\cosh (\Delta )+\mathcal{S}%
_{10}\sinh (\Delta ))\psi =0,  \label{cnmyp}
\end{align}%
with 
\begin{align}
\mathcal{S}_{10}\psi & =\big(-\beta _{1}\Sigma _{1}\cdot p+\epsilon
_{1}\beta _{1}\gamma _{51}+m_{1}\gamma _{51}\big)\psi  \notag \\
\mathcal{S}_{20}\psi & =\big(\beta _{2}\Sigma _{2}\cdot p+\epsilon _{2}\beta
_{2}\gamma _{52}+m_{2}\gamma _{52}\big)\psi  \label{s0}
\end{align}%
and 
\begin{equation}
\Delta ={\frac{1}{2}}\gamma _{51}\gamma _{52}[L(x_{\perp })-\gamma _{1}\cdot
\gamma _{2}\mathcal{G}(x_{\perp })].  \label{del}
\end{equation}%
with $L$ and $\mathcal{G}$ given in Eq. (\ref{me}) (see also \cite{em}). In
matrix form the connection given in Eqs. (\ref{mspn}), (\ref{spn}), and (\ref%
{psin}) between the matrix form of the wave function $\psi $ of (\ref{tbde},%
\ref{cnhyp}) and $\Psi $ is%
\begin{eqnarray}
\mathcal{\psi } &\mathcal{=}&\exp (F)[\cosh K\Psi (\mathbf{r)}-\sinh K%
\boldsymbol{\Sigma }\cdot \mathbf{\hat{r}}\Psi \mathbf{(\mathbf{r)}}%
\boldsymbol{\Sigma }\cdot \mathbf{\hat{r}}  \notag \\
&\equiv &\mathcal{K}\Psi (\mathbf{r)}
\end{eqnarray}%
\ In Appendix C we show that in terms of the matrix wave function $\Psi $
solution (\ref{simp}) to Eq. (\ref{57}) the nomalization condition can be
written as%
\begin{equation}
\int d^{3}xTr\psi ^{\dag }\mathcal{L}\psi =\int d^{3}xTr\left( \mathcal{K}%
\Psi (\mathbf{r)}\right) ^{\dag }\mathcal{LK}\Psi (\mathbf{r)}=1.
\end{equation}%
\ There we also give the matrix form of the operator $\mathcal{L}.$ \ The
deviation of the matrices $\mathcal{K}$ and $\mathcal{L}$ from the unit
matrix will affect the decay rates. The decay amplitude (\ref{dcmp}) in
terms of the matrix wave function $\Psi (\mathbf{r)}$ is%
\begin{eqnarray}
\mathcal{M}_{X\rightarrow 2\gamma } &=&e^{2}\sqrt{\frac{\pi }{2}}\int
d^{3}rTr_{\sigma q}\{\mathcal{K}\Psi (\mathbf{r)}  \notag \\
&&\times \mathbf{[}\exp (-i\mathbf{k\cdot r})q_{3}q_{1}\boldsymbol{\sigma}%
\mathbf{\cdot }\boldsymbol{\epsilon }^{(\alpha _{1})}(m-iq_{3}q_{1}%
\boldsymbol{\sigma}\mathbf{\cdot \nabla )}\frac{\exp (-mr)}{r}q_{3}q_{1}%
\boldsymbol{\sigma}\mathbf{\cdot }\boldsymbol{\epsilon }^{(\alpha _{2})} 
\notag \\
&&+\exp (i\mathbf{k\cdot r)}q_{3}q_{1}\boldsymbol{\sigma}\mathbf{\cdot }%
\boldsymbol{\epsilon }^{(\alpha _{2})}(m-iq_{3}q_{1}\boldsymbol{\sigma}%
\mathbf{\cdot \nabla )}\frac{\exp (-mr)}{r}q_{3}q_{1}\boldsymbol{\sigma}%
\mathbf{\cdot }\boldsymbol{\epsilon }^{(\alpha _{1})}]\}.  \label{dcmpq}
\end{eqnarray}

\subsubsection{Scalar and Vector Wave Functions in Vector Spherical Harmonics%
}

Given the above wave functions we now write down the total 4x4 matrix wave
function in terms of $\mathcal{\psi }_{+}.$ The spin-zero part of the total
wave function is governed by $\mathcal{\psi }_{+0}$, the spin-one portion by 
$\boldsymbol{\psi }_{+}$. \ These wave functions appear in the forms 
\begin{eqnarray}
\mathcal{\psi }_{+} &=&\mathcal{\psi }_{+0}\sigma _{0}+\boldsymbol{\psi }_{+}%
\mathbf{\cdot }\boldsymbol{\sigma},  \notag \\
\mathcal{\psi }_{+0} &=&\frac{u_{j0j}^{+}}{r}Y_{jm},  \notag \\
\boldsymbol{\psi }_{+} &=&\frac{u_{(j+1)1j}^{+}}{r}\mathbf{Y}_{jm+}+\frac{%
u_{(j-1)1j}^{+}}{r}\mathbf{Y}_{jm-}+\frac{u_{j1j}^{+}}{r}\mathbf{X}_{jm},
\label{wvfn}
\end{eqnarray}%
where the labels on the radial wave function $u$ refer to the $lsj$ quantum
numbers of the solutions to Eq. (\ref{57}) and 
\begin{eqnarray}
\mathbf{Y}_{jm+} &=&(a_{+}\mathbf{\hat{r}+}rb_{+}\mathbf{p})Y_{jm},  \notag
\\
\mathbf{Y}_{jm-} &=&(a_{-}\mathbf{\hat{r}+}rb_{-}\mathbf{p})Y_{jm},  \notag
\\
\mathbf{X}_{jm} &=&\frac{\mathbf{L}Y_{jm}}{\sqrt{j(j+1)}},  \label{vsph}
\end{eqnarray}%
are vector spherical harmonic eigenfunctions of $\mathbf{L}^{2}$ with
eigenvalue $l(l+1)$ where $l=j+1,j-1,j$ respectively. The coefficients are 
\begin{eqnarray}
a_{+} &=&-\sqrt{\frac{j+1}{2j+1}};~a_{-}=\sqrt{\frac{j}{2j+1}},  \notag \\
b_{+} &=&\frac{i}{j+1}\sqrt{\frac{j+1}{2j+1}};~b_{-}=\frac{i}{j}\sqrt{\frac{j%
}{2j+1}}.  \label{abj}
\end{eqnarray}%
In our work below, there will be no spin mixing and the unnatural parity
solutions $(u_{j1j}^{+}/r)\mathbf{X}_{jm}(\mathbf{\Omega })$ will not
contribute.

For spin-singlet states ($\boldsymbol{\psi }_{+}$ $=0$), Eqs. (\ref{simp}), (%
\ref{etap}), (\ref{psim}) imply the following combination of scalar and
vector wave functions 
\begin{equation}
\Psi |_{s=0}=\frac{1}{2\sqrt{2}}(\psi _{+0}\sigma _{0}q_{1}+\psi _{-0}\sigma
_{0}iq_{2}+\boldsymbol{\eta }_{+}\cdot \boldsymbol{\sigma}q_{0}).
\label{sing}
\end{equation}%
In Appendix C we show that the TBDN for spin-singlet states is%
\begin{eqnarray}
1 &=&\frac{1}{2}\int d^{3}x\exp (2F)\big([\exp (-2K)(\psi _{+0}^{\dag }\psi
_{+0}+\psi _{-0}^{\dag }\psi _{-0})+\exp (2K)\boldsymbol{\eta }_{+}^{\dag
}\cdot \boldsymbol{\eta }_{+}-2\sinh 2K\boldsymbol{\eta }\mathbf{_{+}^{\dag
}\cdot \hat{r}}\boldsymbol{\eta }\mathbf{_{+}\cdot \hat{r}]}  \notag \\
&&+[2w^{2}\frac{\partial L}{\partial w^{2}}[\exp (-2K)(\psi _{+0}^{\dag
}\psi _{+0}-\psi _{-0}^{\dag }\psi _{-0})-\exp (2K)\boldsymbol{\eta }%
_{+}^{\dag }\cdot \boldsymbol{\eta }_{+}+2\sinh 2K\boldsymbol{\eta }\mathbf{%
_{+}^{\dag }\cdot \hat{r}}\boldsymbol{\eta }\mathbf{_{+}\cdot \hat{r}]} 
\notag \\
&&+2w^{2}\frac{\partial \mathcal{G}}{\partial w^{2}}[2\exp (-2K)(2\psi
_{+0}^{\dag }\psi _{+0}+\psi _{-0}^{\dag }\psi _{-0})]\big)  \label{sinrm}
\end{eqnarray}%
The naive norm (NN) is given by the above two-body Dirac norm by $\mathcal{L}%
,\mathcal{K}\rightarrow 1$ or equivlently by $\exp (F),\exp (K)\rightarrow
1, $ $\partial L/\partial w^{2},$ and $~\partial \mathcal{G}/\partial w^{2}%
\mathcal{\rightarrow }0$ and is%
\begin{equation}
\frac{1}{2}\int d^{3}x\{[\mathcal{\psi }_{+0}^{\dag }\mathcal{\psi }_{+0}+%
\boldsymbol{\eta }_{+}^{\dag }\cdot \boldsymbol{\eta }_{+}+\mathcal{\psi }%
_{-0}^{\dag }\mathcal{\psi }_{-0}]=1.
\end{equation}

Appendix D gives from Eqs. (\ref{etap}) and(\ref{psim}) the needed radial
forms for the contributing wave functions in terms of the radial portions of
the solution to Eq. (\ref{57}). It requires the radial form of Eq. (\ref{57}%
) which is simply 
\begin{equation}
\lbrack -\frac{1}{r}\frac{d^{2}}{dr^{2}}r+\frac{j(j+1)}{r^{2}}+\frac{1}{2}%
\boldsymbol{\nabla }^{2}L-\frac{1}{4}\left( \boldsymbol{\nabla }L\right)
^{2}]\frac{u_{j0j}^{+}}{r}=\mathcal{B}^{2}\exp (-2\mathcal{G})]\frac{%
u_{j0j}^{+}}{r},  \label{rdc}
\end{equation}%
where \cite{em} 
\begin{eqnarray}
\mathcal{B}^{2} &\equiv &E^{2}-M^{2}  \notag \\
-\mathcal{B}^{2}\exp (-2\mathcal{G)} &\mathcal{=}&2m_{w}S+S^{2}+2\varepsilon
_{w}A-A^{2}.  \label{b2}
\end{eqnarray}%
Appendix D gives us the relations between the contributing wave functions $%
\mathcal{\psi }_{+0},\mathcal{\psi }_{-0},$ and $\boldsymbol{\eta }_{+}.$
They are 
\begin{equation}
\mathcal{\psi }_{-0}\equiv \frac{u_{j0j}^{-}}{r}Y_{jm}=\frac{M}{E}\mathcal{%
\psi }_{+0}=\frac{M}{E}\frac{u_{j0j}^{+}}{r}Y_{jm},  \label{ls}
\end{equation}%
and 
\begin{eqnarray}
\boldsymbol{\eta }_{+} &\equiv &i(\frac{\upsilon _{(j-1)1j}^{+}}{r}\mathbf{Y}%
_{jm-}+\frac{\upsilon _{(j+1)1j}^{+}}{r}\mathbf{Y}_{jm+}),  \notag \\
\frac{\upsilon _{(j-1)1j}^{+}}{r} &=&\frac{\exp (\mathcal{G}-2K)}{E}[\exp
(2K)(-\frac{d}{dr}-\frac{L^{\prime }}{2})-\frac{(j+1)}{r}]\frac{u_{j0j}^{+}}{%
r}\sqrt{\frac{j}{2j+1}},  \notag \\
\frac{\upsilon _{(j+1)1j}^{+}}{r} &=&\frac{\exp (\mathcal{G}-2K)}{E}[\exp
(2K)(\frac{d}{dr}+\frac{L^{\prime }}{2})-\frac{j}{r}]\frac{u_{j0j}^{+}}{r}%
\sqrt{\frac{j+1}{2j+1}}.  \label{jl}
\end{eqnarray}%
For the norm we also need%
\begin{equation}
\boldsymbol{\eta }_{+}\cdot \mathbf{\hat{r}}\equiv i(-\frac{\upsilon
_{(j+1)1j}^{+}}{r}\sqrt{\frac{j+1}{2j+1}}+\frac{\upsilon _{(j-1)1j}^{+}}{r}%
\sqrt{\frac{j}{2j+1}})Y_{jm}  \label{lj}
\end{equation}

For spin-triplet states ($\psi _{+0}$ $=0$), Eqs. (\ref{simp}), (\ref{etap}%
), (\ref{etam}), and (\ref{psim}) imply the combination 
\begin{equation}
\Psi |_{s=1}=\frac{1}{2\sqrt{2}}(\boldsymbol{\psi }_{+}\mathbf{\cdot }%
\boldsymbol{\sigma}q_{1}+\boldsymbol{\psi }_{-}\mathbf{\cdot }%
\boldsymbol{\sigma}iq_{2}+\eta _{+0}q_{0}\boldsymbol{+\eta }_{-}\mathbf{%
\cdot }\boldsymbol{\sigma}q_{3}),  \label{trip}
\end{equation}%
and the contributing wave functions are $\boldsymbol{\psi }_{+}\mathbf{,}%
\boldsymbol{\psi }_{-},\mathcal{\eta }_{+0},$ and $\boldsymbol{\eta }_{-}$ .
In Appendix C we show that the TBDN for spin-triplet states is \ 
\begin{eqnarray}
&&\frac{1}{2}\int d^{3}x\exp (2F)\big([\exp (2K)(\boldsymbol{\psi }%
_{+}^{\dag }\mathbf{\cdot }\boldsymbol{\psi }\mathbf{_{+}}+\boldsymbol{\psi }%
_{-}^{\dag }\mathbf{\cdot }\boldsymbol{\psi }_{-}+\boldsymbol{\eta }\mathbf{%
_{-}^{\dag }\cdot }\boldsymbol{\eta }_{-})+\exp (-2K)\eta _{+0}^{\dag }\eta
_{+0}  \notag \\
&&-2\sinh 2K(\boldsymbol{\psi }_{+}^{\dag }\mathbf{\cdot \hat{r}}%
\boldsymbol{\psi }\mathbf{_{+}\mathbf{\cdot }\hat{r}}+\boldsymbol{\psi }%
_{-}^{\dag }\mathbf{\cdot \hat{r}}\boldsymbol{\psi }\mathbf{_{-}}\cdot 
\mathbf{\hat{r}+}\boldsymbol{\eta }\mathbf{_{-}^{\dag }\mathbf{\cdot }\hat{r}%
}\boldsymbol{\eta }\mathbf{_{-}}\cdot \mathbf{\hat{r}})]  \notag \\
&&+\{[2w^{2}\frac{\partial L}{\partial w^{2}}[\exp (2K)(\boldsymbol{\psi }%
_{+}^{\dag }\mathbf{\cdot }\boldsymbol{\psi }\mathbf{_{+}}-\boldsymbol{\psi }%
_{-}^{\dag }\mathbf{\cdot }\boldsymbol{\psi }_{-}+\boldsymbol{\eta }\mathbf{%
_{-}^{\dag }\cdot }\boldsymbol{\eta }_{-})-\exp (-2K)\eta _{+0}^{\dag }\eta
_{+0}  \notag \\
&&-2\sinh 2K(\boldsymbol{\psi }_{+}^{\dag }\mathbf{\cdot \hat{r}}%
\boldsymbol{\psi }\mathbf{_{+}\mathbf{\cdot }\hat{r}}-\boldsymbol{\psi }%
_{-}^{\dag }\mathbf{\cdot \hat{r}}\boldsymbol{\psi }\mathbf{_{-}}\cdot 
\mathbf{\hat{r}+}\boldsymbol{\eta }\mathbf{_{-}^{\dag }\mathbf{\cdot }\hat{r}%
}\boldsymbol{\eta }\mathbf{_{-}}\cdot \mathbf{\hat{r}})]  \notag \\
&&+4w^{2}\frac{\partial \mathcal{G}}{\partial w^{2}}([-\exp (2K)(%
\boldsymbol{\psi }_{-}^{\dag }\mathbf{\cdot }\boldsymbol{\psi }_{-}+%
\boldsymbol{\eta }\mathbf{_{-}^{\dag }\cdot }\boldsymbol{\eta }_{-})+2\exp
(-2K)\eta _{+0}^{\dag }\eta _{+0}  \notag \\
&&+2\sinh 2K[(\boldsymbol{\psi }_{+}^{\dag }\mathbf{\cdot \hat{r}}%
\boldsymbol{\psi }\mathbf{_{+}\mathbf{\cdot }\hat{r}}+\boldsymbol{\psi }%
_{-}^{\dag }\mathbf{\cdot \hat{r}}\boldsymbol{\psi }\mathbf{_{-}\mathbf{%
\cdot }\hat{r}}+\boldsymbol{\eta }\mathbf{\mathbf{_{-}^{\dag }\mathbf{\cdot }%
\hat{r}}}\boldsymbol{\eta }\mathbf{\mathbf{_{-}}\cdot \mathbf{\hat{r}})}]]\}%
\big)=1,  \label{trpnrm}
\end{eqnarray}%
while the naive norm (NN) is%
\begin{equation}
\frac{1}{2}\int d^{3}x(\boldsymbol{\psi }_{+}^{\dag }\mathbf{\cdot }%
\boldsymbol{\psi }\mathbf{_{+}}+\boldsymbol{\psi }_{-}^{\dag }\mathbf{\cdot }%
\boldsymbol{\psi }_{-}+\eta _{+0}^{\dag }\eta _{+0}+\boldsymbol{\eta }%
\mathbf{_{-}^{\dag }\cdot }\boldsymbol{\eta }_{-})=1.
\end{equation}%
In Appendix D we show that Eq. (\ref{psim}) gives $\boldsymbol{\psi }_{-}$ \
from 
\begin{eqnarray}
\boldsymbol{\psi }_{-} &=&\frac{(E^{2}+M^{2})}{2EM}\boldsymbol{\psi }_{+}-%
\frac{\exp (2\mathcal{G})}{2EM}[\mathcal{B}^{2}\exp (-2\mathcal{G)}%
\boldsymbol{\psi }_{+}+\mathcal{J]}  \notag \\
&=&\frac{M}{E}\boldsymbol{\psi }_{+}-\frac{\exp (2\mathcal{G})}{2EM}\mathcal{%
J},  \label{j}
\end{eqnarray}%
\ in which 
\begin{eqnarray}
\mathcal{J} &=&\frac{1}{2j+1}{\biggl (}\{\Phi _{--}-2(j+1)\mathcal{B}%
^{2}\exp (-2\mathcal{G)+}2\sqrt{j(j+1)}\Phi _{+-}+\frac{A_{mm}}{r^{2}}+\frac{%
B_{mm}}{r}+C_{mm}+\frac{F_{mm}}{r}\frac{d}{dr}+G_{mm}\frac{d}{dr}\}\frac{%
u_{-}}{r}\mathbf{Y}_{jm-}  \notag \\
&&+\mathcal{\{}\Phi _{-+}+\sqrt{j(j+1)}[2\Phi _{++}-2\mathcal{B}^{2}\exp (-2%
\mathcal{G)}+\frac{A_{mp}}{r^{2}}+\frac{B_{mp}}{r}+C_{mp}+\frac{F_{mp}}{r}%
\frac{d}{dr}+G_{mp}\frac{d}{dr}]\}\frac{u_{+}}{r}\mathbf{Y}_{jm-}  \notag \\
&&+\{-\Phi _{++}-2j\mathcal{B}^{2}\exp (-2\mathcal{G)}+2\sqrt{j(j+1)}\Phi
_{-+}+\frac{A_{pp}}{r^{2}}+\frac{B_{pp}}{r}+C_{pp}\mathcal{+}\frac{F_{pp}}{r}%
\frac{d}{dr}+G_{pp}\frac{d}{dr}\}\frac{u_{+}}{r}\mathbf{Y}_{jm+}  \notag \\
&&+\{-\Phi _{+-}+\sqrt{j(j+1)}[2\Phi _{--}-2\mathcal{B}^{2}\exp (-2\mathcal{%
G)}+\frac{A_{pm}}{r^{2}}+\frac{B_{pm}}{r}+C_{pm}+\frac{F_{pm}}{r}\frac{d}{dr}%
+G_{pm}\frac{d}{dr}]\}\frac{u_{-}}{r}\mathbf{Y}_{jm+}{\biggr )},  \label{ujp}
\end{eqnarray}%
(see Appendix D for explicit forms of the functions $A_{mm},..,G_{pm},$). \
\ This equation requires the coupled radial wave equations for spin triplet
states that follow from Eq. (\ref{57}). \ They have the form \cite{liu} 
\begin{eqnarray}
\lbrack -\frac{1}{r}\frac{d^{2}}{dr^{2}}r+\frac{(j+2)(j+1)}{r^{2}}+\Phi
_{++}]\frac{u_{(j+1)1j}^{+}}{r}+\Phi _{+-}\frac{u_{(j-1)1j}^{+}}{r} &=&%
\mathcal{B}^{2}\exp (-2\mathcal{G)}\frac{u_{(j+1)1j}^{+}}{r},  \notag \\
\lbrack -\frac{1}{r}\frac{d^{2}}{dr^{2}}r+\frac{j(j-1)}{r^{2}}+\Phi _{--}]%
\frac{u_{(j-1)1j}^{+}}{r}+\Phi _{-+}\frac{u_{(j+1)1j}^{+}}{r} &=&\mathcal{B}%
^{2}\exp (-2\mathcal{G)}\frac{u_{(j-1)1j}^{+}}{r}.
\end{eqnarray}%
(See Appendix D for the explicit forms of $\Phi _{\pm \pm })$. \ Thus with
Eq. (\ref{wvfn}) we have 
\begin{eqnarray}
&&\frac{u_{(j-1)1j}^{-}}{r}=\frac{M}{E}\frac{u_{(j-1)1j}^{+}}{r}-\frac{\exp
(2\mathcal{G})}{2EM\left( 2j+1\right) }\{[\Phi _{--}-2(j+1)\mathcal{B}%
^{2}\exp (-2\mathcal{G)+}2\sqrt{j(j+1)}\Phi _{+-}  \notag \\
&&\mathcal{+}\frac{A_{mm}}{r^{2}}+\frac{B_{mm}}{r}+C_{mm}+(\frac{F_{mm}}{r}%
+G_{mm})\frac{d}{dr}]\frac{u_{(j-1)1j}^{+}}{r}  \notag \\
&&+[\Phi _{-+}+\sqrt{j(j+1)}(2\Phi _{++}-2\mathcal{B}^{2}\exp (-2\mathcal{G)+%
}\frac{A_{mp}}{r^{2}}+\frac{B_{mp}}{r}+C_{mp}+(\frac{F_{mp}}{r}+G_{mp})\frac{%
d}{dr})]\frac{u_{(j+1)1j}^{+}}{r}\},
\end{eqnarray}%
and 
\begin{eqnarray}
&&\frac{u_{(j+1)1j}^{-}}{r}=\frac{M}{E}\frac{u_{(j+1)1j}^{+}}{r}-\frac{\exp
(2\mathcal{G})}{2EM\left( 2j+1\right) }\{[-\Phi _{++}-2j\mathcal{B}^{2}\exp
(-2\mathcal{G)}+2\sqrt{j(j+1)}\Phi _{-+}  \notag \\
&&\mathcal{+}\frac{A_{pp}}{r^{2}}+\frac{B_{pp}}{r}+C_{pp}+(\frac{F_{pp}}{r}%
+G_{pp})\frac{d}{dr}]\frac{u_{(j+1)1j}^{+}}{r}  \notag \\
+ &&[-\Phi _{+-}+\sqrt{j(j+1)}(2\Phi _{--}-2\mathcal{B}^{2}\exp (-2\mathcal{%
G)+}\frac{A_{pm}}{r^{2}}+\frac{B_{pm}}{r}+C_{pm}+(\frac{F_{pm}}{r}+G_{pm})%
\frac{d}{dr})]\frac{u_{(j-1)1j}^{+}}{r}\}.
\end{eqnarray}%
In Appendix D we also show that 
\begin{equation}
\mathcal{\eta }_{+0}=i\frac{\upsilon _{j0j}}{r}Y_{jm},  \label{nu}
\end{equation}%
and 
\begin{equation}
\boldsymbol{\eta }_{-}=i\frac{\upsilon _{j1j}^{-}}{r}\mathbf{X}_{jm},
\label{emi}
\end{equation}%
in which 
\begin{eqnarray}
\frac{\upsilon _{j0j}}{r} &=&\frac{\exp (\mathcal{G}+2K\mathcal{)}}{E}\{[%
\frac{(j-1)-2Q_{m}}{r}-(Q_{m}+1)\frac{d}{dr}+\frac{5}{2}L^{\prime }(Q_{m}+1)]%
\sqrt{\frac{j}{2j+1}}\frac{u_{(j-1)1j}^{+}}{r}  \notag \\
&&+[\frac{(j+2)+2Q_{m}}{r}+(Q_{m}+1)\frac{d}{dr}-\frac{5}{2}L^{\prime
}(Q_{m}+1)]\sqrt{\frac{j+1}{2j+1}}\frac{u_{(j+1)1j}^{+}}{r}\},
\end{eqnarray}%
and 
\begin{eqnarray}
\frac{\upsilon _{j1j}^{-}}{r} &=&-\frac{\exp (\mathcal{G)}}{M}\{[(\frac{d}{dr%
}-\frac{j(Q_{m}+1)-1}{r}-(3\mathcal{G+}\frac{L}{2})^{\prime })]\sqrt{\frac{%
j+1}{2j+1}}\frac{u_{(j-1)1j}^{+}}{r}  \notag \\
&&+[(\frac{d}{dr}+\frac{j(Q_{m}+1)+2+Q_{m}}{r}-(3\mathcal{G+}\frac{L}{2}%
)^{\prime })]\sqrt{\frac{j}{2j+1}}\frac{u_{(j+1)1j}^{+}}{r}.
\end{eqnarray}

We use these wave functions to compute composite $2\gamma $ decay amplitude
Eq.(\ref{dcmpq}) which after performing the $q$ space trace gives 
\begin{eqnarray}
\mathcal{M}_{X\rightarrow 2\gamma } &=&-\frac{e^{2}\sqrt{\pi }}{2}\int
d^{3}r\exp (F)Tr_{\sigma }\big(\exp (-i\mathbf{k\cdot r})\{i[\cosh K\mathcal{%
\psi }_{-}-\sinh K\boldsymbol{\sigma}\cdot \mathbf{\hat{r}}\mathcal{\psi }%
_{-}\boldsymbol{\sigma}\cdot \mathbf{\hat{r}]}\boldsymbol{\sigma}\cdot %
\boldsymbol{\epsilon }^{(\alpha _{1})}(\boldsymbol{\sigma}\mathbf{\cdot
\nabla )}\frac{\exp (-mr)}{r}  \notag \\
&&+[\cosh K\eta _{+}-\sinh K\boldsymbol{\sigma}\cdot \mathbf{\hat{r}}\eta
_{+}\boldsymbol{\sigma}\cdot \mathbf{\hat{r}]}\boldsymbol{\sigma}\cdot %
\boldsymbol{\epsilon }^{(\alpha _{1})}m\frac{\exp (-mr)}{r}\}%
\boldsymbol{\sigma}\cdot \boldsymbol{\epsilon }^{(\alpha _{2})}  \notag \\
&&+\exp (i\mathbf{k\cdot r)\{}i[\cosh K\mathcal{\psi }_{-}-\sinh K%
\boldsymbol{\sigma}\cdot \mathbf{\hat{r}}\mathcal{\psi }_{-}%
\boldsymbol{\sigma}\cdot \mathbf{\hat{r}]}\boldsymbol{\sigma}\cdot %
\boldsymbol{\epsilon }^{(\alpha _{2})}(\boldsymbol{\sigma}\mathbf{\cdot
\nabla )}\frac{\exp (-mr)}{r}  \notag \\
&&+[\cosh K\eta _{+}-\sinh K\boldsymbol{\sigma}\cdot \mathbf{\hat{r}}\eta
_{+}\boldsymbol{\sigma}\cdot \mathbf{\hat{r}]}\boldsymbol{\sigma}\cdot %
\boldsymbol{\epsilon }^{(\alpha _{2})}m\frac{\exp (-mr)}{r}\}%
\boldsymbol{\sigma}\cdot \boldsymbol{\epsilon }^{(\alpha _{1})}\big){\large .%
}  \label{abv}
\end{eqnarray}%
The trace eliminates the contribution of the portion $\mathcal{\psi }_{+}(%
\mathbf{r)}q_{1}~$of the wave function.

\subsubsection{\qquad Decay Amplitude for $^{1}L_{l}$ Composites}

Substituting Eq.(\ref{sing}) into (\ref{abv}), 
\begin{eqnarray}
\mathcal{M}_{^{1}L_{l}\rightarrow 2\gamma } &=&-\frac{e^{2}\sqrt{\pi }}{2}%
\int d^{3}r\exp (F)Tr_{\sigma }\big(\{i\exp (-K)\psi _{-0}  \notag \\
&&\times \mathbf{[}\exp (-i\mathbf{k\cdot r})\boldsymbol{\sigma}\mathbf{%
\cdot }\boldsymbol{\epsilon }^{(\alpha _{1})}(\boldsymbol{\sigma}\mathbf{%
\cdot \nabla )}\boldsymbol{\sigma}\mathbf{\cdot }\boldsymbol{\epsilon }%
^{(\alpha _{2})}+\exp (i\mathbf{k\cdot r)}\boldsymbol{\sigma}\mathbf{\cdot }%
\boldsymbol{\epsilon }^{(\alpha _{2})}(\boldsymbol{\sigma}\mathbf{\cdot
\nabla )}\boldsymbol{\sigma}\mathbf{\cdot }\boldsymbol{\epsilon }^{(\alpha
_{1})}]\}  \notag \\
&&+\{[\exp (K)\boldsymbol{\eta}\mathbf{_{+}\cdot }\boldsymbol{\sigma}-\sinh
(K)2\mathbf{\hat{r}\cdot }\boldsymbol{\eta}\mathbf{_{+}}\boldsymbol{\sigma}%
\mathbf{\cdot \hat{r}]}  \notag \\
&&\times \mathbf{[}\exp (-i\mathbf{k\cdot r})\boldsymbol{\sigma}\mathbf{%
\cdot }\boldsymbol{\epsilon }^{(\alpha _{1})}m\boldsymbol{\sigma}\mathbf{%
\cdot }\boldsymbol{\epsilon }+\exp (i\mathbf{k\cdot r)}\boldsymbol{\sigma}%
\mathbf{\cdot }\boldsymbol{\epsilon }^{(\alpha _{2})}m\boldsymbol{\sigma}%
\mathbf{\cdot }\boldsymbol{\epsilon }^{(\alpha _{1})}]\}\big)\frac{\exp (-mr)%
}{r},
\end{eqnarray}%
performing the remaining trace gives using (\ref{ls},\ref{jl}) gives 
\begin{eqnarray}
\mathcal{M}_{^{1}L_{l}\rightarrow 2\gamma } &=&-\sqrt{\pi }e^{2~}%
\boldsymbol{\epsilon }^{(\alpha _{1})}\times \boldsymbol{\epsilon }^{(\alpha
_{2})}\cdot \int d^{3}r\exp (F)[\exp (-i\mathbf{k\cdot r)-}\exp (+i\mathbf{%
\mathbf{k\cdot r)]}}  \notag \\
&&\times \mathbf{\{}\exp (-K)\frac{u_{j0j}^{-}}{r}Y_{jm}\mathbf{\hat{r}}%
\left( \frac{\exp (-mr)}{r}\right) ^{\prime }  \notag \\
&&+\{-\exp (K)(\frac{\upsilon _{(j-1)1j}^{+}}{r}\mathbf{Y}_{jm-}+\frac{%
\upsilon _{(j+1)1j}^{+}}{r}\mathbf{Y}_{jm+})  \notag \\
&&+2\sinh (K)Y_{jm}\mathbf{\hat{r}}(-\frac{\upsilon _{(j+1)1j}^{+}}{r}\sqrt{%
\frac{j+1}{2j+1}}+\frac{\upsilon _{(j-1)1j}^{+}}{r}\sqrt{\frac{j}{2j+1}})\}m%
\frac{\exp (-mr)}{r}],  \label{lm}
\end{eqnarray}%
with unit vectors defined in terms of the photon decay momenta and
transverse polarization vectors 
\begin{eqnarray}
\mathbf{\hat{z}}\mathbf{=\hat{k},} &&  \notag \\
\frac{\mathbf{(\hat{x}\pm }i\mathbf{\hat{y})}}{\sqrt{2}} &=&%
\boldsymbol{\epsilon }^{(\pm )}.  \label{xiy}
\end{eqnarray}%
The integral forms appearing in Eq. (\ref{lm}) are treated in Appendix E in
which we show (with $g(r)$ appropriately defined) 
\begin{eqnarray}
&&\int d^{3}r\exp (-i\mathbf{k\cdot r)\hat{r}}g(r)Y_{jm}(\mathbf{\Omega }) 
\notag \\
&=&4\pi \sum_{j^{\prime }=|j-1|}^{j+1}(-i)^{j^{\prime }}\sqrt{\frac{(2j+1)}{%
4\pi }}\langle j1;00|j^{\prime }0\rangle \int_{0}^{\infty
}drr^{2}j_{j^{\prime }}(kr)g(r)  \notag \\
&&\times \lbrack \mathbf{\hat{k}}\langle j1;00|j^{\prime }0\rangle \delta
_{m0}-\boldsymbol{\epsilon }^{(-)}\langle j1;-11|j^{\prime }0\rangle \delta
_{m-1}+\boldsymbol{\epsilon }^{(+)}\langle j1;1-11|j^{\prime }0\rangle
\delta _{m1}].  \label{i1}
\end{eqnarray}%
We also show (with $f_{\pm }(r)$ appropriately defined) 
\begin{eqnarray}
&&\int d^{3}r\exp (-i\mathbf{k\cdot r)}f_{\pm }(r)\mathbf{Y}_{jm\pm }(\Omega
)  \notag \\
&=&4\pi \mathbf{\hat{k}}\sqrt{\frac{(2j+1)}{4\pi }}\delta
_{m0}\int_{0}^{\infty }drr^{2}j_{j}(kr)krb_{\pm }f_{\pm }(r)  \notag \\
&&+4\pi \sqrt{\frac{(2j+1)}{4\pi }}\sum_{j^{\prime
}=|j-1|}^{j+1}(-i)^{j^{\prime }}\langle j1;00|j^{\prime }0\rangle
\int_{0}^{\infty }drr^{2}f_{\pm }(r)\mathbf{[(}a_{\pm }-2ib_{\pm
})j_{j^{\prime }}(kr)-ib_{\pm }j_{j^{\prime }}^{\prime }(kr)kr]  \notag \\
&&\times \lbrack \mathbf{\hat{k}}\langle j1;00|j^{\prime }0\rangle \delta
_{m0}-\boldsymbol{\epsilon }^{-}\langle j1;-11|j^{\prime }0\rangle \delta
_{m-1}+\boldsymbol{\epsilon }^{+}\langle j1;1-11|j^{\prime }0\rangle \delta
_{m1}],
\end{eqnarray}%
in which 
\begin{equation}
j_{j}^{\prime }(kr)=\frac{jj_{j-1}(kr)-(j+1)j_{j+1}(kr)}{2j+1}.  \label{jp}
\end{equation}%
Thus 
\begin{eqnarray}
\mathcal{M}_{^{1}L_{l}\rightarrow 2\gamma } &=&-\sqrt{2j+1}%
\boldsymbol{\epsilon }^{(\alpha _{1})}\times \boldsymbol{\epsilon }^{(\alpha
_{2})}\cdot \mathbf{\hat{k}\{}F_{j=l}(1+(-)^{j})\delta _{m0}  \label{samp} \\
&&+\sum_{j^{\prime }=\left\vert j-1\right\vert }^{j+1}G_{j=l}^{(j^{\prime
})}(1-(-)^{j^{\prime }})\frac{(1-(-1)^{j+j^{\prime }})}{2}\langle
j1;00|j^{\prime }0\rangle \langle j1;00|j^{\prime }0\rangle \delta _{m0}\}, 
\notag
\end{eqnarray}%
in which 
\begin{eqnarray}
F_{j=l} &=&-2i\pi e^{2}(-i)^{j}\int_{0}^{\infty }drmr\exp
(-mr)j_{j}(kr)kr\exp (F+K)  \notag \\
&&\times (\frac{1}{j+1}\sqrt{\frac{j+1}{2j+1}}\frac{\upsilon
_{(j+1)1j}^{+}(r)}{r}+\frac{1}{j}\sqrt{\frac{j}{2j+1}}\frac{\upsilon
_{(j-1)1j}^{+}(r)}{r})  \notag \\
G_{j=l}^{(j^{\prime })} &=&-2\pi e^{2}(-i)^{j^{\prime }}\int_{0}^{\infty
}dr\exp (-mr)\exp (F)\big(j_{j^{\prime }}(kr)\{(mr+1)\exp (-K)\frac{%
u_{j0j}^{-}}{r}  \notag \\
&&-2mr\sinh (K)(-\frac{\upsilon _{(j+1)1j}^{+}}{r}\sqrt{\frac{j+1}{2j+1}}+%
\frac{\upsilon _{(j-1)1j}^{+}}{r}\sqrt{\frac{j}{2j+1}})\}  \notag \\
&&+mr\{\mathbf{[(}-1+\frac{2}{j+1})j_{j^{\prime }}(kr)+\frac{1}{j+1}%
j_{j^{\prime }}^{\prime }(kr)kr]\sqrt{\frac{j+1}{2j+1}}\frac{\upsilon
_{(j+1)1j}^{+}(r)}{r}  \notag \\
&&+[\mathbf{(}1+\frac{2}{j})j_{j^{\prime }}(kr)+\frac{1}{j}j_{j^{\prime
}}^{\prime }(kr)kr]\sqrt{\frac{j}{2j+1}}\frac{\upsilon _{(j-1)1j}^{+}(r)}{r}%
\}\exp (K)\big){\large .}  \label{fj}
\end{eqnarray}%
Notice that this amplitude (\ref{samp}) is zero for $j$ odd consistent with
the Landau-Yang Theorem. \ We call this amplitude the two-body Dirac
amplitude (TBDA). What we call the naive amplitudes (NA) would correspond
the use of the naive norm ($\mathcal{K}=\mathcal{L}=1$) together with $\exp
(F),\exp (K)\rightarrow 1$ in Eq. (\ref{fj}).

\subsubsection{ \ \ Decay Amplitude for $^{3}L_{l\pm 1}$ Composites \ \ }

Using Eq. (\ref{trip}) in (\ref{abv}) leaves us with 
\begin{eqnarray}
\mathcal{M}_{^{3}L_{j=l\pm 1}\rightarrow 2\gamma } &=&-\frac{e^{2}}{2}\sqrt{%
\pi }\int d^{3}r\exp (F)Tr_{\sigma }\big(\exp (-i\mathbf{k\cdot r})\{[i\exp K%
\boldsymbol{\psi }_{-}\mathbf{\cdot }\boldsymbol{\sigma}  \notag \\
&&-2\sinh K\boldsymbol{\psi }\mathbf{_{-}\mathbf{\cdot }\hat{r}}%
\boldsymbol{\sigma}\cdot \mathbf{\hat{r}}\mathcal{]}\boldsymbol{\sigma}\cdot %
\boldsymbol{\epsilon }^{(\alpha _{1})}\boldsymbol{\sigma}\mathbf{\cdot
\nabla }+m\exp (-K)\mathcal{\eta }_{+0}\boldsymbol{\sigma}\cdot %
\boldsymbol{\epsilon }^{(\alpha _{1})}]\boldsymbol{\sigma}\cdot %
\boldsymbol{\epsilon }^{(\alpha _{2})}\}  \notag \\
&&+\exp (+i\mathbf{k\cdot r)}\{i[\exp K\boldsymbol{\psi }_{-}\mathbf{\cdot }%
\boldsymbol{\sigma}-2\sinh K\boldsymbol{\psi }\mathbf{_{-}\mathbf{\cdot }%
\hat{r}}\boldsymbol{\sigma}\cdot \mathbf{\hat{r}}\mathcal{]}%
\boldsymbol{\sigma}\cdot \boldsymbol{\epsilon }^{(\alpha _{2})}%
\boldsymbol{\sigma}\mathbf{\cdot \nabla }  \notag \\
&&+m\exp (-K)\mathcal{\eta }_{+0}\boldsymbol{\sigma}\cdot %
\boldsymbol{\epsilon }^{(\alpha _{2})}\}\boldsymbol{\sigma}\cdot %
\boldsymbol{\epsilon }^{(\alpha _{1})}\}\big)\frac{\exp (-mr)}{r}.
\end{eqnarray}%
Notice that only two of the four portions of the triplet wave function (\ref%
{trip}) survive that trace. \ Performing the $\sigma $ space trace and using
Eqs. (\ref{j}) and (\ref{nu}) \ together with%
\begin{equation}
\mathbf{\hat{r}\cdot \psi }_{-}=-\sqrt{\frac{j+1}{2j+1}}\frac{u_{(j+1)1j}^{-}%
}{r}Y_{jm}+\frac{u_{(j-1)1j}^{-}}{r}\sqrt{\frac{j}{2j+1}}Y_{jm},
\end{equation}%
we obtain 
\begin{eqnarray}
&&\mathcal{M}_{^{3}L_{j=l\pm 1}\rightarrow 2\gamma }  \notag \\
&=&-i\sqrt{\pi }e^{2}\int d^{3}r[\exp (-i\mathbf{k\cdot r)+}\exp (i\mathbf{%
k\cdot r)]}\exp (F){\biggl (}\left( \frac{\exp (-mr)}{r}\right) ^{\prime } 
\notag \\
&&\times \exp K\{\frac{u_{(j+1)1j}^{-}}{r}[\mathbf{Y}_{jm+}(\mathbf{\Omega }%
)\cdot \boldsymbol{\epsilon }^{(\alpha _{1})}\mathbf{\hat{r}\cdot }%
\boldsymbol{\epsilon }^{(\alpha _{2})}+\mathbf{Y}_{jm+}(\mathbf{\Omega }%
)\cdot \boldsymbol{\epsilon }^{(\alpha _{2})}\mathbf{\hat{r}\cdot }%
\boldsymbol{\epsilon }^{(\alpha _{1})}-\mathbf{Y}_{jm+}(\mathbf{\Omega }%
)\cdot \mathbf{\hat{r}}\boldsymbol{\epsilon }^{(\alpha _{1})}\mathbf{\cdot }%
\boldsymbol{\epsilon }^{(\alpha _{2})}]  \notag \\
&&+\frac{u_{(j-1)1j}^{-}}{r}[\mathbf{Y}_{jm-}(\mathbf{\Omega })\cdot %
\boldsymbol{\epsilon }^{(\alpha _{1})}\mathbf{\hat{r}\cdot }%
\boldsymbol{\epsilon }^{(\alpha _{2})}+\mathbf{Y}_{jm-}(\mathbf{\Omega }%
)\cdot \boldsymbol{\epsilon }^{(\alpha _{2})}\mathbf{\hat{r}\cdot }%
\boldsymbol{\epsilon }^{(\alpha _{1})}-\mathbf{Y}_{jm-}(\mathbf{\Omega }%
)\cdot \mathbf{\hat{r}}\boldsymbol{\epsilon }^{(\alpha _{1})}\mathbf{\cdot }%
\boldsymbol{\epsilon }^{(\alpha _{2})}]\}  \notag \\
&&-4\sinh K\left( \frac{\exp (-mr)}{r}\right) ^{\prime }[-\sqrt{\frac{j+1}{%
2j+1}}\frac{u_{(j+1)1j}^{-}}{r}+\frac{u_{(j-1)1j}^{-}}{r}\sqrt{\frac{j}{2j+1}%
}]Y_{jm}\mathbf{\hat{r}}\cdot \boldsymbol{ \epsilon }^{(\alpha _{1})}\mathbf{%
\hat{r}\cdot }\boldsymbol{ \epsilon }^{(\alpha _{2})}  \label{amp} \\
&&.+[m\exp (-K)\frac{\upsilon _{j0j}^{+}}{r}-2(m+1/r)\sinh K(-\sqrt{\frac{j+1%
}{2j+1}}\frac{u_{(j+1)1j}^{-}}{r}+\frac{u_{(j-1)1j}^{-}}{r}\sqrt{\frac{j}{%
2j+1}})Y_{jm}(\mathbf{\Omega })\boldsymbol{ \epsilon }^{(\alpha _{1})}\cdot %
\boldsymbol{\epsilon }^{(\alpha _{2})}\frac{\exp (-mr)}{r}{\biggr )}  \notag
\end{eqnarray}%
\ 

With Eq. (\ref{xiy}), the simplest terms in the above expression include
forms like 
\begin{eqnarray}
\int d^{3}r\exp (-i\mathbf{k\cdot r)}g(r)Y_{jm}(\mathbf{\Omega }) &=&4\pi
(-i)^{j}Y_{jm}(\mathbf{\Omega }_{k})\int_{0}^{\infty }drr^{2}j_{j}(kr)g(r) 
\notag \\
&\rightarrow &\sqrt{4\pi (2j+1)}(-i)^{j}\delta _{m0}\int_{0}^{\infty
}drr^{2}j_{j}(kr)g(r).
\end{eqnarray}%
Stepping up in complexity we have the transverse parts of the dyad form%
\begin{equation}
\int d^{3}r\exp (-i\mathbf{k\cdot r)}G(r)\mathbf{\hat{r}\hat{r}}.
\label{rye}
\end{equation}%
In Appendix E we show that transverse portion is%
\begin{eqnarray}
&&\mathbf{(1-\hat{k}\hat{k})\cdot }\int d^{3}r\exp (-i\mathbf{k\cdot r)}G(r)%
\mathbf{\hat{r}\hat{r}}\cdot \mathbf{(1-\hat{k}\hat{k})}  \notag \\
&=&(\boldsymbol{\epsilon }^{(+)}\boldsymbol{\epsilon }^{(-)}+%
\boldsymbol{\epsilon }^{(-)}\boldsymbol{\epsilon }^{(+)}\mathbf{)}\frac{1}{3}%
\sqrt{4\pi (2j+1)}\mathbf{\{}\delta _{m0}\int_{0}^{\infty
}r^{2}dr(-i)^{j}j_{j}(kr)G(r)  \notag \\
&&-\sum_{j^{\prime }=\left\vert j-2\right\vert }^{j+2}\frac{%
(1+(-1)^{j+j^{\prime }})}{2}\langle j2;00|j^{\prime }0\rangle \langle
j2;00|j^{\prime }0\rangle \delta _{m0}\int_{0}^{\infty
}r^{2}dr(-i)^{j^{\prime }}j_{j^{\prime }}(kr)G_{\pm }(r)  \notag \\
&&+\{(\boldsymbol{\epsilon }^{(+)}\boldsymbol{\epsilon }^{(+)}\mathbf{)}%
\sqrt{\frac{8\pi (2j+1)}{3}}\sum_{j^{\prime }=\left\vert j-2\right\vert
}^{j+2}\frac{(1+(-1)^{j+j^{\prime }})}{2}\langle j2;00|j^{\prime }0\rangle
\langle j2;-22|j^{\prime }0\rangle \delta _{m-2}  \notag \\
&&+(\boldsymbol{\epsilon }^{(-)}\boldsymbol{\epsilon }^{(-)}\mathbf{)}\sqrt{%
\frac{8\pi (2j+1)}{3}}\sum_{j^{\prime }=\left\vert j-2\right\vert }^{j+2}%
\frac{(1+(-1)^{j+j^{\prime }})}{2}\langle j2;00|j^{\prime }0\rangle \langle
j2;2-2|j^{\prime }0\rangle \delta _{m2}\}  \notag \\
&&\times \int_{0}^{\infty }r^{2}dr(-i)^{j^{\prime }}j_{j^{\prime }}(kr)G(r).
\end{eqnarray}%
Finally, we need the trace as well as transverse parts of the dyad forms 
\begin{eqnarray}
&&\int d^{3}r\exp (-i\mathbf{k\cdot r)}F_{\pm }(r)\mathbf{Y}_{jm\pm }(\Omega
)\mathbf{\hat{r}}  \notag \\
&=&\int d^{3}r\exp (-i\mathbf{k\cdot r)}[a_{\pm }F_{\pm }(r)\mathbf{\hat{r}}%
Y_{jm}\mathbf{+}b_{\pm }rF_{\pm }(r)\mathbf{p}Y_{jm})\mathbf{\hat{r}.}
\label{sye}
\end{eqnarray}%
In Appendix E we show that with Eq. (\ref{xiy}) the trace portion of Eq. (%
\ref{sye}) is 
\begin{eqnarray}
&&(\boldsymbol{\epsilon }^{(+)}\boldsymbol{\epsilon }^{(-)}+%
\boldsymbol{\epsilon }^{(-)}\boldsymbol{\epsilon }^{(+)})\int d^{3}r\exp (-i%
\mathbf{k\cdot r)}F_{\pm }(r)\mathbf{Y}_{jm\pm }(\Omega )\cdot \mathbf{\hat{r%
}}  \notag \\
&=&(\boldsymbol{\epsilon }^{(+)}\boldsymbol{\epsilon }^{(-)}+%
\boldsymbol{\epsilon }^{(-)}\boldsymbol{\epsilon }^{(+)})a_{\pm }\sqrt{4\pi
(2j+1)}(-i)^{j}\delta _{m0}\int_{0}^{\infty }r^{2}drF_{\pm }(r)j_{j}(kr),
\label{tr}
\end{eqnarray}%
while the\ transverse part is 
\begin{eqnarray}
&&\mathbf{(1-\hat{k}\hat{k})\cdot }\int d^{3}r\exp (-i\mathbf{k\cdot r)}%
F_{\pm }(r)\mathbf{Y}_{jm\pm }(\Omega )\mathbf{\hat{r}}\cdot \mathbf{(1-\hat{%
k}\hat{k})}  \notag \\
&=&(\boldsymbol{\epsilon }^{(+)}\boldsymbol{\epsilon }^{(-)}+%
\boldsymbol{\epsilon }^{(-)}\boldsymbol{\epsilon }^{(+)}\mathbf{)}\frac{1}{3}%
\sqrt{4\pi (2j+1)}\mathbf{\{}\delta _{m0}\int_{0}^{\infty
}r^{2}dr(-i)^{j}j_{j}(kr)[(a_{\pm }+3ib_{\pm })F_{\pm }(r)+ib_{\pm }rF_{\pm
}^{\prime }(r)]  \notag \\
&&-\sum_{j^{\prime }=\left\vert j-2\right\vert }^{j+2}\frac{%
(1+(-1)^{j+j^{\prime }})}{2}\langle j2;00|j^{\prime }0\rangle \langle
j2;00|j^{\prime }0\rangle \delta _{m0}\int_{0}^{\infty
}r^{2}dr(-i)^{j^{\prime }}j_{j^{\prime }}(kr)[a_{\pm }F_{\pm }(r)+ib_{\pm
}rF_{\pm }^{\prime }(r)]\}  \notag \\
&&+\{(\boldsymbol{\epsilon }^{(+)}\boldsymbol{\epsilon }^{(+)}\mathbf{)}%
\sqrt{\frac{8\pi (2j+1)}{3}}\sum_{j^{\prime }=\left\vert j-2\right\vert
}^{j+2}\frac{(1+(-1)^{j+j^{\prime }})}{2}\langle j2;00|j^{\prime }0\rangle
\langle j2;-22|j^{\prime }0\rangle \delta _{m-2}  \notag \\
&&+(\boldsymbol{\epsilon }^{(-)}\boldsymbol{\epsilon }^{(-)}\mathbf{)}\sqrt{%
\frac{8\pi (2j+1)}{3}}\sum_{j^{\prime }=\left\vert j-2\right\vert }^{j+2}%
\frac{(1+(-1)^{j+j^{\prime }})}{2}\langle j2;00|j^{\prime }0\rangle \langle
j2;2-2|j^{\prime }0\rangle \delta _{m2}\}  \notag \\
&&\times \int_{0}^{\infty }r^{2}dr(-i)^{j^{\prime }}j_{j^{\prime
}}(kr)[a_{\pm }F_{\pm }(r)+ib_{\pm }rF_{\pm }^{\prime }(r)].
\end{eqnarray}%
\ After integrations by parts, substitution of values of $a_{\pm },ib_{\pm
}~ $\ and combining with the other portions, we obtain\ 
\begin{eqnarray}
\mathcal{M}_{^{3}L_{j=l\pm 1}\rightarrow 2\gamma } &=&A_{j=l\pm 1}(1+(-)^{j})%
\sqrt{(2j+1)}\boldsymbol{\epsilon }^{(\alpha _{1})}\mathbf{\cdot }%
\boldsymbol{\epsilon }^{(\alpha _{2})}\delta _{m0}  \notag \\
&&+\sum_{j^{\prime }=\left\vert j-2\right\vert }^{j+2}\sqrt{(2j+1)}\langle
j2;00|j^{\prime }0\rangle \frac{(1+(-1)^{j+j^{\prime }})}{2}%
(1+(-)^{j^{\prime }})2B_{j=l\pm 1}^{(j^{\prime })}{\biggl (}%
\boldsymbol{\epsilon }^{(\alpha _{1})}\mathbf{\cdot }\boldsymbol{\epsilon }%
^{(\alpha _{2})}\langle j2;00|j^{\prime }0\rangle \delta _{m0}  \notag \\
&&-\sqrt{6}[\boldsymbol{\epsilon }^{(\alpha _{1})}\cdot (\boldsymbol{%
\epsilon }^{(+)}\boldsymbol{\epsilon }^{(+)}\mathbf{)}\cdot %
\boldsymbol{\epsilon }^{(\alpha _{2})}\langle j2;-22|j^{\prime }0\rangle
\delta _{m-2}  \notag \\
&&+\boldsymbol{\epsilon }^{(\alpha _{1})}\cdot (\boldsymbol{\epsilon }^{(-)}%
\boldsymbol{\epsilon }^{(-)}\mathbf{)}\cdot \boldsymbol{\epsilon }^{(\alpha
_{2})}\langle j2;2-2|j^{\prime }0\rangle \delta _{m2}]{\biggr ),}
\label{mpm}
\end{eqnarray}%
in which 
\begin{eqnarray}
A_{j=l\pm 1} &=&\frac{i2\pi e^{2}}{3}i^{j}\int_{0}^{\infty }dr\exp (-mr)\exp
(F)\big(\{-3j_{j}(kr)\{mr\exp (-K)\frac{\upsilon _{j0j}}{r}  \notag \\
&&-2(mr+1)\sinh K[-\sqrt{\frac{j+1}{2j+1}}\frac{u_{(j+1)1j}^{-}}{r}+\frac{%
u_{(j-1)1j}^{-}}{r}\sqrt{\frac{j}{2j+1}}]\}  \notag \\
&&+(mr+1)\exp K\{[(j_{j}(kr)+\frac{2}{j+1}j_{j}^{\prime }(kr)kr)\sqrt{\frac{%
j+1}{2j+1}}\frac{u_{(j+1)1j}^{-}}{r}  \notag \\
&&+(-j_{j}(kr)+\frac{2}{j}j_{j}^{\prime }(kr)kr)\sqrt{\frac{j}{2j+1}}\frac{%
u_{(j-1)1j}^{-}}{r}]\}  \notag \\
&&-4\sinh K(mr+1)[-\sqrt{\frac{j+1}{2j+1}}\frac{u_{(j+1)1j}^{-}}{r}+\frac{%
u_{(j-1)1j}^{-}}{r}\sqrt{\frac{j}{2j+1}}]j_{j}(kr)  \notag \\
B_{j=l\pm 1}^{(j^{\prime })} &=&-\frac{i2\pi e^{2}}{3}i^{j^{\prime
}}\int_{0}^{\infty }dr\exp (-mr)\exp (F)(mr+1)  \notag \\
&&\times \big(\exp K\{[(\frac{3}{j+1}-1)j_{j^{\prime }}(kr)+\frac{1}{j+1}%
j_{j^{\prime }}^{\prime }(kr)kr]\sqrt{\frac{j+1}{2j+1}}\frac{u_{(j+1)1j}^{-}%
}{r}  \notag \\
&&+[(\frac{3}{j}+1)j_{j^{\prime }}(kr)+\frac{1}{j}j_{j^{\prime }}^{\prime
}(kr)kr]\sqrt{\frac{j}{2j+1}}\frac{u_{(j-1)1j}^{-}}{r}\}  \label{abjpm} \\
&&-2\sinh K(mr+1)[-\sqrt{\frac{j+1}{2j+1}}\frac{u_{(j+1)1j}^{-}}{r}+\frac{%
u_{(j-1)1j}^{-}}{r}\sqrt{\frac{j}{2j+1}}]j_{j}(kr)\big)  \notag
\end{eqnarray}%
As in the singlet case we obtain zero amplitude (\ref{mpm}) for odd $j.~$We
also call these amplitudes the two-body Dirac amplitudes . Again, the
corresponding naive amplitudes would correspond the use of the naive norm ($%
\mathcal{K}=\mathcal{L}=1$) together with $\exp (F),\exp (K)\rightarrow 1$
in Eq. (\ref{abjpm}).

\subsection{ \ \ Decay Rates}

From the above two sets of amplitudes we construct the decay rates. \ In our
present case, we have 
\begin{eqnarray}
\varepsilon _{\gamma 1} &=&\varepsilon _{\gamma 2}=\frac{w}{2},  \notag \\
b &=&\left\vert \mathbf{p}_{\gamma }\right\vert =\frac{w}{2}.
\end{eqnarray}%
Also, we are not interested in the decay of a state with a definite magnetic
quantum number. \ Rather we are interested in the average over all $m$. The
Lagrangian that leads to the Feynman amplitude for the decay process is
Lorentz invariant. Consequently the amplitude and our bound state adaptation
conserves total $j,m$. This implies that we can sum over final states in an
unrestricted way that is most convenient, without picking only special
helicities that one expects to contribute. The details of the amplitude
should do this automatically. Using the general decay rate formula \cite%
{wein} we obtain 
\begin{eqnarray}
\Gamma (X &\rightarrow &2\gamma )=\frac{1}{2!}\frac{1}{(2j+1)w^{2}(2\pi )^{6}%
}\int d\Omega _{k}\frac{2\pi \varepsilon _{\gamma 1}\varepsilon _{\gamma 2}b%
}{w}\sum_{m,\boldsymbol{\epsilon }^{(\alpha _{1})},\boldsymbol{\epsilon }%
^{(\alpha _{2})}}\left\vert \mathcal{M}_{X\rightarrow 2\gamma }\right\vert
^{2}  \notag \\
&=&\frac{1}{(2j+1)16(2\pi )^{5}}\int d\Omega _{k}\sum_{m,\boldsymbol{
\epsilon }^{(\alpha _{1})},\boldsymbol{\epsilon }^{(\alpha _{2})}}\left\vert 
\mathcal{M}_{X\rightarrow 2\gamma }\right\vert ^{2},
\end{eqnarray}%
in which we carry out the initial state $m~$average and final state
polarization sum independently. For spin singlet states with amplitude (\ref%
{samp}) this becomes 
\begin{eqnarray}
\Gamma (^{1}L_{l} &\rightarrow &2\gamma )=\frac{1}{(2j+1)16(2\pi )^{5}}\int
d\Omega _{k}\sum_{m,\boldsymbol{\epsilon }^{(\alpha _{1})},%
\boldsymbol{\epsilon }^{(\alpha _{2})}}|\boldsymbol{\epsilon }^{(\alpha
_{1})}\times \boldsymbol{\epsilon }^{(\alpha _{2})}\cdot \mathbf{\hat{k}|}%
^{2}  \notag \\
&&\times (2j+1)|F_{j=l}(1+(-)^{j})\delta _{m0}+\sum_{j^{\prime }=\left\vert
j-1\right\vert }^{j+1}G_{j=l}^{(j^{\prime })}(1-(-)^{j^{\prime }})\frac{%
(1-(-1)^{j+j^{\prime }})}{2}\langle j1;00|j^{\prime }0\rangle \langle
j1;00|j^{\prime }0\rangle \delta _{m0}|^{2}  \notag \\
&=&\frac{1}{4(2\pi )^{4}}|F_{j=l}(1+(-)^{j})\delta _{m0}+\sum_{j^{\prime
}=\left\vert j-1\right\vert }^{j+1}G_{j=l}^{(j^{\prime })}(1-(-)^{j^{\prime
}})\frac{(1-(-1)^{j+j^{\prime }})}{2}\langle j1;00|j^{\prime }0\rangle
\langle j1;00|j^{\prime }0\rangle \delta _{m0}|^{2}.
\end{eqnarray}%
\ 

We have summed over the following four independent polarization combinations 
\begin{eqnarray}
\boldsymbol{\epsilon }^{(\alpha _{1})},\boldsymbol{\epsilon }^{(\alpha
_{2})} &=&\boldsymbol{\epsilon }^{(\pm )},\boldsymbol{\epsilon }^{(\pm )}, 
\notag \\
\boldsymbol{\epsilon }^{(\alpha _{1})}\cdot \mathbf{\hat{k}} &\mathbf{=}&%
\boldsymbol{\epsilon }^{(\alpha _{2})}\cdot \mathbf{\hat{k}=}0,
\end{eqnarray}%
with 
\begin{equation}
\sum_{\boldsymbol{\epsilon }^{(\alpha _{1})},\boldsymbol{\epsilon }^{(\alpha
_{2})}}|\boldsymbol{\epsilon }^{(\alpha _{1})}\times \boldsymbol{\epsilon }%
^{(\alpha _{2})}\cdot \mathbf{\hat{k}|}^{2}=\sum_{\boldsymbol{\epsilon }%
^{(\alpha _{1})},\boldsymbol{\epsilon }^{(\alpha _{2})}}[1-|%
\boldsymbol{\epsilon }^{(\alpha _{1})}\cdot (\boldsymbol{\epsilon }^{(\alpha
_{2})})^{\ast }|^{2}]=2.
\end{equation}%
\ Note that only the zero helicity states (corresponding to both photons
being either left or right handed polarized) 
\begin{eqnarray}
\boldsymbol{\epsilon }^{(\alpha _{1})},\boldsymbol{\epsilon }^{(\alpha
_{2})} &=&\frac{1}{\sqrt{2}}(\mathbf{\hat{x}}+i\mathbf{\hat{y}}),\frac{1}{%
\sqrt{2}}(\mathbf{\hat{x}}-i\mathbf{\hat{y}})\equiv \boldsymbol{\epsilon }%
^{(+)},\boldsymbol{\epsilon }^{(-)},  \notag \\
\boldsymbol{\epsilon }^{(\alpha _{1})},\boldsymbol{\epsilon }^{(\alpha
_{2})} &=&\frac{1}{\sqrt{2}}(\mathbf{\hat{x}}-i\mathbf{\hat{y}}),\frac{1}{%
\sqrt{2}}(\mathbf{\hat{x}}+i\mathbf{\hat{y}})\equiv \boldsymbol{\epsilon }%
^{(-)},\boldsymbol{\epsilon }^{(+)},  \label{sp0}
\end{eqnarray}%
give non-zero contributions to the rate factor $1-|\boldsymbol{\epsilon }%
^{(\alpha _{1})}\cdot (\boldsymbol{\epsilon }^{(\alpha _{2})})^{\ast
}|^{2}.~ $The total helicity $\pm 2$ \ states \ \ 
\begin{eqnarray}
\boldsymbol{\epsilon }^{(\alpha _{1})},\boldsymbol{\epsilon }^{(\alpha
_{2})} &=&\frac{1}{\sqrt{2}}(\mathbf{\hat{x}}+i\mathbf{\hat{y}}),\frac{1}{%
\sqrt{2}}(\mathbf{\hat{x}}+i\mathbf{\hat{y}})\equiv \boldsymbol{\epsilon }%
^{(+)},\boldsymbol{\epsilon }^{(+)},  \notag \\
\boldsymbol{\epsilon }^{(\alpha _{1})},\boldsymbol{\epsilon }^{(\alpha
_{2})} &=&\frac{1}{\sqrt{2}}(\mathbf{\hat{x}}-i\mathbf{\hat{y}}),\frac{1}{%
\sqrt{2}}(\mathbf{\hat{x}}+i\hat{y})\equiv \boldsymbol{\epsilon }^{(+)},%
\boldsymbol{\epsilon }^{(+)},  \label{sp2}
\end{eqnarray}%
give zero contribution. Performing the angular integration gives \cite{msh} 
\begin{equation}
\Gamma (^{1}S_{0}\rightarrow 2\gamma )=\frac{1}{(2\pi )^{4}}%
|F_{0}+G_{0}^{(1)}\langle 01;00|10\rangle |^{2}=\frac{1}{(2\pi )^{4}}%
|F_{0}+G_{0}^{(1)}|^{2},  \label{singd}
\end{equation}%
and 
\begin{eqnarray}
\Gamma (^{1}D_{2} &\rightarrow &2\gamma )=\frac{1}{4(2\pi )^{4}}%
|2F_{2}+\sum_{j^{\prime }=1,3}^{3}2G_{j=l}^{(j^{\prime })}(1-(-)^{j^{\prime
}})\langle j1;00|j^{\prime }0\rangle \langle j1;00|j^{\prime }0\rangle |^{2}
\notag \\
&=&\frac{1}{(2\pi )^{4}}|F_{2}+G_{2}^{(1)}\langle 21;00|10\rangle
^{2}+G_{2}^{(3)}\langle 21;00|30\rangle ^{2}|^{2}  \notag \\
&=&\frac{1}{(2\pi )^{4}}|F_{2}+\frac{2}{5}G_{2}^{(1)}+\frac{3}{5}%
G_{2}^{(3)}|^{2}.
\end{eqnarray}

Using Eq. (\ref{mpm}) for triplet states $^{3}L_{l\pm 1}$ our rate formula
is 
\begin{eqnarray}
\Gamma (^{3}L_{j=l\pm 1} &\rightarrow &2\gamma )=\frac{1}{8(2\pi )^{4}}%
\sum_{m,\boldsymbol{\epsilon }^{(\alpha _{1})},\boldsymbol{\epsilon }%
^{(\alpha _{2})}}{\bigg |}A_{j=l\pm 1}\boldsymbol{\epsilon }^{(\alpha _{1})}%
\mathbf{\ \cdot }\boldsymbol{\epsilon }^{(\alpha _{2})}(1+(-)^{j})\delta
_{m0}  \notag \\
&&+\sum_{j^{\prime }=\left\vert j-2\right\vert }^{j+2}(1+(-)^{j^{\prime }})%
\frac{(1+(-1)^{j+j^{\prime }})}{2}\langle j2;00|j^{\prime }0\rangle  \notag
\\
&&\times {\biggl (}B_{j=l\pm 1}^{(j^{\prime })}2[\boldsymbol{\epsilon }%
^{(\alpha _{1})}\mathbf{\cdot }\boldsymbol{\epsilon }^{(\alpha _{2})}\langle
j2;00|j^{\prime }0\rangle \delta _{m0}  \notag \\
&&-\sqrt{6}[\boldsymbol{\epsilon }^{(\alpha _{1})}\cdot (\boldsymbol{%
\epsilon }^{(+)}\boldsymbol{\epsilon }^{(+)}\mathbf{)}\cdot %
\boldsymbol{\epsilon }^{(\alpha _{2})}\langle j2;-22|j^{\prime }0\rangle
\delta _{m-2}  \notag \\
&&+\boldsymbol{\epsilon }^{(\alpha _{1})}\cdot (\boldsymbol{\epsilon }^{(-)}%
\boldsymbol{\epsilon }^{(-)}\mathbf{)}\cdot \boldsymbol{\epsilon }^{(\alpha
_{2})}\langle j2;2-2|j^{\prime }0\rangle \delta _{m2}]{\biggr )\bigg |}^{2}.
\end{eqnarray}%
Notice from Eqs. (\ref{sp0}) and (\ref{sp2}) that this rate in general
includes both helicity zero and helicity two contributions.

In the case of $^{3}P_{0}$ decay we have $j=m=0$ and so performing the
polarization sum gives \cite{msh} 
\begin{eqnarray}
\ \Gamma (^{3}P_{0} &\rightarrow &2\gamma )=\frac{1}{2(2\pi )^{4}}\sum_{%
\boldsymbol{\epsilon }^{(\alpha _{1})},\boldsymbol{\epsilon }^{(\alpha
_{2})}}|A_{0}+2B_{0}^{(2)}\langle 02;00|20\rangle ^{2}|^{2}|%
\boldsymbol{\epsilon }^{(\alpha _{1})}\mathbf{\cdot }\boldsymbol{\epsilon }%
^{(\alpha _{2})}|^{2}  \notag \\
&=&\frac{1}{(2\pi )^{4}}|A_{0}+2B_{0}^{(2)}|^{2}.  \label{tpo}
\end{eqnarray}%
This rate includes only helicity zero contributions.

In the case of $^{3}P_{2}$ decay we have \cite{msh} 
\begin{eqnarray}
\Gamma (^{3}P_{2} &\rightarrow &2\gamma )=\frac{1}{2(2\pi )^{4}}\sum_{m,%
\boldsymbol{\epsilon }^{(\alpha _{1})},\boldsymbol{\epsilon }^{(\alpha
_{2})}}{\bigg |}A_{2}\boldsymbol{\epsilon }^{(\alpha _{1})}\mathbf{\cdot }%
\boldsymbol{\epsilon }^{(\alpha _{2})}\delta _{m0}  \notag \\
&&+\sum_{j^{\prime }=0,2,4}\langle 22;00|j^{\prime }0\rangle
2B_{2}^{(j^{\prime })}{\biggl (}\boldsymbol{\epsilon }^{(\alpha _{1})}%
\mathbf{\cdot }\boldsymbol{\epsilon }^{(\alpha _{2})}\langle 22;00|j^{\prime
}0\rangle \delta _{m0}  \notag \\
&&-\sqrt{6}[\boldsymbol{\epsilon }^{(\alpha _{1})}\cdot (\boldsymbol{%
\epsilon }^{(+)}\boldsymbol{\epsilon }^{(+)}\mathbf{)}\cdot %
\boldsymbol{\epsilon }^{(\alpha _{2})}\langle 22;-22|j^{\prime }0\rangle
\delta _{m-2}  \notag \\
&&+\boldsymbol{\epsilon }^{(\alpha _{1})}\cdot (\boldsymbol{\epsilon }^{(-)}%
\boldsymbol{\epsilon }^{(-)}\mathbf{)}\cdot \boldsymbol{\epsilon }^{(\alpha
_{2})}\langle 22;2-2|j^{\prime }0\rangle \delta _{m2}]]{\biggr )\bigg |}^{2}
\notag \\
&=&\frac{1}{(2\pi )^{4}}{\biggl [}|A_{2}-2[B_{2}^{(0)}\langle
22;00|00\rangle ^{2}+B_{2}^{(2)}\langle 22;00|20\rangle
^{2}+B_{2}^{(4)}\langle 22;00|40\rangle ^{2}]|^{2}  \notag \\
&&+12|[B_{2}^{(0)}\langle 22;00|00\rangle \langle 22;-22|00\rangle
+B_{2}^{(2)}\langle 22;00|20\rangle \langle 22;-22|20\rangle
+B_{2}^{(4)}\langle 22;00|40\rangle \langle 22;-22|40\rangle ]|^{2}  \notag
\\
&&+12|[B_{2}^{(0)}\langle 22;00|00\rangle \langle 22;2-2|00\rangle
+B_{2}^{(2)}\langle 22;00|20\rangle \langle 22;2-2|20\rangle
+B_{2}^{(4)}\langle 22;00|40\rangle \langle 22;2-2|40\rangle ]|^{2}{\biggr ]}
\notag \\
&=&\frac{1}{(2\pi )^{4}}{\biggl [}|A_{2}+\frac{2B_{2}^{(0)}}{5}+\frac{%
2B_{2}^{(2)}}{7}+\frac{36B_{2}^{(4)}}{35}|^{2}+24|\frac{B_{2}^{(0)}}{5}-%
\frac{2B_{2}^{(2)}}{7}+\frac{3B_{2}^{(4)}}{35}|^{2}{\biggr ].}  \label{tpt}
\end{eqnarray}

\subsubsection{ Positronium Decays}

For these decays we ignore the effects of the potentials on the norms and
amplitudes since they are relatively weak ($\mathcal{K}=\mathcal{L}=1$ or $%
\exp (F),~\exp (K)\rightarrow 1)$.

\paragraph{$^{1}S_{0}$ Decay}

The amplitude for $^{1}S_{0}$ positronium decay is from Eq. (\ref{samp}) 
\begin{equation}
\mathcal{F}_{^{1}S_{0}}=(F_{0}+G_{0}^{(1)}),
\end{equation}%
where for the weak potentials we expect in QED (with $\exp (F),~\exp
(K)\rightarrow 1$) 
\begin{eqnarray}
F_{0} &=&-2i\pi e^{2}\int_{0}^{\infty }drmr\exp (-mr)j_{0}(kr)kr\frac{%
\upsilon _{110}^{+}(r)}{r} \\
G_{0}^{(1)} &=&i2\pi e^{2}\int_{0}^{\infty }dr\exp (-mr){\biggl (}%
(mr+1)[j_{1}(kr)\frac{u_{000}^{-}(r)}{r}+mr\mathbf{[}j_{1}(kr)-j_{0}^{\prime
}(kr)kr]\frac{\upsilon _{110}^{+}(r)}{r}{\biggr ),}  \notag
\end{eqnarray}%
where we use%
\begin{equation}
\frac{u_{000}^{-}}{r}=\frac{M}{E}\frac{u_{000}^{+}}{r}=\frac{m}{E}\frac{%
u_{000}^{+}}{r}=\frac{2m}{w\sqrt{1+2\alpha /(wr)}}\frac{u_{000}^{+}}{r},
\end{equation}%
and%
\begin{equation}
\frac{\upsilon _{110}^{+}}{r}=\frac{\exp (\mathcal{G})}{E}(\frac{d}{dr}+%
\frac{L^{\prime }}{2})\frac{u_{000}^{+}}{r}=\frac{\exp (\mathcal{G})}{E}%
\frac{d}{dr}\frac{u_{000}^{+}}{r}=\frac{2}{w\left( 1+2\alpha /(wr)\right) }%
\frac{d}{dr}\frac{u_{000}^{+}}{r}.
\end{equation}%
This wave function is one of the small component ones. For positronium, $%
w=2m+O(\alpha ^{2})$ and so 
\begin{equation*}
\frac{M}{E}=\sqrt{\frac{mr}{mr+\alpha }}(1+O(\alpha ^{2}))
\end{equation*}%
and (with $k=m(1+O(\alpha ^{2})$) 
\begin{eqnarray}
F_{0} &=&-2i\pi e^{2}\int_{0}^{\infty }drr^{2}m\frac{\exp (-mr)}{r}%
j_{1}(mr)mr(\frac{r}{mr+\alpha })\frac{d}{dr}\psi _{000})  \notag \\
G_{0}^{(0)} &=&-2\pi e^{2}\int_{0}^{\infty }dr\exp (-mr)\big((mr+1)[j_{1}(mr)%
\sqrt{\frac{mr}{mr+\alpha }}\psi _{000}  \notag \\
&&+mr\mathbf{[}j_{1}(kr)+j_{0}^{\prime }(kr)kr]\frac{1}{m}\frac{d}{dr}\psi
_{000}\big)
\end{eqnarray}%
with the nonrelativistic wave function given by%
\begin{equation}
\psi _{000}=\frac{{(}m\alpha {)^{3/2}}}{\sqrt{8\pi }}\exp (-\alpha mr)=\frac{%
R(r)}{\sqrt{4\pi }}
\end{equation}%
replacing the relativistic one $u_{000}^{+}/r$. \ \ The NN (\ref{nrm})
becomes 
\begin{equation}
\frac{1}{2}\int_{0}^{\infty }drr^{2}[\left( \frac{u_{000}^{+}}{r}\right)
^{2}+\left( \frac{u_{000}^{-}}{r}\right) ^{2}+\left( \frac{\upsilon
_{110}^{+}}{r}\right) ^{2}]=1.
\end{equation}%
In Appendix F we obtain the well known form for the decay rate: 
\begin{equation}
\Gamma =\frac{\left\vert G_{0}^{(1)}\right\vert ^{2}}{(2\pi )^{4}}%
=\left\vert R(0)\right\vert ^{2}\frac{\alpha ^{2}}{m^{2}}=\frac{m\alpha ^{5}%
}{2}.
\end{equation}%
and show that the small component portion $F_{0}$ does not contribute to the
singlet decay rate at this order.

\paragraph{$^{3}P_{0,2}$ Decay}

The branching ratio for these decay have not been measured since the decays
of those states is so largely dominated by the dipole transition to the $%
^{3}S_{1}$ state. \ Nevertheless, it will be of value to determine if our
covariant formalism yields the standard results given in \cite{alx} and \cite%
{beri}. The relevent amplitudes given in Eq. (\ref{abjpm}) for weak
potentials ($\mathcal{L},\mathcal{K}=1$ or $\exp (F)$, $\exp (K)\rightarrow 1
$) are 
\begin{eqnarray}
A_{j=l\pm 1} &=&i\frac{2\pi e^{2}}{3}(-i)^{j}\int_{0}^{\infty }dr\exp (-mr){%
\biggl (}(mr+1)  \notag \\
&&\times \{[j_{j}(kr)+\frac{2}{(j+1)}j_{j}^{\prime }(kr)kr]\sqrt{\frac{j+1}{%
2j+1}}\frac{u_{(j+1)1j}^{-}(r)}{r}+[-j_{j}(kr)+\frac{2}{j}j_{j}^{\prime
}(kr)kr]\sqrt{\frac{j}{2j+1}}\frac{u_{(j-1)1j}^{-}(r)}{r}\}  \notag \\
&&-3j_{j}(kr)mr\frac{\upsilon _{j0j}^{+}(r)}{r}{\biggr ),}  \notag \\
B_{j=l\pm 1}^{(j^{\prime })} &=&i\frac{2\pi e^{2}}{3}(-i)^{j^{\prime
}}\int_{0}^{\infty }dr\exp (-mr)(mr+1)  \notag \\
&&\times \{[(-1+\frac{3}{j+1})j_{j^{\prime }}(kr)+\frac{1}{j+1}j_{j^{\prime
}}^{\prime }(kr)kr]\sqrt{\frac{j+1}{2j+1}}\frac{u_{(j+1)1j}^{-}(r)}{r} 
\notag \\
&&+[(1+\frac{3}{j})j_{j^{\prime }}(kr)+\frac{1}{j}j_{j^{\prime }}^{\prime
}(kr)kr]\sqrt{\frac{j}{2j+1}}\frac{u_{(j-1)1j}^{-}(r)}{r}\}.
\end{eqnarray}%
The connection between the wave functions $u_{(j\pm 1)1j}^{-}$ and $u_{(j\pm
1)1j}^{+}$ (see Eqs. (\ref{j}) and (\ref{ujp}) appears complicated, but
specializing as in the singlet case, we find that the terms beyond the first
include higher order $\alpha $ terms from the various potential. \ 

For the nonrelativistic wave functions we have%
\begin{equation*}
\frac{u_{(j\pm 1)1j}^{+}}{r}=R_{(j\pm 1)1j}(r)=r^{j\pm 1}\chi _{(j\pm
1)1j}(r).
\end{equation*}%
We also need the small component wave function%
\begin{eqnarray}
\frac{\upsilon _{j0j}}{r} &=&\frac{\exp (3\mathcal{G)}}{m}\{[\frac{%
(j-1)-2Q_{m}}{r}-(Q_{m}+1)\frac{d}{dr}]\sqrt{\frac{j}{2j+1}}\frac{%
u_{(j-1)1j}^{+}}{r}  \notag \\
&&+[\frac{(j+2)+2Q_{m}}{r}+(Q_{m}+1)\frac{d}{dr}]\sqrt{\frac{j+1}{2j+1}}%
\frac{u_{(j+1)1j}^{+}}{r}\},
\end{eqnarray}

For the $^{3}P_{0}$ state we have \ 
\begin{eqnarray}
A_{0} &=&\frac{i2\pi e^{2}}{3}\int_{0}^{\infty }dr\exp (-mr)\{-3mrj_{0}(kr)%
\frac{\upsilon _{000}^{+}}{r}+(mr+1)(j_{0}(kr)+2j_{0}^{\prime }(kr)kr)\frac{%
u_{110}^{-}}{r}\},  \notag \\
B_{0}^{(2)} &=&\frac{i2\pi e^{2}}{3}\int_{0}^{\infty }dr\exp
(-mr)(mr+1)(2j_{2}(kr)+j_{2}^{\prime }(kr)kr)\frac{u_{110}^{-}}{r}.
\end{eqnarray}%
and the decay rate (\ref{tpo}) involves the amplitude combination%
\begin{eqnarray}
\mathcal{F}_{^{3}P_{0}} &=&(A_{0}+2B_{0}^{(2)})  \notag \\
&=&\frac{2\pi i}{3}\int_{0}^{\infty }dr\exp (-mr)\{(mr+1)  \notag \\
&&\times \lbrack j_{0}(kr)+2j_{0}^{\prime }(kr)kr+4j_{2}(kr)+2j_{2}^{\prime
}(kr)kr]\frac{u_{110}^{-}(r)}{r}-3j_{0}(kr)mr\frac{\upsilon _{000}^{+}(r)}{r}%
\},  \label{b0}
\end{eqnarray}%
in which (from Appendix D we find for this state that $\Phi _{++}$ cancels
with the remaining portions of $\mathcal{J}$) 
\begin{equation}
\frac{u_{110}^{-}}{r}=\frac{M}{E}\frac{u_{110}^{+}}{r}.
\end{equation}%
\ We also have 
\begin{equation}
\frac{\upsilon _{000}^{+}}{r}=\frac{\exp (\mathcal{G)}}{E}[\frac{2}{r}+\frac{%
d}{dr}]\frac{u_{110}^{+}}{r},
\end{equation}%
and%
\begin{equation}
\frac{d}{dr}R_{110}(r)|_{r=0}=\frac{d}{dr}r\chi _{110}(r)|_{r=0}=\chi
_{110}(0).
\end{equation}%
The NN condition (\ref{nrm}) is 
\begin{equation}
\frac{1}{2}\int_{0}^{\infty }drr^{2}[\left( \frac{u_{110}^{+}}{r}\right)
^{2}+\left( \frac{u_{110}^{-}}{r}\right) ^{2}+\left( \frac{\upsilon
_{000}^{+}}{r}\right) ^{2}]=1.
\end{equation}%
Our multicomponent results uses these relations in Eq. (\ref{b0}). \ \ In
Appendix F we present the details that allows us to obtain the result of%
\begin{equation}
\Gamma (^{3}P_{0}\rightarrow 2\gamma )=\frac{3m\alpha ^{7}}{256},
\end{equation}%
We point out there that in the limit in which the variation of the
positronium wave function is neglected (the nonrelativistic approximation
and single component result) we obtain vanishing amplitude in the $^{3}P_{0}$
$\ $\ case. \ As stressed in \cite{beri} the inclusion of the small
components of the wave functions is essential for this decay.

For the $^{3}P_{2}$ amplitude $j=2,$ $l=j-1=1,~j^{\prime }=0,2,4.$ The
relevant amplitudes are (ignoring angular momentum coupling)%
\begin{equation}
A_{2}=-i\frac{2\pi e^{2}}{3}\int_{0}^{\infty }dr\exp (-mr)\big(%
(mr+1)(-j_{2}(kr)+j_{2}^{\prime }(kr)kr)\sqrt{\frac{2}{5}}\frac{%
u_{112}^{-}(r)}{r}-3j_{2}(kr)mr\frac{\upsilon _{202}^{+}(r)}{r}\big),
\end{equation}%
and%
\begin{eqnarray}
B_{2}^{(0)} &=&-i\frac{2\pi e^{2}}{3}\int_{0}^{\infty }dr\exp (-mr)(mr+1) 
\notag \\
&&\times \{[\frac{5}{2}j_{0}(kr)+\frac{1}{2}j_{0}^{\prime }(kr)kr]\sqrt{%
\frac{2}{5}}\frac{u_{112}^{-}(r)}{r}\},  \notag \\
B_{2}^{(2)} &=&+i\frac{2\pi e^{2}}{3}\int_{0}^{\infty }dr\exp (-mr)(mr+1) 
\notag \\
&&\times \{[\frac{5}{2}j_{2}(kr)+\frac{1}{2}j_{2}^{\prime }(kr)kr]\sqrt{%
\frac{2}{5}}\frac{u_{112}^{-}(r)}{r}\},  \notag \\
B_{2}^{(4)} &=&-i\frac{2\pi e^{2}}{3}\int_{0}^{\infty }dr\exp (-mr)(mr+1) 
\notag \\
&&\times \{[\frac{5}{2}j_{4}(kr)+\frac{1}{2}j_{4}^{\prime }(kr)kr]\sqrt{%
\frac{2}{5}}\frac{u_{112}^{-}(r)}{r}\},
\end{eqnarray}%
with the neglect of orbital mixing where 
\begin{eqnarray}
\frac{u_{112}^{-}}{r} &=&\sqrt{\frac{mr}{mr+\alpha }}\frac{u_{112}^{+}}{r}-%
\frac{12}{5m^{2}}\sqrt{\left( \frac{mr}{mr+\alpha }\right) ^{3}}  \notag \\
&&\times \{[-\frac{1}{r^{2}}(\sqrt{\frac{mr+\alpha }{mr}}-1)+\frac{1}{r}(%
\sqrt{\frac{mr}{mr+\alpha }}-\sqrt{\frac{mr+\alpha }{mr}})\frac{d}{dr}]\frac{%
u_{112}^{+}}{r}\},  \notag \\
\frac{\upsilon _{202}}{r} &=&\frac{1}{m}\sqrt{\left( \frac{mr}{mr+\alpha }%
\right) ^{3}}[\frac{3-2\sqrt{\frac{mr+\alpha }{mr}}}{r}-\sqrt{\frac{%
mr+\alpha }{mr}}\frac{d}{dr}]\sqrt{\frac{2}{5}}\frac{u_{112}^{+}}{r}
\end{eqnarray}%
Using the above expressions for $\upsilon _{202}^{+}$ and $u_{112}^{-}(r)$
with%
\begin{equation}
\frac{u_{112}^{+}}{r}=R_{112}(r)=r\chi _{112}(r),
\end{equation}%
and Eq. (\ref{tpt}) leads to (see Appendix F)%
\begin{equation}
\Gamma (^{3}P_{2}\rightarrow 2\gamma )=\frac{m\alpha ^{7}}{320},
\end{equation}%
and the ratio $\Gamma (^{3}P_{0}\rightarrow 2\gamma )/\Gamma
(^{3}P_{2}\rightarrow 2\gamma )=\frac{15}{4}.$

Even though our approach leads to the earlier results it is of interest to
see how our constraint formalism based approach differs from other
approaches. We first note that in the constraint approach, the general frame
form of the CM amplitude of Eqs. (\ref{3m}) and (\ref{gamp}) is 
\begin{equation}
\int d^{4}pTr\mathbf{\Gamma (}p_{-},p_{+};k_{1},k_{2}\mathbf{)}\delta
(p\cdot \hat{P})\Psi \mathbf{(}p\mathbf{\mathbf{),}}  \label{98}
\end{equation}%
in which 
\begin{equation}
\Gamma (p_{-},p_{+};k_{1},k_{2})\mathbf{=}e^{2}[\boldsymbol{\gamma}\mathbf{%
\cdot }\boldsymbol{\epsilon }^{(\alpha _{1})}\frac{m-\gamma \cdot
(p_{-}-k_{1})}{(p_{-}-k_{1})^{2}+m^{2}}\boldsymbol{\gamma}\mathbf{\cdot }%
\boldsymbol{\epsilon }^{(\alpha _{2})}+\boldsymbol{\gamma}\mathbf{\cdot }%
\boldsymbol{\epsilon }^{(\alpha _{2})}\frac{m-\gamma \cdot (p_{-}-k_{2})}{%
(p_{-}-k_{2})^{2}+m^{2}}\boldsymbol{\gamma}\mathbf{\cdot }%
\boldsymbol{\epsilon }^{(\alpha _{1})}].
\end{equation}%
In the constraint approach, from Eqs. (\ref{tp}) and (\ref{rp}) 
\begin{eqnarray}
p_{-} &=&\frac{\hat{P}}{2}+p,  \notag \\
p_{+} &=&\frac{\hat{P}}{2}-p.
\end{eqnarray}%
The CM form is seen to follow directly from this since there we have $p=(0,%
\mathbf{p)}$ and 
\begin{eqnarray}
p_{-} &=&(\frac{w}{2},\mathbf{p)~};p_{+}=(\frac{w}{2},-\mathbf{p),}  \notag
\\
k_{1} &=&(\frac{w}{2},\mathbf{k)~};k_{2}=(\frac{w}{2},-\mathbf{k).}
\label{msl}
\end{eqnarray}%
This interpretation of the amplitude follows directly from the constraint
formalism and is distinct from that used in other approaches which assume an
on shell form for the amplitude (see e.g. \cite{ack} which uses $p_{i}^{0}=%
\sqrt{m^{2}+\mathbf{p}_{i}^{2}}$). \ The amplitude we use incorporates an
off-mass-shell assumption which is true for constituent particles of the
bound state. \ The\ constraint modification of the off-mass-shell amplitude
in addition places it on energy shell. \ This gives us the Yukawa
modification seen in Eq. (\ref{dcmp}) not appearing in other approaches. \ 

\subsubsection{Meson Decays}

\paragraph{$\protect\eta _{c},\protect\eta _{c^{\prime }\text{ }}$Decays}

For the $\eta _{c}$ the state \ vector is 
\begin{equation}
|\eta _{c}\rangle =\frac{1}{\sqrt{3}}\sum_{r,g,b}|c\bar{c}\rangle ,
\end{equation}%
with the charge of the charmed quark equal to $2e/3$. Since the interaction
is color independent the resultant amplitude is 
\begin{eqnarray}
\mathcal{F}_{\eta _{c}} &=&\frac{4\sqrt{3}}{9}(F_{0}+G_{0}^{(1)})  \notag \\
&=&\frac{4\sqrt{3}e^{2}}{9}2\pi i\int_{0}^{\infty }dr\exp (-mr)\big(%
j_{1}(kr)\exp (F-K)\frac{u_{000}^{-}}{r}](1+mr)  \notag \\
&&+mr\frac{\upsilon _{110}^{+}}{r}\{j_{1}(kr)\exp (F)[\exp (K)+2\sinh (K)] 
\notag \\
&&+kr\exp (F+K)[j_{1}^{\prime }(kr)-j_{0}(kr)]\}\big).  \label{seta}
\end{eqnarray}%
In (\ref{seta}) we take numerical wave functions from the work of \cite%
{crater2}. The remaining parts of our multicomponent wave functions are \ 
\begin{equation}
\frac{u_{000}^{-}}{r}=\frac{M}{E}\frac{u_{000}^{+}}{r}
\end{equation}%
and%
\begin{equation}
\frac{\upsilon _{110}^{+}}{r}=\frac{\exp (\mathcal{G})}{E}(\frac{d}{dr}+%
\frac{L^{\prime }}{2})\frac{u_{000}^{+}}{r},
\end{equation}%
which appear in that equation satisfy the TBDN condition Eq.(\ref{sinrm}).
In the spin singlet state this is (see Appendix C)%
\begin{eqnarray}
&&\frac{1}{2}\int_{0}^{\infty }drr^{2}\exp (2F)\big(\exp (-2K)[\left( \frac{%
u_{000}^{+}}{r}\right) ^{2}+\left( \frac{u_{000}^{-}}{r}\right) ^{2}+\left( 
\frac{\upsilon _{110}^{+}}{r}\right) ^{2}]  \notag \\
&&+2w^{2}\frac{\partial L}{\partial w^{2}}\exp (-2K)[\left( \frac{u_{000}^{+}%
}{r}\right) ^{2}-\left( \frac{u_{000}^{-}}{r}\right) ^{2}-\left( \frac{%
\upsilon _{110}^{+}}{r}\right) ^{2}]  \notag \\
&&+4w^{2}\frac{\partial \mathcal{G}}{\partial w^{2}}\exp (-2K)[2\left( \frac{%
u_{000}^{+}}{r}\right) ^{2}+\left( \frac{u_{000}^{-}}{r}\right) ^{2}]\big)=1.
\label{etanrm}
\end{eqnarray}%
The NN norm condition is 
\begin{equation}
\frac{1}{2}\int_{0}^{\infty }drr^{2}[\left( \frac{u_{000}^{+}}{r}\right)
^{2}+\left( \frac{u_{000}^{-}}{r}\right) ^{2}+\left( \frac{\upsilon
_{110}^{+}}{r}\right) ^{2}]=1,  \label{wetanrm}
\end{equation}%
In that same limit our NA (\ref{seta}) becomes%
\begin{equation}
\mathcal{F}_{\eta _{c}}=\frac{4\sqrt{3}e^{2}}{9}2\pi i\int_{0}^{\infty
}dr\exp (-mr)\big(j_{1}(kr)\frac{u_{000}^{-}}{r}(1+mr)+mr\frac{\upsilon
_{110}^{+}}{r}\{j_{1}(kr)+kr[j_{1}^{\prime }(kr)-j_{0}(kr)]\}\big){\large .}
\label{weta}
\end{equation}%
The multicomponent forms given by the TBDA and TBDN in (\ref{seta}) and (\ref%
{etanrm}) respectively give a decay rate of 9.18 \textrm{keV, }while that
obtained from the corresponding NA and NN forms (\ref{weta}), and (\ref%
{wetanrm}) is 9.15 \textrm{keV}. If we further ignore the small components
in these latter forms by taking $u_{000}^{-}=$\ $u_{000}^{+}$ and $\upsilon
_{110}^{+}=0,$ then the decay rate is 9.09 \textrm{keV. }These are to be
compared with the observed rate of 7.44 $\pm $1.0 \ \textrm{keV}. Including
first order QCD radiative corrections \cite{rsnr} damps these decay rates by
a factor of $(1+\alpha _{s}/\pi (\pi ^{2}/3-20/3))$ giving us 6.20 and 6.18 $%
\mathrm{keV.(}$and 6.14 \textrm{keV }when ignoring small components). \ For
the $\eta _{c}^{\prime }$ our results are 4.81 and 2.79 \textrm{keV} (and
2.68 \textrm{keV) }respectively compared with the observed rate of 1.3$\pm $%
.6 \ \textrm{keV}. The QCD radiative corrections reduce these to 3.36 and
1.95 \textrm{keV (}and 1.87\textrm{\ keV). }The overall additional effects
of using the TBDN and TBDA above that of the NN and NA appear to be very
small for the $\eta _{c}$ but for the $\eta _{c}^{\prime }$ they are
substantial ( but in the wrong direction!). \ It is of interest to trace the
origin of these contrasting behaviors. \ The square root of the norm
(starting with a normed $u_{000}^{+}/r$)$~$for the $\eta _{c}$ in the TBDN
and TBDN case is 1.64, compared with 1.03 in the NN and NA case. \ The
respective raw decay amplitudes (with the norm effects taken out) are 0.252
and 0.160. \ These are both substantial differences. \ However, including
the norm effect in the amplitude cancels out these differences giving us
about a 0.155 amplitude in both cases. This cancelation hides the
substantial effects of both the TBDA and TBDN. Things are different in the
case of the $\eta _{c}^{\prime }$. \ There the square root of the norm in
the TBDN and TBDA case is 1.22, compared with 1.008 in the NN and NA case. \
The respective raw decay amplitudes (with the norm effects taken out) are
-0.138 and -0.087. \ Unlike the case of the $\eta _{c}$ the effect of
including the norm in the amplitude does not cancel out these differences
giving us about a 0.113 amplitude in the first case and a -0.086 amplitude
in the second. The ratio of the TBDA to the NA are 1.58 \ in both cases. \
However, the square root norm ratios are quite different, being 1.59 in the
case of the $\eta _{c}$ but only 1.21 in the case of the $\eta _{c}^{\prime
}.$ This may point to a limitation of the linear confining model used in
working out the wave functions near threshold for the $\eta _{c^{\prime }}$
decay.

\paragraph{$\protect\chi _{0}~$Decay}

The $^{3}P_{0}$ decay amplitudes are from Eqs.(\ref{abjpm}) and in the
combination from (\ref{tpo})%
\begin{eqnarray}
\mathcal{F}_{\chi _{0}} &=&\frac{4\sqrt{3}}{9}(A_{0}+2B_{0}^{(2)})  \notag \\
&=&\frac{4\sqrt{3}e^{2}}{9}\frac{2\pi i}{3}\int_{0}^{\infty }dr\exp
(-mr)\exp (F)\{(mr+1)  \notag \\
&&\times \{\exp K[j_{0}(kr)+2j_{0}^{\prime }(kr)kr+4j_{2}(kr)+2j_{2}^{\prime
}(kr)kr]+2j_{0}(kr)\sinh K\}\frac{u_{110}^{-}(r)}{r}  \notag \\
&&-3\exp (-K)j_{0}(kr)mr\frac{\upsilon _{000}^{+}(r)}{r}\},  \label{chpo}
\end{eqnarray}%
with the same color and flavor factors as before, in which (see Appendix D) 
\begin{equation}
\frac{u_{110}^{-}}{r}=\frac{M}{E}\frac{u_{110}^{+}}{r}.
\end{equation}%
We also have 
\begin{equation}
\frac{\upsilon _{000}^{+}}{r}=\frac{\exp (\mathcal{G)}}{E}[\frac{2}{r}-\frac{%
5}{2}L^{\prime }+\frac{d}{dr}]\frac{u_{110}^{+}}{r}.
\end{equation}%
Our multicomponent TBDA results uses these relations in Eq.(\ref{chpo}). The
TBDN condition (\ref{trpnrm}) becomes (see Appendix C)%
\begin{eqnarray}
&&\frac{1}{2}\int drr^{2}\exp (2F){\biggl (}\exp (-2K)\{\left( \frac{%
u_{110}^{+}}{r}\right) ^{2}+\left( \frac{u_{110}^{-}}{r}\right) ^{2}+\left( 
\frac{v_{000}^{+}}{r}\right) ^{2}  \notag \\
&&+\{2w^{2}\frac{\partial L}{\partial w^{2}}(\left( \frac{u_{110}^{+}}{r}%
\right) ^{2}-\left( \frac{u_{110}^{-}}{r}\right) ^{2}-\left( \frac{%
v_{000}^{+}}{r}\right) ^{2})  \notag \\
&&+4w^{2}\frac{\partial \mathcal{G}}{\partial w^{2}}(-\left( \frac{%
u_{110}^{-}}{r}\right) ^{2}+2\left( \frac{v_{000}^{+}}{r}\right) ^{2})\} 
\notag \\
&&+8w^{2}\frac{\partial \mathcal{G}}{\partial w^{2}}\sinh 2K\left( \frac{%
u_{110}^{+}}{r}\right) ^{2}{\biggr )}=1  \label{chonrm}
\end{eqnarray}%
while the NN condition of Eq. (\ref{nrm}) is 
\begin{equation}
\frac{1}{2}\int drr^{2}[\left( \frac{u_{110}^{+}}{r}\right) ^{2}+\left( 
\frac{u_{110}^{-}}{r}\right) ^{2}+\left( \frac{v_{000}^{+}}{r}\right)
^{2}]=1.  \label{w0rm}
\end{equation}%
In that same limit our decay amplitude (\ref{chpo}) becomes the NA%
\begin{eqnarray}
\mathcal{F}_{\chi _{0}} &=&\frac{4\sqrt{3}e^{2}}{9}\frac{2\pi i}{3}%
\int_{0}^{\infty }dr\exp (-mr)\{(mr+1)[j_{0}(kr)+2j_{0}^{\prime
}(kr)kr+4j_{2}(kr)+2j_{2}^{\prime }(kr)kr]\frac{u_{110}^{-}(r)}{r}  \notag \\
&&-3\exp (-K)j_{0}(kr)mr\frac{\upsilon _{000}^{+}(r)}{r}\}.  \label{whpo}
\end{eqnarray}%
Our multicomponent TBDA and TBDN result from (\ref{chpo}) and (\ref{chonrm})
is 3.90 \textrm{keV, }while that obtained from the corresponding
multicomponent NA and NN result from (\ref{whpo}) and (\ref{w0rm}) is 3.28 
\textrm{keV}. \ If we further ignore the small components in these latter
forms by taking $u_{110}^{-}=$\ $u_{110}^{+}$ and $\upsilon _{000}^{+}=0,$
then the decay rate is 0.646 \textrm{keV. }These are to be compared with the
observed rate of 2.6 \ $\pm 0$.65 \textrm{keV}$.$ The QCD radiative
corrections \cite{rsnr} modify these by a factor of $(1+\alpha _{s}/\pi (\pi
^{2}/3-28/9))$ to 3.96 and 3.34 \textrm{keV (}and 0.656\textrm{\ keV). }The
multicomponent effects are substantial even if we do not include the effects
of the TBDA and TBDN. Those additional effects are small compared with the
effects of including the multicomponents by themselves. \ This parallels
that which occurs in the $^{3}P_{0}$ positronium decay where the amplitude
vanishes without the multicomponent (small) parts of the wave function.

\paragraph{$\protect\chi _{2}~$Decay}

The $^{3}P_{2}$ decay amplitudes (\ref{abjpm}) appear from Eq.(\ref{tpt}) in
the separate combination 
\begin{eqnarray}
\mathcal{F(K)}_{\chi _{2}} &=&\frac{4\sqrt{3}}{9}[A_{2}+\frac{2B_{2}^{(0)}}{5%
}+\frac{2B_{2}^{(2)}}{7}+\frac{36B_{2}^{(4)}}{35})]  \notag \\
&=&\frac{4\sqrt{3}e^{2}}{9}\frac{2\pi i}{3}\int_{0}^{\infty }dr\exp
(-mr)\exp (F)\biggl [{\large -}(mr+1)  \notag \\
&&\times \exp K\{[j_{2}(kr)+\frac{2}{3}j_{2}^{\prime }(kr)kr]\sqrt{\frac{3}{5%
}}\frac{u_{312}^{-}(r)}{r}+\sqrt{\frac{2}{5}}\frac{u_{112}^{-}(r)}{r}%
[-2j_{2}(kr)+j_{2}^{\prime }(kr)kr]\}  \notag \\
&&+j_{2}(kr)\{3mr\exp (-K)\frac{\upsilon _{202}}{r}-2(mr+1)\sinh K[-\sqrt{%
\frac{3}{5}}\frac{u_{312}^{-}}{r}+\frac{u_{112}^{-}}{r}\sqrt{\frac{2}{5}}]\}
\notag \\
&&-\frac{2}{5}(mr+1){\biggl (}\exp K\{\frac{1}{3}j_{0}^{\prime }(kr)kr\sqrt{%
\frac{3}{5}}\frac{u_{312}^{-}(r)}{r}+[\frac{5}{2}j_{0}(kr)+\frac{1}{2}%
j_{0}^{\prime }(kr)kr]\sqrt{\frac{2}{5}}\frac{u_{112}^{-}(r)}{r}\}  \notag \\
&&-2\sinh K(mr+1)[-\sqrt{\frac{3}{5}}\frac{u_{312}^{-}}{r}+\sqrt{\frac{2}{5}}%
\frac{u_{112}^{-}}{r}]j_{2}(kr){\biggr )}  \notag \\
&&+\frac{2}{7}(mr+1){\biggl (}\exp K\{\frac{1}{3}j_{2}^{\prime }(kr)kr\sqrt{%
\frac{3}{5}}\frac{u_{312}^{-}(r)}{r}+[\frac{5}{2}j_{2}(kr)+\frac{1}{2}%
j_{2}^{\prime }(kr)kr]\sqrt{\frac{2}{5}}\frac{u_{112}^{-}(r)}{r}\}  \notag \\
&&-2\sinh K(mr+1)[-\sqrt{\frac{3}{5}}\frac{u_{312}^{-}}{r}+\sqrt{\frac{2}{5}}%
\frac{u_{112}^{-}}{r}]j_{2}(kr){\biggr )}  \notag \\
&&-\frac{36}{35}(mr+1){\biggl (}\exp K\{\frac{1}{3}j_{4}^{\prime }(kr)kr%
\sqrt{\frac{3}{5}}\frac{u_{312}^{-}(r)}{r}+[\frac{5}{2}j_{4}(kr)+\frac{1}{2}%
j_{4}^{\prime }(kr)kr]\sqrt{\frac{2}{5}}\frac{u_{112}^{-}(r)}{r}\}  \notag \\
&&-2\sinh K(mr+1)[-\sqrt{\frac{3}{5}}\frac{u_{312}^{-}}{r}+\sqrt{\frac{2}{5}}%
\frac{u_{112}^{-}}{r}]j_{2}(kr){\biggr )\biggr ],}  \label{b3}
\end{eqnarray}%
and%
\begin{eqnarray}
\mathcal{G(K)}_{\chi _{2}} &=&\frac{4\sqrt{3}}{9}[\frac{B_{2}^{(0)}}{5}-%
\frac{2B_{2}^{(2)}}{7}+\frac{3B_{2}^{(4)}}{35}]  \notag \\
&=&-\frac{4\sqrt{3}e^{2}}{9}\frac{2\pi i}{3}\int_{0}^{\infty }dr\exp
(-mr)\exp (F)(mr+1)  \notag \\
&&\times \biggl [\frac{1}{5}{\biggl (}\exp K\{\frac{1}{3}j_{0}^{\prime
}(kr)kr\sqrt{\frac{3}{5}}\frac{u_{312}^{-}(r)}{r}+[\frac{5}{2}j_{0}(kr)+%
\frac{1}{2}j_{0}^{\prime }(kr)kr]\sqrt{\frac{2}{5}}\frac{u_{112}^{-}(r)}{r}\}
\notag \\
&&-2\sinh K[-\sqrt{\frac{3}{5}}\frac{u_{312}^{-}}{r}+\sqrt{\frac{2}{5}}\frac{%
u_{112}^{-}}{r}]j_{2}(kr){\biggr )}  \notag \\
&&+\frac{2}{7}{\biggl (}\exp K\{[\frac{1}{3}j_{2}^{\prime }(kr)kr]\sqrt{%
\frac{3}{5}}\frac{u_{312}^{-}(r)}{r}+[\frac{5}{2}j_{2}(kr)+\frac{1}{2}%
j_{2}^{\prime }(kr)kr]\sqrt{\frac{2}{5}}\frac{u_{112}^{-}(r)}{r}\}  \notag \\
&&-2\sinh K[-\sqrt{\frac{3}{5}}\frac{u_{312}^{-}}{r}+\sqrt{\frac{2}{5}}\frac{%
u_{112}^{-}}{r}]j_{2}(kr){\biggr )}  \notag \\
&&+\frac{3}{35}{\biggl (}\exp K\{[\frac{1}{3}j_{4}^{\prime }(kr)kr]\sqrt{%
\frac{3}{5}}\frac{u_{312}^{-}(r)}{r}+[\frac{5}{2}j_{4}(kr)+\frac{1}{2}%
j_{4}^{\prime }(kr)kr]\sqrt{\frac{2}{5}}\frac{u_{112}^{-}(r)}{r}\}  \notag \\
&&-2\sinh K[-\sqrt{\frac{3}{5}}\frac{u_{312}^{-}}{r}+\sqrt{\frac{2}{5}}\frac{%
u_{112}^{-}}{r}]j_{2}(kr){\biggr )\biggr ]},  \label{b4}
\end{eqnarray}%
in which%
\begin{eqnarray}
&&\frac{u_{112}^{-}}{r}=\frac{E}{M}\frac{u_{112}^{+}}{r}-\frac{\exp (2%
\mathcal{G})}{10EM}\{[\Phi _{--}+4\exp (2\mathcal{G})(E^{2}-M^{2})\mathcal{-}%
2\sqrt{6}\Phi _{+-}\mathcal{+}\frac{A_{mm}}{r^{2}}+\frac{B_{mm}}{r}+C_{mm}+(%
\frac{F_{mm}}{r}+G_{mm})\frac{d}{dr}]\frac{u_{112}^{+}}{r}  \notag \\
&&+\sqrt{6}[\frac{6\Phi _{-+}}{\sqrt{6}}+2\Phi _{++}-2\exp (2\mathcal{G}%
)(E^{2}-M^{2})\mathcal{+}\frac{A_{mp}}{r^{2}}+\frac{B_{mp}}{r}+C_{mp}+(\frac{%
F_{mp}}{r}+G_{mp})\frac{d}{dr}]\frac{u_{312}^{+}}{r}\},
\end{eqnarray}%
with $\Phi _{--},\Phi _{-+},A_{mm},..,G_{mp}$ given in Appendix D and%
\begin{eqnarray}
&&\frac{u_{312}^{-}}{r}=\frac{M}{E}\frac{u_{312}^{+}}{r}-\frac{\exp (2%
\mathcal{G})}{10EM}\{[-\Phi _{++}-4\exp (2\mathcal{G})(E^{2}-M^{2})+2\sqrt{6}%
\Phi _{-+}\mathcal{+}\frac{A_{pp}}{r^{2}}+\frac{B_{pp}}{r}+C_{pp}+(\frac{%
F_{pp}}{r}+G_{pp})\frac{d}{dr}]\frac{u_{312}^{+}}{r}  \notag \\
&&+\sqrt{6}[-\frac{\Phi _{+-}}{\sqrt{6}}+2\Phi _{--}-2\exp (2\mathcal{G}%
)(E^{2}-M^{2})\mathcal{+}\frac{A_{pm}}{r^{2}}+\frac{B_{pm}}{r}+C_{pm}+(\frac{%
F_{pm}}{r}+G_{pm})\frac{d}{dr}]\frac{u_{112}^{+}}{r}\}  \label{u312}
\end{eqnarray}%
with the expression for $\Phi _{++},\Phi _{+-},A_{pp},..,G_{pm}$ also in the
Appendix D$.$ The other radial wave functions are%
\begin{eqnarray}
\frac{\upsilon _{202}}{r} &=&\frac{\exp (\mathcal{G}+2K\mathcal{)}}{E}\{[%
\frac{1-2Q_{m}}{r}-(Q_{m}+1)\frac{d}{dr}-\frac{5L}{2}^{\prime }(Q_{m}+1)]%
\sqrt{\frac{2}{5}}\frac{u_{112}^{+}}{r}  \notag \\
&&+[\frac{4+2Q_{m}}{r}+(Q_{m}+1)\frac{d}{dr}-\frac{5L}{2}^{\prime }(Q_{m}+1)]%
\sqrt{{\frac{3}{5}}}\frac{u_{312}^{+}}{r}\}  \label{v202}
\end{eqnarray}%
and 
\begin{eqnarray}
\frac{\upsilon _{212}^{-}}{r} &=&-\frac{\exp (\mathcal{G)}}{M}\{[(\frac{d}{dr%
}-\frac{1}{r}-\frac{2Q_{m}}{r}+\frac{(L+6\mathcal{G})}{2}^{\prime })]\sqrt{%
\frac{3}{5}}\frac{u_{112}^{+}}{r}  \notag \\
&&+[(\frac{d}{dr}+\frac{4}{r}+\frac{3Q_{m}}{r}+\frac{(L+6\mathcal{G})}{2}%
^{\prime })]\sqrt{\frac{2}{5}}\frac{u_{312}^{+}}{r}.  \label{v212}
\end{eqnarray}%
Only the first of these latter two wave functions contributes to the decay
amplitude. All wave functions contribute to the\ TBDN condition (\ref{trpnrm}%
) which has the form (see Appendix C)\ 
\begin{eqnarray}
1 &=&\frac{1}{2}\int drr^{2}\exp (2F){\biggl (}\left( \frac{u_{112}^{+}}{r}%
\right) ^{2}\{[\exp (2K)-\frac{4}{5}\sinh 2K][1+2w^{2}\frac{\partial L}{%
\partial w^{2}}]+\frac{4}{5}\sinh 2K(4w^{2}\frac{\partial \mathcal{G}}{%
\partial w^{2}})\}  \notag \\
&&+\left( \frac{u_{112}^{-}}{r}\right) ^{2}\{[\exp (2K)-\frac{4}{5}\sinh
2K][1-2w^{2}\frac{\partial L}{\partial w^{2}}]+\frac{4}{5}\sinh 2K(4w^{2}%
\frac{\partial \mathcal{G}}{\partial w^{2}})\}  \notag \\
&&+\left( \frac{u_{312}^{+}}{r}\right) ^{2}\{[\exp (2K)-\frac{6}{5}\sinh
2K][1+2w^{2}\frac{\partial L}{\partial w^{2}}]+\frac{6}{5}\sinh 2K(8w^{2}%
\frac{\partial \mathcal{G}}{\partial w^{2}})\}  \notag \\
&&+\left( \frac{u_{312}^{-}}{r}\right) ^{2}\{[\exp (2K)-\frac{6}{5}\sinh
2K][1-2w^{2}\frac{\partial L}{\partial w^{2}}-4w^{2}\frac{\partial \mathcal{G%
}}{\partial w^{2}}]\}  \notag \\
&&+\left( \frac{u_{312}^{+}}{r}\right) \left( \frac{u_{112}^{+}}{r}\right) 
\frac{4\sqrt{6}}{5}\sinh 2K[1+2w^{2}\frac{\partial L}{\partial w^{2}}-4w^{2}%
\frac{\partial \mathcal{G}}{\partial w^{2}}]  \notag \\
&&+\left( \frac{u_{312}^{-}}{r}\right) \left( \frac{u_{112}^{-}}{r}\right) 
\frac{4\sqrt{6}}{5}\sinh 2K[1-2w^{2}\frac{\partial L}{\partial w^{2}}-4w^{2}%
\frac{\partial \mathcal{G}}{\partial w^{2}}]  \notag \\
&&+\left( \frac{\upsilon _{202}^{+}}{r}\right) ^{2}\exp (-2K)(1-2w^{2}\frac{%
\partial L}{\partial w^{2}}+8w^{2}\frac{\partial \mathcal{G}}{\partial w^{2}}%
)  \notag \\
&&+\left( \frac{\upsilon _{212}^{-}}{r}\right) ^{2}\exp (2K)((1+2w^{2}\frac{%
\partial L}{\partial w^{2}}+4w^{2}\frac{\partial \mathcal{G}}{\partial w^{2}}%
){\biggr ).}  \label{chp2nrm}
\end{eqnarray}%
Our multicomponent results uses these relations in Eq.(\ref{b3},\ref{b4}). \
The NN is 
\begin{equation}
\frac{1}{2}\int drr^{2}[\left( \frac{u_{112}^{+}}{r}\right) ^{2}+\left( 
\frac{u_{112}^{-}}{r}\right) ^{2}+\left( \frac{u_{312}^{+}}{r}\right)
^{2}+\left( \frac{u_{312}^{-}}{r}\right) ^{2}+\left( \frac{\upsilon
_{202}^{+}}{r}\right) ^{2}+\left( \frac{\upsilon _{212}^{-}}{r}\right)
^{2}]=1.  \label{w2rm0}
\end{equation}%
In that same limit our decay amplitudes (\ref{b3},\ref{b4}) become the NAs%
\begin{eqnarray}
\mathcal{F(K)}_{\chi _{2}} &=&\frac{4\sqrt{3}e^{2}}{9}\frac{2\pi i}{3}%
\int_{0}^{\infty }dr\exp (-mr)\biggl [{\large -}(mr+1)  \notag \\
&&\times \{[j_{2}(kr)+\frac{2}{3}j_{2}^{\prime }(kr)kr]\sqrt{\frac{3}{5}}%
\frac{u_{312}^{-}(r)}{r}+\sqrt{\frac{2}{5}}\frac{u_{112}^{-}(r)}{r}%
[-2j_{2}(kr)+j_{2}^{\prime }(kr)kr]\}+j_{2}(kr)\{3mr\frac{\upsilon _{202}}{r}%
]\}  \notag \\
&&-\frac{2}{5}(mr+1){\biggl (}\{\frac{1}{3}j_{0}^{\prime }(kr)kr\sqrt{\frac{3%
}{5}}\frac{u_{312}^{-}(r)}{r}+[\frac{5}{2}j_{0}(kr)+\frac{1}{2}j_{0}^{\prime
}(kr)kr]\sqrt{\frac{2}{5}}\frac{u_{112}^{-}(r)}{r}\}{\biggr )}  \notag \\
&&+\frac{2}{7}(mr+1){\biggl (}\{\frac{1}{3}j_{2}^{\prime }(kr)kr\sqrt{\frac{3%
}{5}}\frac{u_{312}^{-}(r)}{r}+[\frac{5}{2}j_{2}(kr)+\frac{1}{2}j_{2}^{\prime
}(kr)kr]\sqrt{\frac{2}{5}}\frac{u_{112}^{-}(r)}{r}\}{\biggr )}  \notag \\
&&-\frac{36}{35}(mr+1){\biggl (}\{\frac{1}{3}j_{4}^{\prime }(kr)kr\sqrt{%
\frac{3}{5}}\frac{u_{312}^{-}(r)}{r}+[\frac{5}{2}j_{4}(kr)+\frac{1}{2}%
j_{4}^{\prime }(kr)kr]\sqrt{\frac{2}{5}}\frac{u_{112}^{-}(r)}{r}\}{\biggr )%
\biggr ]}{\large ,}  \label{b30}
\end{eqnarray}%
and%
\begin{eqnarray}
\mathcal{G(K)}_{\chi _{2}} &=&-\frac{4\sqrt{3}e^{2}}{9}\frac{2\pi i}{3}%
\int_{0}^{\infty }dr\exp (-mr)(mr+1)  \notag \\
&&\times \biggl [\frac{1}{5}{\biggl (}\{\frac{1}{3}j_{0}^{\prime }(kr)kr%
\sqrt{\frac{3}{5}}\frac{u_{312}^{-}(r)}{r}+[\frac{5}{2}j_{0}(kr)+\frac{1}{2}%
j_{0}^{\prime }(kr)kr]\sqrt{\frac{2}{5}}\frac{u_{112}^{-}(r)}{r}\}{\biggr )}
\notag \\
&&+\frac{2}{7}{\biggl (}\{[\frac{1}{3}j_{2}^{\prime }(kr)kr]\sqrt{\frac{3}{5}%
}\frac{u_{312}^{-}(r)}{r}+[\frac{5}{2}j_{2}(kr)+\frac{1}{2}j_{2}^{\prime
}(kr)kr]\sqrt{\frac{2}{5}}\frac{u_{112}^{-}(r)}{r}\}{\biggr )}  \notag \\
&&+\frac{3}{35}{\biggl (}\{[\frac{1}{3}j_{4}^{\prime }(kr)kr]\sqrt{\frac{3}{5%
}}\frac{u_{312}^{-}(r)}{r}+[\frac{5}{2}j_{4}(kr)+\frac{1}{2}j_{4}^{\prime
}(kr)kr]\sqrt{\frac{2}{5}}\frac{u_{112}^{-}(r)}{r}\}{\biggr )\biggr ]}%
{\large .}  \label{b40}
\end{eqnarray}%
\ Our strong potential, multicomponent result from (\ref{b3}), (\ref{b4}%
),and (\ref{chp2nrm}) is 1.43 \textrm{keV, }while that obtained from the
corresponding weak potential forms of (\ref{b30}) and (\ref{b40}) is 0.836 
\textrm{keV}. \ Our multicomponent TBDA\ and TBDN result from (\ref{b3}), (%
\ref{b4}),and (\ref{chp2nrm}) is 1.43 \textrm{keV, }while that obtained from
the corresponding multicomponent NA and NN result of (\ref{b30}), (\ref{b40}%
), and (\ref{w2rm0}) is 0.836 \textrm{keV}. \ \ If we further ignore the
small and tensor coupled components in these latter forms by taking $%
u_{112}^{-}=$\ $u_{112}^{+}$ and $u_{312}^{-}=$\ $u_{312}^{+}=\upsilon
_{202}^{+}=\upsilon _{212}^{+}=0,$then the decay rate is 0.033 \textrm{keV. }%
These are to be compared with the observed rate of 0.528 $\pm $.09\ \textrm{%
keV}$.$The QCD radiative corrections \cite{rsnr} modify these by a factor of 
$(1-16\alpha _{s}/\pi )$ to 0.743 and 0.435 \textrm{keV }(0.017 \textrm{keV)}%
. Full tensor couplings are included in the first two results. As with the $%
^{3}P_{0}$ decay the NA and NN multicomponent effects are substantial even
if we do not include those of the TBDA and TBDN. \ Those effects are
themselves significantly larger than the effects of the NA and NN. \ 

\paragraph{$\protect\pi ^{0}$ Decay}

The $\pi ^{0}$ state vector is 
\begin{equation}
|\pi ^{0}\rangle =\sum_{c=r,g,b}\frac{1}{\sqrt{2}}(|\bar{u}u\rangle -|\bar{d}%
d\rangle )_{c}\frac{1}{\sqrt{3}},
\end{equation}%
where the charge of the $u$ is $+2e/3$ that of the $d$ is $-e/3.$ Thus, the
amplitude for its annihilation is modified by a factor of $\sqrt{3}%
[(2/3)^{2}-(-1/3)^{2}]/\sqrt{2}.~$Otherwise the \ wave function discussion
is the same as in the section on $\eta _{c}$ decay. So we obtain 
\begin{eqnarray}
\mathcal{F}_{\pi _{0}} &=&\sqrt{\frac{3}{2}}\frac{e^{2}}{3}2\pi
i\int_{0}^{\infty }dr\exp (-mr){\biggl (}j_{1}(kr)\exp (F-K)\frac{u_{000}^{-}%
}{r}](1+mr)  \notag \\
&&+mr\frac{\upsilon _{110}^{+}}{r}\{j_{1}(kr)\exp (F)[\exp (K)+2\sinh (K)] 
\notag \\
&&+kr\exp (F+K)[j_{1}^{\prime }(kr)-j_{0}(kr)]\}{\biggr )}.
\end{eqnarray}%
Otherwise the norm and amplitude discussion is the same as in the section on 
$\eta _{c}$ decay. Our multicomponent TBDA\ and TBDN result from this is
24.7 \textrm{eV, }while that obtained from the multicomponent NA\ and NN
result is 94.4 \textrm{eV}. If we further ignore the small components in the
weak potential form by taking $u_{000}^{-}=$\ $u_{000}^{+}$ and $\upsilon
_{110}^{+}=0,$ then the decay rate is 89.5 \textrm{eV. }These are to be
compared with the observed rate of 7.72$\pm $.04 \textrm{eV}. \ QCD
radiative corrections modify these to 8.73 \textrm{eV }and 33.5 \textrm{eV (}%
31.5 \textrm{eV)}$.$ The influence of including the TBDA and TBDN
multicomponent effects in the norm and the amplitude are substantial when
compared to that of including just the NA and NN effects and bring our pion
decay rate reasonably close to the observed rate.

\paragraph{$\protect\pi _{2}$ Decay}

For this spin singlet decay the relevant amplitude is%
\begin{eqnarray}
\mathcal{F}_{\pi _{2}} &=&\sqrt{\frac{3}{2}}\frac{1}{3}(F_{2}+\frac{2}{5}%
G_{2}^{(1)}+\frac{3}{5}G_{2}^{(3)})  \notag \\
&=&i\sqrt{\frac{3}{2}}\frac{e^{2}}{3}2\pi \int_{0}^{\infty }drmr\exp (-mr) 
\notag \\
&&\times {\biggl [}\exp (F+K)mrj_{2}(kr)kr(\frac{1}{3}\sqrt{\frac{3}{5}}%
\frac{\upsilon _{312}^{+}(r)}{r}+\frac{1}{2}\sqrt{\frac{2}{5}}\frac{\upsilon
_{112}^{+}(r)}{r})  \notag \\
&&+\frac{2}{5}{\biggl (}(mr+1)j_{1}(kr)\exp (F-K)\frac{u_{202}^{-}(r)}{r}%
+\exp (F+K)mr\{\mathbf{[-}\frac{j_{1}(kr)}{3}+\frac{1}{3}j_{1}^{\prime
}(kr)kr]\sqrt{\frac{3}{5}}\frac{\upsilon _{312}^{+}(r)}{r}  \notag \\
&&+[2j_{1}(kr)+\frac{1}{2}j_{1}^{\prime }(kr)kr]\sqrt{\frac{2}{5}}\frac{%
\upsilon _{112}^{+}(r)}{r}\}-2mr\exp (F)\sinh (K)j_{1}(kr)(-\frac{\upsilon
_{312}^{+}}{r}\sqrt{\frac{3}{5}}+\frac{\upsilon _{112}^{+}}{r}\sqrt{\frac{2}{%
5}}){\biggr )}  \notag \\
&&-\frac{3}{5}{\biggl (}(mr+1)j_{3}(kr)\exp (F-K)\frac{u_{202}^{-}(r)}{r}%
+\exp (F+K)mr\{\mathbf{[-}\frac{j_{3}(kr)}{3}+\frac{1}{3}j_{3}^{\prime
}(kr)kr]\sqrt{\frac{3}{5}}\frac{\upsilon _{312}^{+}(r)}{r}  \notag \\
&&+[2j_{3}(kr)+\frac{1}{2}j_{3}^{\prime }(kr)kr]\sqrt{\frac{2}{5}}\frac{%
\upsilon _{112}^{+}(r)}{r}\}-2mr\exp (F)\sinh (K)j_{3}(kr)(-\frac{\upsilon
_{312}^{+}}{r}\sqrt{\frac{3}{5}}+\frac{\upsilon _{112}^{+}}{r}\sqrt{\frac{2}{%
5}}){\biggr )\biggr ]}{\large ,}
\end{eqnarray}%
where%
\begin{equation}
\frac{u_{202}^{-}}{r}=\frac{M}{E}\frac{u_{202}^{+}}{r},
\end{equation}%
and%
\begin{eqnarray}
\frac{\upsilon _{112}^{+}}{r} &=&\frac{\exp (\mathcal{G}-2K)}{E}[\exp (2K)(-%
\frac{d}{dr}-\frac{L^{\prime }}{2})-\frac{3}{r}]\frac{u_{202}^{+}}{r}\sqrt{%
\frac{2}{5}}  \notag \\
\frac{\upsilon _{312}^{+}}{r} &=&\frac{\exp (\mathcal{G}-2K)}{E}[\exp (2K)(%
\frac{d}{dr}+\frac{L^{\prime }}{2})-\frac{2}{r}]\frac{u_{202}^{+}}{r}\sqrt{%
\frac{3}{5}},
\end{eqnarray}%
together with the normalization condition (\ref{sinrm}) (see Appendix C)%
\begin{eqnarray}
&&\frac{1}{2}\int_{0}^{\infty }drr^{2}\exp (2F){\biggl (}\exp (-2K)[\left( 
\frac{u_{202}^{+}}{r}\right) ^{2}+\left( \frac{u_{202}^{-}}{r}\right)
^{2}]+\exp (2K)[\left( \frac{\upsilon _{112}^{+}}{r}\right) ^{2}+\left( 
\frac{\upsilon _{312}^{+}}{r}\right) ^{2}]  \notag \\
&&-2\sinh 2K(-\frac{\upsilon _{312}^{+}}{r}\sqrt{\frac{3}{5}}+\frac{\upsilon
_{112}^{+}}{r}\sqrt{\frac{2}{5}})^{2}  \notag \\
&&+2w^{2}\frac{\partial L}{\partial w^{2}}[\exp (-2K)[\left( \frac{%
u_{202}^{+}}{r}\right) ^{2}-\left( \frac{u_{202}^{-}}{r}\right) ^{2}]-\exp
(2K)[\left( \frac{\upsilon _{112}^{+}}{r}\right) ^{2}+\left( \frac{\upsilon
_{312}^{+}}{r}\right) ^{2}]  \notag \\
&&+2\sinh 2K(-\frac{\upsilon _{312}^{+}}{r}\sqrt{\frac{3}{5}}+\frac{\upsilon
_{112}^{+}}{r}\sqrt{\frac{2}{5}})^{2}\mathbf{]}  \notag \\
&&+4w^{2}\frac{\partial \mathcal{G}}{\partial w^{2}}\exp (-2K)[2\left( \frac{%
u_{202}^{+}}{r}\right) ^{2}+\left( \frac{u_{202}^{-}}{r}\right) ^{2}]{\biggr
)}=1.
\end{eqnarray}%
Our multicomponent TBDA\ and TBDN result from this is results is 180 eV
while the NA and NN result is 142 eV. If we further ignore the small
components in the weak potential form by taking $u_{202}^{-}=$\ $u_{202}^{+}$
and $\upsilon _{112}^{+}=\upsilon _{112}^{+}=0,$ then the decay rate is 4.66 
\textrm{eV.}\ The experimental situation is unclear. \ \ Earlier results had
very large widths on the order of 1 keV. \ In the latest compilation, one
result is listed as $<$70 eV and one at $<$190 eV both at the 90\% \
confidence level. \ In any event, \textrm{\ }the multicomponent effects here
are substantial which ever result we use. The difference between the results
are small compared with the effects of including the multicomponents by
themselves.

\paragraph{$^{1}D_{2}(3872)$ Decay}

The quark-content of this state is unsure. \ If we assume this is a $%
^{1}D_{2}$ spin singlet, then the relevant decay amplitude has the same wave
function structures as with the $\pi _{2}$ except for the flavor factor. Our
multicomponent TBDA\ and TBDN result from this is results is 65.7 eV while
the NA and NN result is 73.2 eV. \ If we further ignore the small components
in the latter form by taking $u_{202}^{-}=$\ $u_{202}^{+}$ and $\upsilon
_{112}^{+}=\upsilon _{112}^{+}=0,$ then the decay rate is 168. \textrm{keV.}%
\ The experimental situation is unclear. \ \ Using ratios given in the
latest table (for the $(c\bar{c},l=2,j=2$ state) we take the observed value
to be 435 \textrm{eV. }As with the $\pi _{2}$ decay, \textrm{\ }both
multicomponent effects here are substantial. The difference between them are
small compared with the effects of including the multicomponents by
themselves. However, unlike the $\pi _{2}$ the effects are to reduce rather
than enhance the rate.

\paragraph{$a_{2}~$Decay}

Except for the quark content (same as with $\pi _{0},\pi _{2})$ this
particle has a $^{3}P_{2}$ decay amplitude and wave functions given as with
the $\chi _{2}$.\ Our multicomponent TBDA and TBDN decay rate result is 31.5 
\textrm{keV} reducing to 9.02 \textrm{keV~}when QCD radiative effects are
included. The corresponding NA and NN rate result is 10.9 \textrm{keV }%
reducing to 3.12 \textrm{keV~}when QCD radiative effects are included. The
observed rate of 1.00 $\pm $.06\ \textrm{keV}$.$ Including TBDN and TBDA
effects in the norm and the amplitude are substantial and produce too large
a decay rate.

\paragraph{$f_{2}^{\prime }~$Decay}

With an $s\bar{s}$ quark content, this particle has a $^{3}P_{2}$ decay
amplitude otherwise similar to that of the above $a_{2}$. \ Our strong
potential, multicomponent decay rate is 2.36 \textrm{keV }reducing to 760 
\textrm{eV} when QCD effects are included . The corresponding weak component
rates is 1.08 \textrm{keV }reducing to 348 \textrm{eV~}when QCD radiative
effects are included. The observed rate is 81 $\pm $9.6\ \textrm{eV}$.$ As
with the $a_{2}$, including TBDN and TBDA effects in the norm and the
amplitude are substantial and produce too large a decay rate.

\section{Discussion and Earlier Results}

\subsection{Charmonium}

\ \ The table below (units are in \textrm{keV}) compares our results (both
the ones that come from TBDA and NA multicomponent results) with a variety
of other quark models (ones that have not yet been subjected to the tests
imposed on the TBDE and which for the most part do not include the light
mesons in their spectroscopic calculations).

\ \ \ \ \ \ \ \ \ \ \ \ \ \ \ \ \ \ \ \ \ 
\begin{tabular}{llllllll}
& Expt & TBDE-TBDA & TBDE-NA & \cite{ack} & \cite{gupt} & \cite{ebrt} & \cite%
{munz} \\ 
$\eta _{c}(1^{1}S_{0}-2976)$ & 7.4$\pm $1.0 & 6.20 & 6.18 & 4.8 & 10.94,10.81
& 5.5 & 3.50 \\ 
$\eta _{c}(2^{1}S_{0}-3263)$ & 1.3$\pm $0.6 & 3.36 & 1.95 & 3.7 & - & 1.8 & 
1.38 \\ 
$\chi _{0}(1^{3}P_{0}-3415)$ & 2.6$\pm $.65 & 3.96 & 3.34 & - & 6.38,8.13 & 
2.9 & 1.39 \\ 
$\chi _{2}(1^{3}P_{2}-3556)$ & 0.53$\pm .$09 & 0.743 & 0.435 & - & 0.57,1.14
& 0.50 & .440%
\end{tabular}

Ackleh and Barnes \cite{ack} independently developed a similar approach to
the one we developed for positronium decay \cite{posit} and then applied it
to spin singlet quarkonium decay into two gammas.\ As in our approach, they
include the effects of the bound state wave function on the initial decaying
particle. \ Gupta, Johnson, and Repko \cite{gupt} follow a similar approach.
The two numbers displayed in their column correspond to two distinct
approaches used in incorporating off shell effects. \ The first is similar
to that used in \cite{ack} where the energy factors which arise from the
Feynman amplitude are treated on mass shell ($E=\sqrt{\mathbf{p}^{2}+m^{2}};$
energy conservation, which would have $E=w/2,$ is not used) while the second
set of numbers come from treating the particle on energy shell ($E=w/2$ but
with $m^{2}=E^{2}-\mathbf{p}^{2})$. Our approach is different from both of
these in that it is on energy shell, $E=w/2,$ but with $m^{2}\neq E^{2}-%
\mathbf{p}^{2}$ it is off-mass-shell. The energy factors that appear in our
equations are those required from the way in which the constraint formalism
eliminates the CM relative energy- see Eq. (\ref{rle}, \ref{rp})). (See also
our discussion in our section on $^{3}P_{0}$ positronium.) To be more
explicit, the portion of the Feynman propagator $(p_{-}-k_{1})^{2}+m^{2}-i0$
in the approaches of \cite{ack} and the first of \cite{gupt} is treated as $2%
\mathbf{p\cdot k+}w\sqrt{\mathbf{p}^{2}+m^{2}}$, in the second of \cite{gupt}
as $2\mathbf{p\cdot k+}w^{2}/2$ and in the constraint approach as $\mathbf{%
(p-k)}^{2}+m^{2}$. \ The treatment of the spin-dependent aspects of the wave
function in \cite{ack}, \cite{gupt}, and \cite{ebrt} is similar to that
appearing in our earlier paper on positronium decay in \cite{posit}. \ We
point out, however, that in our paper here, the spin dependent aspects of
the wave function do not arise from the free spinor factors in the Feynman
decay amplitude, but rather from the multicomponent structure of the
interacting TBDE. \ The treatment appearing in \cite{ebrt} uses a
quasipotential wave equation that gives a Schr\"{o}dinger-like equation for
the bound states. Their amplitude treatment is otherwise similar to that of 
\cite{ack} except that they include (as we do here) QCD radiative
corrections. Another treatment is that of \cite{hng}. \ They list decay
rates of 5.5 and 2.1 \textrm{keV }for the $\eta _{c}$ and $\eta _{c}^{\prime
}$ respectively, similar to the results of \cite{ebrt} (see also recent
result of \cite{land}). Their treatment of the spin-dependent aspects of the
wave function appearing is more like ours except that they use the Salpeter
truncation of the Bethe-Salpeter equation but with energy factors in the
amplitude treated on shell as in \cite{ack}. \ \ The treatment in \cite{munz}
is similar to that of \cite{hng} except that it involves the
four-dimensional Bethe-Salpeter amplitude constructed from the Salpeter
solution. \ As with \cite{hng} they use a combination of scalar and
time-like confining potentials. \ Unlike \cite{hng} and \cite{ebrt}, and
like the treatment of \cite{ack} and the present one, \cite{munz} then goes
on to treat the light quark pseudoscalar decays. For the $^{1}D_{2}(3872)$
particle, Ackleh and Barnes obtain a result of 20 \ \textrm{eV }close to
ours of 27 \ \textrm{eV. }None of the other authors include this particle.

\subsection{Light Quark Mesons}

Our formalism at this stage does not include the effects of flavor mixing
and consequently we do not \ compute the rates for $\eta ,\eta ^{\prime
}\rightarrow 2\gamma $. \ This leaves us with the 2 $\gamma $ decays of $\pi
^{0}$, $\pi _{2},a_{2},f_{2}^{\prime }$. \ We present the results of the
decay width in the table below, including those approaches above that give
predictions for some of these decays. The units are in \textrm{eV}. 
\begin{equation*}
\begin{tabular}{llllll}
& Expt. & TBDE-TBDA & TBDE-NA & \cite{ack} & \cite{munz} \\ 
$\pi _{0}(^{1}S_{0}-0.135)$ & 7.22$\pm $.04 & 8.73 & 33.5. & 3.4$\rightarrow 
$6.4 & 3.81,5.07 \\ 
$\pi _{2}(^{1}D_{2}-1.670)$ & $<$70,190 & 181 & 142 & 110$\rightarrow $270 & 
73.2,129 \\ 
$a_{2}(^{3}P_{2}-1.318)$ & 1000$\pm $60 & 9020 & 3120 & - & 766,900 \\ 
$f_{2}^{\prime }(^{3}P_{2}-1.525)$ & 81$\pm $10 & 760 & 348 & - & -%
\end{tabular}%
\end{equation*}%
\ Our pion rate is comparable to the others. However, the assumptions of 
\cite{ack} are quite different from ours. First they use a non-relativistic
potential model for the wave function. As pointed out by \cite{munz}, \cite%
{eft} , and \cite{gui} standard approaches to the pion decay fail miserably
for such models, typically too large by three orders of magnitude (by
comparison our result is only off by \ 20\%). \ Ackleh and Barnes, however,
included, as did Hayne and Isgur \cite{hne} in an earlier paper, a
phenomenological resonance mass factor motivated by an effective field
theory Lagrangian ($\sim \frac{1}{2}g\phi F_{\mu \nu }\tilde{F}_{\mu \nu }$)
which greatly suppresses the \textquotedblleft bare\textquotedblright\ rate.
\ The approach we have taken above did not include such a factor.\ Of
course, as they point out, the factor implied by that effective field theory
is not contained within the positronium-like model that they and we use. \
(The range of values in their column correspond to a \ range of assumed
constituent masses). The approach taken by M\"{u}nz \cite{munz} is much
closer in spirit to the one we employ. \ He uses the framework of the
Salpeter equation for the formulation of two photon decays and finds that
including relativistic effects, and the negative energy components of the
wave function, is important even for heavy quarkonia. \ In addition, unlike
our approach, which in the CM frame would have momentum space wave function
dependence only on the relative three momentum, he works out a decay matrix
element which includes relative four-momentum dependence (including relative
energy dependence in the CM). \ It is his claim that in this way, not only
does the amplitude depend on off-mass-shell annihilating quark pairs
(through the wave function) but also the exchanged quark within the diagram
that are both off mass shell and off energy shell. In contrast to our Eq. (%
\ref{3m}) his amplitude involves an additional integral over the relative
energy. However, he finds it necessary to introduce a cutoff factor in his
spectral analysis for the one-gluon exchange. \ In addition he finds that he
must assume not only a different confinement mechanism for the light and
heavy quarks, but different confinement strengths. \ The spectral results we
obtain do not treat the heavy and light quark bound states differently. \
Further, in the two models that he considers, he finds that it is not
possible to formulate the one gluon exchange gauge invariantly and so uses
the Feynman gauge in one parameter set for a semirelativistic model and the
Coulomb gauge in the other. By contrast, the constraint approach displays
gauge invariance, and, for simplicity uses the Feynman gauge.

There are other approaches that develop formalisms with natural suppression
of the $\pi ^{0}\rightarrow 2\gamma $ width in the quark model. Guisasu and
Koniuk \cite{gui}, using a multi-pair structure in the context of the
Bethe-Salpeter formalism, show how the extremely bound highly relativistic
nature of the pion suppress the decay rate. \ The authors of \cite{eft} also
show how the assumption of a completely diagonalized QCD Hamiltonian (with
meson eigenstates predominantly $q\bar{q}$), implies bound state effects can
greatly suppress the width.

\section{Concluding Remarks}

The Two-Body Dirac equations are based on Dirac's constraint formalism and a
minimal interaction structure for the effective particle of relative motion
(first used by Todorov) confirmed by both classical \cite{fw} and quantum
field theory \cite{saz97}. This formalism displays spectral results with
flavor independent interactions in very good experimental agreement for most
of the meson spectra. At the same time, and we have stressed the importance
of this in a recent publication \cite{crater2}, the formalism when treated
in a nonperturbative manner naturally accounts for the perturbative results
of QED bound states. So far this has not been fully replicated in any other
approach. In a natural way it leads not only to good singlet-triplet ground
state\ splittings for the light meson, but also a Goldstone behavior for the
pion. \ This we showed is tied to the same relativistic structures that
account for the nonperturbative positronium and muonium results. \ Based on
this and the successful off shell treatment of positronium two photon\ decay
we had reason to anticipate that the quarkonium decays to two photons would
be reasonable. \ We have found this to be particularly true for the $\pi
_{0} $ and $\eta _{c}$. There we found relativistic effects, including most
importantly the full multicomponent wave function and the influence of the
TBDN and TBDA, to be of crucial importance. \ The results compare favorably
with models based on two-body formalisms not tested as extensively as that
of the Two-Body Dirac equations. \ Our pion results are unlike some of the
competing approaches in that no additional effective field theory
assumptions were made that go beyond the relativistic potential model
approach, and spectral results for all mesons are obtained in a flavor
independent way. \ In light of this our results are not too unreasonable.
Still one may speculate on assumptions made in the constraint approach which
may be relaxed. \ For example, \ it may very well be that even though
spectral results are independent of the method by which the relative time is
controlled in the constraint formalism \cite{cra87}, the decay and other
amplitudes may depend on this effect. \ \ The work in \cite{cra87} (see
Appendix A in that paper), allows one to show how the relative energy
restriction Eq. (\ref{rle}) which in quantum form is $P\cdot p\psi =0$ or $%
\psi (p)=\delta (p\cdot \hat{P})\psi (p_{\perp })$ and this could be
replaced in the amplitude Eq. (\ref{98}) by a more general form, say 
\begin{equation*}
\psi (p)=\delta (p\cdot \hat{P})\psi (p_{\perp })\rightarrow \Delta (p\cdot 
\hat{P})\psi (p_{\perp })
\end{equation*}%
in which $\Delta $ is a distribution with non zero support \cite{zh}. \ In
future work, having shown that the meson wave functions of \cite{crater2}
used in this paper give in most circumstances reasonable results, \ one will
now consider applications of them to the meson-meson scattering process such
as discussed in the beginning of this paper.

\ 

The authors would like to thank Profs. T. Barnes and E. S. Swanson for
helpful discussions. They also thank Prof. Paul Vetter for information about
\ positronium decay experiments done with Gammasphere at LBL. This research
was supported by the National Science Foundation under Contract No.\
NSF-PHY-0244819 at the University of Tennessee Space Institute and by the
Division of Nuclear Physics, U.S.D.O.E., under Contract No.\
DE-AC05-00OR22725, managed by UT-Battelle, LLC and by the National Science
Foundation under contract NSF-Phy-0244786 at the University of Tennessee.

\appendix

\section{Dirac Matrices for the Two-Body Dirac Equations}

\begin{equation*}
\beta _{1}=%
\begin{bmatrix}
1_{8} & 0 \\ 
0 & -1_{8}%
\end{bmatrix}%
,\ \ \gamma _{51}=%
\begin{bmatrix}
0 & 1_{8} \\ 
1_{8} & 0%
\end{bmatrix}%
,\ \ \beta _{1}\gamma _{51}\equiv \rho _{1}=%
\begin{bmatrix}
0 & 1_{8} \\ 
-1_{8} & 0%
\end{bmatrix}%
,
\end{equation*}%
\begin{equation*}
\beta _{2}=%
\begin{bmatrix}
\beta & 0 \\ 
0 & \beta%
\end{bmatrix}%
,\ \beta =%
\begin{bmatrix}
1_{4} & 0 \\ 
0 & -1_{4}%
\end{bmatrix}%
,
\end{equation*}%
\begin{equation*}
\gamma _{52}=%
\begin{bmatrix}
\gamma _{5} & 0 \\ 
0 & \gamma _{5}%
\end{bmatrix}%
,\ \gamma _{5}=%
\begin{bmatrix}
0 & 1_{4} \\ 
1_{4} & 0%
\end{bmatrix}%
,
\end{equation*}%
\begin{equation*}
\beta _{2}\gamma _{52}\equiv \rho _{2}=%
\begin{bmatrix}
\rho & 0 \\ 
0 & \rho%
\end{bmatrix}%
,\ \rho =%
\begin{bmatrix}
0 & 1_{4} \\ 
-1_{4} & 0%
\end{bmatrix}%
,
\end{equation*}%
\begin{equation*}
\gamma _{51}\gamma _{52}=%
\begin{bmatrix}
0 & \gamma _{5} \\ 
\gamma _{5} & 0%
\end{bmatrix}%
,\rho _{1}\rho _{2}=%
\begin{bmatrix}
0 & \rho \\ 
-\rho & 0%
\end{bmatrix}%
\end{equation*}%
\begin{equation*}
\beta _{1}\gamma _{51}\gamma _{52}=%
\begin{bmatrix}
0 & \gamma _{5} \\ 
-\gamma _{5} & 0%
\end{bmatrix}%
,\beta _{2}\gamma _{52}\gamma _{51}=%
\begin{bmatrix}
0 & \rho \\ 
\rho & 0%
\end{bmatrix}%
,
\end{equation*}%
\begin{equation*}
\beta _{i}=-\gamma _{i}\cdot \hat{P},
\end{equation*}%
\begin{equation}
\Sigma _{i}=\gamma _{5i}\beta _{i}\gamma _{\perp i}.
\end{equation}

\section{Relations Between Four Component Dirac Subspinors}

We rewrite the TBDE Eqs. (\ref{ctbde}) in terms of the subspinors 
\begin{eqnarray}
\phi _{\pm } &=&\psi _{1}\pm \psi _{4},  \notag \\
\chi _{\pm } &=&\psi _{2}\pm \psi _{3}.
\end{eqnarray}%
With Appendix A these equations then lead to 
\begin{eqnarray}
D_{1}^{++}\phi _{+} &=&E_{1}\chi _{+}-M_{1}\chi _{-},  \notag \\
-D_{2}^{++}\phi _{+} &=&E_{2}\chi _{+}+M_{2}\chi _{-},
\end{eqnarray}%
and so 
\begin{eqnarray}
\chi _{+} &=&{\frac{1}{\mathcal{D}}}(M_{2}D_{1}^{++}-M_{1}D_{2}^{++})\phi
_{+},  \notag \\
\chi _{-} &=&-{\frac{1}{\mathcal{D}}}(E_{2}D_{1}^{++}+E_{1}D_{2}^{++})\phi
_{+},  \label{fg}
\end{eqnarray}%
and 
\begin{eqnarray}
D_{1}^{-+}\chi _{+} &=&E_{1}\phi _{+}-M_{1}\phi _{-},  \notag \\
D_{1}^{--}\chi _{-} &=&M_{1}\phi _{+}-E_{1}\phi _{-},  \label{gf} \\
-D_{2}^{-+}\chi _{+} &=&E_{2}\phi _{+}-M_{2}\phi _{-},  \notag \\
-D_{2}^{--}\chi _{-} &=&-M_{2}\phi _{+}+E_{2}\phi _{-},  \notag
\end{eqnarray}%
with the $D_{i}^{\pm \pm }$ given in Eqs. (\ref{d1}) and(\ref{d2}). \ These
lead directly to Eqs. (\ref{chip}) and (\ref{chim}). \ From Eq. (\ref{gf})
we \ find using Eqs. (\ref{chip}) and (\ref{chim}) leads to \ ~ Eq. (\ref{54}%
).

Using Eqs. (\ref{chip}), (\ref{chim}) and (\ref{fch}),(\ref{smtrx}) we
obtain 
\begin{eqnarray}
\chi _{+} &=&\frac{\exp (\mathcal{G)}}{\mathcal{D}}\{M_{2}[%
\boldsymbol{\sigma}_{1}\cdot \mathbf{p}-\frac{i}{2}\boldsymbol{\sigma}%
_{2}\cdot \boldsymbol{\nabla }(J-L+\mathcal{G}\boldsymbol{\sigma}_{1}\cdot %
\boldsymbol{\sigma}_{2})]  \notag \\
&&-M_{1}[\boldsymbol{\sigma}_{2}\cdot \mathbf{p}-\frac{i}{2}%
\boldsymbol{\sigma}_{1}\cdot \boldsymbol{\nabla }(J-L+\mathcal{G}%
\boldsymbol{\sigma}_{1}\cdot \boldsymbol{\sigma}_{2})]\}\phi _{+}  \notag \\
&\rightarrow &\frac{\exp (\mathcal{G)}}{\mathcal{D}}\{(M_{2}+M_{1})[(\mathbf{%
\ p}+\frac{i}{2}\boldsymbol{\nabla }(J-L+\mathcal{G}))\cdot \boldsymbol{\phi
}_{+}1  \notag \\
&&+(\mathbf{p}+\frac{i}{2}\boldsymbol{\nabla }(J-L-3\mathcal{G}))\cdot %
\boldsymbol{\sigma}\mathcal{\phi }_{+0}]+(M_{2}-M_{1})[i\mathbf{p}+\frac{1}{2%
}\boldsymbol{\nabla }(J-L+\mathcal{G})]\times \boldsymbol{\phi }_{+}\cdot %
\boldsymbol{\sigma}  \notag \\
&\equiv &\mathcal{\chi }_{+0}1+\boldsymbol{\chi }_{+}\cdot \sigma .
\end{eqnarray}%
Similarly 
\begin{eqnarray}
\chi _{-} &=&-\frac{\exp (\mathcal{G)}}{\mathcal{D}}\{E_{2}[%
\boldsymbol{\sigma}_{1}\cdot \mathbf{p}-\frac{i}{2}\boldsymbol{\sigma}%
_{2}\cdot \boldsymbol{\nabla }(J-L+\mathcal{G}\boldsymbol{\sigma}_{1}\cdot %
\boldsymbol{\sigma}_{2})]  \notag \\
&&+E_{1}[\boldsymbol{\sigma}_{2}\cdot \mathbf{p}-\frac{i}{2}%
\boldsymbol{\sigma}_{1}\cdot \boldsymbol{\nabla }(J-L+\mathcal{G}%
\boldsymbol{\sigma}_{1}\cdot \boldsymbol{\sigma}_{2})]\}\phi _{+}  \notag \\
&\rightarrow &-\frac{\exp (\mathcal{G)}}{\mathcal{D}}\{(E_{2}-E_{1})[(%
\mathbf{p}+\frac{i}{2}\boldsymbol{\nabla }(J-L+\mathcal{G}))\cdot %
\boldsymbol{\phi }_{+}1  \notag \\
&&+(\mathbf{p}+\frac{i}{2}\boldsymbol{\nabla }(J-L-3\mathcal{G}))\mathcal{\
\phi }_{+0}\cdot \boldsymbol{\sigma}\mathbf{]}+(E_{2}+E_{1})[i\mathbf{p}+%
\frac{1}{2}\boldsymbol{\nabla }(J-L+\mathcal{G})]\times \boldsymbol{\phi }%
_{+}\cdot \sigma \}  \notag \\
&\equiv &\mathcal{\chi }_{-0}1+\boldsymbol{\chi }_{-}\cdot %
\boldsymbol{\sigma}.
\end{eqnarray}

For equal mass and electromagnetic-like interactions ($J=-\mathcal{G}$) we
find 
\begin{eqnarray}
3\mathcal{G}^{\prime }-L^{\prime } &\mathcal{=}&\mathcal{-}2F^{\prime }, 
\notag \\
-(\mathcal{G+}L)^{\prime } &=&-2K^{\prime },
\end{eqnarray}
and 
\begin{equation}
\chi _{+}\rightarrow \frac{\exp (\mathcal{G)}}{E}\{(\mathbf{p}-\frac{i}{2} %
\boldsymbol{\nabla }L)\cdot \boldsymbol{\phi }_{+}1+[\mathbf{p}-\frac{i}{2} 
\mathbf{\nabla }(L+4\mathcal{G})]\cdot \boldsymbol{\sigma}\mathcal{\phi }
_{+0}]\},
\end{equation}
so that with the use of Eq. (\ref{sie}) 
\begin{eqnarray}
\mathcal{\chi }_{0+} &=&\exp (F-K)\mathcal{\eta }_{+0}=\frac{\exp (\mathcal{%
\ G)}}{E}(\mathbf{p}-\frac{i}{2}\boldsymbol{\nabla }L)\cdot \boldsymbol{\phi
} _{+}  \notag \\
&=&\frac{\exp (\mathcal{G}+F+K\mathcal{)}}{E}[\mathbf{p}-\frac{i}{2}\mathbf{%
\ \nabla }(L+2F+2K)]\cdot \lbrack \mathbf{1}+Q_{m}\mathbf{\hat{r}\hat{r}}%
\cdot ]\boldsymbol{\psi }_{+}  \notag \\
~\ \boldsymbol{\chi }_{+} &=&\exp F(\exp K+[\exp (-K)-\exp (K)]\mathbf{\hat{%
r }\hat{r}}\cdot )\boldsymbol{\eta }_{+}  \notag \\
&=&\frac{\exp (\mathcal{G)}}{E}\{[\mathbf{p}-\frac{i}{2}\boldsymbol{\nabla }
(L+4\mathcal{G})]\mathcal{\phi }_{+0}\}  \notag \\
&=&\frac{\exp (F-K+\mathcal{G)}}{E}\{[\mathbf{p}-\frac{i}{2} %
\boldsymbol{\nabla }(L+4\mathcal{G+}2F-2K)]\mathcal{\psi }_{+0}\},
\end{eqnarray}
and so inverting the exponentials and dyads we find Eq. (\ref{etap}).
Similarly, in this case, 
\begin{equation}
\chi _{-}\rightarrow -\frac{\exp (\mathcal{G)}}{M}(i\mathbf{p}-\frac{1}{2} %
\boldsymbol{\nabla }L)\times \boldsymbol{\phi }_{+}\cdot \boldsymbol{\sigma},
\end{equation}
so that 
\begin{eqnarray}
\mathcal{\chi }_{0-} &=&0  \notag \\
\boldsymbol{\chi }_{-} &=&\exp (F+K)(\mathbf{1}-(1-\exp (-2K))\mathbf{\hat{r}
\hat{r}})\cdot \boldsymbol{\eta }_{-}  \notag \\
&=&-\frac{\exp (\mathcal{G+}F+K\mathcal{)}}{M}[i\mathbf{p}-\frac{1}{2} %
\boldsymbol{\nabla }(L-2F-2K)]\times (\mathbf{1}+Q_{m}\mathbf{\hat{r}\hat{r}}
)\cdot \boldsymbol{\psi }_{+},
\end{eqnarray}
giving us Eq. (\ref{etam}) again by inverting the exponentials and dyads.

Next we work with Eq. (\ref{54}). \ We transform this equation involving
relations between spinors into one involving relations between the scalar
and vector wave functions $\mathcal{\psi }_{-0},\boldsymbol{\psi }_{-}$ and $
\mathcal{\psi }_{+0},\boldsymbol{\psi }_{+}$ . \ We \ begin by transforming
the portion 
\begin{eqnarray}
&&(E_{2}D_{1}^{-+}-E_{1}D_{2}^{-+})\chi _{+}  \notag \\
&=&\exp (\mathcal{G)}\{E_{2}[\boldsymbol{\sigma}_{1}\cdot \mathbf{p}+\frac{i 
}{2}\boldsymbol{\sigma}_{2}\cdot \boldsymbol{\nabla }(-J-L-\mathcal{G} %
\boldsymbol{\sigma}_{1}\cdot \boldsymbol{\sigma}_{2})]  \notag \\
&&-E_{1}[\boldsymbol{\sigma}_{2}\cdot \mathbf{p}+\frac{i}{2} %
\boldsymbol{\sigma}_{1}\cdot \boldsymbol{\nabla }(-J-L-\mathcal{G} %
\boldsymbol{\sigma}_{1}\cdot \boldsymbol{\sigma}_{2})]\}\chi _{+}  \notag \\
&\rightarrow &\exp (\mathcal{G)}\{(E_{2}+E_{1})[(\mathbf{p}+\frac{i}{2} %
\boldsymbol{\nabla }(J+L+\mathcal{G}))\cdot \boldsymbol{\chi }_{+}1  \notag
\\
&&+(\mathbf{p}+\frac{i}{2}\boldsymbol{\nabla }(J+L-3\mathcal{G}))\mathcal{\
\chi }_{+0}\mathbf{\cdot }\boldsymbol{\sigma}]+(E_{2}-E_{1})[i\mathbf{p}+ 
\frac{1}{2}\boldsymbol{\nabla }(J+L+\mathcal{G})]\times \boldsymbol{\chi }
_{+}\mathbf{\cdot }\boldsymbol{\sigma}\},
\end{eqnarray}
and then next 
\begin{eqnarray}
&&(M_{2}D_{1}^{--}+M_{1}D_{2}^{--})\chi _{-}  \notag \\
&=&\exp (\mathcal{G)}\{M_{2}[\boldsymbol{\sigma}_{1}\cdot \mathbf{p}+\frac{i 
}{2}\boldsymbol{\sigma}_{2}\cdot \boldsymbol{\nabla }(J-L-\mathcal{G} %
\boldsymbol{\sigma}_{1}\cdot \boldsymbol{\sigma}_{2})]  \notag \\
&&+M_{1}[\boldsymbol{\sigma}_{2}\cdot \mathbf{p}+\frac{i}{2} %
\boldsymbol{\sigma}_{1}\cdot \boldsymbol{\nabla }(J-L-\mathcal{G} %
\boldsymbol{\sigma}_{1}\cdot \boldsymbol{\sigma}_{2})]\}\chi _{-}  \notag \\
&\rightarrow &\exp (\mathcal{G)}\{(M_{2}-M_{1})[(\mathbf{p}+\frac{i}{2} %
\boldsymbol{\nabla }(-J+L+\mathcal{G}))\cdot \boldsymbol{\chi }_{-}1  \notag
\\
&&+(\mathbf{p}+\frac{i}{2}\boldsymbol{\nabla }(-J+L-3\mathcal{G}))\mathcal{\
\chi }_{+0}\mathbf{\cdot }\boldsymbol{\sigma}]+(M_{2}+M_{1})[i\mathbf{p}+ 
\frac{1}{2}\boldsymbol{\nabla }(-J+L+\mathcal{G})]\times \boldsymbol{\chi }
_{-}\mathbf{\cdot }\boldsymbol{\sigma}\}.
\end{eqnarray}
Thus Eq. (\ref{54}) becomes 
\begin{eqnarray}
\mathcal{\phi }_{-} &=&\frac{(E_{2}E_{1}+M_{2}M_{1})}{\mathcal{D}}\mathcal{\
\phi }_{+}  \notag \\
&&-\frac{\exp (\mathcal{G)}}{2\mathcal{D}}\{(E_{2}+E_{1})[(\mathbf{p}+\frac{%
i }{2}\boldsymbol{\nabla }(J+L+\mathcal{G}))\cdot \boldsymbol{\chi }%
_{+}\sigma _{0}  \notag \\
&&+(\mathbf{p}+\frac{i}{2}\boldsymbol{\nabla }(J+L-3\mathcal{G}))\mathcal{\
\chi }_{+0}\mathbf{\cdot }\boldsymbol{\sigma}]+(E_{2}-E_{1})[i\mathbf{p}+ 
\frac{1}{2}\boldsymbol{\nabla }(J+L+\mathcal{G})]\times \boldsymbol{\chi }
_{+}\mathbf{\cdot }\boldsymbol{\sigma}  \notag \\
&&-(M_{2}-M_{1})[(\mathbf{p}+\frac{i}{2}\boldsymbol{\nabla }(-J+L+\mathcal{G}
))\cdot \boldsymbol{\chi }_{-}\sigma _{0}  \notag \\
&&+(\mathbf{p}+\frac{i}{2}\boldsymbol{\nabla }(-J+L-3\mathcal{G}))\mathcal{\
\chi }_{+0}\mathbf{\cdot }\boldsymbol{\sigma}]-(M_{2}+M_{1})[i\mathbf{p}+ 
\frac{1}{2}\boldsymbol{\nabla }(-J+L+\mathcal{G})]\times \boldsymbol{\chi }
_{-}\mathbf{\cdot }\boldsymbol{\sigma}\}.  \label{fim}
\end{eqnarray}
For $\mathcal{G=-}J$ and equal mass the matrix wave function $\mathcal{\phi }
_{-}$ \ becomes, with the use of Eq. (\ref{mspn}), 
\begin{eqnarray}
\mathcal{\phi }_{-} &=&\frac{(E^{2}+M^{2})}{2EM}\{\exp (F-K)\mathcal{\psi }
_{+0}\sigma _{0}+\exp (F+K)[(\mathbf{1}-Q_{m}\mathbf{\hat{r}\hat{r}})\cdot %
\boldsymbol{\psi }_{+}]\cdot \boldsymbol{\sigma}\}  \notag \\
&&-\frac{\exp (2\mathcal{G+}F-K\mathcal{)}}{2ME}[\mathbf{p}+\frac{i}{2} %
\boldsymbol{\nabla }(L-4\mathcal{G}-2F+2K)]\cdot \lbrack \mathbf{p}-\frac{i}{
2}\boldsymbol{\nabla }(L+4\mathcal{G+}2F-2K)]\mathcal{\psi }_{+0}\sigma _{0}
\notag \\
&&-\frac{\exp (F+K+2\mathcal{G})}{2ME}[(\mathbf{p}+\frac{i}{2}\mathbf{\nabla 
}(L-2F-2K-8\mathcal{G}))([\mathbf{p}-\frac{i}{2}\boldsymbol{\nabla }
(L+2F+2K))]  \notag \\
&&\cdot \lbrack \mathbf{1}+Q_{m}\mathbf{\hat{r}\hat{r}}\cdot ] %
\boldsymbol{\psi }_{+})]\cdot \boldsymbol{\sigma}  \notag \\
&&+\frac{\exp (2\mathcal{G}+F+K)}{2EM}[i\mathbf{p}+\frac{1}{2}\mathbf{\nabla 
}(-L+4\mathcal{G+}2F+2K)]\times \{[i\mathbf{p}-\frac{1}{2}\boldsymbol{\nabla
}(L-2F-2K)]  \notag \\
&&\times \lbrack 1+Q_{m}\mathbf{\hat{r}\hat{r}}]\cdot \boldsymbol{\psi }
_{+}\}\cdot \boldsymbol{\sigma}  \notag \\
&=&\exp (F-K)\mathcal{\psi }_{-0}\sigma _{0}+\exp (F+K)[(\mathbf{1}+Q_{m} 
\mathbf{\hat{r}\hat{r}})\cdot \boldsymbol{\psi }_{-}]\cdot %
\boldsymbol{\sigma},  \label{wow}
\end{eqnarray}
and so 
\begin{eqnarray}
\mathcal{\psi }_{-0} &=&\{\frac{(E^{2}+M^{2})}{2EM}-\frac{\exp (2\mathcal{G)}
}{2ME}[\mathbf{p}+\frac{i}{2}\boldsymbol{\nabla }(L-4\mathcal{G}
-2F+2K)]\cdot \lbrack \mathbf{p}-\frac{i}{2}\boldsymbol{\nabla }(L+4\mathcal{%
\ G+}2F-2K)\}\mathcal{\psi }_{+0}  \notag \\
\boldsymbol{\psi }_{-} &=&\frac{(E^{2}+M^{2})}{2EM}\boldsymbol{\psi }_{+} 
\notag \\
&&+\frac{\exp (2\mathcal{G})}{2EM}[\mathbf{1}+Q_{p}\mathbf{\hat{r}\hat{r}}
]\cdot {\biggl (-}[\mathbf{p}+\frac{i}{2}\boldsymbol{\nabla }(L-8\mathcal{G-}
2F-2K)][\mathbf{p}-\frac{i}{2}\boldsymbol{\nabla }(L+2F+2K)]  \notag \\
&&\cdot \lbrack \mathbf{1}+Q_{m}\mathbf{\hat{r}\hat{r}}]\cdot %
\boldsymbol{\psi }_{+}  \notag \\
&&-[\mathbf{p}-\frac{i}{2}\boldsymbol{\nabla }(-L+4\mathcal{G+}2F+2K)]\times
\{[\mathbf{p}+\frac{i}{2}\boldsymbol{\nabla }(L-2F-2K)]\times \lbrack 
\mathbf{1}+Q_{m}\mathbf{\hat{r}\hat{r}}]\cdot \boldsymbol{\psi }_{+}\} {%
\biggr ).}  \label{oh}
\end{eqnarray}
Using 
\begin{eqnarray}
2K-2F &=&4\mathcal{G},  \notag \\
2K+2F &=&2L-2\mathcal{G},
\end{eqnarray}
then gives us Eq. (\ref{psim}).

As a check on our formalism consider the case of no potential, so that for
the singlet we have 
\begin{equation}
\Psi |_{s=0}=\frac{1}{2\sqrt{2}}(\psi _{+0}\sigma _{0}q_{1}+\psi _{-0}\sigma
_{0}iq_{2}+\boldsymbol{\eta }_{+}\cdot \boldsymbol{\sigma}q_{0}).
\end{equation}
Using the results in Eqs. (\ref{etap}), (\ref{etam}), and (\ref{psim}) we
obtain 
\begin{eqnarray*}
\psi _{-0} &=&\frac{m}{\varepsilon }\psi _{+0}, \\
\boldsymbol{\eta }_{+} &=&\frac{1}{\varepsilon }\mathbf{p}\psi _{+0},
\end{eqnarray*}
The total matrix wave function is 
\begin{equation}
\Psi |_{s=0}=\frac{1}{2\sqrt{2}}\psi _{+0}(\sigma _{0}q_{1}+\frac{m}{
\varepsilon }\sigma _{0}iq_{2}+\frac{1}{\varepsilon }\mathbf{p}\cdot %
\boldsymbol{\sigma}q_{0}).  \label{sng}
\end{equation}
We compare that with the free wave function 
\begin{equation}
\frac{1}{\sqrt{2}}[u^{(s_{-})}(\mathbf{p})\bar{v}^{(s_{+})}(-\mathbf{p)}
-u^{(s_{+})}(\mathbf{p})\bar{v}^{(s_{-})}(-\mathbf{p)],}
\end{equation}
in which 
\begin{eqnarray}
u^{(s_{-})}(\mathbf{p}) &=&\frac{(\varepsilon +m+\boldsymbol{\alpha}\mathbf{%
\ \cdot p)}}{\sqrt{2m(\varepsilon +m)}} 
\begin{bmatrix}
0 \\ 
1 \\ 
0 \\ 
0%
\end{bmatrix}
,  \notag \\
u^{(s_{+})}(\mathbf{p}) &=&\frac{(\varepsilon +m+\boldsymbol{\alpha}\mathbf{%
\ \cdot p)}}{\sqrt{2m(\varepsilon +m)}} 
\begin{bmatrix}
1 \\ 
0 \\ 
0 \\ 
0%
\end{bmatrix}
,  \notag \\
\bar{v}^{(s_{+})}(-\mathbf{p)} &\mathbf{=}& 
\begin{bmatrix}
0 & 0 & 0 & -1%
\end{bmatrix}
\frac{(\varepsilon +m+\boldsymbol{\alpha}\mathbf{\cdot p)}}{\sqrt{
2m(\varepsilon +m)}},  \notag \\
\bar{v}^{(s_{-})}(-\mathbf{p)} &\mathbf{=}& 
\begin{bmatrix}
0 & 0 & 1 & 0%
\end{bmatrix}
\frac{(\varepsilon +m+\boldsymbol{\alpha}\mathbf{\cdot p)}}{\sqrt{
2m(\varepsilon +m)}}.
\end{eqnarray}
Thus, we find 
\begin{eqnarray}
&&\frac{1}{\sqrt{2}}[u^{(s_{-})}(\mathbf{p})\bar{v}^{(s_{+})}(-\mathbf{p)}
-u^{(s_{+})}(\mathbf{p})\bar{v}^{(s_{-})}(-\mathbf{p)]}  \notag \\
&=&\frac{1}{2\sqrt{2}m(\varepsilon +m)}(\varepsilon +m+q_{1}\sigma \mathbf{\
\cdot p)} 
\begin{bmatrix}
0 & 0 & -1 & 0 \\ 
0 & 0 & 0 & -1 \\ 
0 & 0 & 0 & 0 \\ 
0 & 0 & 0 & 0%
\end{bmatrix}
(\varepsilon +m+q_{1}\sigma \mathbf{\cdot p)}  \notag \\
&=&\frac{1}{2\sqrt{2}m(\varepsilon +m)}(\varepsilon +m+q_{1}\sigma \mathbf{\
\cdot p)}(-\frac{1}{2})(q_{1}+iq_{2})(\varepsilon +m+q_{1}\sigma \mathbf{\
\cdot p)}  \notag \\
&=&-\frac{\varepsilon }{2\sqrt{2}m}[q_{1}+\frac{miq_{2}}{\varepsilon }+\frac{
\sigma \mathbf{\cdot p}}{\varepsilon }],
\end{eqnarray}
which has the same form as in Eq. (\ref{sng}).

For the triplet case things are a bit more complex. For the triplet we have 
\begin{equation}
\Psi |_{s=1}=\frac{1}{2\sqrt{2}}(\boldsymbol{\psi }_{+}\mathbf{\cdot }%
\boldsymbol{\sigma}q_{1}+\boldsymbol{\psi }_{-}\mathbf{\cdot }%
\boldsymbol{\sigma}iq_{2}+\eta _{+0}q_{0}\boldsymbol{+\eta }_{-}\mathbf{%
\cdot }\boldsymbol{\sigma}q_{3}),
\end{equation}%
in which from (\ref{psim}) 
\begin{equation}
\boldsymbol{\psi }_{-}=\frac{(\varepsilon ^{2}+m^{2})}{2\varepsilon m}%
\boldsymbol{\psi }_{+}-\frac{1}{2\varepsilon m}{\biggl (}\mathbf{pp}\cdot %
\boldsymbol{\psi }_{+}+\mathbf{p}\times (\mathbf{p}\times \boldsymbol{\psi }%
_{+}){\biggr ),}
\end{equation}%
or 
\begin{eqnarray}
\boldsymbol{\psi }_{-} &=&\frac{1}{2\varepsilon m}[(\varepsilon ^{2}+m^{2})+%
\mathbf{p}^{2}{\large -}2\mathbf{pp}\cdot ]\boldsymbol{\psi }_{+}  \notag \\
&=&\frac{\varepsilon }{m}\boldsymbol{\psi }_{+}-\frac{1}{\varepsilon m}%
\mathbf{pp}\cdot \boldsymbol{\psi }_{+},
\end{eqnarray}%
and from (\ref{etam}) 
\begin{equation}
\boldsymbol{\eta }_{-}=-\frac{1}{m}i\mathbf{p}\times \boldsymbol{\psi }_{+},
\end{equation}%
and (\ref{etap}) 
\begin{equation*}
\mathcal{\eta }_{+0}=\frac{1}{\varepsilon }\mathbf{p}\cdot \boldsymbol{\psi }%
_{+},
\end{equation*}%
and thus 
\begin{equation}
\Psi |_{s=1}=\frac{\boldsymbol{\psi }_{+}}{2\sqrt{2}}\mathbf{\cdot }[%
\boldsymbol{\sigma}q_{1}+\mathbf{(}\frac{\varepsilon }{m}-\frac{1}{%
\varepsilon m}\mathbf{pp}\cdot )\boldsymbol{\sigma}iq_{2}+\frac{1}{%
\varepsilon }\mathbf{p}q_{0}+\frac{1}{m}i\mathbf{p}\times \boldsymbol{\sigma}%
q_{3}].  \label{psi}
\end{equation}%
Compare this with the free spinor solutions as in the singlet case. 
\begin{eqnarray}
u^{(s_{1})}(\mathbf{p)}\bar{v}^{(s_{2})}(-\mathbf{p)} &\mathbf{=}&\mathbf{(}%
\varepsilon +m+\boldsymbol{\alpha}\mathbf{\cdot p)}u^{(s_{2})}(\mathbf{0)}%
u^{(s_{2})}(\mathbf{0)}^{T}i\alpha _{y}\mathbf{(}\varepsilon +m+%
\boldsymbol{\alpha}\mathbf{\cdot p)}  \notag \\
&=&\mathbf{(}\varepsilon +m+q_{1}\boldsymbol{\sigma}\mathbf{\cdot p)}%
u^{(s_{1})}(\mathbf{0)}u^{(s_{2})}(\mathbf{0)}^{T}iq_{1}\sigma
_{2}(\varepsilon +m+q_{1}\boldsymbol{\sigma}\mathbf{\cdot p),}
\end{eqnarray}%
in which 
\begin{equation}
u^{(s)}(\mathbf{0)}^{T}i\alpha _{y}=%
\begin{bmatrix}
\chi ^{T(s)} & 0%
\end{bmatrix}%
\begin{bmatrix}
0 & i\sigma _{y} \\ 
i\sigma _{y} & 0%
\end{bmatrix}%
=%
\begin{bmatrix}
0 & \chi ^{T(s)}i\sigma _{y}%
\end{bmatrix}%
.
\end{equation}%
Thus 
\begin{equation}
u^{(s_{1})}(\mathbf{p)}\bar{v}^{(s_{2})}(-\mathbf{p)=(}\varepsilon +m+q_{1}%
\boldsymbol{\sigma}\mathbf{\cdot p)}%
\begin{bmatrix}
\chi ^{(s_{1})} \\ 
0%
\end{bmatrix}%
\begin{bmatrix}
0 & \chi ^{T(s_{2})}i\sigma _{y}%
\end{bmatrix}%
(\varepsilon +m+q_{1}\boldsymbol{\sigma}\mathbf{\cdot p).}
\end{equation}%
Thus, 
\begin{eqnarray}
&&\frac{1}{\sqrt{2}}[u^{(s_{-})}(\mathbf{p})\bar{v}^{(s_{+})}(-\mathbf{p)}%
+u^{(s_{+})}(\mathbf{p})\bar{v}^{(s_{-})}(-\mathbf{p)]}  \notag \\
&=&\frac{1}{4\sqrt{2}m(\varepsilon +m)}(\varepsilon +m+q_{1}%
\boldsymbol{\sigma}\mathbf{\cdot p)(}q_{1}+iq_{2}\mathbf{)}\sigma
_{3}(\varepsilon +m+q_{1}\boldsymbol{\sigma}\mathbf{\cdot p),}
\end{eqnarray}%
and 
\begin{eqnarray}
&&u^{(s_{-})}(\mathbf{p})\bar{v}^{(s_{-})}(-\mathbf{p)}  \notag \\
&=&\frac{1}{4m(\varepsilon +m)}(\varepsilon +m+q_{1}\sigma \mathbf{\cdot p)(}%
q_{1}+iq_{2}\mathbf{)(}\sigma _{1}-i\sigma _{2})(\varepsilon +m+q_{1}%
\boldsymbol{\sigma}\mathbf{\cdot p),}
\end{eqnarray}%
and 
\begin{eqnarray}
&&u^{(s_{+})}(\mathbf{p})\bar{v}^{(s_{+})}(-\mathbf{p)}  \notag \\
&=&\frac{1}{4m(\varepsilon +m)}(\varepsilon +m+q_{1}\sigma \mathbf{\cdot p)(}%
q_{1}+iq_{2}\mathbf{)(-}\sigma _{1}-i\sigma _{2})(\varepsilon +m+q_{1}%
\boldsymbol{\sigma}\mathbf{\cdot p),}
\end{eqnarray}%
or in general 
\begin{eqnarray}
&&\frac{1}{4\sqrt{2}m(\varepsilon +m)}(\varepsilon +m+q_{1}\sigma \mathbf{\
\cdot p)(}q_{1}+iq_{2}\mathbf{)}\boldsymbol{\phi }\mathbf{_{+}\mathbf{\cdot }%
}\boldsymbol{\sigma}(\varepsilon +m+q_{1}\boldsymbol{\sigma}\mathbf{\cdot p)}
\notag \\
&=&\frac{1}{2\sqrt{2}}\{q_{1}\boldsymbol{\phi }\mathbf{_{+}\mathbf{\cdot }}%
\boldsymbol{\sigma}\mathbf{+}\frac{\varepsilon }{m}iq_{2}\boldsymbol{\phi }%
\mathbf{_{+}\mathbf{\cdot }}\boldsymbol{\sigma}+\frac{1}{m}\boldsymbol{\phi }%
\mathbf{_{+}\mathbf{\cdot }}[\mathbf{p}-iq_{3}\boldsymbol{\sigma}\mathbf{\
\times p]}+\frac{\mathbf{(}q_{1}-iq_{2}\mathbf{)}}{m(\varepsilon +m)}%
\boldsymbol{\phi }\mathbf{_{+}\mathbf{\cdot }p}\boldsymbol{\sigma}\mathbf{%
\cdot p\}.}  \label{phi}
\end{eqnarray}%
Consider the special case of particle and antiparticle at rest. \ In that
case the above reduces to 
\begin{equation}
\frac{\mathbf{(}q_{1}+iq_{2}\mathbf{)}\boldsymbol{\phi }\mathbf{_{+}\mathbf{%
\ \cdot }}\boldsymbol{\sigma}}{2\sqrt{2}}.
\end{equation}%
This agrees in essential terms with 
\begin{equation}
\Psi |_{s=1}=\frac{\boldsymbol{\psi }_{+}\mathbf{\cdot }\boldsymbol{\sigma}}{%
2\sqrt{2}}(q_{1}+iq_{2}),
\end{equation}%
provided that in that limit 
\begin{equation}
\boldsymbol{\phi }\mathbf{_{+}=}\boldsymbol{\psi}_{+}.  \label{fs}
\end{equation}%
But what about the general case? One can show that letting 
\begin{equation}
\boldsymbol{\psi }_{+}=\frac{\varepsilon m}{(\varepsilon +m)m}(b\mathbf{\
1\cdot -}a\mathbf{pp\cdot +}c\mathbf{p\times )}\boldsymbol{\phi}\mathbf{_{+},%
}
\end{equation}%
with 
\begin{eqnarray}
c &=&\frac{i(q_{0}-q_{3})\left( q_{1}m+iq_{2}\varepsilon \right) }{\left(
\varepsilon +m\right) \varepsilon },  \notag \\
b &=&2-\frac{(\varepsilon -m)}{\varepsilon }q_{3},  \notag \\
a &=&-\frac{\varepsilon q_{0}+mq_{3}}{m\varepsilon (\varepsilon +m)},
\end{eqnarray}%
the expressions (\ref{phi}) and (\ref{psi}) are equivalent. \ 

\section{Interaction Dependent Modifications of the Norm}

In order to obtain the intereraction dependent modification%
\begin{equation}
\int d^{3}xTr[\psi ^{\dag }(1+4w^{2}\beta _{1}\beta _{2}\frac{\partial
\Delta }{\partial w^{2}})\psi ]=1
\end{equation}%
of Eq. (\ref{nrm2}) we first need the matrix connection 
\begin{eqnarray}
\psi &=&%
\begin{bmatrix}
\psi _{1} \\ 
\psi _{2} \\ 
\psi _{3} \\ 
\psi _{4}%
\end{bmatrix}%
=\frac{1}{2}(\beta _{1}+\gamma _{51}\gamma _{52})%
\begin{bmatrix}
\phi _{+} \\ 
\chi _{+} \\ 
\chi _{-} \\ 
\phi _{-}%
\end{bmatrix}
\notag \\
&=&\frac{1}{2}(\beta _{1}+\gamma _{51}\gamma _{52})\exp F(\cosh K+\sinh K%
\boldsymbol{\Sigma}_{1}\mathbf{\cdot \hat{r}}\boldsymbol{\Sigma}_{2}\mathbf{%
\cdot \hat{r}})%
\begin{bmatrix}
\psi _{+} \\ 
\eta _{+} \\ 
\eta _{-} \\ 
\psi _{-}%
\end{bmatrix}
\notag \\
&\equiv &\mathcal{L}_{0}%
\begin{bmatrix}
\psi _{+} \\ 
\eta _{+} \\ 
\eta _{-} \\ 
\psi _{-}%
\end{bmatrix}%
,  \label{abv1}
\end{eqnarray}%
between the Dirac spinor solutions of Eqs. (\ref{tbde}) and those of (\ref%
{57}). \ 

The transformation between the 16 component column vector form of the wave
function that satisfies our quasipotential equation (\ref{57}) and the one
which satisfies the Two-Body Dirac equation in hyperbolic form is given in
Eq.(\ref{abv1}). \ The corresponding 4x4 matrix form is%
\begin{eqnarray}
\mathcal{\psi } &\mathcal{=}&\frac{\exp (F)}{2\sqrt{2}}[\cosh K(\mathcal{%
\psi }_{+}q_{1}+\mathcal{\psi }_{-}iq_{2}+\eta _{+}q_{0}\mathbf{+}\eta
_{-}q_{3})  \notag \\
&&-\sinh K\boldsymbol{\Sigma}\cdot \mathbf{\hat{r}}(\mathcal{\psi }_{+}q_{1}+%
\mathcal{\psi }_{-}iq_{2}+\eta _{+}q_{0}\mathbf{+}\eta _{-}q_{3})%
\boldsymbol{\Sigma}\cdot \mathbf{\hat{r}]}  \notag \\
&\equiv &\mathcal{K}\Psi (\mathbf{r)}  \label{ksi}
\end{eqnarray}%
where%
\begin{equation}
\Psi (\mathbf{r)}\mathcal{=}\frac{1}{2\sqrt{2}}(\mathcal{\psi }_{+}q_{1}+%
\mathcal{\psi }_{-}iq_{2}+\eta _{+}q_{0}\mathbf{+}\eta _{-}q_{3}).
\end{equation}%
Whereas the normalization condition (\ref{nrm2}) in 16 component form is 
\begin{eqnarray}
&&\int d^{3}x[\psi ^{\dag }(1+4w^{2}\beta _{1}\beta _{2}\frac{\partial
\Delta }{\partial w^{2}})\psi ]  \notag \\
&=&\int d^{3}x\{%
\begin{bmatrix}
\psi _{+}^{\dag } & \eta _{+}^{\dag } & \eta _{-}^{\dag } & \psi _{-}^{\dag }%
\end{bmatrix}%
\frac{1}{2}(\beta _{1}+\gamma _{51}\gamma _{52})\left( A+B\boldsymbol{\Sigma}%
_{1}\cdot \mathbf{\hat{r}}\boldsymbol{\psi}_{2}\cdot \mathbf{\hat{r}}\right)
\notag \\
&&\times \left( A+B\boldsymbol{\Sigma}_{1}\cdot \mathbf{\hat{r}}%
\boldsymbol{\Sigma}_{2}\cdot \mathbf{\hat{r}}\right) \frac{1}{2}(\beta
_{1}+\gamma _{51}\gamma _{52})%
\begin{bmatrix}
\psi _{+} \\ 
\eta _{+} \\ 
\eta _{-} \\ 
\psi _{-}%
\end{bmatrix}%
\}  \notag \\
&&+4w^{2}\int d^{3}x\{%
\begin{bmatrix}
\psi _{+}^{\dag } & \eta _{+}^{\dag } & \eta _{-}^{\dag } & \psi _{-}^{\dag }%
\end{bmatrix}%
\frac{1}{2}(\beta _{1}+\gamma _{51}\gamma _{52})\left( A+B\boldsymbol{\Sigma}%
_{1}\cdot \mathbf{\hat{r}}\boldsymbol{\Sigma}_{2}\cdot \mathbf{\hat{r}}%
\right)  \notag \\
&&\times {\frac{1}{2}}[\rho _{1}\rho _{2}\frac{\partial L}{\partial w^{2}}%
+(\gamma _{51}\gamma _{52}-\boldsymbol{\Sigma}_{1}\mathbf{\cdot }%
\boldsymbol{\Sigma}_{2})\frac{\partial \mathcal{G}}{\partial w^{2}}]  \notag
\\
&&\times \left( A+B\boldsymbol{\Sigma}_{1}\cdot \mathbf{\hat{r}}%
\boldsymbol{\Sigma}_{2}\cdot \mathbf{\hat{r}}\right) \frac{1}{2}(\beta
_{1}+\gamma _{51}\gamma _{52})%
\begin{bmatrix}
\psi _{+} \\ 
\eta _{+} \\ 
\eta _{-} \\ 
\psi _{-}%
\end{bmatrix}%
\}
\end{eqnarray}%
since the matrix form of 
\begin{eqnarray}
(1+4w^{2}\beta _{1}\beta _{2}\frac{\partial \Delta }{\partial w^{2}})\psi
&=&[1+2w^{2}(\rho _{1}\rho _{2}\frac{\partial L}{\partial w^{2}}+(\gamma
_{51}\gamma _{52}-\boldsymbol{\Sigma}_{1}\mathbf{\cdot }\boldsymbol{\Sigma}%
_{2})\frac{\partial \mathcal{G}}{\partial w^{2}})]\psi  \notag \\
&\rightarrow &\mathcal{K}\Psi (\mathbf{r)+}[-2w^{2}\frac{\partial L}{%
\partial w^{2}}\rho \mathcal{K}\Psi (\mathbf{r)}\rho +2w^{2}\frac{\partial 
\mathcal{G}}{\partial w^{2}}(\gamma _{5}\mathcal{K}\Psi (\mathbf{r)}\gamma
_{5}+\boldsymbol{\Sigma}\mathcal{K}\Psi (\mathbf{r)\cdot }\boldsymbol{\Sigma}%
)]  \notag \\
&=&\mathcal{K}\Psi (\mathbf{r)+}[2w^{2}\frac{\partial L}{\partial w^{2}}%
iq_{2}\mathcal{K}\Psi (\mathbf{r)}iq_{2}+2w^{2}\frac{\partial \mathcal{G}}{%
\partial w^{2}}(q_{1}\mathcal{K}\Psi (\mathbf{r)}q_{1}+\boldsymbol{\Sigma}%
\mathcal{K}\Psi (\mathbf{r)\cdot }\boldsymbol{\Sigma})]
\end{eqnarray}%
in terms of matrix wave functions the norm condition (\ref{nrm2}) is%
\begin{eqnarray}
1 &=&\int d^{3}xTr\psi ^{\dag }\mathcal{L}\psi =\int d^{3}xTr\left( \mathcal{%
K}\Psi (\mathbf{r)}\right) ^{\dag }\mathcal{LK}\Psi (\mathbf{r)}\equiv \int
d^{3}xTr\{(\mathcal{K}\Psi (\mathbf{r)}\mathcal{)}^{\dag }\mathcal{K}\Psi (%
\mathbf{r)}  \notag \\
&&+(\mathcal{K}\Psi (\mathbf{r)}\mathcal{)}^{\dag }[2w^{2}\frac{\partial L}{%
\partial w^{2}}iq_{2}\mathcal{K}\Psi (\mathbf{r)}iq_{2}+2w^{2}\frac{\partial 
\mathcal{G}}{\partial w^{2}}(q_{1}\mathcal{K}\Psi (\mathbf{r)}q_{1}+%
\boldsymbol{\Sigma}\mathcal{K}\Psi (\mathbf{r)\cdot }\boldsymbol{\Sigma})]\}.
\label{ks1}
\end{eqnarray}%
Substituting Eq.(\ref{ksi}) and its conjugate this norm condition becomes

\begin{eqnarray}
1 &=&\frac{1}{8}\int d^{3}x\exp (2F)Tr_{\sigma q}\{[\cosh K(\mathcal{\psi }%
_{+}^{\dag }q_{1}-\mathcal{\psi }_{-}^{\dag }iq_{2}+\eta _{+}^{\dag }q_{0}%
\mathbf{+}\eta _{-}^{\dag }q_{3})  \notag \\
&&-\sinh K\boldsymbol{\Sigma}\cdot \mathbf{\hat{r}(}\psi _{+}^{\dag
}q_{1}-\psi _{-}^{\dag }iq_{2}+\eta _{+}^{\dag }q_{0}+\eta _{-}^{\dag }q_{3}%
\mathbf{)}\boldsymbol{\Sigma}\cdot \mathbf{\hat{r}]}  \notag \\
&&\times \lbrack \cosh K(\mathcal{\psi }_{+}q_{1}+\mathcal{\psi }%
_{-}iq_{2}+\eta _{+}q_{0}\mathbf{+}\eta _{-}q_{3})  \notag \\
&&-\sinh K\boldsymbol{\Sigma}\cdot \mathbf{\hat{r}}(\mathcal{\psi }_{+}q_{1}+%
\mathcal{\psi }_{-}iq_{2}+\eta _{+}q_{0}\mathbf{+}\eta _{-}q_{3})%
\boldsymbol{\Sigma}\cdot \mathbf{\hat{r}]}  \notag \\
&&+[\cosh K(\mathcal{\psi }_{+}^{\dag }q_{1}-\mathcal{\psi }_{-}^{\dag
}iq_{2}+\eta _{+}^{\dag }q_{0}\mathbf{+}\eta _{-}^{\dag }q_{3})  \notag \\
&&-\sinh K\boldsymbol{\Sigma}\cdot \mathbf{\hat{r}}(\mathcal{\psi }%
_{+}^{\dag }q_{1}-\mathcal{\psi }_{-}^{\dag }iq_{2}+\eta _{+}^{\dag }q_{0}%
\mathbf{+}\eta _{-}^{\dag }q_{3})\boldsymbol{\Sigma}\cdot \mathbf{\hat{r}]} 
\notag \\
&&\times \{[2w^{2}\frac{\partial L}{\partial w^{2}}iq_{2}[\cosh K(\mathcal{%
\psi }_{+}q_{1}+\mathcal{\psi }_{-}iq_{2}+\eta _{+}q_{0}\mathbf{+}\eta
_{-}q_{3})  \notag \\
&&-\sinh K\boldsymbol{\Sigma}\cdot \mathbf{\hat{r}}(\mathcal{\psi }_{+}q_{1}+%
\mathcal{\psi }_{-}iq_{2}+\eta _{+}q_{0}\mathbf{+}\eta _{-}q_{3})%
\boldsymbol{\Sigma}\cdot \mathbf{\hat{r}]}iq_{2}  \notag \\
&&+2w^{2}\frac{\partial \mathcal{G}}{\partial w^{2}}(q_{1}[\cosh K(\mathcal{%
\psi }_{+}q_{1}+\mathcal{\psi }_{-}iq_{2}+\eta _{+}q_{0}\mathbf{+}\eta
_{-}q_{3})  \notag \\
&&-\sinh K\boldsymbol{\Sigma}\cdot \mathbf{\hat{r}}(\mathcal{\psi }_{+}q_{1}+%
\mathcal{\psi }_{-}iq_{2}+\eta _{+}q_{0}\mathbf{+}\eta _{-}q_{3})%
\boldsymbol{\Sigma}\cdot \mathbf{\hat{r}]}q_{1}  \notag \\
&&+\boldsymbol{\Sigma}[\cosh K(\mathcal{\psi }_{+}q_{1}+\mathcal{\psi }%
_{-}iq_{2}+\eta _{+}q_{0}\mathbf{+}\eta _{-}q_{3})  \notag \\
&&-\sinh K\boldsymbol{\Sigma}\cdot \mathbf{\hat{r}}(\mathcal{\psi }_{+}q_{1}+%
\mathcal{\psi }_{-}iq_{2}+\eta _{+}q_{0}\mathbf{+}\eta _{-}q_{3})%
\boldsymbol{\Sigma}\cdot \mathbf{\hat{r}]\cdot }\boldsymbol{\Sigma})]\}\}
\end{eqnarray}%
\begin{eqnarray}
&=&\frac{1}{4}\int d^{3}x\exp (2F)Tr_{\sigma }\big(\cosh 2K(\mathcal{\psi }%
_{+}^{\dag }\mathcal{\psi }_{+}+\mathcal{\psi }_{-}^{\dag }\mathcal{\psi }%
_{-}+\eta _{+}^{\dag }\eta _{+}\mathbf{+}\eta _{-}^{\dag }\eta _{-})  \notag
\\
&&-\sinh 2K(\psi _{+}^{\dag }\boldsymbol{\sigma}\cdot \mathbf{\hat{r}}\psi
_{+}\boldsymbol{\sigma}\cdot \mathbf{\hat{r}}+\psi _{-}^{\dag }%
\boldsymbol{\sigma}\cdot \mathbf{\hat{r}}\psi _{-}\boldsymbol{\sigma}\cdot 
\mathbf{\hat{r}}+\eta _{+}^{\dag }\boldsymbol{\sigma}\cdot \mathbf{\hat{r}}%
\eta _{+}\boldsymbol{\sigma}\cdot \mathbf{\hat{r}+}\eta _{-}^{\dag }%
\boldsymbol{\sigma}\cdot \mathbf{\hat{r}}\eta _{-}\boldsymbol{\sigma}\cdot 
\mathbf{\hat{r}})  \notag \\
&&+\{[2w^{2}\frac{\partial L}{\partial w^{2}}[\cosh 2K(\mathcal{\psi }%
_{+}^{\dag }\mathcal{\psi }_{+}-\mathcal{\psi }_{-}^{\dag }\mathcal{\psi }%
_{-}-\eta _{+}^{\dag }\eta _{+}\mathbf{+}\eta _{-}^{\dag }\eta _{-})  \notag
\\
&&-\sinh 2K(\psi _{+}^{\dag }\boldsymbol{\sigma}\cdot \mathbf{\hat{r}}\psi
_{+}\boldsymbol{\sigma}\cdot \mathbf{\hat{r}}-\psi _{-}^{\dag }%
\boldsymbol{\sigma}\cdot \mathbf{\hat{r}}\psi _{-}\boldsymbol{\sigma}\cdot 
\mathbf{\hat{r}}-\eta _{+}^{\dag }\boldsymbol{\sigma}\cdot \mathbf{\hat{r}}%
\eta _{+}\boldsymbol{\sigma}\cdot \mathbf{\hat{r}+}\eta _{-}^{\dag }%
\boldsymbol{\sigma}\cdot \mathbf{\hat{r}}\eta _{-}\boldsymbol{\sigma}\cdot 
\mathbf{\hat{r}})\mathbf{]}  \notag \\
&&+2w^{2}\frac{\partial \mathcal{G}}{\partial w^{2}}\{[\cosh 2K(\mathcal{%
\psi }_{+}^{\dag }\mathcal{\psi }_{+}-\mathcal{\psi }_{-}^{\dag }\mathcal{%
\psi }_{-}+\eta _{+}^{\dag }\eta _{+}\mathbf{-}\eta _{-}^{\dag }\eta _{-}) 
\notag \\
&&-\sinh 2K(\psi _{+}^{\dag }\boldsymbol{\sigma}\cdot \mathbf{\hat{r}}\psi
_{+}\boldsymbol{\sigma}\cdot \mathbf{\hat{r}}-\psi _{-}^{\dag }%
\boldsymbol{\sigma}\cdot \mathbf{\hat{r}}\psi _{-}\boldsymbol{\sigma}\cdot 
\mathbf{\hat{r}}+\eta _{+}^{\dag }\boldsymbol{\sigma}\cdot \mathbf{\hat{r}}%
\eta _{+}\boldsymbol{\sigma}\cdot \mathbf{\hat{r}-}\eta _{-}^{\dag }%
\boldsymbol{\sigma}\cdot \mathbf{\hat{r}}\eta _{-}\boldsymbol{\sigma}\cdot 
\mathbf{\hat{r}})\mathbf{]}  \notag \\
&&+\cosh ^{2}K(\psi _{+}^{\dag }\boldsymbol{\sigma}\psi _{+}\mathbf{\cdot }%
\boldsymbol{\sigma}+\psi _{-}^{\dag }\boldsymbol{\sigma}\psi _{-}\mathbf{%
\cdot }\boldsymbol{\sigma}+\eta _{+}^{\dag }\boldsymbol{\sigma}\eta _{+}%
\mathbf{\cdot }\boldsymbol{\sigma}\mathbf{+}\eta _{-}^{\dag }%
\boldsymbol{\sigma}\eta _{-}\mathbf{\cdot }\boldsymbol{\sigma})  \notag \\
&&+\sinh ^{2}K[\boldsymbol{\sigma}\cdot \mathbf{\hat{r}}\psi _{+}^{\dag }%
\boldsymbol{\sigma}\cdot \mathbf{\hat{r}}\boldsymbol{\sigma}\mathbf{(}%
\boldsymbol{\sigma}\cdot \mathbf{\hat{r}}\psi _{+}\boldsymbol{\sigma}\cdot 
\mathbf{\hat{r})\cdot }\boldsymbol{\sigma}\mathbf{+}\boldsymbol{\sigma}\cdot 
\mathbf{\hat{r}}\psi _{-}^{\dag }\boldsymbol{\sigma}\cdot \mathbf{\hat{r}}%
\boldsymbol{\sigma}\mathbf{(}\boldsymbol{\sigma}\cdot \mathbf{\hat{r}}\psi
_{-}\boldsymbol{\sigma}\cdot \mathbf{\hat{r})\cdot }\boldsymbol{\sigma} 
\notag \\
&&+\boldsymbol{\sigma}\cdot \mathbf{\hat{r}}\eta _{+}^{\dag }%
\boldsymbol{\sigma}\cdot \mathbf{\hat{r}}\boldsymbol{\sigma}\mathbf{(}%
\boldsymbol{\sigma}\cdot \mathbf{\hat{r}}\eta _{+}\boldsymbol{\sigma}\cdot 
\mathbf{\hat{r})\cdot }\boldsymbol{\sigma}\mathbf{+}\boldsymbol{\sigma}\cdot 
\mathbf{\hat{r}}\eta _{-}^{\dag }\boldsymbol{\sigma}\cdot \mathbf{\hat{r}}%
\boldsymbol{\sigma}\mathbf{(}\boldsymbol{\sigma}\cdot \mathbf{\hat{r}}\eta
_{-}\boldsymbol{\sigma}\cdot \mathbf{\hat{r}\cdot }\boldsymbol{\sigma}%
\mathbf{)}]  \notag \\
&&-\cosh K\sinh K[\mathcal{\psi }_{+}^{\dag }\boldsymbol{\sigma}\mathbf{(}%
\boldsymbol{\sigma}\cdot \mathbf{\hat{r}}\mathcal{\psi }_{+}%
\boldsymbol{\sigma}\cdot \mathbf{\hat{r}}+\mathcal{\psi }_{-}^{\dag }%
\boldsymbol{\sigma}\boldsymbol{\sigma}\cdot \mathbf{\hat{r}}\mathcal{\psi }%
_{-}\boldsymbol{\sigma}\cdot \mathbf{\hat{r}}+\eta _{+}^{\dag }%
\boldsymbol{\sigma}\boldsymbol{\sigma}\cdot \mathbf{\hat{r}}\eta \mathbf{_{+}%
}\boldsymbol{\sigma}\mathbf{\cdot \mathbf{\hat{r}}+}\eta _{-}^{\dag }%
\boldsymbol{\sigma}\boldsymbol{\sigma}\cdot \mathbf{\hat{r}}\eta %
\boldsymbol{\sigma}\cdot \mathbf{\hat{r})\cdot }\boldsymbol{\sigma}  \notag
\\
&&+\mathbf{(}\boldsymbol{\sigma}\cdot \mathbf{\hat{r}}\psi _{+}^{\dag }%
\boldsymbol{\sigma}\cdot \mathbf{\hat{r}}\boldsymbol{\sigma}\mathcal{\psi }%
_{+}+\boldsymbol{\sigma}\cdot \mathbf{\hat{r}}\psi _{-}^{\dag }%
\boldsymbol{\sigma}\cdot \mathbf{\hat{r}}\boldsymbol{\sigma}\mathcal{\psi }%
_{-}+\boldsymbol{\sigma}\cdot \mathbf{\hat{r}}\eta _{+}^{\dag }%
\boldsymbol{\sigma}\cdot \mathbf{\hat{r}}\boldsymbol{\sigma}\eta _{+}+%
\boldsymbol{\sigma}\cdot \mathbf{\hat{r}}\eta _{-}^{\dag }\boldsymbol{\sigma}%
\cdot \mathbf{\hat{r}}\boldsymbol{\sigma}\eta _{-})\mathbf{\cdot }%
\boldsymbol{\sigma}]\}\big)
\end{eqnarray}%
\ \ 

For the spin singlet case (\ref{sing}) 
\begin{equation}
\mathcal{\psi }_{+}=\psi _{+0}\sigma _{0};~\mathcal{\psi }_{-}=\psi
_{-0}\sigma _{0};~\eta _{+}=\boldsymbol{\eta}_{+}\cdot \boldsymbol{\sigma}%
\mathbf{;~}\eta _{-}=0.
\end{equation}%
Substitution and performing the remaining trace using matrix identities such
as%
\begin{eqnarray}
\boldsymbol{\sigma}\mathbf{(}\boldsymbol{\sigma}\cdot \mathbf{\hat{r}}%
\boldsymbol{\eta}\mathbf{_{+}\cdot }\boldsymbol{\sigma}\boldsymbol{\sigma}%
\mathbf{\cdot \mathbf{\hat{r}})\cdot }\boldsymbol{\sigma} &\mathbf{=}&%
\boldsymbol{\sigma}\mathbf{(}2\boldsymbol{\eta}\mathbf{_{+}\cdot \mathbf{%
\hat{r}}}\boldsymbol{\sigma}\mathbf{\cdot \mathbf{\hat{r}-}}\boldsymbol{\eta}%
\mathbf{_{+}\cdot }\boldsymbol{\sigma}\mathbf{)\cdot }\boldsymbol{\sigma} 
\notag \\
&=&-\mathbf{(}2\boldsymbol{\eta}\mathbf{_{+}\cdot \mathbf{\hat{r}}}%
\boldsymbol{\sigma}\mathbf{\cdot \mathbf{\hat{r}-}}\boldsymbol{\eta}\mathbf{%
_{+}\cdot }\boldsymbol{\sigma}\mathbf{)}  \notag \\
-\boldsymbol{\eta}\mathbf{_{+}^{\dag }\cdot }\boldsymbol{\sigma}\mathbf{(}2%
\boldsymbol{\eta}\mathbf{_{+}\cdot \mathbf{\hat{r}}}\boldsymbol{\sigma}%
\mathbf{\cdot \mathbf{\hat{r}-}}\boldsymbol{\eta}\mathbf{_{+}\cdot }%
\boldsymbol{\sigma}\mathbf{)} &\mathbf{=}&-2\boldsymbol{\eta}\mathbf{%
_{+}^{\dag }\cdot \mathbf{\hat{r}}}\boldsymbol{\eta}\mathbf{\mathbf{_{+}}%
\cdot \mathbf{\hat{r}+}}\boldsymbol{\eta}_{+}^{\dag }\cdot \boldsymbol{\eta}%
\mathbf{_{+}}
\end{eqnarray}%
leads to Eq. (\ref{sinrm}) in the text. From that equation and Eqs.(\ref{ls}%
) and (\ref{jl}) we obtain the general radial form of%
\begin{eqnarray}
&&\frac{1}{2}\int_{0}^{\infty }drr^{2}\exp (2F)\big(\exp (-2K)[\left( \frac{%
u_{j0j}^{+}}{r}\right) ^{2}+\left( \frac{u_{j0j}^{-}}{r}\right) ^{2}]+\exp
(2K)[\left( \frac{\upsilon _{(j-1)1j}^{+}}{r}\right) ^{2}+\left( \frac{%
\upsilon _{(j+1)1j}^{+}}{r}\right) ^{2}]  \notag \\
&&-2\sinh 2K(-\frac{\upsilon _{(j+1)1j}^{+}}{r}\sqrt{\frac{j+1}{2j+1}}+\frac{%
\upsilon _{(j-1)1j}^{+}}{r}\sqrt{\frac{j}{2j+1}})^{2}  \notag \\
&&+2w^{2}\frac{\partial L}{\partial w^{2}}\{\exp (-2K)[[\left( \frac{%
u_{j0j}^{+}}{r}\right) ^{2}-\left( \frac{u_{j0j}^{-}}{r}\right) ^{2}-\exp
(2K)[\left( \frac{\upsilon _{(j-1)1j}^{+}}{r}\right) ^{2}+\left( \frac{%
\upsilon _{(j+1)1j}^{+}}{r}\right) ^{2}]  \notag \\
&&+2\sinh 2K(-\frac{\upsilon _{(j+1)1j}^{+}}{r}\sqrt{\frac{j+1}{2j+1}}+\frac{%
\upsilon _{(j-1)1j}^{+}}{r}\sqrt{\frac{j}{2j+1}})^{2}\}  \notag \\
&&+4w^{2}\frac{\partial \mathcal{G}}{\partial w^{2}}\exp (-2K)[2\left( \frac{%
u_{j0j}^{+}}{r}\right) ^{2}+\left( \frac{u_{j0j}^{-}}{r}\right) ^{2}]\big)=1
\end{eqnarray}

For the spin triplet case (\ref{trip})%
\begin{equation}
\mathcal{\psi }_{+}=\boldsymbol{\psi}_{+}\mathbf{\cdot }\boldsymbol{\sigma};~%
\mathcal{\psi }_{-}=\boldsymbol{\psi}_{-}\mathbf{\cdot }\boldsymbol{\sigma}%
;~\eta _{+}=\mathcal{\eta }_{+0}\sigma _{0}\mathbf{;~}\eta _{-}=%
\boldsymbol{\eta}_{-}\mathbf{\cdot }\boldsymbol{\sigma}.
\end{equation}%
and performing the trace gives Eq. (\ref{trpnrm}) in the text. From that
equation and Eqs.(\ref{wvfn}), (\ref{nu}), and (\ref{emi}) and%
\begin{equation}
\boldsymbol{\psi}_{-}=\frac{u_{(j+1)1j}^{-}}{r}\mathbf{Y}_{jm+}+\frac{%
u_{(j-1)1j}^{-}}{r}\mathbf{Y}_{jm-}
\end{equation}%
we obtain the general radial form of%
\begin{eqnarray}
1 &=&\frac{1}{2}\int drr^{2}\exp (2F)\big([\exp (2K)(\left( \frac{%
u_{(j+1)1j}^{+}}{r}\right) ^{2}+\left( \frac{u_{(j-1)1j}^{+}}{r}\right)
^{2}+\left( \frac{u_{(j+1)1j}^{-}}{r}\right) ^{2}+\left( \frac{%
u_{(j-1)1j}^{-}}{r}\right) ^{2}+\left( \frac{\upsilon _{j1j}^{-}}{r}\right)
^{2})  \notag \\
&&+\exp (-2K)\left( \frac{v_{j0j}^{+}}{r}\right) ^{2}  \notag \\
&&-2\sinh 2K[(-\frac{u_{(j+1)1j}^{+}}{r}\sqrt{\frac{j+1}{2j+1}}+\frac{%
u_{(j-1)1j}^{+}}{r}\sqrt{\frac{j}{2j+1}})^{2}+(-\frac{u_{(j+1)1j}^{-}}{r}%
\sqrt{\frac{j+1}{2j+1}}+\frac{u_{(j-1)1j}^{-}}{r}\sqrt{\frac{j}{2j+1}})^{2}]
\notag \\
&&+\{[2w^{2}\frac{\partial L}{\partial w^{2}}[\exp (2K)(\left( \frac{%
u_{(j+1)1j}^{+}}{r}\right) ^{2}+\left( \frac{u_{(j-1)1j}^{+}}{r}\right)
^{2}-\left( \frac{u_{(j+1)1j}^{-}}{r}\right) ^{2}-\left( \frac{%
u_{(j-1)1j}^{-}}{r}\right) ^{2}+\left( \frac{\upsilon _{j1j}^{-}}{r}\right)
^{2})  \notag \\
&&-\exp (-2K)\left( \frac{v_{j0j}^{+}}{r}\right) ^{2}  \notag \\
&&-2\sinh 2K[(-\frac{u_{(j+1)1j}^{+}}{r}\sqrt{\frac{j+1}{2j+1}}+\frac{%
u_{(j-1)1j}^{+}}{r}\sqrt{\frac{j}{2j+1}})^{2}-(-\frac{u_{(j+1)1j}^{-}}{r}%
\sqrt{\frac{j+1}{2j+1}}+\frac{u_{(j-1)1j}^{-}}{r}\sqrt{\frac{j}{2j+1}})^{2}]
\notag \\
&&+4w^{2}\frac{\partial \mathcal{G}}{\partial w^{2}}([-\exp (2K)(\left( 
\frac{u_{(j+1)1j}^{-}}{r}\right) ^{2}+\left( \frac{\upsilon _{j1j}^{-}}{r}%
\right) ^{2})+2\exp (-2K)\left( \frac{v_{j0j}^{+}}{r}\right) ^{2}  \notag \\
&&+2\sinh 2K[(-\frac{u_{(j+1)1j}^{+}}{r}\sqrt{\frac{j+1}{2j+1}}+\frac{%
u_{(j-1)1j}^{+}}{r}\sqrt{\frac{j}{2j+1}})^{2}+(-\frac{u_{(j+1)1j}^{-}}{r}%
\sqrt{\frac{j+1}{2j+1}}+\frac{u_{(j-1)1j}^{-}}{r}\sqrt{\frac{j}{2j+1}}%
)^{2}]]\}\big)
\end{eqnarray}

\section{Relations Between Radial Parts of Scalar and Vector Wave Functions}

\subsection{The Singlet Wave Function}

We start with scalar part of Eq. (\ref{psim}) 
\begin{equation}
\mathcal{\psi }_{-0}=\{\frac{(E^{2}+M^{2})}{2EM}-\frac{\exp (2\mathcal{G)}}{%
2ME}[\mathbf{p}+\frac{i}{2}\boldsymbol{\nabla }L]\cdot \lbrack \mathbf{p}-%
\frac{i}{2}\boldsymbol{\nabla }L]\}\mathcal{\psi }_{+0}.
\end{equation}%
Use 
\begin{equation}
\lbrack \mathbf{p}+\frac{i}{2}\boldsymbol{\nabla }L]\cdot \lbrack \mathbf{p}-%
\frac{i}{2}\boldsymbol{\nabla }L]=\mathbf{p}^{2}-\frac{1}{2}%
\boldsymbol{\nabla }^{2}L+\frac{1}{4}\left( \boldsymbol{\nabla }L\right)
^{2},
\end{equation}%
and so 
\begin{equation}
\mathcal{\psi }_{-0}=\{\frac{(E^{2}+M^{2})}{2EM}-\frac{\exp (2\mathcal{G)}}{%
2ME}[\mathbf{p}^{2}-\frac{1}{2}\boldsymbol{\nabla }^{2}L+\frac{1}{4}\left( %
\boldsymbol{\nabla }L\right) ^{2}]\}\mathcal{\psi }_{+0}.
\end{equation}%
By converting Eq. (\ref{57}) to matrix form and taking the scalar part one
can show (see \cite{liu}) for an alternative approach) 
\begin{equation}
-\mathbf{p}^{2}\psi =[\frac{1}{4}\left( \boldsymbol{\nabla }L\right) ^{2}-%
\mathcal{B}^{2}\exp (-2\mathcal{G})-\frac{1}{2}\boldsymbol{\nabla }%
^{2}L]\psi .
\end{equation}%
and so 
\begin{eqnarray}
\mathcal{\psi }_{-0} &=&\{\frac{(E^{2}+M^{2})}{2EM}-\frac{\exp (2\mathcal{G)}%
}{2ME}\mathcal{B}^{2}\exp (-2\mathcal{G})\mathcal{\psi }_{+0}  \notag \\
&=&\frac{M}{E}\mathcal{\psi }_{-0}=\frac{M}{E}\frac{u_{j0j}^{+}}{r}Y_{jm}.
\end{eqnarray}

The other wave function contributing is from Eq. (\ref{etap}). \ 
\begin{eqnarray}
\boldsymbol{\eta }_{+} &=&\frac{\exp (\mathcal{G-}2K)}{E}\{(\mathbf{p}-\frac{%
i}{2}\boldsymbol{\nabla }L)+[\exp (2K)-1](\mathbf{\hat{r}\hat{r}}\cdot 
\mathbf{p}-\frac{i}{2}\boldsymbol{\nabla }L)\}\mathcal{\psi }_{+0}  \notag \\
&=&\frac{\exp (\mathcal{G-}2K)}{E}\{(\mathbf{p}-\frac{iL^{\prime }}{2}%
\mathbf{\hat{r}})+[\exp (2K)-1](-i\frac{d}{dr}-\frac{iL^{\prime }}{2})%
\mathbf{\hat{r}}\}\frac{u_{j0j}^{+}}{r}Y_{jm}  \notag \\
&=&i\frac{\exp (\mathcal{G-}2K)}{E}\{[\exp (2K)(-\frac{d}{dr}-\frac{%
L^{\prime }}{2})-\frac{(j+1)}{r}]\frac{u_{j0j}^{+}}{r}\sqrt{{\frac{j}{2j+1}}}%
\mathbf{Y}_{-}  \notag \\
&&+[-\exp (2K)(-\frac{d}{dr}-\frac{L^{\prime }}{2})-\frac{j}{r}]\frac{%
u_{j0j}^{+}}{r}\sqrt{{\frac{j+1}{2j+1}}}\mathbf{Y}_{+}\},
\end{eqnarray}%
in which we have used Eq. (\ref{vsph}) in the form 
\begin{eqnarray}
\mathbf{\hat{r}}Y_{jm} &=&\sqrt{\frac{j}{2j+1}}\mathbf{Y}_{-}-\sqrt{\frac{j+1%
}{2j+1}}\mathbf{Y}_{+},  \notag \\
\mathbf{p}Y_{jm} &=&-\frac{i}{r}[(j+1)\sqrt{\frac{j}{2j+1}}\mathbf{Y}_{-}+j%
\sqrt{\frac{j+1}{2j+1}}\mathbf{Y}_{+}].
\end{eqnarray}

\subsection{The Triplet Wave Functions}

In this appendix we evaluate the details of the wave functions below
contributing to the triplet state. Beginning with Eqs. (\ref{psim}), (\ref%
{etap}), and (\ref{etam}) we find the relations between the radial parts of
the wave functions for $\boldsymbol{\psi }_{-},\mathcal{\eta }_{+0},%
\boldsymbol{\eta }_{-}$ and those of $\boldsymbol{\psi }_{+}$. For \
simplicity of notation we define 
\begin{eqnarray}
A &=&\frac{(L+2\mathcal{G})}{2},  \notag \\
B &=&\frac{(L-2\mathcal{G})}{2},  \notag \\
C &=&\frac{(L+6\mathcal{G})}{2},  \notag \\
D &=&\frac{(3L-2\mathcal{G})}{2},
\end{eqnarray}%
and so Eqs. (\ref{psim}), (\ref{etap}), and (\ref{etam}) become 
\begin{eqnarray}
\boldsymbol{\psi }_{-} &=&\frac{(E^{2}+M^{2})}{2EM}\boldsymbol{\psi }_{+}-%
\frac{\exp (2\mathcal{G})}{2EM}[\mathbf{1}+Q_{p}\mathbf{\hat{r}\hat{r}}] 
\notag \\
&&\cdot {\biggl (}[\mathbf{p}-i\boldsymbol{\nabla }A]\times \{[\mathbf{p}-i%
\boldsymbol{\nabla }B]\times \lbrack \mathbf{1}+Q_{m}\mathbf{\hat{r}\hat{r}}%
]\cdot \boldsymbol{\psi }_{+}\}  \notag \\
&&{\large +}[\mathbf{p}-i\boldsymbol{\nabla }C][\mathbf{p}-i%
\boldsymbol{\nabla }D]\cdot \lbrack \mathbf{1}+Q_{m}\mathbf{\hat{r}\hat{r}}%
]\cdot \boldsymbol{\psi }_{+}{\biggr )}  \notag \\
\mathcal{\eta }_{+0} &=&\frac{\exp (\mathcal{G}+2K\mathcal{)}}{E}[\mathbf{p}%
-i\boldsymbol{\nabla }D]\cdot \lbrack \mathbf{1}+Q_{m}\mathbf{\hat{r}\hat{r}}%
]\cdot \boldsymbol{\psi }_{+}  \notag \\
\boldsymbol{\eta }_{-} &=&-\frac{\exp (\mathcal{G)}}{M}[\mathbf{1}+Q_{p}%
\mathbf{\hat{r}\hat{r}}\cdot ][i\mathbf{p}+\boldsymbol{\nabla }B]\times
\lbrack \mathbf{1}+Q_{m}\mathbf{\hat{r}\hat{r}}]\boldsymbol{\psi }_{+}.
\end{eqnarray}

\subsubsection{The Wave Function $\boldsymbol{\psi }_{-}$}

\ \ In this section we consider $\boldsymbol{\psi }_{-}$ and in the
subsequent sections $\mathcal{\eta }_{+0}$ and\textbf{\ }$\boldsymbol{\eta }%
_{-}$. \ The first portion of the first term of $\boldsymbol{\psi }_{-}$
involves 
\begin{eqnarray}
&&[\mathbf{p}-i\boldsymbol{\nabla }A]\times \{[\mathbf{p}-i%
\boldsymbol{\nabla }B]\times \lbrack \mathbf{1}+Q_{m}\mathbf{\hat{r}\hat{r}}%
]\cdot \boldsymbol{\psi }_{+}\}  \notag \\
&=&[\mathbf{p}-iA^{\prime }\mathbf{\hat{r}]}\times \{[\mathbf{p}-iB^{\prime }%
\mathbf{\hat{r}]}\times \lbrack \mathbf{1}+Q_{m}\mathbf{\hat{r}\hat{r}}%
]\cdot \boldsymbol{\psi }_{+}\}.
\end{eqnarray}%
The inner portion is 
\begin{eqnarray}
&&[\mathbf{p}-iB^{\prime }\mathbf{\hat{r}]}\times \lbrack \mathbf{1}+Q_{m}%
\mathbf{\hat{r}\hat{r}}]\cdot \boldsymbol{\psi }_{+}  \notag \\
&=&\mathbf{p}\times \lbrack \mathbf{1}+Q_{m}\mathbf{\hat{r}\hat{r}}]\cdot %
\boldsymbol{\psi }_{+}-iB^{\prime }\mathbf{\hat{r}\times }[\mathbf{1}+Q_{m}%
\mathbf{\hat{r}\hat{r}}]\cdot \boldsymbol{\psi }_{+}.
\end{eqnarray}%
The first part of this is 
\begin{eqnarray}
&&\mathbf{p}\times \lbrack \mathbf{1}+Q_{m}\mathbf{\hat{r}\hat{r}}]\cdot %
\boldsymbol{\psi }_{+}  \notag \\
&=&\mathbf{p}\times \boldsymbol{\psi }_{+}-\frac{Q_{m}}{r}\mathbf{L\hat{r}}%
\cdot \boldsymbol{\psi }_{+},
\end{eqnarray}%
and the second part is 
\begin{eqnarray}
&&-iB^{\prime }\mathbf{\hat{r}\times }[\mathbf{1}+Q_{m}\mathbf{\hat{r}\hat{r}%
}]\cdot \boldsymbol{\psi }_{+}  \notag \\
&=&-iB^{\prime }\mathbf{\hat{r}\times }\boldsymbol{\psi }_{+}.
\end{eqnarray}%
Thus the inner portion is 
\begin{eqnarray}
&&[\mathbf{p}-iB^{\prime }\mathbf{\hat{r}]}\times \lbrack \mathbf{1}+Q_{m}%
\mathbf{\hat{r}\hat{r}}]\cdot \boldsymbol{\psi }_{+}  \notag \\
&=&\mathbf{p}\times \boldsymbol{\psi }_{+}-\frac{Q_{m}}{r}\mathbf{L\hat{r}}%
\cdot \boldsymbol{\psi }_{+}-iB^{\prime }\mathbf{\hat{r}\times }{%
\boldsymbol{\psi}}_{+}.  \label{ip2}
\end{eqnarray}%
Hence, the first term involves 
\begin{eqnarray}
&&(\mathbf{p}-iA^{\prime }\mathbf{\hat{r}})\times (\mathbf{p}\times %
\boldsymbol{\psi }_{+}-\frac{Q_{m}}{r}\mathbf{L\hat{r}}\cdot %
\boldsymbol{\psi }_{+}-iB^{\prime }\mathbf{\hat{r}\times }{\boldsymbol{ \psi}%
}_{+})  \notag \\
&=&\mathbf{p}\times (\mathbf{p}\times \boldsymbol{\psi }_{+}-\frac{Q_{m}}{r}%
\mathbf{L\hat{r}}\cdot \boldsymbol{\psi }_{+}-iB^{\prime }\mathbf{\hat{r}%
\times }{\boldsymbol{ \psi}}_{+})  \notag \\
&&-iA^{\prime }\mathbf{\hat{r}\times }(\mathbf{p}\times \boldsymbol{\psi }%
_{+}-\frac{Q_{m}}{r}\mathbf{L\hat{r}}\cdot \boldsymbol{\psi }_{+}-iB^{\prime
}\mathbf{\hat{r}\times }{\boldsymbol{ \psi}}_{+}).
\end{eqnarray}%
The first part is 
\begin{eqnarray}
&&\mathbf{p}\times (\mathbf{p}\times \boldsymbol{\psi }_{+}-\frac{Q_{m}}{r}%
\mathbf{L\hat{r}}\cdot \boldsymbol{\psi }_{+}-iB^{\prime }\mathbf{\hat{r}%
\times }{\boldsymbol{ \psi}}_{+})  \notag \\
&=&\mathbf{pp}\cdot \boldsymbol{\psi }_{+}-\mathbf{p}^{2}\boldsymbol{\psi }%
_{+}-i[\frac{Q_{m}}{r^{2}}+\frac{2K^{\prime }\exp (-2K)}{r}]\mathbf{\hat{r}%
\times L\hat{r}}\cdot \boldsymbol{\psi }_{+}-\frac{Q_{m}}{r}\mathbf{p}\times 
\mathbf{L\hat{r}}\cdot \boldsymbol{\psi }_{+}  \notag \\
&&-B^{\prime \prime }\mathbf{\hat{r}\times }(\mathbf{\hat{r}\times }{%
\boldsymbol{\psi}}_{+})-iB^{\prime }\mathbf{p}\times (\mathbf{\hat{r}\times }%
{\boldsymbol{ \psi}}_{+}).
\end{eqnarray}%
The second is 
\begin{eqnarray}
&&-iA^{\prime }\mathbf{\hat{r}\times }(\mathbf{p}\times \boldsymbol{\psi }%
_{+}-\frac{Q_{m}}{r}\mathbf{L\hat{r}}\cdot \boldsymbol{\psi }_{+}-iB^{\prime
}\mathbf{\hat{r}\times }{\boldsymbol{\psi}}_{+})  \notag \\
&=&-iA^{\prime }\mathbf{\hat{r}\times }(\mathbf{p}\times \boldsymbol{\psi }%
_{+})+\frac{Q_{m}}{r}A^{\prime }i(\mathbf{\hat{r}\times L})\mathbf{\hat{r}}%
\cdot \boldsymbol{\psi }_{+}-A^{\prime }B^{\prime }\mathbf{\hat{r}\times }(%
\mathbf{\hat{r}\times }{\boldsymbol{ \psi}}_{+}).
\end{eqnarray}%
Combining, the first portion of the first term is 
\begin{eqnarray*}
&&(\mathbf{p}-iA^{\prime }\mathbf{\hat{r})}\times (\mathbf{p}\times %
\boldsymbol{\psi }_{+}-\frac{Q_{m}}{r}\mathbf{L\hat{r}}\cdot %
\boldsymbol{\psi }_{+}-iB^{\prime }\mathbf{\hat{r}\times }{\boldsymbol{\psi}}%
_{+}) \\
&=&\mathbf{pp}\cdot \boldsymbol{\psi }_{+}-\mathbf{p}^{2}\boldsymbol{\psi }%
_{+}-[\frac{Q_{m}}{r^{2}}+\frac{(2K-A)^{\prime }\exp (-2K)+A^{\prime }}{r}]i%
\mathbf{\hat{r}\times L\hat{r}}\cdot \boldsymbol{\psi }_{+} \\
&&-\frac{Q_{m}}{r}\mathbf{p}\times \mathbf{L\hat{r}}\cdot \boldsymbol{\psi }%
_{+}-(B^{\prime \prime }+A^{\prime }B^{\prime })\mathbf{\hat{r}\times }(%
\mathbf{\hat{r}\times }{\boldsymbol{\psi}}_{+})-iB^{\prime }\mathbf{p}\times
(\mathbf{\hat{r}\times }{\boldsymbol{\psi}}_{+})-iA^{\prime }\mathbf{\hat{r}%
\times }(\mathbf{p}\times \boldsymbol{\psi }_{+}).
\end{eqnarray*}%
The second portion of the first term involves 
\begin{eqnarray}
&&Q_{p}\mathbf{\hat{r}\hat{r}}\cdot {\biggl (}[\mathbf{p}-i%
\boldsymbol{\nabla }A]\times \{[\mathbf{p}-i\boldsymbol{\nabla }B]\times
\lbrack \mathbf{1}+Q_{m}\mathbf{\hat{r}\hat{r}}]\cdot \boldsymbol{\psi }%
_{+}\}  \notag \\
&=&Q_{p}\{\mathbf{\hat{r}\hat{r}}\cdot \mathbf{p}^{2}\boldsymbol{\psi }_{+}-%
\mathbf{\hat{r}\hat{r}}\cdot \mathbf{pp}\cdot \boldsymbol{\psi }_{+}-\frac{%
Q_{m}}{r^{2}}\mathbf{\hat{r}L}\cdot \mathbf{L\hat{r}}\cdot \boldsymbol{\psi }%
_{+}  \notag \\
&&-\frac{B^{\prime }}{r}\mathbf{\hat{r}}i\mathbf{L\times \hat{r}}\cdot %
\boldsymbol{\psi }_{+}\}.
\end{eqnarray}%
Thus, the entire first term involves 
\begin{eqnarray}
&&[\mathbf{1}+Q_{p}\mathbf{\hat{r}\hat{r}}]\cdot {\biggl (}[\mathbf{p}-i%
\boldsymbol{\nabla }A]\times \{[\mathbf{p}-i\boldsymbol{\nabla }B]\times
\lbrack \mathbf{1}+Q_{m}\mathbf{\hat{r}\hat{r}}]\cdot \boldsymbol{\psi }%
_{+}\}{\biggr )}  \notag \\
&=&\mathbf{pp}\cdot \boldsymbol{\psi }_{+}-\mathbf{p}^{2}\boldsymbol{\psi }%
_{+}-[\frac{Q_{m}}{r^{2}}+\frac{(2K-A)^{\prime }\exp (-2K)+A^{\prime }}{r}]i%
\mathbf{\hat{r}\times L\hat{r}}\cdot \boldsymbol{\psi }_{+}  \notag \\
&&-\frac{Q_{m}}{r}\mathbf{p}\times \mathbf{L\hat{r}}\cdot \boldsymbol{\psi }%
_{+}-(B^{\prime \prime }+A^{\prime }B^{\prime })\mathbf{\hat{r}\times }(%
\mathbf{\hat{r}\times }{\boldsymbol{\psi}}_{+})-iB^{\prime }\mathbf{p}\times
(\mathbf{\hat{r}\times }{\boldsymbol{\psi}}_{+})-iA^{\prime }\mathbf{\hat{r}%
\times }(\mathbf{p}\times \boldsymbol{\psi }_{+})  \notag \\
&&+Q_{p}\{+\mathbf{\hat{r}\hat{r}}\cdot \mathbf{pp}\cdot \boldsymbol{\psi }%
_{+}-\mathbf{\hat{r}\hat{r}}\cdot \mathbf{p}^{2}\boldsymbol{\psi }_{+}-\frac{%
Q_{m}}{r^{2}}\mathbf{\hat{r}L}^{2}\mathbf{\hat{r}}\cdot \boldsymbol{\psi }%
_{+}-\frac{B^{\prime }}{r}\mathbf{\hat{r}}i\mathbf{L\times \hat{r}}\cdot %
\boldsymbol{\psi }_{+}\}.
\end{eqnarray}

The first portion of the second term involves

\begin{equation*}
\lbrack \mathbf{p}-i\boldsymbol{\nabla }C][\mathbf{p}-i\boldsymbol{\nabla }%
D]\cdot \lbrack \mathbf{1}+Q_{m}\mathbf{\hat{r}\hat{r}}]\cdot %
\boldsymbol{\psi }_{+}.
\end{equation*}%
The inner portion is 
\begin{eqnarray}
&&[\mathbf{p}-iD^{\prime }\mathbf{\hat{r}]}\cdot \lbrack \mathbf{1}+Q_{m}%
\mathbf{\hat{r}\hat{r}}]\cdot \boldsymbol{\psi }_{+}  \notag \\
&=&\mathbf{p}\cdot \lbrack \mathbf{1}+Q_{m}\mathbf{\hat{r}\hat{r}}]\cdot %
\boldsymbol{\psi }_{+}  \notag \\
&&-iD^{\prime }\mathbf{\hat{r}}\cdot \lbrack 1+Q_{m}\mathbf{\hat{r}\hat{r}]}%
\cdot \boldsymbol{\psi }.  \label{ip}
\end{eqnarray}%
The first part is 
\begin{eqnarray}
&&\mathbf{p}\cdot \lbrack \mathbf{1}+Q_{m}\mathbf{\hat{r}\hat{r}}]\cdot %
\boldsymbol{\psi }_{+}  \notag \\
&=&\mathbf{p}\cdot \boldsymbol{\psi }_{+}+i2K^{\prime }\exp (-2K\mathbf{)%
\hat{r}}\cdot \boldsymbol{\psi }_{+}+Q_{m}\mathbf{p}\cdot \mathbf{\hat{r}%
\hat{r}}\cdot \boldsymbol{\psi }_{+},
\end{eqnarray}%
while the second part is 
\begin{eqnarray*}
&&-iD^{\prime }\mathbf{\hat{r}}\cdot \lbrack \mathbf{1}+Q_{m}\mathbf{\hat{r}%
\hat{r}]}\cdot \boldsymbol{\psi } \\
&=&-i\exp (-2K)D^{\prime }\mathbf{\hat{r}\cdot }{\boldsymbol{\psi}.}
\end{eqnarray*}%
The inner portion is thus 
\begin{eqnarray}
&&[\mathbf{p}-iD^{\prime }\mathbf{\hat{r}]}\cdot \lbrack \mathbf{1}+Q_{m}%
\mathbf{\hat{r}\hat{r}}]\cdot \boldsymbol{\psi }_{+}  \notag \\
&=&\mathbf{p}\cdot \boldsymbol{\psi }_{+}+i(2K-D)^{\prime }\exp (-2K\mathbf{)%
\hat{r}}\cdot \boldsymbol{\psi }_{+}+Q_{m}\mathbf{p}\cdot \mathbf{\hat{r}%
\hat{r}}\cdot \boldsymbol{\psi }_{+}.  \label{ip1}
\end{eqnarray}%
The first portion of the second term thus involves 
\begin{eqnarray}
&&[\mathbf{p}-iC^{\prime }\mathbf{\hat{r}]}[\mathbf{p}\cdot \boldsymbol{\psi
}_{+}+i(2K-D)^{\prime }\exp (-2K\mathbf{)\hat{r}}\cdot \boldsymbol{\psi }%
_{+}+Q_{m}\mathbf{p}\cdot \mathbf{\hat{r}\hat{r}}\cdot \boldsymbol{\psi }%
_{+}]  \notag \\
&=&\mathbf{p}[\mathbf{p}\cdot \boldsymbol{\psi }_{+}+i(2K-D)^{\prime }\exp
(-2K\mathbf{)\hat{r}}\cdot \boldsymbol{\psi }_{+}+Q_{m}\mathbf{p}\cdot 
\mathbf{\hat{r}\hat{r}}\cdot \boldsymbol{\psi }_{+}]  \notag \\
&&{\large -}iC^{\prime }\mathbf{\hat{r}}[\mathbf{p}\cdot \boldsymbol{\psi }%
_{+}+i(2K-D)^{\prime }\exp (-2K\mathbf{)\hat{r}}\cdot \boldsymbol{\psi }%
_{+}+Q_{m}\mathbf{p}\cdot \mathbf{\hat{r}\hat{r}}\cdot \boldsymbol{\psi }%
_{+}].
\end{eqnarray}%
The first part of this is 
\begin{eqnarray}
&&\mathbf{p}[\mathbf{p}\cdot \boldsymbol{\psi }_{+}+i(2K-D)^{\prime }\exp
(-2K\mathbf{)\hat{r}}\cdot \boldsymbol{\psi }_{+}+Q_{m}\mathbf{p}\cdot 
\mathbf{\hat{r}\hat{r}}\cdot \boldsymbol{\psi }_{+}]  \notag \\
&=&\mathbf{pp}\cdot \boldsymbol{\psi }_{+}+[(2K-D)^{\prime \prime
}-2K^{\prime }(2K-D)^{\prime }]\exp (-2K)\mathbf{\hat{r}\hat{r}}\cdot %
\boldsymbol{\psi }_{+}{\large +}(2K-D)^{\prime }\exp (-2K)i\mathbf{p\hat{r}}%
\cdot \boldsymbol{\psi }_{+}  \notag \\
&&+2K^{\prime }\exp (-2K)i\mathbf{\hat{r}p}\cdot \mathbf{\hat{r}\hat{r}}%
\cdot \boldsymbol{\psi }_{+}+Q_{m}\mathbf{pp}\cdot \mathbf{\hat{r}\hat{r}}%
\cdot \boldsymbol{\psi }_{+},
\end{eqnarray}%
while the second part is simply 
\begin{eqnarray}
&&-iC^{\prime }\mathbf{\hat{r}}[\mathbf{p}\cdot \boldsymbol{\psi }%
_{+}+i(2K-D)^{\prime }\exp (-2K\mathbf{)\hat{r}}\cdot \boldsymbol{\psi }%
_{+}+Q_{m}\mathbf{p}\cdot \mathbf{\hat{r}\hat{r}}\cdot \boldsymbol{\psi }%
_{+}]  \notag \\
&=&-C^{\prime }i\mathbf{\hat{r}p}\cdot \boldsymbol{\psi }_{+}+C^{\prime
}(2K-D)^{\prime }\exp (-2K)\mathbf{\hat{r}\hat{r}}\cdot \boldsymbol{\psi }%
_{+}-Q_{m}C^{\prime }i\mathbf{\hat{r}}\left( \mathbf{p}\cdot \mathbf{\hat{r}}%
\right) \mathbf{\hat{r}}\cdot \boldsymbol{\psi }_{+}.
\end{eqnarray}%
This combines with the first part to give. 
\begin{eqnarray}
&&[\mathbf{p}-iC^{\prime }\mathbf{\hat{r}]}[\mathbf{p}\cdot \boldsymbol{\psi
}_{+}+i(2K-D)^{\prime }\exp (-2K\mathbf{)\hat{r}}\cdot \boldsymbol{\psi }%
_{+}+Q_{m}\mathbf{p}\cdot \mathbf{\hat{r}\hat{r}}\cdot \boldsymbol{\psi }%
_{+}]  \notag \\
&=&\mathbf{pp}\cdot \boldsymbol{\psi }_{+}+Q_{m}\mathbf{pp}\cdot \mathbf{%
\hat{r}\hat{r}}\cdot \boldsymbol{\psi }_{+}+[(2K-D)^{\prime \prime
}-(2K-C)^{\prime }(2K-D)^{\prime }](Q_{m}+1)\mathbf{\hat{r}\hat{r}}\cdot %
\boldsymbol{\psi }_{+}  \notag \\
&&+(2K-D)^{\prime }(Q_{m}+1)i\mathbf{p\hat{r}\cdot }\boldsymbol{\psi }%
_{+}+2K^{\prime }(Q_{m}+1)i\mathbf{\hat{r}}\left( \mathbf{p}\cdot \mathbf{%
\hat{r}}\right) \mathbf{\hat{r}}\cdot \boldsymbol{\psi }_{+}-C^{\prime }i%
\mathbf{\hat{r}(p}\cdot \boldsymbol{\psi }_{+}+Q_{m}\left( \mathbf{p}\cdot 
\mathbf{\hat{r}}\right) \mathbf{\hat{r}}\cdot \boldsymbol{\psi }_{+}).
\end{eqnarray}%
The second portion of the second term is 
\begin{eqnarray}
&&+Q_{p}\mathbf{\hat{r}\hat{r}}\cdot \lbrack \mathbf{p}-iC^{\prime }\mathbf{%
\hat{r}]}[\mathbf{p}\cdot \boldsymbol{\psi }_{+}+i(2K-D)^{\prime }(Q_{m}+1)%
\mathbf{\hat{r}}\cdot \boldsymbol{\psi }_{+}+Q_{m}\mathbf{p}\cdot \mathbf{%
\hat{r}\hat{r}}\cdot \boldsymbol{\psi }_{+}]  \notag \\
&=&Q_{p}\mathbf{\hat{r}}\{\mathbf{\hat{r}}\cdot \mathbf{pp}\cdot %
\boldsymbol{\psi }_{+}+Q_{m}\mathbf{\hat{r}}\cdot \mathbf{pp}\cdot \mathbf{%
\hat{r}\hat{r}}\cdot \boldsymbol{\psi }_{+}  \notag \\
&&+[(2K-D)^{\prime \prime }-(2K-C)^{\prime }(2K-D)^{\prime }](Q_{m}+1)%
\mathbf{\hat{r}}\cdot \boldsymbol{\psi }_{+}  \notag \\
&&{\large +}(2K-D)^{\prime }(Q_{m}+1)i\mathbf{\hat{r}\cdot p\hat{r}\cdot }%
\boldsymbol{\psi }_{+}  \notag \\
&&+2K^{\prime }(Q_{m}+1)i\mathbf{p}\cdot \mathbf{\hat{r}\hat{r}}\cdot %
\boldsymbol{\psi }_{+}-C^{\prime }i(\mathbf{p}\cdot \boldsymbol{\psi }%
_{+}+Q_{m}\left( \mathbf{p}\cdot \mathbf{\hat{r}}\right) \mathbf{\hat{r}}%
\cdot \boldsymbol{\psi }_{+})\}.
\end{eqnarray}%
Combining the two portions gives 
\begin{eqnarray}
&&[\mathbf{1}+Q_{p}\mathbf{\hat{r}\hat{r}}\cdot ][\mathbf{p}-i\mathbf{\nabla 
}C][\mathbf{p}-i\boldsymbol{\nabla }D]\cdot \lbrack \mathbf{1}+Q_{m}\mathbf{%
\hat{r}\hat{r}}]\cdot \boldsymbol{\psi }_{+}  \notag \\
&=&\mathbf{pp}\cdot \boldsymbol{\psi }_{+}+Q_{m}\mathbf{pp}\cdot \mathbf{%
\hat{r}\hat{r}}\cdot \boldsymbol{\psi }_{+}+Q_{p}\mathbf{\hat{r}\hat{r}}%
\cdot \mathbf{pp}\cdot \boldsymbol{\psi }_{+}+Q_{p}Q_{m}\mathbf{\hat{r}\hat{r%
}}\cdot \mathbf{pp}\cdot \mathbf{\hat{r}\hat{r}}\cdot \boldsymbol{\psi }_{+}
\notag \\
&&+[(2K-D)^{\prime \prime }-(2K-C)^{\prime }(2K-D)^{\prime }]\mathbf{\hat{r}%
\hat{r}}\cdot \boldsymbol{\psi }_{+}  \notag \\
&&+(2K-D)^{\prime }(Q_{m}+1)i(\mathbf{p\hat{r}\cdot }\boldsymbol{\psi }%
_{+}+Q_{p}\mathbf{\hat{r}\hat{r}\cdot p\hat{r}\cdot }\boldsymbol{\psi }_{+})
\notag \\
&&+2K^{\prime }i\mathbf{\hat{r}}\left( \mathbf{p}\cdot \mathbf{\hat{r}}%
\right) \mathbf{\hat{r}}\cdot \boldsymbol{\psi }_{+}-(Q_{p}+1)C^{\prime }i%
\mathbf{\hat{r}(p}\cdot \boldsymbol{\psi }_{+}+Q_{m}\left( \mathbf{p}\cdot 
\mathbf{\hat{r}}\right) \mathbf{\hat{r}}\cdot \boldsymbol{\psi }_{+}).
\end{eqnarray}

Now $\ $we combine the first and second terms \ to give 
\begin{eqnarray}
&&[\mathbf{1}+Q_{p}\mathbf{\hat{r}\hat{r}}]\cdot {\biggl (}[\mathbf{p}-\frac{
i}{2}\boldsymbol{\nabla }(L+2\mathcal{G})]\times \{[\mathbf{p}-\frac{i}{2} %
\boldsymbol{\nabla }(L-2\mathcal{G})]\times \lbrack \mathbf{1}+Q_{m}\mathbf{%
\ \hat{r}\hat{r}}]\cdot \boldsymbol{\psi }_{+}\}  \notag \\
&&+[\mathbf{p}-\frac{i}{2}\boldsymbol{\nabla }(L+6\mathcal{G})][\mathbf{p}- 
\frac{i}{2}\boldsymbol{\nabla }(3L-2\mathcal{G})]\cdot \lbrack \mathbf{1}
+Q_{m}\mathbf{\hat{r}\hat{r}}]\cdot \boldsymbol{\psi }_{+}{\biggr )}  \notag
\\
&=&\mathbf{pp}\cdot \boldsymbol{\psi }_{+}-\mathbf{p}^{2}\boldsymbol{\psi }
_{+}-[\frac{Q_{m}}{r^{2}}+\frac{(2K-A)^{\prime }\exp (-2K)+A^{\prime }}{r}]i 
\mathbf{\hat{r}\times L\hat{r}}\cdot \boldsymbol{\psi }_{+}  \notag \\
&&-\frac{Q_{m}}{r}\mathbf{p}\times \mathbf{L\hat{r}}\cdot \boldsymbol{\psi }
_{+}-(B^{\prime \prime }+A^{\prime }B^{\prime })\mathbf{\hat{r}\times }( 
\mathbf{\hat{r}\times }\boldsymbol{\psi }_{+})-iB^{\prime }\mathbf{p}\times
( \mathbf{\hat{r}\times }\boldsymbol{\psi }_{+})-iA^{\prime }\mathbf{\hat{r}
\times }(\mathbf{p}\times \boldsymbol{\psi }_{+})  \notag \\
&&+Q_{p}\{\mathbf{\hat{r}\hat{r}}\cdot \mathbf{pp}\cdot \boldsymbol{\psi }
_{+}-\mathbf{\hat{r}\hat{r}}\cdot \mathbf{p}^{2}\boldsymbol{\psi }_{+}-\frac{
Q_{m}}{r^{2}}\mathbf{\hat{r}L}^{2}\mathbf{\hat{r}}\cdot \boldsymbol{\psi }
_{+}-\frac{B^{\prime }}{r}\mathbf{\hat{r}}i\mathbf{L\times \hat{r}}\cdot %
\boldsymbol{\psi }_{+}\}  \notag \\
&&+\mathbf{pp}\cdot \boldsymbol{\psi }_{+}+Q_{m}\mathbf{pp}\cdot \mathbf{\ 
\hat{r}\hat{r}}\cdot \boldsymbol{\psi }_{+}+Q_{p}\mathbf{\hat{r}\hat{r}}
\cdot \mathbf{pp}\cdot \boldsymbol{\psi }_{+}+Q_{p}Q_{m}\mathbf{\hat{r}\hat{%
r }}\cdot \mathbf{pp}\cdot \mathbf{\hat{r}\hat{r}}\cdot \boldsymbol{\psi }%
_{+}  \notag \\
&&+[(2K-D)^{\prime \prime }-(2K-C)^{\prime }(2K-D)^{\prime }]\mathbf{\hat{r} 
\hat{r}}\cdot \boldsymbol{\psi }_{+}  \notag \\
&&+(2K-D)^{\prime }(Q_{m}+1)i(\mathbf{p\hat{r}\cdot }\boldsymbol{\psi }
_{+}+Q_{p}\mathbf{\hat{r}\hat{r}\cdot p\hat{r}\cdot }\boldsymbol{\psi }_{+})
\notag \\
&&+2K^{\prime }i\mathbf{\hat{r}}\left( \mathbf{p}\cdot \mathbf{\hat{r}}
\right) \mathbf{\hat{r}}\cdot \boldsymbol{\psi }_{+}-(Q_{p}+1)C^{\prime }i 
\mathbf{\hat{r}(p}\cdot \boldsymbol{\psi }_{+}+Q_{m}\left( \mathbf{p}\cdot 
\mathbf{\hat{r}}\right) \mathbf{\hat{r}}\cdot \boldsymbol{\psi }_{+}).
\end{eqnarray}

We make the following abbreviations 
\begin{eqnarray}
G &\equiv &[\frac{Q_{m}}{r^{2}}+\frac{(2K-A)^{\prime }(Q_{m}+1)+A^{\prime }}{%
r}],  \notag \\
H &\equiv &(B^{\prime \prime }+A^{\prime }B^{\prime }),  \notag \\
I &\equiv &[(2K-D)^{\prime \prime }-(2K-C)^{\prime }(2K-D)^{\prime }], 
\notag \\
N &\equiv &(2K-D)^{\prime }(Q_{m}+1),
\end{eqnarray}%
and so 
\begin{eqnarray}
&&[\mathbf{1}+Q_{p}\mathbf{\hat{r}\hat{r}}]\cdot {\biggl (}[\mathbf{p}-\frac{%
i}{2}\boldsymbol{\nabla }(L+2\mathcal{G})]\times \{[\mathbf{p}-\frac{i}{2}%
\boldsymbol{\nabla }(L-2\mathcal{G})]\times \lbrack \mathbf{1}+Q_{m}\mathbf{%
\hat{r}\hat{r}}]\cdot \boldsymbol{\psi }_{+}\}  \notag \\
&&{\large +}[\mathbf{p}-\frac{i}{2}\boldsymbol{\nabla }(L+6\mathcal{G})][%
\mathbf{p}-\frac{i}{2}\boldsymbol{\nabla }(3L-2\mathcal{G})]\cdot \lbrack 
\mathbf{1}+Q_{m}\mathbf{\hat{r}\hat{r}}]\cdot \boldsymbol{\psi }_{+}{\biggr )%
}  \notag \\
&=&+2\mathbf{pp}\cdot \boldsymbol{\psi }_{+}-\mathbf{p}^{2}\boldsymbol{\psi }%
_{+}+2Q_{p}\mathbf{\hat{r}\hat{r}}\cdot \mathbf{pp}\cdot \boldsymbol{\psi }%
_{+}+Q_{m}\mathbf{pp}\cdot \mathbf{\hat{r}\hat{r}}\cdot \boldsymbol{\psi }%
_{+}+Q_{p}Q_{m}\mathbf{\hat{r}\hat{r}}\cdot \mathbf{pp}\cdot \mathbf{\hat{r}%
\hat{r}}\cdot \boldsymbol{\psi }_{+}-Q_{p}\mathbf{\hat{r}\hat{r}}\cdot 
\mathbf{p}^{2}\boldsymbol{\psi }_{+}  \notag \\
&&-Gi\mathbf{\hat{r}\times L\hat{r}}\cdot \boldsymbol{\psi }_{+}-\frac{Q_{m}%
}{r}\mathbf{p}\times \mathbf{L\hat{r}}\cdot \boldsymbol{\psi }_{+}-H\mathbf{%
\hat{r}\times }(\mathbf{\hat{r}\times }{\boldsymbol{\psi}}_{+})-iB^{\prime }%
\mathbf{p}\times (\mathbf{\hat{r}\times }{\boldsymbol{\psi}}_{+})-iA^{\prime
}\mathbf{\hat{r}\times }(\mathbf{p}\times \boldsymbol{\psi }_{+})  \notag \\
&&+Q_{p}\{-\frac{Q_{m}}{r^{2}}\mathbf{\hat{r}L}^{2}\mathbf{\hat{r}}\cdot %
\boldsymbol{\psi }_{+}-\frac{B^{\prime }}{r}\mathbf{\hat{r}}i\mathbf{L\times 
\hat{r}}\cdot \boldsymbol{\psi }_{+}\}+I\mathbf{\hat{r}\hat{r}}\cdot %
\boldsymbol{\psi }_{+}  \notag \\
&&+Ni(\mathbf{p\hat{r}\cdot }\boldsymbol{\psi }_{+}+Q_{p}\mathbf{\hat{r}\hat{%
r}\cdot p\hat{r}\cdot }\boldsymbol{\psi }_{+})+2K^{\prime }i\hat{r}\mathbf{p}%
\cdot \mathbf{\hat{r}\hat{r}}\cdot \boldsymbol{\psi }_{+}-(Q_{p}+1)C^{\prime
}i\mathbf{\hat{r}(p}\cdot \boldsymbol{\psi }_{+}+Q_{m}\left( \mathbf{p}\cdot 
\mathbf{\hat{r}}\right) \mathbf{\hat{r}}\cdot \boldsymbol{\psi }_{+}).
\end{eqnarray}

To facilitate possible cancellations expand out the triple cross products
and use $\mathbf{p}\cdot \mathbf{\hat{r}=\hat{r}}\cdot \mathbf{p}-\frac{2i}{%
r }$. \ We need

\begin{equation*}
i\mathbf{\hat{r}\times L\hat{r}}\cdot \boldsymbol{\psi }_{+}=ir\mathbf{\hat{%
r }\hat{r}}\cdot \mathbf{p\hat{r}\cdot }\boldsymbol{\psi }_{+}-i\left\vert
r\right\vert \mathbf{p\hat{r}\cdot }\boldsymbol{\psi }_{+},
\end{equation*}
or 
\begin{equation}
i\mathbf{p\hat{r}\cdot }\boldsymbol{\psi }_{+}=i\mathbf{\hat{r}\hat{r}}\cdot 
\mathbf{p\hat{r}\cdot }\boldsymbol{\psi }_{+}+\frac{1}{r}i\mathbf{\hat{r}
\times L\hat{r}}\cdot \boldsymbol{\psi }_{+},
\end{equation}
and 
\begin{equation}
\mathbf{L\times \hat{r}}\cdot \boldsymbol{\psi }_{+}=\left( -\mathbf{\hat{r}%
r }\cdot \mathbf{p}+\left\vert r\right\vert \mathbf{p}+2i\mathbf{\hat{r}}
\right) \cdot \boldsymbol{\psi }_{+},
\end{equation}
and 
\begin{eqnarray}
i\mathbf{\hat{r}p}\cdot \mathbf{\hat{r}\hat{r}}\cdot \boldsymbol{\psi }_{+}
&=&\mathbf{\hat{r}(}i\mathbf{\hat{r}}\cdot \mathbf{p}+\frac{2}{r}\mathbf{) 
\hat{r}\cdot }\boldsymbol{\psi }_{+},  \notag \\
\mathbf{pp}\cdot \mathbf{\hat{r}\hat{r}}\cdot \boldsymbol{\psi }_{+} &=& 
\mathbf{p}(\mathbf{\hat{r}}\cdot \mathbf{p}-\frac{2i}{r}\mathbf{)\hat{r}
\cdot }\boldsymbol{\psi }_{+},  \notag \\
i\mathbf{\hat{r}}\left( \mathbf{p}\cdot \mathbf{\hat{r}}\right) \mathbf{\hat{
r}}\cdot \boldsymbol{\psi }_{+} &=&\mathbf{\hat{r}}\left( i\mathbf{\hat{r}}
\cdot \mathbf{p}+\frac{2}{r}\right) \hat{r}\cdot \boldsymbol{\psi }_{+}
\end{eqnarray}
and 
\begin{equation}
\mathbf{\hat{r}\hat{r}}\cdot \mathbf{p}\left( \mathbf{p}\cdot \mathbf{\hat{r}
}\right) \mathbf{\hat{r}}\cdot \boldsymbol{\psi }_{+}=\mathbf{\hat{r}}\left( 
\mathbf{\hat{r}}\cdot \mathbf{p\hat{r}\cdot p}+\frac{2}{r^{2}}-\frac{2}{r}i 
\mathbf{\hat{r}}\cdot \mathbf{p}\right) \hat{r}\cdot \boldsymbol{\psi }_{+}.
\end{equation}
Use these to simplify 
\begin{eqnarray}
&&+Ni(\mathbf{p\hat{r}\cdot }\boldsymbol{\psi }_{+}+Q_{p}\mathbf{\hat{r}\hat{
r}\cdot p\hat{r}\cdot }\boldsymbol{\psi }_{+})+2K^{\prime }i\mathbf{\hat{r}p}
\cdot \mathbf{\hat{r}\hat{r}}\cdot \boldsymbol{\psi }_{+}-(Q_{p}+1)C^{\prime
}i\mathbf{\hat{r}(p}\cdot \boldsymbol{\psi }_{+}+Q_{m}\left( \mathbf{p}\cdot 
\mathbf{\hat{r}}\right) \mathbf{\hat{r}}\cdot \boldsymbol{\psi }_{+})  \notag
\\
&&+Q_{m}\mathbf{pp}\cdot \mathbf{\hat{r}\hat{r}}\cdot \boldsymbol{\psi }
_{+}+Q_{p}Q_{m}\mathbf{\hat{r}\hat{r}}\cdot \mathbf{pp}\cdot \mathbf{\hat{r} 
\hat{r}}\cdot \boldsymbol{\psi }_{+}  \notag \\
&=&(\frac{4K^{\prime }-2Q_{m}(Q_{p}+1)C^{\prime }}{r}+\frac{
2Q_{p}Q_{m}+2Q_{m}}{r^{2}})\mathbf{\hat{r}\hat{r}}\cdot \boldsymbol{\psi }
_{+}  \notag \\
&&+(N(1+Q_{p})+2K^{\prime }-(Q_{p}+1)Q_{m}C^{\prime }-\frac{%
2Q_{p}Q_{m}-Q_{m} }{r})\mathbf{\hat{r}}i\mathbf{\hat{r}}\cdot \mathbf{p\hat{r%
}\cdot } \boldsymbol{\psi }_{+}  \notag \\
&&-\frac{Ni}{r}\mathbf{\hat{r}\times L\hat{r}}\cdot \boldsymbol{\psi }
_{+}-(Q_{p}+1)C^{\prime }\mathbf{\hat{r}}i\mathbf{p}\cdot \boldsymbol{\psi }
_{+}+Q_{m}(\mathbf{\hat{r}}\cdot \mathbf{pp}-\frac{3i}{r}\mathbf{p?)\hat{r}
\cdot }\boldsymbol{\psi }_{+}  \notag \\
&&+Q_{p}Q_{m}\mathbf{\hat{r}}(\mathbf{\hat{r}}\cdot \mathbf{p\hat{r}\cdot p}%
) \mathbf{\hat{r}}\cdot \boldsymbol{\psi }_{+}  \notag \\
&\equiv &\frac{M}{r}\mathbf{\hat{r}\hat{r}}\cdot \boldsymbol{\psi }_{+}+O 
\mathbf{\hat{r}}i\mathbf{\hat{r}}\cdot \mathbf{p\hat{r}\cdot } %
\boldsymbol{\psi }_{+}-\frac{Ni}{r}\mathbf{\hat{r}\times L\hat{r}}\cdot %
\boldsymbol{\psi }_{+}-(Q_{p}+1)C^{\prime }\mathbf{\hat{r}}i\mathbf{p}\cdot %
\boldsymbol{\psi }_{+}  \notag \\
&&+Q_{m}(\mathbf{\hat{r}}\cdot \mathbf{pp}-\frac{3i}{r}\mathbf{p)\hat{r}
\cdot }\boldsymbol{\psi }_{+}+Q_{p}Q_{m}\mathbf{\hat{r}}(\mathbf{\hat{r}}
\cdot \mathbf{p\hat{r}\cdot p})\mathbf{\hat{r}}\cdot \boldsymbol{\psi }_{+},
\end{eqnarray}
where we have used 
\begin{equation}
\mathbf{p\hat{r}}\cdot \mathbf{p=\hat{r}}\cdot \mathbf{pp-}\frac{i}{r} 
\mathbf{p+}\frac{i}{r}\mathbf{\hat{r}\hat{r}}\cdot \mathbf{p,}
\end{equation}
and defined 
\begin{eqnarray}
\frac{M}{r} &\equiv &(\frac{4K^{\prime }-2Q_{m}(Q_{p}+1)C^{\prime }}{r}+ 
\frac{2Q_{p}Q_{m}+2Q_{m}}{r^{2}})  \notag \\
&=&(\frac{4K^{\prime }+2Q_{p}C^{\prime }}{r}-\frac{2Q_{p}}{r^{2}}),  \notag
\\
O &\equiv &N(1+Q_{p})+2K^{\prime }-(Q_{p}+1)Q_{m}C^{\prime }-\frac{
2Q_{p}Q_{m}-Q_{m}}{r}  \notag \\
&=&N(1+Q_{p})+2K^{\prime }+Q_{p}C^{\prime }+\frac{2Q_{p}+3Q_{m}}{r}.
\end{eqnarray}
Furthermore using 
\begin{equation}
\mathbf{\hat{r}\times }(\mathbf{\hat{r}\times }\boldsymbol{\psi }_{+})= 
\mathbf{\hat{r}\hat{r}}\cdot \boldsymbol{\psi }_{+}-\boldsymbol{\psi }_{+},
\end{equation}
we obtain 
\begin{eqnarray}
&&[\mathbf{1}+Q_{p}\mathbf{\hat{r}\hat{r}}]\cdot {\biggl (}[\mathbf{p}-\frac{
i}{2}\boldsymbol{\nabla }(L+2\mathcal{G})]\times \{[\mathbf{p}-\frac{i}{2} %
\boldsymbol{\nabla }(L-2\mathcal{G})]\times \lbrack \mathbf{1}+Q_{m}\mathbf{%
\ \hat{r}\hat{r}}]\cdot \boldsymbol{\psi }_{+}\}  \notag \\
&&{\large +}[\mathbf{p}-\frac{i}{2}\boldsymbol{\nabla }(L+6\mathcal{G})][ 
\mathbf{p}-\frac{i}{2}\boldsymbol{\nabla }(3L-2\mathcal{G})]\cdot \lbrack 
\mathbf{1}+Q_{m}\mathbf{\hat{r}\hat{r}}]\cdot \boldsymbol{\psi }_{+}{\biggr
) }  \notag \\
&=&(H-\mathbf{p}^{2})\boldsymbol{\psi }_{+}+2\mathbf{pp}\cdot %
\boldsymbol{\psi }_{+}+Q_{m}(\mathbf{\hat{r}}\cdot \mathbf{pp}-\frac{3i}{r} 
\mathbf{p)\hat{r}\cdot }\boldsymbol{\psi }_{+}+Q_{p}Q_{m}\mathbf{\hat{r}}( 
\mathbf{\hat{r}}\cdot \mathbf{p\hat{r}\cdot p})\mathbf{\hat{r}}\cdot %
\boldsymbol{\psi }_{+}  \notag \\
&&-(G+\frac{N}{r})i\mathbf{\hat{r}\times L\hat{r}}\cdot \boldsymbol{\psi }
_{+}-\frac{Q_{m}}{r}\mathbf{p}\times \mathbf{L\hat{r}}\cdot \boldsymbol{\psi
}_{+}+\left( \frac{M}{r}-H+I\right) \mathbf{\hat{r}\hat{r}}\cdot %
\boldsymbol{\psi }_{+} +O\mathbf{\hat{r}}i\mathbf{\hat{r}}\cdot \mathbf{p%
\hat{r}\cdot } \boldsymbol{\psi }_{+}-iB^{\prime }\mathbf{p}\times (\mathbf{%
\hat{r}\times } \boldsymbol{\psi }_{+})  \notag \\
&&-iA^{\prime }\mathbf{\hat{r}\times }(\mathbf{p}\times \boldsymbol{\psi }
_{+})+Q_{p}\{+2\mathbf{\hat{r}\hat{r}}\cdot \mathbf{pp}\cdot %
\boldsymbol{\psi }_{+}-\mathbf{\hat{r}\hat{r}}\cdot \mathbf{p}^{2} %
\boldsymbol{\psi }_{+}-\frac{Q_{m}}{r^{2}}\mathbf{\hat{r}L}^{2}\mathbf{\hat{%
r }}\cdot \boldsymbol{\psi }_{+}-\frac{B^{\prime }}{r}\mathbf{\hat{r}}i%
\mathbf{\ L\times \hat{r}}\cdot \boldsymbol{\psi }_{+}\}  \notag \\
&&-(Q_{p}+1)C^{\prime }\mathbf{\hat{r}}i\mathbf{p}\cdot \boldsymbol{\psi }
_{+}.
\end{eqnarray}

Of further use are 
\begin{eqnarray}
i\mathbf{p}\times (\mathbf{\hat{r}\times }{\boldsymbol{\psi}}_{+}) &=&-\frac{%
1}{r}\boldsymbol{\psi }_{+}-\frac{1}{r}\mathbf{\hat{r}\hat{r}}\cdot %
\boldsymbol{\psi }_{+}-i\mathbf{\hat{r}}\cdot \mathbf{p}\boldsymbol{\psi }%
_{+}+i\mathbf{\hat{r}p\cdot }{\boldsymbol{\psi}}_{+},  \notag \\
i\mathbf{\hat{r}\times }(\mathbf{p}\times \boldsymbol{\psi }_{+}) &=&-\frac{1%
}{r}\boldsymbol{\psi }_{+}+\frac{1}{r}\mathbf{\hat{r}\hat{r}}\cdot %
\boldsymbol{\psi }_{+}-i\mathbf{\hat{r}}\cdot \mathbf{p}\boldsymbol{\psi }%
_{+}+i\mathbf{p\hat{r}\cdot }{\boldsymbol{\psi}}_{+},
\end{eqnarray}%
and 
\begin{eqnarray}
i\mathbf{\hat{r}\times L\hat{r}}\cdot \boldsymbol{\psi }_{+} &=&ir\mathbf{%
\hat{r}\hat{r}}\cdot \mathbf{p\hat{r}\cdot }{\boldsymbol{\psi}}%
_{+}-i\left\vert r\right\vert \mathbf{p\hat{r}\cdot }{\boldsymbol{\psi}}_{+},
\notag \\
i\mathbf{L\times \hat{r}}\cdot \boldsymbol{\psi }_{+} &=&\left( -i\mathbf{%
\hat{r}r}\cdot \mathbf{p}+i\left\vert r\right\vert \mathbf{p}-2\mathbf{\hat{r%
}}\right) \cdot \boldsymbol{\psi }_{+}.
\end{eqnarray}%
So using these we have 
\begin{eqnarray}
&&[\mathbf{1}+Q_{p}\mathbf{\hat{r}\hat{r}}]\cdot {\biggl (}[\mathbf{p}-\frac{%
i}{2}\boldsymbol{\nabla }(L+2\mathcal{G})]\times \{[\mathbf{p}-\frac{i}{2}%
\boldsymbol{\nabla }(L-2\mathcal{G})]\times \lbrack \mathbf{1}+Q_{m}\mathbf{%
\hat{r}\hat{r}}]\cdot \boldsymbol{\psi }_{+}\}  \notag \\
&&{\large +}[\mathbf{p}-\frac{i}{2}\boldsymbol{\nabla }(L+6\mathcal{G})][%
\mathbf{p}-\frac{i}{2}\boldsymbol{\nabla }(3L-2\mathcal{G})]\cdot \lbrack 
\mathbf{1}+Q_{m}\mathbf{\hat{r}\hat{r}}]\cdot \boldsymbol{\psi }_{+}{\biggr )%
}  \notag \\
&=&(H-\mathbf{p}^{2})\boldsymbol{\psi }_{+}+2\mathbf{pp}\cdot %
\boldsymbol{\psi }_{+}+Q_{m}(\mathbf{\hat{r}}\cdot \mathbf{pp}-\frac{3i}{r}%
\mathbf{p)\hat{r}\cdot }\boldsymbol{\psi }_{+}+Q_{p}Q_{m}\mathbf{\hat{r}}(%
\mathbf{\hat{r}}\cdot \mathbf{p\hat{r}\cdot p})\mathbf{\hat{r}}\cdot %
\boldsymbol{\psi }_{+}  \notag \\
&&-(rG+N)(i\mathbf{\hat{r}\hat{r}}\cdot \mathbf{p\hat{r}\cdot }{%
\boldsymbol{\psi}}_{+}-i\mathbf{p\hat{r}\cdot }{\boldsymbol{\psi}}_{+})-%
\frac{Q_{m}}{r}\mathbf{p}\times \mathbf{L\hat{r}}\cdot \boldsymbol{\psi }%
_{+}+\left( \frac{M}{r}-H+I\right) \mathbf{\hat{r}\hat{r}}\cdot %
\boldsymbol{\psi }_{+}  \notag \\
&&-B^{\prime }\mathbf{(-}\frac{1}{r}\mathbf{\boldsymbol{\psi }_{+}-}\frac{1}{%
r}\mathbf{\mathbf{\hat{r}\hat{r}}\cdot \boldsymbol{\psi }_{+}-}i\mathbf{%
\mathbf{\hat{r}}\cdot \mathbf{p}\boldsymbol{\psi }_{+}+}i\mathbf{\mathbf{%
\hat{r}p\cdot }{\boldsymbol{\psi}}_{+})}-A^{\prime }\mathbf{(-}\frac{1}{r}%
\mathbf{\boldsymbol{\psi }_{+}+}\frac{1}{r}\mathbf{\mathbf{\hat{r}\hat{r}}%
\cdot \boldsymbol{\psi }_{+}-}i\mathbf{\mathbf{\hat{r}}\cdot \mathbf{p}%
\boldsymbol{\psi }_{+}+}i\mathbf{\mathbf{p\hat{r}\cdot }{\boldsymbol{\psi}}%
_{+})}  \notag \\
&&+Q_{p}\{+2\mathbf{\hat{r}\hat{r}}\cdot \mathbf{pp}\cdot \boldsymbol{\psi }%
_{+}-\mathbf{\hat{r}\hat{r}}\cdot \mathbf{p}^{2}\boldsymbol{\psi }_{+}-\frac{%
Q_{m}}{r^{2}}\mathbf{\hat{r}L}^{2}\mathbf{\hat{r}}\cdot \boldsymbol{\psi }%
_{+}  \notag \\
&&-\frac{B^{\prime }}{r}\mathbf{\hat{r}}\left( -i\mathbf{\hat{r}r}\cdot 
\mathbf{p}+i\left\vert r\right\vert \mathbf{p}-2\mathbf{\hat{r}}\right)
\cdot \boldsymbol{\psi }_{+}\}+O\mathbf{\hat{r}}i\mathbf{\hat{r}}\cdot 
\mathbf{p\hat{r}\cdot }\boldsymbol{\psi }_{+}  \notag \\
&&-(Q_{p}+1)C^{\prime }\mathbf{\hat{r}}i\mathbf{p}\cdot \boldsymbol{\psi }%
_{+}  \notag \\
&\equiv &(R-\mathbf{p}^{2})\boldsymbol{\psi }_{+}+2\mathbf{pp}\cdot %
\boldsymbol{\psi }_{+}+Q_{m}\mathbf{\hat{r}}\cdot \mathbf{pp\hat{r}\cdot }%
\boldsymbol{\psi }_{+}+Q_{p}Q_{m}(\mathbf{\hat{r}}\cdot \mathbf{p\hat{r}%
\cdot p})\mathbf{\hat{r}\hat{r}}\cdot \boldsymbol{\psi }_{+}-\frac{Q_{m}}{r}%
\mathbf{p}\times \mathbf{L\hat{r}}\cdot \boldsymbol{\psi }_{+}  \notag \\
&&+Q_{p}[2\mathbf{\hat{r}\hat{r}}\cdot \mathbf{pp}\cdot \boldsymbol{\psi }%
_{+}-\mathbf{\hat{r}\hat{r}}\cdot \mathbf{p}^{2}\boldsymbol{\psi }_{+}-\frac{%
Q_{m}}{r^{2}}\mathbf{\hat{r}L}^{2}\mathbf{\hat{r}}\cdot \boldsymbol{\psi }%
_{+}]  \notag \\
&&+S\mathbf{\hat{r}\hat{r}}\cdot \boldsymbol{\psi }_{+}+(A+B)^{\prime }i%
\mathbf{\hat{r}\cdot p}{\boldsymbol{\psi}}_{+}+Ti\mathbf{\hat{r}}\cdot 
\mathbf{p\hat{r}\hat{r}}\cdot \boldsymbol{\psi }_{+}+Ui\mathbf{p\hat{r}\cdot 
}{\boldsymbol{\psi}}_{+}+Vi\mathbf{\hat{r}p}\cdot \boldsymbol{\psi }_{+},
\label{imt}
\end{eqnarray}%
where we use the abbreviations 
\begin{eqnarray}
R &=&H+\frac{(A+B)^{\prime }}{r},  \notag \\
S &=&\frac{M}{r}-H+I+\frac{(3B-A)^{\prime }}{r},  \notag \\
T &=&(O{\large -}N+B^{\prime }-Gr),  \notag \\
U &=&-\frac{3Q_{m}}{r}+rG+N-A^{\prime },  \notag \\
V &=&-(Q_{p}+1)C^{\prime }-2B^{\prime }.
\end{eqnarray}%
Taking the above expression term by term we need (using the abbreviations $%
\mathbf{Y}_{\pm }=\mathbf{Y}_{jm\pm }$) 
\begin{eqnarray}
\mathbf{L}^{2}\mathbf{Y}_{+} &=&\left[ (j+1)(j+2)\right] \mathbf{Y}_{+}, 
\notag \\
\mathbf{L}^{2}\mathbf{Y}_{-} &=&(j-1)j\mathbf{Y}_{-},
\end{eqnarray}%
and so 
\begin{eqnarray}
\mathbf{p}^{2}\boldsymbol{\psi }_{+} &=&-\left( \frac{d^{2}}{dr^{2}}+\frac{2%
}{r}\frac{d}{dr}-\frac{(j+1)(j+2)}{r}\right) \frac{u_{+}}{r}\mathbf{Y}_{+} 
\notag \\
&&-\left( \frac{d^{2}}{dr^{2}}+\frac{2}{r}\frac{d}{dr}-\frac{(j-1)j}{r}%
\right) \frac{u_{-}}{r}\mathbf{Y}_{-}.
\end{eqnarray}%
Also needed are 
\begin{eqnarray}
\mathbf{\hat{r}(\hat{r}\cdot }{\boldsymbol{\psi}}_{+}) &=&\frac{1}{2j+1}[(j%
\frac{u_{-}}{r}-\sqrt{j(j+1)}\frac{u_{+}}{r})\mathbf{Y}_{-}+((j+1)\frac{u_{+}%
}{r}-\sqrt{j(j+1)}\frac{u_{-}}{r})\mathbf{Y}_{+}],  \notag \\
i\mathbf{\hat{r}\cdot p}{\boldsymbol{\psi}}_{+} &=&\frac{d}{dr}\frac{u_{-}}{r%
}\mathbf{Y}_{-}+\frac{d}{dr}\frac{u_{+}}{r}\mathbf{Y}_{+},  \notag \\
i\mathbf{\hat{r}}\cdot \mathbf{p\hat{r}\hat{r}}\cdot \boldsymbol{\psi }_{+}
&=&\frac{1}{2j+1}[(j\frac{d}{dr}\frac{u_{-}}{r}-\sqrt{j(j+1)}\frac{d}{dr}%
\frac{u_{+}}{r})\mathbf{Y}_{-}+((j+1)\frac{d}{dr}\frac{u_{+}}{r}-\sqrt{j(j+1)%
}\frac{d}{dr}\frac{u_{-}}{r})\mathbf{Y}_{+}],  \notag \\
\mathbf{\hat{r}L}^{2}\mathbf{\hat{r}\cdot }{\boldsymbol{\psi}}_{+} &=&\frac{1%
}{2j+1}[j(j+1)(j\frac{u_{-}}{r}-\sqrt{j(j+1)}\frac{u_{+}}{r})\mathbf{Y}%
_{-}+j(j+1)((j+1)\frac{u_{+}}{r}-\sqrt{j(j+1)}\frac{u_{-}}{r})\mathbf{Y}%
_{+}],
\end{eqnarray}%
and 
\begin{eqnarray}
\mathbf{p}\times \mathbf{L\hat{r}}\cdot \boldsymbol{\psi }_{+} &=&\frac{j+1}{%
2j+1}[(\frac{j(j+1)}{r}+j\frac{d}{dr})\frac{u_{-}}{r}-\sqrt{j(j+1)}(\frac{%
(j+1)}{r}+\frac{d}{dr})\frac{u_{+}}{r}]\mathbf{Y}_{-},  \notag \\
&&+\frac{j}{2j+1}[(\frac{j(j+1)}{r}-(j+1{)}\frac{d}{dr})\frac{u_{+}}{r}+%
\sqrt{j(j+1)}(-\frac{j}{r}+\frac{d}{dr})\frac{u_{-}}{r}]\mathbf{Y}_{+} 
\notag \\
i\mathbf{p\hat{r}\cdot }{\boldsymbol{\psi}} &=&\frac{1}{2j+1}\{[(\frac{j(j+1)%
}{r}+j\frac{d}{dr})\frac{u_{-}}{r}+\sqrt{j(j+1)}(-\frac{(j+1)}{r}-\frac{d}{dr%
})\frac{u_{+}}{r}]\mathbf{Y}_{-}  \notag \\
&&+[(-\frac{j(j+1)}{r}+(j+1)\frac{d}{dr})\frac{u_{+}}{r}+\sqrt{j(j+1)}(-%
\frac{j}{r}-\frac{d}{dr})\frac{u_{-}}{r}]\mathbf{Y}_{+}\},
\end{eqnarray}%
\begin{eqnarray}
i\mathbf{\hat{r}p}\cdot \boldsymbol{\psi }_{+} &=&\frac{1}{2j+1}\{[j(-\frac{%
(j-1)}{r}+\frac{d}{dr})\frac{u_{-}}{r}+\sqrt{j(j+1)}(-\frac{(j+2)}{r}-\frac{d%
}{dr})\frac{u_{+}}{r}]\mathbf{Y}_{-}  \notag \\
&&+[(j+1)(\frac{(j+2)}{r}+\frac{d}{dr})\frac{u_{+}}{r}+\sqrt{j(j+1)}(\frac{%
(j-1)}{r}-\frac{d}{dr})\frac{u_{-}}{r}]\mathbf{Y}_{+}\},  \notag \\
\mathbf{p\hat{r}\cdot p\hat{r}\cdot }{\boldsymbol{\psi}}_{+} &=&\frac{1}{2j+1%
}\{[(+\frac{j(j+1)}{r^{2}}-j\frac{d^{2}}{dr^{2}})\frac{u_{-}}{r}+\sqrt{j(j+1)%
}(-\frac{(j+1)}{r^{2}}+\frac{d^{2}}{dr^{2}})\frac{u_{+}}{r}]\mathbf{Y}_{-} 
\notag \\
&&+[(-\frac{j(j+1)}{r^{2}}-(j+1)\frac{d^{2}}{dr^{2}})\frac{u_{+}}{r}+\sqrt{%
j(j+1)}(-\frac{j}{r^{2}}+\frac{d^{2}}{dr^{2}})\frac{u_{-}}{r}]\mathbf{Y}_{+},
\end{eqnarray}%
and 
\begin{eqnarray}
\mathbf{\hat{r}\cdot pp\hat{r}\cdot }{\boldsymbol{\psi}} &=&\frac{1}{2j+1}%
\{[(\frac{j(j+1)}{r^{2}}-\frac{j(j+1)}{r}\frac{d}{dr}-j\frac{d^{2}}{dr^{2}})%
\frac{u_{-}}{r}+\sqrt{j(j+1)}(-\frac{(j+1)}{r^{2}}+\frac{(j+1)}{r}\frac{d}{dr%
}+\frac{d^{2}}{dr^{2}})\frac{u_{+}}{r}]\mathbf{Y}_{-}  \notag \\
&&+[(-\frac{j(j+1)}{r^{2}}+\frac{j(j+1)}{r}\frac{d}{dr}-(j+1)\frac{d^{2}}{%
dr^{2}})\frac{u_{+}}{r}+\sqrt{j(j+1)}(\frac{j}{r^{2}}-\frac{j}{r}\frac{d}{dr}%
+\frac{d^{2}}{dr^{2}})\frac{u_{-}}{r}]\mathbf{Y}_{+}\mathbf{\},}
\end{eqnarray}%
\begin{eqnarray}
&&\mathbf{\hat{r}\hat{r}}\cdot \mathbf{p}^{2}\boldsymbol{\psi }  \notag \\
&=&\frac{1}{2j+1}{\biggl (}\{[-j\left( \frac{d^{2}}{dr^{2}}+\frac{2}{r}\frac{%
d}{dr}-\frac{(j-1)j}{r}\right) ]\frac{u_{-}}{r}+\sqrt{j(j+1)}[\frac{d^{2}}{%
dr^{2}}+\frac{2}{r}\frac{d}{dr}-\frac{(j+1)(j+2)}{r}]\frac{u_{+}}{r}\}%
\mathbf{Y}_{-}  \notag \\
&&+\{[-(j+1)\left( \frac{d^{2}}{dr^{2}}+\frac{2}{r}\frac{d}{dr}-\frac{%
(j+1)(j+2)}{r}\right) ]\frac{u_{+}}{r}+\sqrt{j(j+1)}[\frac{d^{2}}{dr^{2}}+%
\frac{2}{r}\frac{d}{dr}-\frac{(j-1)j}{r}]\frac{u_{-}}{r}\}\mathbf{Y}_{+}{%
\biggr ),}
\end{eqnarray}%
\begin{equation}
\mathbf{\hat{r}}\cdot \mathbf{p\hat{r}\cdot p\hat{r}\hat{r}}\cdot %
\boldsymbol{\psi }_{+}=\frac{1}{2j+1}[(-j\frac{d^{2}}{dr^{2}}\frac{u_{-}}{r}+%
\sqrt{j(j+1)}\frac{d^{2}}{dr^{2}}\frac{u_{+}}{r})\mathbf{Y}_{-}+(-(j+1)\frac{%
d^{2}}{dr^{2}}\frac{u_{+}}{r}+\sqrt{j(j+1)}\frac{d^{2}}{dr^{2}}\frac{u_{-}}{r%
})\mathbf{Y}_{+}],
\end{equation}%
\begin{eqnarray}
\mathbf{pp\cdot }{\boldsymbol{\psi}}_{+} &=&{\frac{1}{2j+1}}\{-j\left( \frac{%
d^{2}}{dr^{2}}+\frac{2}{r}\frac{d}{dr}-\frac{j(j-1)}{r^{2}}\right) \frac{%
u_{-}}{r}  \notag \\
&&+\sqrt{j(j+1)}[\left( \frac{d^{2}}{dr^{2}}+\frac{2}{r}\frac{d}{dr}-\frac{%
(j+2)(j+1)}{r^{2}}\right) +\frac{(2j+1)}{r}\frac{d}{dr}\frac{u_{+}}{r}+\frac{%
(2j+1)(j+2)}{r^{2}}]\frac{u_{+}}{r}\}\mathbf{Y}_{-}  \notag \\
&&+{\frac{1}{2j+1}}\{-(j+1)\left( \frac{d^{2}}{dr^{2}}+\frac{2}{r}\frac{d}{dr%
}-\frac{(j+2)(j+1)}{r^{2}}\right) \frac{u_{+}}{r}  \notag \\
&&+\sqrt{j(j+1)}[\left( \frac{d^{2}}{dr^{2}}+\frac{2}{r}\frac{d}{dr}-\frac{%
j(j-1)}{r^{2}}\right) -\frac{2j+1}{r}\frac{d}{dr}\frac{u_{-}}{r}+\frac{%
(2j-1)(j-1)}{r^{2}}]\frac{u_{-}}{r}\}\mathbf{Y}_{+},
\end{eqnarray}%
\begin{eqnarray}
\mathbf{\hat{r}\hat{r}\cdot pp\cdot }{\boldsymbol{\psi}}_{+} &=&[\sqrt{\frac{%
j}{2j+1}}(\frac{(j-1)}{r}\frac{d}{dr}\frac{u_{-}}{r}-\frac{(j-1)}{r^{2}}%
\frac{u_{-}}{r}-\frac{d^{2}}{dr^{2}}\frac{u_{-}}{r})  \notag \\
&&+\sqrt{\frac{j+1}{2j+1}}(\frac{(j+2)}{r}\frac{d}{dr}\frac{u_{+}}{r}-\frac{%
(j+2)}{r^{2}}\frac{u_{+}}{r}+\frac{d^{2}}{dr^{2}}\frac{u_{+}}{r})](\sqrt{{%
\frac{j}{2j+1}}}\mathbf{Y}_{-}-\sqrt{{\frac{j+1}{2j+1}}}\mathbf{Y}_{+}),
\end{eqnarray}%
Using the simplification 
\begin{equation}
Q_{p}+Q_{m}+Q_{p}Q_{m}=0,
\end{equation}%
and the above identities in Eq. (\ref{imt}) gives 
\begin{eqnarray*}
&&[\mathbf{1}+Q_{p}\mathbf{\hat{r}\hat{r}}]\cdot {\biggl (}[\mathbf{p}-\frac{%
i}{2}\boldsymbol{\nabla }(L+2\mathcal{G})]\times \{[\mathbf{p}-\frac{i}{2}%
\boldsymbol{\nabla }(L-2\mathcal{G})]\times \lbrack \mathbf{1}+Q_{m}\mathbf{%
\hat{r}\hat{r}}]\cdot \boldsymbol{\psi }_{+}\} \\
&&+[\mathbf{p}-\frac{i}{2}\boldsymbol{\nabla }(L+6\mathcal{G})][\mathbf{p}-%
\frac{i}{2}\boldsymbol{\nabla }(3L-2\mathcal{G})]\cdot \lbrack \mathbf{1}%
+Q_{m}\mathbf{\hat{r}\hat{r}}]\cdot \boldsymbol{\psi }_{+}{\biggr )}
\end{eqnarray*}%
\begin{eqnarray}
&=&\frac{1}{2j+1}{\biggl (}\{\Phi _{--}-\mathcal{B}^{2}\exp (-2\mathcal{G)+}2%
\sqrt{j(j+1)}\Phi _{+-}+\frac{A_{mm}}{r^{2}}+\frac{B_{mm}}{r}+C_{mm}+\frac{%
F_{mm}}{r}\frac{d}{dr}+G_{mm}\frac{d}{dr}\}\frac{u_{-}}{r}\mathbf{Y}_{-} 
\notag \\
&&+\mathcal{\{}\Phi _{-+}+\sqrt{j(j+1)}[2\Phi _{++}-2\mathcal{B}^{2}\exp (-2%
\mathcal{G)}+\frac{A_{mp}}{r^{2}}+\frac{B_{mp}}{r}+C_{mp}+\frac{F_{mp}}{r}%
\frac{d}{dr}+G_{mp}\frac{d}{dr}]\}\frac{u_{+}}{r}\mathbf{Y}_{-}  \notag \\
&&+\{-\Phi _{++}+\mathcal{B}^{2}\exp (-2\mathcal{G)}+2\sqrt{j(j+1)}\Phi
_{-+}+\frac{A_{pp}}{r^{2}}+\frac{B_{pp}}{r}+C_{pp}\mathcal{+}\frac{F_{mp}}{r}%
\frac{d}{dr}+G_{pp}\frac{d}{dr}\}\frac{u_{+}}{r}\mathbf{Y}_{+}  \notag \\
&&\{-\Phi _{+-}+\sqrt{j(j+1)}[2\Phi _{--}-2\mathcal{B}^{2}\exp (-2\mathcal{G)%
}+\frac{A_{pm}}{r^{2}}+\frac{B_{pm}}{r}+C_{pm}+\frac{F_{pm}}{r}\frac{d}{dr}%
+G_{pm}\frac{d}{dr}]\}\frac{u_{-}}{r}\mathbf{Y}_{+}{\biggr )}.  \label{damp}
\end{eqnarray}

The quasipotentials $\Phi _{\pm \pm }$ come from \cite{liu} or \cite{long}
or alternatively from taking the vector part of the matrix form of Eq. (\ref%
{57})) 
\begin{eqnarray}
\lbrack -\frac{1}{r}\frac{d^{2}}{dr^{2}}r+\frac{(j+2)(j+1)}{r^{2}}+\Phi
_{++}]\frac{u_{+}}{r}+\Phi _{+-}\frac{u_{-}}{r} &=&\mathcal{B}^{2}\exp (-2 
\mathcal{G)}\frac{u_{+}}{r},  \notag \\
\lbrack -\frac{1}{r}\frac{d^{2}}{dr^{2}}r+\frac{j(j-1)}{r^{2}}+\Phi _{--}] 
\frac{u_{-}}{r}+\Phi _{-+}\frac{u_{+}}{r} &=&\mathcal{B}^{2}\exp (-2\mathcal{%
\ G)}\frac{u_{-}}{r},
\end{eqnarray}
in which we find 
\begin{eqnarray}
\text{ }\Phi _{++} &=&(\frac{5}{2}+\frac{3}{2\left( 2j+1\right) })\mathcal{G}
^{\prime 2}-(\frac{1}{2\left( 2j+1\right) }+\frac{3}{2})\mathcal{G}^{\prime
}L^{\prime }+\frac{L^{\prime 2}}{4}+\frac{L^{\prime }}{2r}(1+\frac{3}{2j+1})
\notag \\
&&+(\frac{3}{2}+\frac{1}{2\left( 2j+1\right) })\nabla ^{2}\mathcal{G-}\frac{
\nabla ^{2}L}{2}+\frac{(j+2)}{r}(-3\mathcal{G}^{\prime }+L^{\prime }-\frac{( 
\mathcal{G}+L)^{\prime }}{\left( 2j+1\right) })  \notag \\
&&-[(\cosh 2K-1)(3\mathcal{G}^{\prime }-L^{\prime }-\frac{2}{r})+(\mathcal{G}
+L)^{\prime }\sinh 2K]\frac{2j(j+1)}{(2j+1)r},
\end{eqnarray}
\begin{eqnarray}
\text{ }\Phi _{--} &=&(\frac{5}{2}-\frac{3}{2(2j+1)})\mathcal{G}^{\prime
2}+( \frac{1}{2\left( 2j+1\right) }-\frac{3}{2})\mathcal{G}^{\prime
}L^{\prime }+ \frac{L^{\prime 2}}{4}-\frac{L^{\prime }}{2r}(-1+\frac{3}{2j+1}%
)  \notag \\
&&+(\frac{3}{2}-\frac{1}{2\left( 2j+1\right) })\nabla ^{2}\mathcal{G-}\frac{
\nabla ^{2}L}{2}+\frac{(j-1)}{r}(3\mathcal{G}^{\prime }-L-\frac{(\mathcal{G+}
L)^{\prime }}{\left( 2j+1\right) })  \notag \\
&&+[(\cosh 2K-1)(2\mathcal{G}^{\prime }-(L-\mathcal{G})^{\prime }-\frac{2}{r}
)+(\mathcal{G}+L)^{\prime }\sinh 2K]\frac{2j(j+1)}{(2j+1)r},
\end{eqnarray}
\begin{eqnarray}
&&\Phi _{+-}=\sqrt{j(j+1)}[\frac{1}{2j+1}\left( \frac{3}{r}\mathcal{G}
^{\prime }+\frac{3(\mathcal{G+}L)^{\prime }}{r}-\nabla ^{2}\mathcal{G}-2 
\mathcal{G}^{\prime }\left( \frac{3\mathcal{G}^{\prime }}{2}-\frac{L^{\prime
}}{2}\right) \right) +\frac{(\frac{E}{M}-2)}{r^{2}}  \notag \\
&&-\frac{\left( \frac{3\mathcal{G}^{\prime }}{2}-\frac{L^{\prime }}{2}
\right) }{r}(\frac{\frac{E}{M}+\frac{M}{E}-2}{2\left( 2j+1\right) }+\frac{ 
\frac{E}{M}-\frac{M}{E}}{2})+\frac{(\mathcal{G+}L)^{\prime }}{2r}(\frac{E}{M}
+\frac{M}{E}+\frac{\frac{E}{M}-\frac{M}{E}}{(2j+1)})],
\end{eqnarray}
\begin{eqnarray}
&&\Phi _{-+}=\sqrt{j(j+1)}[\frac{1}{2j+1}\left( \frac{3}{r}\mathcal{G}
^{\prime }+\frac{3(\mathcal{G+}L)^{\prime }}{r}-\nabla ^{2}\mathcal{G}-2 
\mathcal{G}^{\prime }\left( \frac{3\mathcal{G}^{\prime }}{2}-\frac{L^{\prime
}}{2}\right) \right) +\frac{(\frac{M}{E}-2)}{r^{2}}  \notag \\
&&-\frac{\left( \frac{3\mathcal{G}^{\prime }}{2}-\frac{L^{\prime }}{2}
\right) }{r}(\frac{\frac{E}{M}+\frac{M}{E}-2}{2\left( 2j+1\right) }-\frac{ 
\frac{E}{M}-\frac{M}{E}}{2})-\frac{(\mathcal{G+}L)^{\prime }}{2r}(\frac{E}{M}
+\frac{M}{E}+\frac{\frac{E}{M}-\frac{M}{E}}{(2j+1)})].
\end{eqnarray}
The other functions are 
\begin{eqnarray}
A_{mm} &=&-2j(j+1)Q_{m},  \notag \\
B_{mm} &=&j(j+1)(Q_{p}+Q_{m})(\frac{L^{\prime }}{2}+3\mathcal{G}^{\prime
})+(2j^{2}+j-1)L^{\prime }+\mathcal{G}^{\prime }(4j^{2}-2j+2),  \notag \\
C_{mm} &=&\frac{1}{2}\partial ^{2}L+\partial ^{2}\mathcal{G}(j-1)+\frac{1}{4}
(2j+1)L^{\prime 2}-2jL^{\prime }\mathcal{G}^{\prime }+(3j-1)\mathcal{G}
^{\prime 2},  \notag \\
F_{mm} &=&2j(j+1)(Q_{p}-Q_{m}),  \notag \\
G_{mm} &=&(j+1)L^{\prime },
\end{eqnarray}
\begin{eqnarray}
A_{mp} &=&2[(2j+1)(j+2)+(j+1)Q_{m}],  \notag \\
B_{mp} &=&[\frac{1}{2}j(Q_{p}-Q_{m})+j-\frac{1}{2}(5+Q_{m})]L^{\prime
}+[-3(j+1)Q_{m}+3jQ_{p}-2j+7]\mathcal{G}^{\prime },  \notag \\
C_{mp} &=&\partial ^{2}L-3\partial ^{2}\mathcal{G}-5\mathcal{G}^{\prime
2}+2L^{\prime }\mathcal{G}^{\prime },  \notag \\
F_{mp} &=&[2(2j+1)+2(Q_{m}+Q_{p})(j+1)-2Q_{p}],  \notag \\
G_{mp} &=&L^{\prime },
\end{eqnarray}
\begin{eqnarray}
A_{pp} &=&2j(j+1)Q_{m},  \notag \\
B_{pp} &=&-(j+1)j(Q_{p}+Q_{m})(\frac{L}{2}+3\mathcal{G)}^{\prime
}+(-2j^{2}-2j+1)L^{\prime }-(4j^{2}+6j)\mathcal{G}^{\prime },  \notag \\
C_{pp} &=&-\frac{1}{2}\partial ^{2}L+(2+j)\partial ^{2}\mathcal{G}+\frac{1}{%
4 }(2j+1)L^{\prime 2}-2(j+1)L^{\prime }\mathcal{G}^{\prime }+(3j+4)\mathcal{G%
} ^{\prime 2},  \notag \\
F_{pp} &=&2j(j+1)(Q_{m}-Q_{p})~\ ,  \notag \\
G_{pp} &=&jL^{\prime },
\end{eqnarray}
\begin{eqnarray}
A_{pm} &=&2(j-1)(2j-1)-3jQ_{m},  \notag \\
B_{pm} &=&[-\frac{1}{2}(j+1)Q_{p}+\frac{1}{2}jQ_{m}-j-\frac{7}{2}]L^{\prime
}+[3jQ_{m}-3(j+1)Q_{p}+9+2j]\mathcal{G}^{\prime },  \notag \\
C_{pm} &=&\partial ^{2}L-3\partial ^{2}\mathcal{G}-5\mathcal{G}^{\prime
2}+2L^{\prime }\mathcal{G}^{\prime },  \notag \\
F_{pm} &=&[-2(1+2j)-(2j+1)Q_{p}-2jQ_{m}],  \notag \\
G_{pm} &=&L^{\prime },
\end{eqnarray}

For the $^{3}P_{0}$ states we need 
\begin{equation}
\text{ }\Phi _{++}=4\mathcal{G}^{\prime 2}-2\mathcal{G}^{\prime }L^{\prime
}+ \frac{L^{\prime 2}}{4}+2\nabla ^{2}\mathcal{G-}\frac{\nabla ^{2}L}{2}+%
\frac{ (2L-8\mathcal{G)}^{\prime }}{r},
\end{equation}
and 
\begin{eqnarray}
A_{pp} &=&0,  \notag \\
B_{pp} &=&2L^{\prime }-8\mathcal{G}^{\prime },  \notag \\
C_{pp} &=&-\frac{1}{2}\partial ^{2}L+2\partial ^{2}\mathcal{G}+\frac{1}{4}
(L^{\prime }-4\mathcal{G}^{\prime })^{2},  \notag \\
F_{pp} &=&0,  \notag \\
G_{pp} &=&0.
\end{eqnarray}
Note that these coefficients are such as to cancel the effects of $\Phi
_{++} $ in Eq. (\ref{damp}) exactly!

For the $^{3}P_{2}$ states we require 
\begin{eqnarray}
\Phi _{--} &=&\frac{11}{5}\mathcal{G}^{\prime 2}-\frac{7}{5}\mathcal{G}
^{\prime }L^{\prime }+\frac{L^{\prime 2}}{4}+\frac{L^{\prime }}{5r}+\frac{7}{
5}\nabla ^{2}\mathcal{G-}\frac{\nabla ^{2}L}{2}+\frac{1}{r}(3\mathcal{G}
^{\prime }-L-\frac{(\mathcal{G+}L)^{\prime }}{5})  \notag \\
&&+\frac{12}{5r}[(\cosh 2K-1)(2\mathcal{G}^{\prime }-(L-\mathcal{G})^{\prime
}-\frac{2}{r})+(\mathcal{G}+L)^{\prime }\sinh 2K],
\end{eqnarray}
and 
\begin{eqnarray}
&&\Phi _{-+}=\sqrt{6}[\frac{1}{5}\left( \frac{3}{r}\mathcal{G}^{\prime }+ 
\frac{3(\mathcal{G+}L)^{\prime }}{r}-\nabla ^{2}\mathcal{G}-2\mathcal{G}
^{\prime }\left( \frac{3\mathcal{G}^{\prime }}{2}-\frac{L^{\prime }}{2}
\right) \right) +\frac{(\exp (2K)-2)}{r^{2}}  \notag \\
&&-\frac{1}{r}\left( \frac{3\mathcal{G}^{\prime }}{2}-\frac{L^{\prime }}{2}
\right) (\frac{\cosh 2K-1}{5}+\sinh 2K)-\frac{(\mathcal{G+}L)^{\prime }}{r}
(\cosh 2K-\frac{\sinh 2K}{5})],
\end{eqnarray}
\begin{eqnarray}
A_{mm} &=&-6Q_{m},  \notag \\
B_{mm} &=&(Q_{p}+Q_{m})(3L^{\prime }+18\mathcal{G}^{\prime })+9L^{\prime
}+14 \mathcal{G}^{\prime },  \notag \\
C_{mm} &=&\frac{1}{2}\partial ^{2}L+\partial ^{2}\mathcal{G}+\frac{5}{4}
L^{\prime 2}-4L^{\prime }\mathcal{G}^{\prime }+5\mathcal{G}^{\prime 2}, 
\notag \\
F_{mm} &=&12Q_{p}-12Q_{m},  \notag \\
G_{mm} &=&3L^{\prime },
\end{eqnarray}
\begin{eqnarray}
A_{mp} &=&2[20+3Q_{m}],  \notag \\
B_{mp} &=&[(Q_{p}-\frac{3}{2}Q_{m})-\frac{1}{2}]L^{\prime
}+[-9Q_{m}+6Q_{p}+3]\mathcal{G}^{\prime },  \notag \\
C_{mp} &=&\partial ^{2}L-3\partial ^{2}\mathcal{G}-5\mathcal{G}^{\prime
2}+2L^{\prime }\mathcal{G}^{\prime },  \notag \\
F_{mp} &=&10+6Q_{m}+4Q_{p},  \notag \\
G_{mp} &=&L^{\prime },
\end{eqnarray}
and 
\begin{eqnarray}
\Phi _{++} &=&\frac{14}{5}\mathcal{G}^{\prime 2}-\frac{8}{5}\mathcal{G}
^{\prime }L^{\prime }+\frac{L^{\prime 2}}{4}+\frac{4L^{\prime }}{5r}  \notag
\\
&&+\frac{8}{5}\nabla ^{2}\mathcal{G-}\frac{\nabla ^{2}L}{2}+\frac{4}{r}(-3 
\mathcal{G}^{\prime }+L^{\prime }-\frac{(\mathcal{G}+L)^{\prime }}{5}) 
\notag \\
&&-\frac{12}{5r}[(\cosh 2K-1)(3\mathcal{G}^{\prime }-L^{\prime }-\frac{2}{r}
)+(\mathcal{G}+L)^{\prime }\sinh 2K],
\end{eqnarray}
and 
\begin{eqnarray}
&&\Phi _{+-}=\sqrt{6}[\frac{1}{5}\left( \frac{3}{r}\mathcal{G}^{\prime }+ 
\frac{3(\mathcal{G+}L)^{\prime }}{r}-\nabla ^{2}\mathcal{G}-2\mathcal{G}
^{\prime }\left( \frac{3\mathcal{G}^{\prime }}{2}-\frac{L^{\prime }}{2}
\right) \right) +\frac{(\exp (2K)-2)}{r^{2}}  \notag \\
&&-\frac{1}{r}\left( \frac{3\mathcal{G}^{\prime }}{2}-\frac{L^{\prime }}{2}
\right) (\frac{\cosh 2K-1}{5}-\sinh 2K)+\frac{(\mathcal{G+}L)^{\prime }}{r}
(\cosh 2K-\frac{\sinh 2K}{5})],
\end{eqnarray}
\begin{eqnarray*}
A_{pp} &=&12Q_{m}, \\
B_{pp} &=&-(Q_{p}+Q_{m})(3L+18\mathcal{G)}^{\prime }-11L^{\prime }-28 
\mathcal{G}^{\prime }, \\
C_{pp} &=&-\frac{1}{2}\partial ^{2}L+4\partial ^{2}\mathcal{G}+\frac{5}{4}
L^{\prime 2}-6L^{\prime }\mathcal{G}^{\prime }+10\mathcal{G}^{\prime 2}, \\
F_{pp} &=&12(Q_{m}-Q_{p}), \\
G_{pp} &=&2L^{\prime },
\end{eqnarray*}
\begin{eqnarray}
A_{pm} &=&6-6Q_{m},  \notag \\
B_{pm} &=&[-\frac{3}{2}Q_{p}+Q_{m}-\frac{11}{2}]L^{\prime
}+[6Q_{m}-9Q_{p}+13]\mathcal{G}^{\prime },  \notag \\
C_{pm} &=&\partial ^{2}L-3\partial ^{2}\mathcal{G}-5\mathcal{G}^{\prime
2}+2L^{\prime }\mathcal{G}^{\prime },  \notag \\
F_{pm} &=&[-10-5Q_{p}-4Q_{m}],  \notag \\
G_{pm} &=&L^{\prime }.
\end{eqnarray}

\subsubsection{The Wave Function $\mathcal{\protect\eta }_{+0}$}

For this wave function (see Eq. (\ref{etap})) we need the single term
already evaluated in Eq. (\ref{ip1}) 
\begin{eqnarray}
&&[\mathbf{p}-iD^{\prime }\mathbf{\hat{r}]}\cdot \lbrack \mathbf{1}+Q_{m} 
\mathbf{\hat{r}\hat{r}}]\cdot \boldsymbol{\psi }_{+}  \notag \\
&=&\mathbf{p}\cdot \boldsymbol{\psi }_{+}+i(2K-D)^{\prime }(Q_{m}+1\mathbf{) 
\hat{r}\cdot }\boldsymbol{\psi }_{+}+Q_{m}\mathbf{p}\cdot \mathbf{\hat{r} 
\hat{r}}\cdot \boldsymbol{\psi }_{+}.
\end{eqnarray}
Use 
\begin{equation}
\mathbf{p}\cdot \mathbf{\hat{r}\hat{r}}\cdot \boldsymbol{\psi }_{+}=(\mathbf{%
\ \hat{r}}\cdot \mathbf{p}-\frac{2i}{r}\mathbf{)\hat{r}\cdot }%
\boldsymbol{\psi }_{+},
\end{equation}
and we obtain 
\begin{eqnarray}
&&[\mathbf{p}-iD^{\prime }\mathbf{\hat{r}]}\cdot \lbrack \mathbf{1}+Q_{m} 
\mathbf{\hat{r}\hat{r}}]\cdot \boldsymbol{\psi }_{+}  \notag \\
&=&\mathbf{p}\cdot \boldsymbol{\psi }_{+}+i[(2K-D)^{\prime }(Q_{m}+1)-\frac{
2Q_{m}}{r}-\frac{d}{dr}]\mathbf{\hat{r}}\cdot \boldsymbol{\psi }_{+},
\end{eqnarray}
with 
\begin{eqnarray}
\mathbf{p}\cdot \boldsymbol{\psi }_{+} &=&i[\sqrt{\frac{j}{2j+1}}(\frac{%
(j-1) }{r}-\frac{d}{dr})\frac{u_{-}}{r}+\sqrt{\frac{j+1}{2j+1}}(\frac{(j+2)}{%
r}+ \frac{d}{dr})\frac{u_{+}}{r}]Y_{jm},  \notag \\
\mathbf{\hat{r}}\cdot \boldsymbol{\psi }_{+} &=&[-\sqrt{{\frac{j+1}{2j+1}}} 
\frac{u_{+}}{r}+\sqrt{{\frac{j}{2j+1}}}\frac{u_{-}}{r}]Y_{jm},
\end{eqnarray}
Thus, 
\begin{eqnarray}
\mathcal{\eta }_{+0} &=&i\frac{\exp (\mathcal{G}+2K\mathcal{)}}{E}\{[\frac{
(j-1)-2Q_{m}}{r}-(Q_{m}+1)\frac{d}{dr}+(2K-D)^{\prime }(Q_{m}+1)]\sqrt{\frac{
j}{2j+1}}\frac{u_{-}}{r}  \notag \\
&&+[\frac{(j+2)+2Q_{m}}{r}+(Q_{m}+1)\frac{d}{dr}-(2K-D)^{\prime }(Q_{m}+1)] 
\sqrt{\frac{j+1}{2j+1}}\frac{u_{+}}{r}\}Y_{jm}.
\end{eqnarray}

\subsubsection{ \protect\bigskip The Wave Function $\boldsymbol{\protect%
\eta
} _{-}$}

This wave function involves the term

\begin{equation}
\lbrack \mathbf{1}+Q_{p}\mathbf{\hat{r}\hat{r}}]\cdot \lbrack i\mathbf{p}
+C^{\prime }\mathbf{\hat{r}]}\times \lbrack \mathbf{1}+Q_{m}\mathbf{\hat{r} 
\hat{r}}]\cdot \boldsymbol{\psi }_{+}.
\end{equation}
Eq. (\ref{ip2}) gives the inner portion of this term as 
\begin{eqnarray}
&&[i\mathbf{p}+C^{\prime }\mathbf{\hat{r}]}\times \lbrack \mathbf{1}+Q_{m} 
\mathbf{\hat{r}\hat{r}}]\cdot \boldsymbol{\psi }_{+}  \notag \\
&=&i\mathbf{p}\times \boldsymbol{\psi }_{+}-i\frac{Q_{m}}{r}\mathbf{L\hat{r}}
\cdot \boldsymbol{\psi }_{+}+C^{\prime }\mathbf{\hat{r}\times } %
\boldsymbol{\psi }_{+}.
\end{eqnarray}
Multiplying by $[\mathbf{1}+Q_{p}\mathbf{\hat{r}\hat{r}}]\cdot $ gives 
\begin{eqnarray}
&&[\mathbf{1}+Q_{p}\mathbf{\hat{r}\hat{r}}]\cdot \lbrack i\mathbf{p}
+C^{\prime }\mathbf{\hat{r}]}\times \lbrack \mathbf{1}+Q_{m}\mathbf{\hat{r} 
\hat{r}}]\cdot \boldsymbol{\psi }_{+}  \notag \\
&=&i\mathbf{p}\times \boldsymbol{\psi }_{+}+\frac{i}{r}\hat{r}L\cdot %
\boldsymbol{\psi }_{+} -i\frac{Q_{m}}{r}\mathbf{L\hat{r}}\cdot %
\boldsymbol{\psi } _{+}+C^{\prime }\mathbf{\hat{r}\times }\boldsymbol{\psi }%
_{+}.
\end{eqnarray}
With the definition 
\begin{equation}
\mathbf{X}_{jm}=\frac{{\boldsymbol{L}}Y_{jm}}{\sqrt{j(j+1)}},
\end{equation}
we have 
\begin{eqnarray}
\mathbf{p}\times \boldsymbol{\psi }_{+} &=&[(\frac{d}{dr}-\frac{j-1}{r}) 
\sqrt{\frac{j+1}{2j+1}}\frac{u_{-}(r)}{r}+(\frac{d}{dr}+\frac{j+2}{r})\sqrt{ 
\frac{j}{2j+1}}\frac{u_{+}(r)}{r}]\mathbf{X}_{jm},  \notag \\
\mathbf{L}\cdot \boldsymbol{\psi }_{+} &=&0,  \notag \\
\mathbf{L\hat{r}}\cdot \boldsymbol{\psi }_{+} &=&\sqrt{j(j+1)}[-\sqrt{\frac{
j+1}{2j+1}}\frac{u_{+}}{r}+\sqrt{\frac{j}{2j+1}}\frac{u_{-}}{r}]\mathbf{X}
_{jm},  \notag \\
\mathbf{\hat{r}\times }\boldsymbol{\psi }_{+} &=&i[\frac{u_{-}(r)}{r}\sqrt{ 
\frac{j+1}{2j+1}}+\frac{u_{+}(r)}{r}\sqrt{\frac{j}{2j+1}}]\mathbf{X}_{jm},
\end{eqnarray}
and so 
\begin{eqnarray}
&&[\mathbf{1}+Q_{p}\mathbf{\hat{r}\hat{r}}\cdot ][i\mathbf{p}+C^{\prime } 
\mathbf{\hat{r}]}\times \lbrack 1+Q_{m}\mathbf{\hat{r}\hat{r}}\cdot ] %
\boldsymbol{\psi }_{+}  \notag \\
&=&i\{[(\frac{d}{dr}-\frac{j-1}{r}-\frac{jQ_{m}}{r}+C^{\prime })]\sqrt{\frac{
j+1}{2j+1}}\frac{u_{-}(r)}{r}  \notag \\
&&+[(\frac{d}{dr}+\frac{j+2}{r}+\frac{(j+1)Q_{m}}{r}+C^{\prime })]\sqrt{ 
\frac{j}{2j+1}}\frac{u_{+}}{r}\}\mathbf{X}_{jm}.
\end{eqnarray}

\section{Plane Wave Integrals}

\subsection{States with $j=l$}

For general $j=l$ states we consider separately integrals from Eq. (\ref{lm}%
) of the form 
\begin{eqnarray}
&&\int d^{3}r\exp (-i\mathbf{k\cdot r)\hat{r}}g(r)Y_{jm}(\mathbf{\Omega }), 
\notag \\
&&\int d^{3}r\exp (-i\mathbf{k\cdot r)}f_{\pm }(r)\mathbf{Y}_{jm\pm }(%
\mathbf{\Omega }).
\end{eqnarray}%
Consider first 
\begin{equation}
\int d^{3}r\exp (-i\mathbf{k\cdot r)\hat{r}}g(r)Y_{jm}(\mathbf{\Omega }).
\end{equation}%
We have 
\begin{eqnarray}
\mathbf{\hat{r}} &=&\sin \theta (\cos \phi \mathbf{\hat{x}}+\sin \theta 
\mathbf{\hat{y})+}\cos \theta \mathbf{\hat{z}}  \notag \\
&=&\sqrt{\frac{4\pi }{3}}Y_{10}(\mathbf{\Omega )\hat{z}}-\sqrt{\frac{2\pi }{3%
}}Y_{11}(\mathbf{\Omega )(\hat{x}-}i\mathbf{\hat{y})+}\sqrt{\frac{2\pi }{3}}%
Y_{1-1}(\mathbf{\Omega )(\hat{x}+}i\mathbf{\hat{y}).}
\end{eqnarray}%
Here $\mathbf{\hat{x},\hat{y},\hat{z}}$ are arbitrary unit vectors fixed in
space. \ And so 
\begin{eqnarray}
&&\int d^{3}r\exp (-i\mathbf{k\cdot r)\hat{r}}g(r)Y_{jm}(\mathbf{\Omega }) 
\notag \\
&=&\mathbf{\hat{z}}\sqrt{\frac{4\pi }{3}}\int d^{3}r\exp (-i\mathbf{k\cdot r)%
}g(r)Y_{10}(\mathbf{\Omega )}Y_{jm}(\mathbf{\Omega })  \notag \\
&&-\mathbf{(\hat{x}-}i\mathbf{\hat{y})}\sqrt{\frac{2\pi }{3}}\int d^{3}r\exp
(-i\mathbf{k\cdot r)\hat{r}}g(r)Y_{11}(\mathbf{\Omega )}Y_{jm}(\mathbf{\
\Omega })  \notag \\
&&+\mathbf{(\hat{x}+}i\mathbf{\hat{y})}\sqrt{\frac{2\pi }{3}}\int d^{3}r\exp
(-i\mathbf{k\cdot r)\hat{r}}g(r)Y_{1-1}(\mathbf{\Omega )}Y_{jm}(\mathbf{\
\Omega }).
\end{eqnarray}%
Now with 
\begin{equation}
\exp (-i\mathbf{k\cdot r}y\mathbf{)=}4\pi \sum_{j^{\prime }=0}^{\infty
}\sum_{m^{\prime }=-j^{\prime }}^{+j^{\prime }}(-i)^{j^{\prime
}}j_{j^{\prime }}(kr)Y_{j^{\prime }m^{\prime }}^{\ast }(\mathbf{\Omega }%
)Y_{j^{\prime }m^{\prime }}(\mathbf{\Omega }_{k}),  \label{plnwv}
\end{equation}%
in which we define the angles $\mathbf{\Omega }_{k}(\theta _{k},\phi _{k})$
relative to the fixed unit vectors $\mathbf{\hat{x},\hat{y},\hat{z},}~$we
use 
\begin{eqnarray}
\int d\Omega Y_{j^{\prime }m^{\prime }}^{\ast }Y_{1m^{\prime \prime }}Y_{jm}
&=&\sqrt{\frac{(2j+1)3}{4\pi (2j^{\prime }+1)}}\langle j1;00|j^{\prime
}0\rangle \langle j1;mm^{\prime \prime }|j^{\prime }m+m^{\prime \prime
}\rangle ;\text{ }j^{\prime }+j+1\text{ even}  \notag \\
&=&0;\text{ }j^{\prime }+j+1\text{ odd,}  \label{jodd}
\end{eqnarray}%
and find 
\begin{eqnarray}
&&\int d^{3}r\exp (-i\mathbf{k\cdot r)\hat{r}}g(r)Y_{jm}(\mathbf{\Omega }) 
\notag \\
&=&4\pi \sum_{j^{\prime }=|j-1|}^{j+1}(-i)^{j^{\prime }}\frac{%
(1-(-1)^{j+j^{\prime }})}{2}\sqrt{\frac{(2j+1)}{(2j^{\prime }+1)}}\langle
j1;00|j^{\prime }0\rangle \int_{0}^{\infty }drr^{2}j_{j^{\prime
}}(kr)g(r)\times \lbrack \mathbf{\hat{z}}Y_{j^{\prime }m}(\mathbf{\Omega }%
_{k})\langle j1;m0|j^{\prime }m\rangle  \notag \\
&&-\frac{\mathbf{(\hat{x}-}i\mathbf{\hat{y})}}{\sqrt{2}}Y_{j^{\prime }m+1}(%
\mathbf{\Omega }_{k})\langle j1;m1|j^{\prime }m+1\rangle +\frac{\mathbf{(%
\hat{x}+}i\mathbf{\hat{y})}}{\sqrt{2}}Y_{j^{\prime }m-1}(\mathbf{\Omega }%
_{k})\langle j1;m-1|j^{\prime }m-1\rangle ].  \label{kr}
\end{eqnarray}

Next we examine 
\begin{eqnarray}
&&\int d^{3}r\exp (-i\mathbf{k\cdot r)}f_{\pm }(r\mathbf{\mathbf{)}Y}_{jm\pm
}(\mathbf{\Omega })  \notag \\
&=&\int d^{3}r\exp (-i\mathbf{k\cdot r)}f_{\pm }(r\mathbf{)[}a_{\pm }\mathbf{%
\ \hat{r}}Y_{jm}\mathbf{+}rb_{\pm }\mathbf{p}Y_{jm}].
\end{eqnarray}%
Use 
\begin{eqnarray}
&&\int d^{3}r\exp (-i\mathbf{k\cdot r)}f_{\pm }(r)\mathbf{p}Y_{jm}(\Omega ) 
\notag \\
&=&\mathbf{k}\int d^{3}r\exp (-i\mathbf{k\cdot r)}f_{\pm }(r)Y_{jm}(\Omega
)+i\int d^{3}r\exp (-i\mathbf{k\cdot r)\hat{r}}f_{\pm }^{\prime
}(r)Y_{jm}(\Omega ),
\end{eqnarray}%
and Eq. (\ref{kr}) 
\begin{eqnarray}
&&\int d^{3}r\exp (-i\mathbf{k\cdot r)}f_{\pm }(r)\mathbf{Y}_{jm\pm }(\Omega
)  \notag \\
&=&\int d^{3}r\exp (-i\mathbf{k\cdot r)[(}a_{\pm }f_{\pm }(r)\mathbf{+}%
ib_{\pm }(rf_{\pm }^{\prime }(r)+f_{\pm }(r))\mathbf{\hat{r}}Y_{jm}\mathbf{+}%
rb_{\pm }f_{\pm }(r)\mathbf{k}Y_{jm}]  \notag \\
&=&4\pi \mathbf{\hat{k}}Y_{jm}(\mathbf{\Omega }_{k})(-i)^{j}\int_{0}^{\infty
}drr^{2}j_{j}(kr)krb_{\pm }f_{\pm }(r)  \notag \\
&&+4\pi \sum_{j^{\prime }=|j-1|}^{j+1}(-i)^{j^{\prime }}\frac{%
(1-(-1)^{j+j^{\prime }})}{2}\sqrt{\frac{(2j+1)}{(2j^{\prime }+1)}}\langle
j1;00|j^{\prime }0\rangle  \notag \\
&&\int_{0}^{\infty }drr^{2}f_{\pm }(r)\mathbf{[(}a_{\pm }-2ib_{\pm
})j_{j^{\prime }}(kr)-ib_{\pm }j_{j^{\prime }}^{\prime }(kr)kr][\mathbf{\hat{%
z}}Y_{j^{\prime }m}(\mathbf{\Omega }_{k})\langle j1;m0|j^{\prime }m\rangle
\label{pkr} \\
&&-\frac{\mathbf{(\hat{x}-}i\mathbf{\ \hat{y})}}{\sqrt{2}}Y_{j^{\prime }m+1}(%
\mathbf{\Omega }_{k})\langle j1;m1|j^{\prime }m+1\rangle +\frac{\mathbf{(%
\hat{x}+}i\mathbf{\hat{y})}}{\sqrt{2}}Y_{j^{\prime }m-1}(\mathbf{\Omega }%
_{k})\langle j1;m-1|j^{\prime }m-1\rangle ].  \notag
\end{eqnarray}%
Likewise 
\begin{eqnarray}
&&\int d^{3}r\exp (+i\mathbf{k\cdot r)}f_{\pm }(r)\mathbf{Y}_{jm\pm }(\Omega
)  \notag \\
&=&-4\pi \mathbf{\hat{k}}Y_{jm}(\mathbf{\Omega }_{k})(-i)^{j}(-)^{j}%
\int_{0}^{\infty }drr^{2}j_{j}(kr)krb_{\pm }f_{\pm }(r)  \notag \\
&&+4\pi \sum_{j^{\prime }=|j-1|}^{j+1}(-i)^{j^{\prime }}(-)^{j^{\prime }}%
\frac{(1-(-1)^{j+j^{\prime }})}{2}\sqrt{\frac{(2j+1)}{(2j^{\prime }+1)}}%
\langle j1;00|j^{\prime }0\rangle  \notag \\
&&\times \int_{0}^{\infty }drr^{2}f_{\pm }(r)\mathbf{[(}a_{\pm }-2ib_{\pm
})j_{j^{\prime }}(kr)-ib_{\pm }j_{j^{\prime }}^{\prime }(kr)kr][\mathbf{\hat{%
z}}Y_{j^{\prime }m}(\mathbf{\Omega }_{k})\langle j1;m0|j^{\prime }m\rangle 
\notag \\
&&-\frac{\mathbf{(\hat{x}-}i\mathbf{\ \hat{y})}}{\sqrt{2}}Y_{j^{\prime }m+1}(%
\mathbf{\Omega }_{k})\langle j1;m1|j^{\prime }m+1\rangle +\frac{\mathbf{(%
\hat{x}+}i\mathbf{\hat{y})}}{\sqrt{2}}Y_{j^{\prime }m-1}(\mathbf{\Omega }%
_{k})\langle j1;m-1|j^{\prime }m-1\rangle ].
\end{eqnarray}

We choose our unit vectors to be defined in terms of the photon decay
momenta and transverse polarization vectors, 
\begin{eqnarray}
\mathbf{\hat{z}}\mathbf{=\hat{k}} &&  \notag \\
\frac{\mathbf{(\hat{x}\pm }i\mathbf{\hat{y})}}{\sqrt{2}} &=&%
\boldsymbol{\epsilon }^{(\pm )}.  \label{zxy}
\end{eqnarray}%
With this choice, the spherical harmonic reduces to just 
\begin{equation}
Y_{jm}(\mathbf{\Omega }_{k})=\delta _{m0}\sqrt{\frac{2j+1}{4\pi }}.
\label{smp}
\end{equation}%
Then Eq. (\ref{kr}) becomes 
\begin{eqnarray}
&&\int d^{3}r\exp (-i\mathbf{k\cdot r)\hat{r}}g(r)Y_{jm}(\mathbf{\Omega }) 
\notag \\
&=&\sqrt{4\pi (2j+1)}\sum_{j^{\prime }=|j-1|}^{j+1}(-i)^{j^{\prime }}\frac{%
(1-(-1)^{j+j^{\prime }})}{2}\langle j1;00|j^{\prime }0\rangle
\int_{0}^{\infty }drr^{2}j_{j^{\prime }}(kr)g(r)  \notag \\
&&\times \lbrack \mathbf{\hat{k}}\langle j1;00|j^{\prime }0\rangle \delta
_{m0}-\boldsymbol{\epsilon }^{(-)}\langle j1;-11|j^{\prime }0\rangle \delta
_{m-1}+\boldsymbol{\epsilon }^{(+)}\langle j1;1-11|j^{\prime }0\rangle
\delta _{m1}],  \label{ry}
\end{eqnarray}%
and Eq. (\ref{pkr}) becomes 
\begin{eqnarray}
&&\int d^{3}r\exp (-i\mathbf{k\cdot r)}f_{\pm }(r)\mathbf{Y}_{jm\pm }(\Omega
)  \notag \\
&=&\sqrt{4\pi (2j+1)}\mathbf{\hat{k}}\delta _{m0}(-i)^{j}\int_{0}^{\infty
}drr^{2}j_{j}(kr)krb_{\pm }f_{\pm }(r)  \notag \\
&&+\sqrt{4\pi (2j+1)}\sum_{j^{\prime }=|j-1|}^{j+1}(-i)^{j^{\prime }}\frac{%
(1-(-1)^{j+j^{\prime }})}{2}\langle j1;00|j^{\prime }0\rangle
\int_{0}^{\infty }drr^{2}f_{\pm }(r)\mathbf{[(}a_{\pm }-2ib_{\pm
})j_{j^{\prime }}(kr)-ib_{\pm }j_{j^{\prime }}^{\prime }(kr)kr]  \notag \\
&&\times \lbrack \mathbf{\hat{k}}\langle j1;00|j^{\prime }0\rangle \delta
_{m0}-\boldsymbol{\epsilon }^{(-)}\langle j1;-11|j^{\prime }0\rangle \delta
_{m-1}+\boldsymbol{\epsilon }^{(+)}\langle j1;1-11|j^{\prime }0\rangle
\delta _{m1}].  \label{y}
\end{eqnarray}%
In the expression Eq. (\ref{lm}) the prefactor $\boldsymbol{\epsilon }%
^{(\alpha _{1})}\times \boldsymbol{\epsilon }^{(\alpha _{2})}$ is orthogonal
to $\boldsymbol{\epsilon }^{\left( \pm \right) }$. \ Hence, only the $%
\mathbf{\hat{k}}$ terms in Eqs. (\ref{ry}) and (\ref{y}) will contribute and
including forms for $g(r)$ and $f_{\pm }(r)$

\begin{eqnarray}
&&\sqrt{\pi }e^{2}\boldsymbol{\epsilon }^{(\alpha _{1})}\times %
\boldsymbol{\epsilon }^{(\alpha _{2})}\cdot \int d^{3}r\exp (-i\mathbf{\
k\cdot r)}\exp (F)\mathbf{[\hat{r}}\frac{u_{j0j}^{-}(r)}{r}\exp
(K)Y_{jm}(\Omega )\left( \frac{\exp (-mr)}{r}\right) ^{\prime }  \notag \\
&&-\exp (-K)(\frac{\upsilon _{(j-1)1j}^{+}(r)}{r}\mathbf{Y}_{-}+\frac{%
\upsilon _{(j+1)1j}^{+}(r)}{r}\mathbf{Y}_{+})m\frac{\exp (-mr)}{r}]  \notag
\\
&=&2\pi e^{2}\boldsymbol{\epsilon }^{(\alpha _{1})}\times %
\boldsymbol{\epsilon }^{(\alpha _{2})}\cdot \mathbf{\hat{k}}\sqrt{(2j+1)}{%
\biggl (}\sum_{j^{\prime }=|j-1|}^{j+1}(-i)^{j^{\prime }}\frac{%
(1-(-1)^{j+j^{\prime }})}{2}\langle j1;00|j^{\prime }0\rangle \langle
j1;00|j^{\prime }0\rangle \delta _{m0}  \notag \\
&&\times \int_{0}^{\infty }drr^{2}\exp (F+K)\left( \frac{\exp (-mr)}{r}%
\right) ^{\prime }j_{j^{\prime }}(kr)\frac{u_{j0j}^{-}(r)}{r}  \notag \\
&&-\delta _{m0}(-i)^{j}\int_{0}^{\infty }drr^{2}m\frac{\exp (-mr)}{r}%
j_{j}(kr)kr\exp (F-K)(b_{+}\frac{\upsilon _{(j+1)1j}^{+}(r)}{r}+b_{-}\frac{%
\upsilon _{(j-1)1j}^{+}(r)}{r})  \notag \\
&&-\sum_{j^{\prime }=|j-1|}^{j+1}(-i)^{j^{\prime }}\frac{(1-(-1)^{j+j^{%
\prime }})}{2}\langle j1;00|j^{\prime }0\rangle \langle j1;00|j^{\prime
}0\rangle \delta _{m0}  \notag \\
&&\times \int_{0}^{\infty }drr^{2}m\frac{\exp (-mr)}{r}\mathbf{[(}%
a_{+}-2ib_{+})j_{j^{\prime }}(kr)-ib_{+}j_{j^{\prime }}^{\prime }(kr)kr]\exp
(F-K)\frac{\upsilon _{(j+1)1j}^{+}(r)}{r}  \notag \\
&&+[\mathbf{(}a_{-}-2ib_{-})j_{j^{\prime }}(kr)-ib_{-}j_{j^{\prime
}}^{\prime }(kr)kr]\exp (F-K)\frac{\upsilon _{(j-1)1j}^{+}(r)}{r}]{\biggr )},
\end{eqnarray}%
and with Eq. (\ref{abj}) this leads to 
\begin{eqnarray}
&&\sqrt{\pi }e^{2}\boldsymbol{\epsilon }^{(\alpha _{1})}\times %
\boldsymbol{\epsilon }^{(\alpha _{2})}\cdot \int d^{3}r[\exp (-i\mathbf{\
k\cdot r)-}\exp (+i\mathbf{\mathbf{k\cdot r)]}}  \notag \\
&&\times \exp (F)\mathbf{[\hat{r}}\frac{u_{j0j}^{-}(r)}{r}\exp
(K)Y_{jm}(\Omega )\left( \frac{\exp (-mr)}{r}\right) ^{\prime }-\exp (-K)(%
\frac{\upsilon _{(j-1)1j}^{+}(r)}{r}\mathbf{Y}_{-}+\frac{\upsilon
_{(j+1)1j}^{+}(r)}{r}\mathbf{Y}_{+})m\frac{\exp (-mr)}{r}]  \notag \\
&=&2\pi e^{2}\boldsymbol{\epsilon }^{(\alpha _{1})}\times %
\boldsymbol{\epsilon }^{(\alpha _{2})}\cdot \mathbf{\hat{k}}\sqrt{2j+1} 
\notag \\
&&\times {\biggl (}\sum_{j^{\prime }=|j-1|}^{j+1}(-i)^{j^{\prime
}}(1-(-)^{j^{\prime }})\frac{(1-(-1)^{j+j^{\prime }})}{2}\langle
j1;00|j^{\prime }0\rangle \langle j1;00|j^{\prime }0\rangle \delta
_{m0}\int_{0}^{\infty }drr^{2}\left( \frac{\exp (-mr)}{r}\right) ^{\prime
}j_{j^{\prime }}(kr)\exp (F+K)\frac{u_{j0j}^{-}(r)}{r}  \notag \\
&&-(1+(-)^{j})\delta _{m0}(-i)^{j}\int_{0}^{\infty }drr^{2}m\frac{\exp (-mr)%
}{r}j_{j}(kr)kr\exp (F-K)(\frac{i}{j+1}\sqrt{\frac{j+1}{2j+1}}\frac{\upsilon
_{(j+1)1j}^{+}(r)}{r}+\frac{i}{j}\sqrt{\frac{j}{2j+1}}\frac{\upsilon
_{(j-1)1j}^{+}(r)}{r})  \notag \\
&&-\sum_{j^{\prime }=|j-1|}^{j+1}(-i)^{j^{\prime }}\langle j1;00|j^{\prime
}0\rangle (1-(-)^{j^{\prime }})\frac{(1-(-1)^{j+j^{\prime }})}{2}\langle
j1;00|j^{\prime }0\rangle \delta _{m0}  \notag \\
&&\times \int_{0}^{\infty }drr^{2}m\frac{\exp (-mr)}{r}\mathbf{[(}-1+\frac{2%
}{j+1})j_{j^{\prime }}(kr)+\frac{1}{j+1}j_{j^{\prime }}^{\prime }(kr)kr]\exp
(F-K)\sqrt{\frac{j+1}{2j+1}}\frac{\upsilon _{(j+1)1j}^{+}(r)}{r}  \notag \\
&&+[\mathbf{(}1+\frac{2}{j})j_{j^{\prime }}(kr)+\frac{1}{j}j_{j^{\prime
}}^{\prime }(kr)kr]\sqrt{\frac{j}{2j+1}}\exp (F-K)\frac{\upsilon
_{(j-1)1j}^{+}(r)}{r}]{\biggr )},
\end{eqnarray}%
and to Eq. (\ref{samp}).

\subsection{States with $j=l\pm 1$}

We seek from Eq. (\ref{amp}) the evaluation of the dyadic integral of the
forms 
\begin{eqnarray}
&&\int d^{3}r\exp (-i\mathbf{k\cdot r)}G(r)\mathbf{\hat{r}\hat{r},}  \notag
\\
&&\int d^{3}r\exp (-i\mathbf{k\cdot r)}F_{\pm }(r)\mathbf{Y}_{jm\pm }(%
\mathbf{\Omega })\mathbf{\hat{r}}  \notag \\
&=&\int d^{3}r\exp (-i\mathbf{k\cdot r)}F_{\pm }(r)[(a_{\pm }\mathbf{\hat{r}+%
}b_{\pm }r\mathbf{p})Y_{jm}(\mathbf{\Omega })]\mathbf{\hat{r}.}  \label{yab}
\end{eqnarray}%
Consider the portion 
\begin{equation}
\int d^{3}r\exp (-i\mathbf{k\cdot r)}F_{\pm }(r)[\mathbf{p}Y_{jm}(\mathbf{\
\Omega })]\mathbf{\hat{r}.}
\end{equation}%
Integration by parts gives 
\begin{eqnarray}
&&\int d^{3}r\exp (-i\mathbf{k\cdot r)}rF_{\pm }(r)[\mathbf{p}Y_{jm}(\mathbf{%
\ \Omega })]\mathbf{\hat{r}}  \notag \\
&=&\int d^{3}rY_{jm}(\mathbf{\Omega })\exp (-i\mathbf{k\cdot r)[k\hat{r}}%
rF_{\pm }(r)+iF_{\pm }(r)\mathbf{1}+i\mathbf{\hat{r}\hat{r}}rF_{\pm
}^{\prime }(r)\mathbf{].}
\end{eqnarray}%
Its trace is 
\begin{eqnarray}
&&\int d^{3}rY_{jm}(\mathbf{\Omega })\mathbf{[k\cdot \hat{r}}rF_{\pm
}(r)+3iF_{\pm }(r)+irF_{\pm }^{\prime }(r)\mathbf{]}\exp (-i\mathbf{k\cdot r)%
}  \notag \\
&=&\int d^{3}rY_{jm}(\mathbf{\Omega })\mathbf{[}F_{\pm }(r)i\frac{d}{dy}%
+3iF_{\pm }(r)+irF_{\pm }^{\prime }(r)\mathbf{]}\exp (-iy\mathbf{k\cdot r)|}%
_{y=1}  \notag \\
&=&4\pi (-i)^{j}Y_{jm}(\mathbf{\Omega }_{k})\int_{0}^{\infty }r^{2}dr\mathbf{%
\ [}F_{\pm }(r)i\frac{d}{dy}+3iF_{\pm }(r)+irF_{\pm }^{\prime }(r)\mathbf{]}%
j_{j}(ykr)\mathbf{|}_{y=1}  \notag \\
&=&4\pi (-i)^{j}Y_{jm}(\mathbf{\Omega }_{k})\int_{0}^{\infty }r^{2}dr\mathbf{%
\ [}F_{\pm }(r)ij_{j}^{\prime }(kr)kr+3iF_{\pm }(r)j_{j}(kr)-3iF_{\pm
}(r)j_{j}(kr)-irF_{\pm }(r)kj_{j}^{\prime }(kr)\mathbf{]}  \notag \\
&\mathbf{=}&0.
\end{eqnarray}%
Hence the trace 
\begin{eqnarray}
&&\int d^{3}r\exp (-i\mathbf{k\cdot r)}F_{\pm }(r)\mathbf{Y}_{jm\pm }(%
\mathbf{\Omega })\cdot \mathbf{\hat{r}}  \notag \\
&=&4\pi (-i)^{j}Y_{jm}(\mathbf{\Omega }_{k})\int r^{2}drj_{j}(kr)a_{\pm
}F_{\pm }(r),  \label{at}
\end{eqnarray}%
and with Eq. (\ref{smp}) leads to Eq. (\ref{tr}) upon using Eq. (\ref{smp}).
Continuing with the dyad, note that 
\begin{equation}
\int d^{3}rY_{jm}(\mathbf{\Omega })\exp (-i\mathbf{k\cdot r)k\hat{r}}rF_{\pm
}(r),
\end{equation}%
will give zero contribution to the amplitude Eq. (\ref{amp}) since it is
sandwiched between transverse polarization vectors. \ The remaining part is 
\begin{equation}
4\pi i(-i)^{j}Y_{jm}(\mathbf{\Omega }_{k})\mathbf{1}\int_{0}^{\infty
}r^{2}drF_{\pm }(r)j_{j}(kr)+i\int_{0}^{\infty }\mathbf{\hat{r}\hat{r}}%
d^{3}r\exp (-i\mathbf{k\cdot r)}rF_{\pm }^{\prime }(r)Y_{jm},  \label{pr}
\end{equation}%
\ which because of the integration by parts is symmetric. \ And so,
effectively we have 
\begin{eqnarray}
&&\int d^{3}r\exp (-i\mathbf{k\cdot r)}F_{\pm }(r)\mathbf{Y}_{jm\pm }(\Omega
)\mathbf{\hat{r}}  \notag \\
&=&\int d^{3}r\exp (-i\mathbf{k\cdot r)}[a_{\pm }F_{\pm }(r)+ib_{\pm
}rF_{\pm }^{\prime }(r)]\mathbf{\hat{r}\hat{r}}Y_{jm}\mathbf{+}b_{\pm }4\pi
i(-i)^{j}Y_{jm}(\mathbf{\Omega }_{k})\mathbf{1}\int_{0}^{\infty
}r^{2}drF_{\pm }(r)j_{j}(kr)].
\end{eqnarray}

Since this dyad is sandwiched between polarization vectors transverse to $%
\mathbf{k}$ we need only consider the portions of $\mathbf{\hat{r}\hat{r}}$
and $\mathbf{1}$ orthogonal to $\mathbf{k.}$ \ Hence we replace $\mathbf{1}$
by $\mathbf{1-\hat{k}\hat{k}}$ and $\mathbf{\hat{r}\hat{r}}$ by $\mathbf{(1-%
\hat{k}\hat{k})\cdot \hat{r}\hat{r}\cdot (1-\hat{k}\hat{k})}$. \ As in the
above section we choose the unit vectors in Eq. (\ref{zxy})$.$ So, with 
\begin{eqnarray}
\mathbf{(1-\hat{k}\hat{k})\cdot \hat{r}\hat{r}\cdot (1-\hat{k}\hat{k})} &%
\mathbf{=}&\sin \theta \mathbf{(}\cos \phi \mathbf{\hat{x}+}\sin \phi 
\mathbf{\hat{y})}\sin \theta \mathbf{(}\cos \phi \mathbf{\hat{x}}+\sin \phi 
\mathbf{\hat{y})}  \notag \\
&=&\frac{1}{3}[\sqrt{4\pi }Y_{00}-\sqrt{\frac{4\pi }{5}}Y_{20}](%
\boldsymbol{\epsilon }^{(+)}\boldsymbol{\epsilon }^{(-)}+\boldsymbol{
\epsilon }^{(-)}\boldsymbol{\epsilon }^{(+)}\mathbf{)}  \notag \\
&&\mathbf{+}\sqrt{\frac{8\pi }{15}}Y_{2-2}\boldsymbol{\epsilon }^{(+)}%
\boldsymbol{\epsilon }^{(+)}\mathbf{+}\sqrt{\frac{8\pi }{15}}Y_{22}%
\boldsymbol{\epsilon }^{(-)}\boldsymbol{\epsilon }^{(-)},
\end{eqnarray}%
we find 
\begin{eqnarray}
&&\mathbf{(1-\hat{k}\hat{k})\cdot }\int d^{3}r\exp (-i\mathbf{k\cdot r)}G(r)%
\mathbf{\hat{r}\hat{r}}Y_{jm}\mathbf{\cdot (1-\hat{k}\hat{k})}  \notag \\
&=&(\boldsymbol{\epsilon }^{(+)}\boldsymbol{\epsilon }^{(-)}+%
\boldsymbol{\epsilon }^{(-)}\boldsymbol{\epsilon }^{(+)}\mathbf{)}\int
d^{3}r\exp (-i\mathbf{k\cdot r)}F_{\pm }(r)\frac{1}{3}[\sqrt{4\pi }Y_{00}-%
\sqrt{\frac{4\pi }{5}}Y_{20}]Y_{jm}  \notag \\
&&+\boldsymbol{\epsilon }^{(+)}\boldsymbol{\epsilon }^{(+)}\sqrt{\frac{8\pi 
}{15}}\int d^{3}r\exp (-i\mathbf{k\cdot r)}G(r)Y_{2-2}Y_{jm}  \notag \\
&&+\boldsymbol{\epsilon }^{(-)}\boldsymbol{\epsilon }^{(-)}\sqrt{\frac{8\pi 
}{15}}\int d^{3}r\exp (-i\mathbf{k\cdot r)}G(r)Y_{22}Y_{jm}.
\end{eqnarray}%
Using Eq. (\ref{plnwv}), Eq. (\ref{smp}) and 
\begin{eqnarray}
\int d\Omega Y_{j^{\prime }m^{\prime }}^{\ast }Y_{2m^{\prime \prime }}Y_{jm}
&=&\sqrt{\frac{(2j+1)5}{4\pi (2j^{\prime }+1)}}\langle j2;00|j^{\prime
}0\rangle \langle j2;mm^{\prime \prime }|j^{\prime }m+m^{\prime \prime
}\rangle ;~j+j^{\prime }+2\text{ even}  \notag \\
&=&0;~j+j^{\prime }+2\text{ odd,}
\end{eqnarray}%
\ we have \ \ 
\begin{eqnarray}
&&\mathbf{(1-\hat{k}\hat{k})\cdot }\int d^{3}r\exp (-i\mathbf{k\cdot r)}G(r)%
\mathbf{\hat{r}\hat{r}}Y_{jm}\cdot \mathbf{(1-\hat{k}\hat{k})}  \notag \\
&=&\sqrt{\frac{(2j+1)}{4\pi }}(\boldsymbol{\epsilon }^{(+)}%
\boldsymbol{\epsilon }^{(-)}+\boldsymbol{\epsilon }^{(-)}\boldsymbol{
\epsilon }^{(+)}\mathbf{)[}\frac{4\pi }{3}\delta _{m0}\int_{0}^{\infty
}r^{2}dr(-i)^{j}j_{j}(kr)G(r)  \notag \\
&&-\frac{4\pi }{3}\sum_{j^{\prime }=\left\vert j-2\right\vert }^{j+2}\frac{%
(1+(-1)^{j+j^{\prime }})}{2}\langle j2;00|j^{\prime }0\rangle \langle
j2;00|j^{\prime }0\rangle \delta _{m0}\int_{0}^{\infty
}r^{2}dr(-i)^{j^{\prime }}j_{j^{\prime }}(kr)G(r)]  \notag \\
&&+\sqrt{\frac{(2j+1)}{4\pi }}(\boldsymbol{\epsilon }^{(+)}%
\boldsymbol{\epsilon }^{(+)}\mathbf{)}4\pi \sqrt{\frac{2}{3}}\sum_{j^{\prime
}=\left\vert j-2\right\vert }^{j+2}\frac{(1+(-1)^{j+j^{\prime }})}{2}\langle
j2;00|j^{\prime }0\rangle \langle j2;-22|j^{\prime }0\rangle \delta
_{m-2}\int r^{2}dr(-i)^{j^{\prime }}j_{j^{\prime }}(kr)G(r)  \notag \\
&&+\sqrt{\frac{(2j+1)}{4\pi }}(\boldsymbol{\epsilon }^{(-)}%
\boldsymbol{\epsilon }^{(-)}\mathbf{)}4\pi \sqrt{\frac{2}{3}}\sum_{j^{\prime
}=\left\vert j-2\right\vert }^{j+2}\frac{(1+(-1)^{j+j^{\prime }})}{2}\langle
j2;00|j^{\prime }0\rangle \langle j2;2-3|j^{\prime }0\rangle \delta
_{m2}\int r^{2}dr(-i)^{j^{\prime }}j_{j^{\prime }}(kr)G(r),  \label{str}
\end{eqnarray}%
and so 
\begin{eqnarray}
&&\mathbf{(1-\hat{k}\hat{k})\cdot }\int d^{3}r\exp (-i\mathbf{k\cdot r)}%
F_{\pm }(r)\mathbf{Y}_{jm\pm }(\Omega )\mathbf{\hat{r}}\cdot \mathbf{(1-\hat{%
k}\hat{k})}  \notag \\
&=&\mathbf{(1-\hat{k}\hat{k})\cdot }\int d^{3}r\exp (-i\mathbf{k\cdot r)\hat{%
r}\hat{r}}Y_{jm}[a_{\pm }F_{\pm }(r)+ib_{\pm }rF_{\pm }^{\prime }(r)]\cdot 
\mathbf{(1-\hat{k}\hat{k})}  \notag \\
&&+4\pi ib_{\pm }(-i)^{j}(\boldsymbol{\epsilon }^{(+)}\boldsymbol{\epsilon }%
^{(-)}+\boldsymbol{\epsilon }^{(-)}\boldsymbol{\epsilon }^{(+)}\mathbf{)}%
\int_{0}^{\infty }r^{2}drF_{\pm }(r)j_{j}(kr)  \notag \\
&=&(\boldsymbol{\epsilon }^{(+)}\boldsymbol{\epsilon }^{(-)}+%
\boldsymbol{\epsilon }^{(-)}\boldsymbol{\epsilon }^{(+)}\mathbf{)}\frac{1}{3}%
\sqrt{4\pi (2j+1)}\mathbf{\{}\delta _{m0}\int_{0}^{\infty
}r^{2}dr(-i)^{j}j_{j}(kr)[(a_{\pm }+3ib_{\pm })F_{\pm }(r)+ib_{\pm }rF_{\pm
}^{\prime }(r)]  \notag \\
&&-\sum_{j^{\prime }=\left\vert j-2\right\vert }^{j+2}\frac{%
(1+(-1)^{j+j^{\prime }})}{2}\langle j2;00|j^{\prime }0\rangle \langle
j2;00|j^{\prime }0\rangle \delta _{m0}\int_{0}^{\infty
}r^{2}dr(-i)^{j^{\prime }}j_{j^{\prime }}(kr)[a_{\pm }F_{\pm }(r)+ib_{\pm
}rF_{\pm }^{\prime }(r)]\}  \notag \\
&&+(\boldsymbol{\epsilon }^{(+)}\boldsymbol{\epsilon }^{(+)}\mathbf{)}\sqrt{%
\frac{8\pi (2j+1)}{3}}\sum_{j^{\prime }=\left\vert j-2\right\vert }^{j+2}%
\frac{(1+(-1)^{j+j^{\prime }})}{2}\langle j2;00|j^{\prime }0\rangle \langle
j2;-22|j^{\prime }0\rangle \delta _{m-2}  \notag \\
&&\times \int_{0}^{\infty }r^{2}dr(-i)^{j^{\prime }}j_{j^{\prime
}}(kr)[a_{\pm }F_{\pm }(r)+ib_{\pm }rF_{\pm }^{\prime }(r]  \notag \\
&&+(\boldsymbol{\epsilon }^{(-)}\boldsymbol{\epsilon }^{(-)}\mathbf{)}\sqrt{%
\frac{8\pi (2j+1)}{3}}\sum_{j^{\prime }=\left\vert j-2\right\vert }^{j+2}%
\frac{(1+(-1)^{j+j^{\prime }})}{2}\langle j2;00|j^{\prime }0\rangle \langle
j2;2-2|j^{\prime }0\rangle \delta _{m2}  \notag \\
&&\times \int_{0}^{\infty }r^{2}dr(-i)^{j^{\prime }}j_{j^{\prime
}}(kr)[a_{\pm }F_{\pm }(r)+ib_{\pm }rF_{\pm }^{\prime }(r)].
\end{eqnarray}%
Since these expressions involve symmetric dyads we find together with the
trace part 
\begin{eqnarray}
&&-(\boldsymbol{\epsilon }^{(+)}\boldsymbol{\epsilon }^{(-)}+%
\boldsymbol{\epsilon }^{(-)}\boldsymbol{\epsilon }^{(+)}\mathbf{)}\int
d^{3}r\exp (-i\mathbf{k\cdot r)}F_{\pm }(r)\mathbf{Y}_{jm\pm }(\Omega )\cdot 
\mathbf{\hat{r}}  \notag \\
&=&-(\boldsymbol{\epsilon }^{(+)}\boldsymbol{\epsilon }^{(-)}+%
\boldsymbol{\epsilon }^{(-)}\boldsymbol{\epsilon }^{(+)}\mathbf{)}4\pi
(-i)^{j}\sqrt{\frac{2j+1}{4\pi }}\delta _{m0}\int r^{2}drj_{j}(kr)a_{\pm
}F_{\pm }(r),
\end{eqnarray}%
that 
\begin{eqnarray}
&&-i\sqrt{\pi }e^{2}\int d^{3}r\exp (-i\mathbf{k\cdot r)}F_{\pm }(r)[\mathbf{%
\ Y}_{jm\pm }(\Omega )\cdot \boldsymbol{\epsilon }^{(\alpha _{1})}\mathbf{%
\hat{r}}\cdot \boldsymbol{\epsilon }^{(\alpha _{2})}+\mathbf{Y}_{jm\pm
}(\Omega )\cdot \boldsymbol{\epsilon }^{(\alpha _{2})}\mathbf{\hat{r}}\cdot %
\boldsymbol{\epsilon }^{(\alpha _{1})}  \notag \\
&&-\mathbf{Y}_{jm\pm }(\Omega )\mathbf{\cdot \hat{r}}\boldsymbol{\epsilon }%
^{(\alpha _{1})}\cdot \boldsymbol{\epsilon }^{(\alpha _{2})}]  \notag \\
&=&-i\frac{2\pi e^{2}}{3}\sqrt{(2j+1)}\{2\delta
_{m0}(-i)^{j}\int_{0}^{\infty }r^{2}drj_{j}(kr)[(a_{\pm }+3ib_{\pm })F_{\pm
}(r)+ib_{\pm }rF_{\pm }^{\prime }(r)]  \notag \\
&&-3\delta _{m0}(-i)^{j}\int_{0}^{\infty }r^{2}drj_{j}(kr)a_{\pm }F_{\pm }(r)
\notag \\
&&-2\sum_{j^{\prime }=\left\vert j-2\right\vert }^{j+2}\frac{%
(1+(-1)^{j+j^{\prime }})}{2}\langle j2;00|j^{\prime }0\rangle \langle
j2;00|j^{\prime }0\rangle \delta _{m0}\int_{0}^{\infty
}r^{2}dr(-i)^{j^{\prime }}j_{j^{\prime }}(kr)[a_{\pm }F_{\pm }(r)+ib_{\pm
}rF_{\pm }^{\prime }(r)]\}(\boldsymbol{\epsilon }\mathbf{^{(\alpha
_{1})}\cdot }\boldsymbol{\epsilon }\mathbf{^{(\alpha _{2})})}  \notag \\
&&-i4\pi e^{2}\sqrt{\frac{2(2j+1)}{3}}\sum_{j^{\prime }=\left\vert
j-2\right\vert }^{j+2}\frac{(1+(-1)^{j+j^{\prime }})}{2}\langle
j2;00|j^{\prime }0\rangle \langle j2;-22|j^{\prime }0\rangle \delta _{m-2} 
\notag \\
&&\times \int_{0}^{\infty }r^{2}dr(-i)^{j^{\prime }}j_{j^{\prime
}}(kr)(a_{\pm }F_{\pm }(r)+ib_{\pm }rF_{\pm }^{\prime }(r)]%
\boldsymbol{\epsilon }^{(\alpha _{1})}\cdot (\boldsymbol{\epsilon }^{(+)}%
\boldsymbol{\epsilon }^{(+)}\mathbf{)}\cdot \boldsymbol{\epsilon }^{(\alpha
_{2})}  \notag \\
&&-i4\pi e^{2}\sqrt{\frac{2(2j+1)}{3}}\sum_{j^{\prime }=\left\vert
j-2\right\vert }^{j+2}\frac{(1+(-1)^{j+j^{\prime }})}{2}\langle
j2;00|j^{\prime }0\rangle \langle j2;2-2|j^{\prime }0\rangle \delta _{m2} 
\notag \\
&&\times \int_{0}^{\infty }r^{2}(-i)^{j^{\prime }}drj_{j^{\prime
}}(kr)(a_{\pm }F_{\pm }(r)+ib_{\pm }rF_{\pm }^{\prime }(r)]%
\boldsymbol{\epsilon }^{(\alpha _{1})}\cdot (\boldsymbol{\epsilon }^{(-)}%
\boldsymbol{\epsilon }^{(-)}\mathbf{)}\cdot \boldsymbol{\epsilon }^{(\alpha
_{2})}.
\end{eqnarray}%
Integration by parts gives 
\begin{eqnarray}
&&-i\sqrt{\pi }e^{2}\int d^{3}r\exp (-i\mathbf{k\cdot r)}F_{\pm }(r)[\mathbf{%
\ Y}_{jm\pm }(\Omega )\cdot \boldsymbol{\epsilon }^{(\alpha _{1})}\mathbf{%
\hat{r}}\cdot \boldsymbol{\epsilon }^{(\alpha _{2})}+\mathbf{Y}_{jm\pm
}(\Omega )\cdot \boldsymbol{\epsilon }^{(\alpha _{2})}\mathbf{\hat{r}}\cdot %
\boldsymbol{\epsilon }^{(\alpha _{1})}  \notag \\
&&-\mathbf{Y}_{jm\pm }(\Omega )\mathbf{\cdot \hat{r}}\boldsymbol{\epsilon }%
^{(\alpha _{1})}\cdot \boldsymbol{\epsilon }^{(\alpha _{2})}]  \notag \\
&=&-i\frac{2\pi e^{2}}{3}\sqrt{(2j+1)}\{(-i)^{j}\delta _{m0}\int_{0}^{\infty
}r^{2}dr[-j_{j}(kr)a_{\pm }F_{\pm }(r)-2ib_{\pm }j_{j}^{\prime }(kr)krF_{\pm
}(r)]  \notag \\
&&-2\sum_{j^{\prime }=\left\vert j-2\right\vert }^{j+2}\frac{%
(1+(-1)^{j+j^{\prime }})}{2}\langle j2;00|j^{\prime }0\rangle \langle
j2;00|j^{\prime }0\rangle \delta _{m0}  \notag \\
&&\times \int_{0}^{\infty }r^{2}dr(-i)^{j^{\prime }}j_{j^{\prime
}}(kr)[(a_{\pm }-3ib_{\pm })j_{j^{\prime }}(kr)F_{\pm }(r)-ib_{\pm
}j_{j^{\prime }}^{\prime }(kr)krF_{\pm }(r)]\}(\boldsymbol{\epsilon }\mathbf{%
\ ^{(\alpha _{1})}\cdot }\boldsymbol{\epsilon }\mathbf{^{(\alpha _{2})})} 
\notag \\
&&-i4\pi e^{2}\sqrt{\frac{2(2j+1)}{3}}\sum_{j^{\prime }=\left\vert
j-2\right\vert }^{j+2}\frac{(1+(-1)^{j+j^{\prime }})}{2}\langle
j2;00|j^{\prime }0\rangle \langle j2;-22|j^{\prime }0\rangle \delta _{m-2} 
\notag \\
&&\times \int_{0}^{\infty }r^{2}dr(-i)^{j^{\prime }}[(a_{\pm }-3ib_{\pm
})j_{j^{\prime }}(kr)F_{\pm }(r)-ib_{\pm }j_{j^{\prime }}^{\prime
}(kr)krF_{\pm }(r)]\boldsymbol{\epsilon }^{(\alpha _{1})}\cdot (%
\boldsymbol{\epsilon }^{(+)}\boldsymbol{\epsilon }^{(+)}\mathbf{)}\cdot %
\boldsymbol{\epsilon }^{(\alpha _{2})}  \notag \\
&&-i4\pi e^{2}\sqrt{\frac{2(2j+1)}{3}}\sum_{j^{\prime }=\left\vert
j-2\right\vert }^{j+2}\frac{(1+(-1)^{j+j^{\prime }})}{2}\langle
j2;00|j^{\prime }0\rangle \langle j2;2-2|j^{\prime }0\rangle \delta _{m2} 
\notag \\
&&\times \int_{0}^{\infty }r^{2}dr(-i)^{j^{\prime }}[(a_{\pm }-3ib_{\pm
})j_{j^{\prime }}(kr)F_{\pm }(r)-ib_{\pm }j_{j^{\prime }}^{\prime
}(kr)krF_{\pm }(r)]\boldsymbol{\epsilon }^{(\alpha _{1})}\cdot (%
\boldsymbol{\epsilon }^{(-)}\boldsymbol{\epsilon }^{(-)}\mathbf{)}\cdot %
\boldsymbol{\epsilon }^{(\alpha _{2})},
\end{eqnarray}%
and simplifying 
\begin{eqnarray}
&&-i\sqrt{\pi }e^{2}\int d^{3}r\exp (-i\mathbf{k\cdot r)}F_{\pm }(r)[\mathbf{%
\ Y}_{jm\pm }(\Omega )\cdot \boldsymbol{\epsilon }^{(\alpha _{1})}\mathbf{%
\hat{r}}\cdot \boldsymbol{\epsilon }^{(\alpha _{2})}+\mathbf{Y}_{jm\pm
}(\Omega )\cdot \boldsymbol{\epsilon }^{(\alpha _{2})}\mathbf{\hat{r}}\cdot %
\boldsymbol{\epsilon }^{(\alpha _{1})}  \notag \\
&&-\mathbf{Y}_{jm\pm }(\Omega )\mathbf{\cdot \hat{r}}\boldsymbol{\epsilon }%
^{(\alpha _{1})}\cdot \boldsymbol{\epsilon }^{(\alpha _{2})}]  \notag \\
&=&-i\frac{2\pi e^{2}}{3}\sqrt{(2j+1)}\{(-i)^{j}\int_{0}^{\infty
}r^{2}dr[-j_{j}(kr)a_{\pm }-2ib_{\pm }j_{j}^{\prime }(kr)kr]F_{\pm }(r) 
\notag \\
&&-2\sum_{j^{\prime }=\left\vert j-2\right\vert }^{j+2}(-i)^{j^{\prime
}}\langle j2;00|j^{\prime }0\rangle \langle j2;00|j^{\prime }0\rangle \delta
_{m0}  \notag \\
&&\times \int_{0}^{\infty }r^{2}dr[(a_{\pm }-3ib_{\pm })j_{j^{\prime
}}(kr)-ib_{\pm }j_{j^{\prime }}^{\prime }(kr)kr]F_{\pm }(r)\}(%
\boldsymbol{\epsilon }\mathbf{^{(\alpha _{1})}\cdot }\boldsymbol{\epsilon }%
\mathbf{^{(\alpha _{2})})}  \notag \\
&&-i4\pi e^{2}\sqrt{\frac{2(2j+1)}{3}}\sum_{j^{\prime }=\left\vert
j-2\right\vert }^{j+2}(-i)^{j^{\prime }}\int_{0}^{\infty }r^{2}dr[(a_{\pm
}-3ib_{\pm })j_{j^{\prime }}(kr)-ib_{\pm }j_{j^{\prime }}^{\prime
}(kr)kr]F_{\pm }(r)  \notag \\
&&\times \frac{(1+(-1)^{j+j^{\prime }})}{2}[\langle j2;00|j^{\prime
}0\rangle \langle j2;-22|j^{\prime }0\rangle \delta _{m-2}]%
\boldsymbol{\epsilon }^{(\alpha _{1})}\cdot (\boldsymbol{\epsilon }^{(+)}%
\boldsymbol{\epsilon }^{(+)}\mathbf{)}\cdot \boldsymbol{\epsilon }^{(\alpha
_{2})}  \notag \\
&&+\langle j2;00|j^{\prime }0\rangle \langle j2;2-2|j^{\prime }0\rangle
\delta _{m2}\boldsymbol{\epsilon }^{(\alpha _{1})}\cdot (\boldsymbol{
\epsilon }^{(-)}\boldsymbol{\epsilon }^{(-)}\mathbf{)}\cdot %
\boldsymbol{\epsilon }^{(\alpha _{2})}].
\end{eqnarray}%
From Eq. (\ref{str})%
\begin{eqnarray}
&&-i\sqrt{\pi }e^{2}\int d^{3}r\exp (-i\mathbf{k\cdot r)}G(r)\mathbf{\hat{r}}%
\cdot \boldsymbol{\epsilon }^{(\alpha _{1})}\mathbf{\hat{r}}\cdot %
\boldsymbol{\epsilon }^{(\alpha _{2})}  \notag \\
&=&-i\frac{2\pi e^{2}}{3}\sqrt{(2j+1)}\{\mathbf{[}\delta
_{m0}\int_{0}^{\infty }r^{2}dr(-i)^{j}j_{j}(kr)G(r)  \notag \\
&&-\sum_{j^{\prime }=\left\vert j-2\right\vert }^{j+2}\frac{%
(1+(-1)^{j+j^{\prime }})}{2}\langle j2;00|j^{\prime }0\rangle \langle
j2;00|j^{\prime }0\rangle \delta _{m0}\int_{0}^{\infty
}r^{2}dr(-i)^{j^{\prime }}j_{j^{\prime }}(kr)G(r)](\boldsymbol{\epsilon }%
\mathbf{^{(\alpha _{1})}\cdot }\boldsymbol{\epsilon }\mathbf{^{(\alpha
_{2})})}  \notag \\
&&+\sqrt{6}\sum_{j^{\prime }=\left\vert j-2\right\vert }^{j+2}\frac{%
(1+(-1)^{j+j^{\prime }})}{2}\langle j2;00|j^{\prime }0\rangle \langle
j2;-22|j^{\prime }0\rangle \delta _{m-2}\int r^{2}dr(-i)^{j^{\prime
}}j_{j^{\prime }}(kr)G(r)\boldsymbol{\epsilon }^{(\alpha _{1})}\cdot (%
\boldsymbol{\epsilon }^{(+)}\boldsymbol{\epsilon }^{(+)}\mathbf{)}\cdot %
\boldsymbol{\epsilon }^{(\alpha _{2})}  \notag \\
&&+\sqrt{6}\sum_{j^{\prime }=\left\vert j-2\right\vert }^{j+2}\frac{%
(1+(-1)^{j+j^{\prime }})}{2}\langle j2;00|j^{\prime }0\rangle \langle
j2;2-3|j^{\prime }0\rangle \delta _{m2}\int r^{2}dr(-i)^{j^{\prime
}}j_{j^{\prime }}(kr)G(r)\boldsymbol{\epsilon }^{(\alpha _{1})}\cdot (%
\boldsymbol{ \epsilon }^{(-)}\boldsymbol{\epsilon }^{(-)}\mathbf{)}\cdot %
\boldsymbol{\epsilon }^{(\alpha _{2})}\}
\end{eqnarray}

Combining and including for $g(r),F_{\pm }(r)$ and $G(r)$%
\begin{eqnarray}
&&\mathcal{M}_{^{3}L_{j=l\pm 1}\rightarrow 2\gamma }  \notag \\
&=&-i\sqrt{\pi }e^{2}\int d^{3}r[\exp (-i\mathbf{k\cdot r)+}\exp (i\mathbf{%
k\cdot r)]}\exp (F){\biggl (}\left( \frac{\exp (-mr)}{r}\right) ^{\prime } 
\notag \\
&&\times \exp K\{\frac{u_{(j+1)1j}^{-}}{r}[\mathbf{Y}_{jm+}(\mathbf{\Omega }%
)\cdot \boldsymbol{\epsilon }^{(\alpha _{1})}\mathbf{\hat{r}\cdot }%
\boldsymbol{\epsilon }^{(\alpha _{2})}+\mathbf{Y}_{jm+}(\mathbf{\Omega }%
)\cdot \boldsymbol{\epsilon }^{(\alpha _{2})}\mathbf{\hat{r}\cdot }%
\boldsymbol{\epsilon }^{(\alpha _{1})}-\mathbf{Y}_{jm+}(\mathbf{\Omega }%
)\cdot \mathbf{\hat{r}}\boldsymbol{\epsilon }^{(\alpha _{1})}\mathbf{\cdot }%
\boldsymbol{\epsilon }^{(\alpha _{2})}]  \notag \\
&&+\frac{u_{(j-1)1j}^{-}}{r}[\mathbf{Y}_{jm-}(\mathbf{\Omega })\cdot %
\boldsymbol{\epsilon }^{(\alpha _{1})}\mathbf{\hat{r}\cdot }%
\boldsymbol{\epsilon }^{(\alpha _{2})}+\mathbf{Y}_{jm-}(\mathbf{\Omega }%
)\cdot \boldsymbol{\epsilon }^{(\alpha _{2})}\mathbf{\hat{r}\cdot }%
\boldsymbol{\epsilon }^{(\alpha _{1})}-\mathbf{Y}_{jm-}(\mathbf{\Omega }%
)\cdot \mathbf{\hat{r}}\boldsymbol{\epsilon }^{(\alpha _{1})}\mathbf{\cdot }%
\boldsymbol{\epsilon }^{(\alpha _{2})}]\}  \notag \\
&&-4\sinh K\left( \frac{\exp (-mr)}{r}\right) ^{\prime }[-\sqrt{\frac{j+1}{%
2j+1}}\frac{u_{(j+1)1j}^{-}}{r}+\frac{u_{(j-1)1j}^{-}}{r}\sqrt{\frac{j}{2j+1}%
}]Y_{jm}\mathbf{\hat{r}}\cdot \boldsymbol{ \epsilon }^{(\alpha _{1})}\mathbf{%
\hat{r}\cdot }\boldsymbol{ \epsilon }^{(\alpha _{2})} \\
&&.+[m\exp (-K)\frac{\upsilon _{j0j}^{+}}{r}-2(m+1/r)\sinh K(-\sqrt{\frac{j+1%
}{2j+1}}\frac{u_{(j+1)1j}^{-}}{r}+\frac{u_{(j-1)1j}^{-}}{r}\sqrt{\frac{j}{%
2j+1}})]Y_{jm}(\mathbf{\Omega })\boldsymbol{ \epsilon }^{(\alpha _{1})}\cdot %
\boldsymbol{\epsilon }^{(\alpha _{2})}\frac{\exp (-mr)}{r}{\biggr )}  \notag
\end{eqnarray}%
\begin{eqnarray}
&=&-i\frac{2\pi e^{2}}{3}\sqrt{(2j+1)}{\biggl [(}\{(1+(-)^{j})(-i)^{j}%
\int_{0}^{\infty }r^{2}dr\left( \frac{\exp (-mr)}{r}\right) ^{\prime }\exp
(F+K)  \notag \\
&&\times \{[j_{j}(kr)+\frac{2}{(j+1)}j_{j}^{\prime }(kr)kr]\sqrt{\frac{j+1}{%
2j+1}}\frac{u_{(j+1)1j}^{-}(r)}{r}+[-j_{j}(kr)+\frac{2}{j}j_{j}^{\prime
}(kr)kr]\sqrt{\frac{j}{2j+1}}\frac{u_{(j-1)1j}^{-}(r)}{r}\}  \notag \\
&&-2\sum_{j^{\prime }=\left\vert j-2\right\vert }^{j+2}(1+(-)^{j^{\prime
}})(-i)^{j^{\prime }}\frac{(1+(-1)^{j+j^{\prime }})}{2}\langle
j2;00|j^{\prime }0\rangle \langle j2;00|j^{\prime }0\rangle \delta _{m0} 
\notag \\
&&\times \lbrack \int_{0}^{\infty }r^{2}dr\left( \frac{\exp (-mr)}{r}\right)
^{\prime }\exp (F+K)\{[(-1+\frac{3}{j+1})j_{j^{\prime }}(kr)+\frac{1}{j+1}%
j_{j^{\prime }}^{\prime }(kr)kr]\sqrt{\frac{j+1}{2j+1}}\frac{%
u_{(j+1)1j}^{-}(r)}{r}  \notag \\
&&+[(1+\frac{3}{j})j_{j^{\prime }}(kr)+\frac{1}{j}j_{j^{\prime }}^{\prime
}(kr)kr]\sqrt{\frac{j}{2j+1}}\frac{u_{(j-1)1j}^{-}(r)}{r}\}{\biggr )}%
\boldsymbol{\epsilon }^{(\alpha _{1})}\cdot \boldsymbol{\epsilon }^{(\alpha
_{2})}  \notag \\
&&+2\sqrt{6}\sum_{j^{\prime }=\left\vert j-2\right\vert
}^{j+2}(1+(-)^{j^{\prime }})\frac{(1+(-1)^{j+j^{\prime }})}{2}%
(-i)^{j^{\prime }}\int_{0}^{\infty }r^{2}dr\left( \frac{\exp (-mr)}{r}%
\right) ^{\prime }  \notag \\
&&\times \exp (F+K)\{[(-1+\frac{3}{j+1})j_{j^{\prime }}(kr)+\frac{1}{j+1}%
j_{j^{\prime }}^{\prime }(kr)kr]\sqrt{\frac{j+1}{2j+1}}\frac{%
u_{(j+1)1j}^{-}(r)}{r}  \notag \\
&&+[(1+\frac{3}{j})j_{j^{\prime }}(kr)+\frac{1}{j}j_{j^{\prime }}^{\prime
}(kr)kr]\sqrt{\frac{j}{2j+1}}\frac{u_{(j-1)1j}^{-}(r)}{r}\}  \notag \\
&&\times \lbrack \langle j2;00|j^{\prime }0\rangle \langle j2;-22|j^{\prime
}0\rangle \delta _{m-2}]\boldsymbol{\epsilon }^{(\alpha _{1})}\cdot (%
\boldsymbol{\epsilon }^{(+)}\boldsymbol{\epsilon }^{(+)}\mathbf{)}\cdot %
\boldsymbol{\epsilon }^{(\alpha _{2})}  \notag \\
&&+\langle j2;00|j^{\prime }0\rangle \langle j2;2-2|j^{\prime }0\rangle
\delta _{m2}\boldsymbol{\epsilon }^{(\alpha _{1})}\cdot (\boldsymbol{
\epsilon }^{(-)}\boldsymbol{\epsilon }^{(-)}\mathbf{)}\cdot %
\boldsymbol{\epsilon }^{(\alpha _{2})}]  \notag \\
&&+\{\delta _{m0}(-i)^{j}(1+(-)^{j})\int_{0}^{\infty
}r^{2}drj_{j}(kr)[-4\sinh K\left( \frac{\exp (-mr)}{r}\right) ^{\prime }[-%
\sqrt{\frac{j+1}{2j+1}}\frac{u_{(j+1)1j}^{-}}{r}+\frac{u_{(j-1)1j}^{-}}{r}%
\sqrt{\frac{j}{2j+1}}]  \notag \\
&&-\sum_{j^{\prime }=\left\vert j-2\right\vert }^{j+2}\frac{%
(1+(-1)^{j+j^{\prime }})}{2}\langle j2;00|j^{\prime }0\rangle \langle
j2;00|j^{\prime }0\rangle \delta _{m0}(1+(-)^{j^{\prime }})  \notag \\
&&\times \int_{0}^{\infty }r^{2}dr(-i)^{j^{\prime }}j_{j^{\prime
}}(kr)[-4\sinh K\left( \frac{\exp (-mr)}{r}\right) ^{\prime }[-\sqrt{\frac{%
j+1}{2j+1}}\frac{u_{(j+1)1j}^{-}}{r}+\frac{u_{(j-1)1j}^{-}}{r}\sqrt{\frac{j}{%
2j+1}}]\}(\boldsymbol{\epsilon }\mathbf{^{(\alpha _{1})}\cdot }%
\boldsymbol{\epsilon }\mathbf{^{(\alpha _{2})})}  \notag \\
&&+\sqrt{6}\sum_{j^{\prime }=\left\vert j-2\right\vert }^{j+2}\frac{%
(1+(-1)^{j+j^{\prime }})}{2}\langle j2;00|j^{\prime }0\rangle \langle
j2;-22|j^{\prime }0\rangle \delta _{m-2}(1+(-)^{j^{\prime }})\int
r^{2}dr(-i)^{j^{\prime }}j_{j^{\prime }}(kr)  \notag \\
&&\times \lbrack -4\sinh K\left( \frac{\exp (-mr)}{r}\right) ^{\prime }[-%
\sqrt{\frac{j+1}{2j+1}}\frac{u_{(j+1)1j}^{-}}{r}+\frac{u_{(j-1)1j}^{-}}{r}%
\sqrt{\frac{j}{2j+1}}]\boldsymbol{\epsilon }^{(\alpha _{1})}\cdot (%
\boldsymbol{\epsilon }^{(+)}\boldsymbol{\epsilon }^{(+)}\mathbf{)}\cdot %
\boldsymbol{\epsilon }^{(\alpha _{2})}  \notag \\
&&+\sqrt{6}\sum_{j^{\prime }=\left\vert j-2\right\vert }^{j+2}\frac{%
(1+(-1)^{j+j^{\prime }})}{2}\langle j2;00|j^{\prime }0\rangle \langle
j2;2-2|j^{\prime }0\rangle \delta _{m2}(1+(-)^{j^{\prime }})\int
r^{2}dr(-i)^{j^{\prime }}j_{j^{\prime }}(kr)  \notag \\
&&\times \lbrack -4\sinh K\left( \frac{\exp (-mr)}{r}\right) ^{\prime }[-%
\sqrt{\frac{j+1}{2j+1}}\frac{u_{(j+1)1j}^{-}}{r}+\frac{u_{(j-1)1j}^{-}}{r}%
\sqrt{\frac{j}{2j+1}}]\boldsymbol{\epsilon }^{(\alpha _{1})}\cdot (%
\boldsymbol{\epsilon }^{(+)}\boldsymbol{\epsilon }^{(+)}\mathbf{)}\cdot %
\boldsymbol{\epsilon }^{(\alpha _{2})}  \notag \\
&&+(1+(-)^{j})(-i)^{j}\int_{0}^{\infty }\exp (-mr)\exp (F)rj_{j}(kr)3[m\exp
(-K)\frac{\upsilon _{j0j}^{+}}{r}  \notag \\
&&-2(m+1/r)\sinh K(-\sqrt{\frac{j+1}{2j+1}}\frac{u_{(j+1)1j}^{-}}{r}+\frac{%
u_{(j-1)1j}^{-}}{r}\sqrt{\frac{j}{2j+1}})]\boldsymbol{\epsilon }^{(\alpha
_{1})}\cdot \boldsymbol{\epsilon }^{(\alpha _{2})}{\biggr ]},
\end{eqnarray}%
or 
\begin{eqnarray}
&&\mathcal{M}_{^{3}L_{j=l\pm 1}\rightarrow 2\gamma }  \notag \\
&=&i\frac{2\pi e^{2}}{3}\sqrt{(2j+1)}{\biggl [}(1+(-)^{j})(-i)^{j}\delta
_{m0}\int_{0}^{\infty }dr\exp (-mr)\exp (F){\biggl (}-3j_{j}(kr)[mr\exp (-K)%
\frac{\upsilon _{j0j}^{+}(r)}{r}  \notag \\
&&-2(mr+1)\sinh K(-\sqrt{\frac{j+1}{2j+1}}\frac{u_{(j+1)1j}^{-}}{r}+\frac{%
u_{(j-1)1j}^{-}}{r}\sqrt{\frac{j}{2j+1}})]  \notag \\
&&+(mr+1)\exp (K)\{[(j_{j}(kr)+\frac{2}{(j+1)}j_{j}^{\prime }(kr)kr)\sqrt{%
\frac{j+1}{2j+1}}\frac{u_{(j+1)1j}^{-}(r)}{r}  \notag \\
&&+(-j_{j}(kr)+\frac{2}{j}j_{j}^{\prime }(kr)kr)\sqrt{\frac{j}{2j+1}}\frac{%
u_{(j-1)1j}^{-}(r)}{r}]\}  \notag \\
&&-4\sinh K(mr+1)[-\sqrt{\frac{j+1}{2j+1}}\frac{u_{(j+1)1j}^{-}}{r}+\frac{%
u_{(j-1)1j}^{-}}{r}\sqrt{\frac{j}{2j+1}}]j_{j}(kr){\biggr )}\boldsymbol{
\epsilon }^{(\alpha _{1})}\cdot \boldsymbol{\epsilon }^{(\alpha _{2})} 
\notag \\
&&+\sum_{j^{\prime }=\left\vert j-2\right\vert }^{j+2}\frac{%
(1+(-1)^{j+j^{\prime }})}{2}\langle j2;00|j^{\prime }0\rangle
(1+(-)^{j^{\prime }})\frac{(1+(-1)^{j+j^{\prime }})}{2}(-i)^{j^{\prime }}{%
\biggl (2}\int_{0}^{\infty }dr\exp (-mr)\exp (F)(mr+1)  \notag \\
&&-\exp (K)\{[(-1+\frac{3}{j+1})j_{j^{\prime }}(kr)+\frac{1}{j+1}%
j_{j^{\prime }}^{\prime }(kr)kr]\sqrt{\frac{j+1}{2j+1}}\frac{%
u_{(j+1)1j}^{-}(r)}{r}+[(1+\frac{3}{j})j_{j^{\prime }}(kr)+\frac{1}{j}%
j_{j^{\prime }}^{\prime }(kr)kr]\sqrt{\frac{j}{2j+1}}\frac{u_{(j-1)1j}^{-}(r)%
}{r}\}  \notag \\
&&+j_{j^{\prime }}(kr)[2\sinh K(mr+1)[-\sqrt{\frac{j+1}{2j+1}}\frac{%
u_{(j+1)1j}^{-}}{r}+\frac{u_{(j-1)1j}^{-}}{r}\sqrt{\frac{j}{2j+1}}]{\biggr )}
\notag \\
&&\times {\biggl (}(\boldsymbol{\epsilon }\mathbf{^{(\alpha _{1})}\cdot }%
\boldsymbol{\epsilon }\mathbf{^{(\alpha _{2})})}\langle j2;00|j^{\prime
}0\rangle \delta _{m0}  \notag \\
&&-\sqrt{6}[\boldsymbol{\epsilon }^{(\alpha _{1})}\cdot (\boldsymbol{%
\epsilon }^{(+)}\boldsymbol{\epsilon }^{(+)}\mathbf{)}\cdot %
\boldsymbol{\epsilon }^{(\alpha _{2})}\langle j2;-22|j^{\prime }0\rangle
\delta _{m-2}  \notag \\
&&-\sqrt{6}\boldsymbol{\epsilon }^{(\alpha _{1})}\cdot (\boldsymbol{\epsilon
}^{(-)}\boldsymbol{\epsilon }^{(-)}\mathbf{)}\cdot \boldsymbol{\epsilon }%
^{(\alpha _{2})}\langle j2;2-2|j^{\prime }0\rangle \delta _{m2}\}{\biggr )%
\biggr ]}
\end{eqnarray}%
From this we obtain the result Eq. (\ref{mpm}).

\section{ Positronium Decay}

In this appendix we focus on positronium decay amplitudes assuming weak
potentials ($\mathcal{L}$, $\mathcal{K}=1$). \ We wish to verify that our
relativistic formalism gives the standard results for the $%
^{1}S_{0},^{3}P_{0}$, and $^{3}P_{2}$ decays.

\subsection{$^{1}L_{l}$ amplitudes}

\ The $^{1}L_{l}$ amplitude is 
\begin{equation}
\mathcal{M}_{^{1}L_{l}\rightarrow 2\gamma }=-\sqrt{2j+1}\boldsymbol{\epsilon
}^{(\alpha _{1})}\times \boldsymbol{\epsilon }^{(\alpha _{2})}\cdot \mathbf{%
\ \hat{k}\{}F_{j=l}(1-(-)^{j})\delta _{m0}+\sum_{j^{\prime }=\left\vert
j-1\right\vert }^{j+1}G_{j=l}^{(j^{\prime })}(1-(-)^{j^{\prime }})\langle
j1;00|j^{\prime }0\rangle \langle j1;00|j^{\prime }0\rangle \delta _{m0}\},
\end{equation}%
in which 
\begin{eqnarray}
F_{j=l} &=&-2i\pi e^{2}(-i)^{j}\int_{0}^{\infty }drr^{2}m\frac{\exp (-mr)}{r}%
j_{j}(kr)kr(\frac{1}{j+1}\sqrt{\frac{j+1}{2j+1}}\frac{\upsilon
_{(j+1)1j}^{+}(r)}{r}+\frac{1}{j}\sqrt{\frac{j}{2j+1}}\frac{\upsilon
_{(j-1)1j}^{+}(r)}{r}),  \notag \\
G_{j=l}^{(j^{\prime })} &=&-2\pi e^{2}(-i)^{j^{\prime }}\int_{0}^{\infty
}dr\exp (-mr){\biggl (}(mr+1)[j_{j^{\prime }}(kr)\frac{u_{j0j}^{-}(r)}{r} 
\notag \\
&&+mr\mathbf{[(}-1+\frac{2}{j+1})j_{j^{\prime }}(kr)+\frac{1}{j+1}%
j_{j^{\prime }}^{\prime }(kr)kr]\sqrt{\frac{j+1}{2j+1}}\frac{\upsilon
_{(j+1)1j}^{+}(r)}{r}  \notag \\
&&+[\mathbf{(}1+\frac{2}{j})j_{j^{\prime }}(kr)+\frac{1}{j}j_{j^{\prime
}}^{\prime }(kr)kr]\sqrt{\frac{j}{2j+1}}\frac{\upsilon _{(j-1)1j}^{+}(r)}{r}{%
\biggr ).}
\end{eqnarray}%
We tabluate here the portions of the wave function that will contribute to
the same order in $\alpha $ to the decay amplitude. \ We consider the
relation between the various contributing components of the wave function. \
The relation between $u_{j0j}^{-}$ and $u_{j0j}^{+}$ from Eq.(\ref{ls}) is$%
~u_{j0j}^{-}=\frac{M}{E}u_{j0j}^{+}$. \ For electromagnetic interactions
only (see \cite{em}) 
\begin{equation}
M=m,E=\frac{w}{2}\sqrt{1-\frac{2A}{w}}
\end{equation}%
and for positronium $A=-\alpha /r$. \ For positronium, $w=m+O(\alpha ^{2})$
and so 
\begin{equation}
\frac{M}{E}=\sqrt{\frac{mr}{mr+\alpha }}(1+O(\alpha ^{2}))
\end{equation}%
$\ $Next consider the contribution of the small component wave functions $%
\upsilon _{(j-1)1j}^{+}$ and $\upsilon _{(j+1)1j}^{+}.$ From Eq.(\ref{jl})
for positronium we have%
\begin{eqnarray}
\frac{\upsilon _{(j-1)1j}^{+}}{r} &=&\frac{1}{m}[-\frac{d}{dr}-\frac{(j+1)}{r%
}]\psi _{j0j}\sqrt{\frac{j}{2j+1}}\text{ }+O(\alpha ^{2}),  \notag \\
\frac{\upsilon _{(j+1)1j}^{+}}{r} &=&\frac{1}{m}[\frac{d}{dr}-\frac{j}{r}%
]\psi _{j0j}\sqrt{\frac{j+1}{2j+1}}+O(\alpha ^{2}).
\end{eqnarray}%
We focus on the decay amplitude for $^{1}S_{0}$ positronium ($j=0,j^{\prime
}=1$). In that case, our amplitudes are%
\begin{eqnarray}
F_{0} &=&-2i\pi e^{2}\int_{0}^{\infty }drr^{2}m\frac{\exp (-mr)}{r}%
j_{1}(kr)kr(\frac{1}{m}\frac{d}{dr}\psi _{000}),  \notag \\
G_{0}^{(0)} &=&-2\pi e^{2}\int_{0}^{\infty }dr\exp (-mr)\big((mr+1)[j_{1}(kr)%
\sqrt{\frac{mr}{mr+\alpha }}\psi _{000}  \notag \\
&&+mr\mathbf{[}j_{1}(kr)+j_{0}^{\prime }(kr)kr]\frac{1}{m}\frac{d}{dr}\psi
_{000}\big),
\end{eqnarray}%
with the nonrelativistic wave function given by%
\begin{equation}
\psi _{000}=\frac{{(}m\alpha {)^{3/2}}}{\sqrt{8\pi }}\exp (-\alpha mr)=\frac{%
R(r)}{\sqrt{4\pi }}.
\end{equation}%
Because of the short ranged $\exp (-mr)$ factor, the exponential wave
function can be replaced with its value at the origin. Hence (with $%
k=m(1+O(\alpha ^{2}))$ we have 
\begin{eqnarray}
F_{0} &=&-2i\pi e^{2}(-\alpha )\frac{{(}m\alpha {)^{3/2}}}{\sqrt{8\pi }}%
\int_{0}^{\infty }drr^{2}m\frac{\exp (-mr)}{r}j_{1}(mr)mr  \notag \\
&=&-2i\pi (4\pi \alpha )(-\alpha )\frac{{(}m\alpha {)^{3/2}}}{m\sqrt{8\pi }}%
\int_{0}^{\infty }dxx^{2}\exp (-x)j_{1}(x)\sim m^{1/2}\alpha ^{7/2},  \notag
\\
G_{0}^{(1)} &=&-2\pi e^{2}\frac{{(}m\alpha {)^{3/2}}}{\sqrt{8\pi }}%
\int_{0}^{\infty }dr\exp (-mr)\big((mr+1)[j_{1}(mr)\sqrt{\frac{mr}{mr+\alpha 
}}  \notag \\
&&+mr\mathbf{[}j_{1}(mr)+j_{1}^{\prime }(mr)mr](-\alpha )\big)  \notag \\
&=&-2\pi (4\pi \alpha )\frac{{(}m\alpha {)^{3/2}}}{m\sqrt{8\pi }}%
\int_{0}^{\infty }dx\exp (-x)\big((x+1)[j_{1}(x)\sqrt{\frac{x}{x+\alpha }}%
-\alpha x\cos x\mathbf{[}j_{1}(x)+j_{1}^{\prime }(x)x]\big)  \notag \\
&=&-2\pi (4\pi \alpha )\frac{{(}m\alpha {)^{3/2}}}{m\sqrt{8\pi }}%
\int_{0}^{\infty }dx\exp (-x)(x+1)j_{1}(x)\sqrt{\frac{x}{x+\alpha }}%
+O(m^{1/2}\alpha ^{7/2}).
\end{eqnarray}%
So, the small components do not contribute to the singlet decay rate at the
order $m^{1/2}\alpha ^{5/2}$. The term that does is%
\begin{eqnarray}
G_{0}^{(1)} &=&\frac{{(}m\alpha {)^{3/2}}}{\sqrt{2}}\frac{8\pi ^{2}\alpha i}{%
m}g(\alpha ), \\
g(\alpha ) &=&\int_{0}^{\infty }dxj_{1}(x)\exp (-x)(1+x)\sqrt{\frac{x}{%
x+\alpha }}  \notag \\
&=&g(0)+\alpha g^{\prime }(0)+...,  \notag
\end{eqnarray}%
with 
\begin{equation}
g(0)=\int_{0}^{\infty }dxj_{1}(x)\exp (-x)(1+x)=\frac{1}{2},
\end{equation}%
and%
\begin{equation}
g^{\prime }(0)=-\frac{1}{2}\int_{0}^{\infty }dxj_{1}(x)\exp (-x)(1+x)\frac{1%
}{x},
\end{equation}%
which is finite (the integrand behaves like a constant at the origin). \
Hence to lowest order we find%
\begin{equation}
G_{0}^{(1)}=m^{1/2}\alpha ^{5/2}2\sqrt{8}\pi ^{2}\times \frac{1}{2}=2\sqrt{2}%
m^{1/2}\alpha ^{5/2}\pi ^{2},
\end{equation}%
which gives the correct form for the decay rate:%
\begin{equation}
\Gamma =\frac{\left\vert G_{0}^{(1)}\right\vert ^{2}}{(2\pi )^{4}}%
=\left\vert R(0)\right\vert ^{2}\frac{\alpha ^{2}}{m^{2}}=\frac{m\alpha ^{5}%
}{2}.
\end{equation}

For later use define 
\begin{equation}
K_{j,k}=\int_{0}^{\infty }dxj_{j}(x)\exp (-x)x^{j+k};~2j+\ k>-1,
\end{equation}%
and 
\begin{equation}
I_{jj^{\prime }}=\frac{1}{j!}K_{j^{\prime },j-j^{\prime }.}
\end{equation}%
Now we have%
\begin{equation}
j_{n}(x)=2^{n}x^{n}\sum_{k=0}^{\infty }\frac{(-)^{k}(k+n)!}{k!(2k+2n+1)!}%
x^{2k},
\end{equation}%
and so%
\begin{eqnarray}
K_{j,k} &=&2^{j}\sum_{l=0}^{\infty }\frac{(-)^{l}(l+j)!}{l!(2l+2j+1)!}%
\int_{0}^{\infty }dx\exp (-x)x^{j+k+2l+j}  \notag \\
&=&\lim_{z\rightarrow 1}2^{j}\sum_{l=0}^{\infty }\frac{%
(-)^{l}(l+j)!(k+2j+2l)!}{l!(2l+2j+1)!}z^{l}.  \label{kjk}
\end{eqnarray}%
We evaluate numerous related forms of these summations for several sets of
needed indices $j,k$ in the section following the one below.

$^{3}L_{j=l\pm 1}$ amplitudes

\ The $^{3}L_{j=l\pm 1}$ amplitudes are 
\begin{eqnarray}
A_{j=l\pm 1} &=&i\frac{2\pi e^{2}}{3}(-i)^{j}\int_{0}^{\infty }dr\exp (-mr){%
\biggl (}(mr+1)  \notag \\
&&\times \{[j_{j}(kr)+\frac{2}{(j+1)}j_{j}^{\prime }(kr)kr]\sqrt{\frac{j+1}{%
2j+1}}\frac{u_{(j+1)1j}^{-}(r)}{r}+[-j_{j}(kr)+\frac{2}{j}j_{j}^{\prime
}(kr)kr]\sqrt{\frac{j}{2j+1}}\frac{u_{(j-1)1j}^{-}(r)}{r}\}  \notag \\
&&-3j_{j}(kr)mr\frac{\upsilon _{j0j}^{+}(r)}{r}{\biggr ),}  \notag \\
B_{j=l\pm 1}^{(j^{\prime })} &=&i\frac{2\pi e^{2}}{3}(-i)^{j^{\prime
}}\int_{0}^{\infty }dr\exp (-mr)(mr+1)  \notag \\
&&\times \{[(-1+\frac{3}{j+1})j_{j^{\prime }}(kr)+\frac{1}{j+1}j_{j^{\prime
}}^{\prime }(kr)kr]\sqrt{\frac{j+1}{2j+1}}\frac{u_{(j+1)1j}^{-}(r)}{r} 
\notag \\
&&+[(1+\frac{3}{j})j_{j^{\prime }}(kr)+\frac{1}{j}j_{j^{\prime }}^{\prime
}(kr)kr]\sqrt{\frac{j}{2j+1}}\frac{u_{(j-1)1j}^{-}(r)}{r}\}.
\end{eqnarray}%
We neglect the effects of orbital angular momentum coupling. Then 
\begin{equation}
A_{j=l-1}=i\frac{2\pi e^{2}}{3}(-i)^{j}\int_{0}^{\infty }dr\exp (-mr)\big(%
(mr+1)(j_{j}(kr)+\frac{2}{j+1}j_{j}^{\prime }(kr)kr)\sqrt{\frac{j+1}{2j+1}}%
\frac{u_{(j+1)1j}^{-}(r)}{r}-3j_{j}(kr)mr\frac{\upsilon _{j0j}^{+}(r)}{r}%
\big){\large ,}
\end{equation}%
and%
\begin{equation}
A_{j=l+1}=i\frac{2\pi e^{2}}{3}(-i)^{j}\int_{0}^{\infty }dr\exp (-mr)\big(%
(mr+1)(-j_{j}(kr)+\frac{2}{j}j_{j}^{\prime }(kr)kr)\sqrt{\frac{j}{2j+1}}%
\frac{u_{(j-1)1j}^{-}(r)}{r}-3j_{j}(kr)mr\frac{\upsilon _{j0j}^{+}(r)}{r}%
\big),
\end{equation}%
The connection between the wave functions $u_{(j\pm 1)1j}^{-}$ and $u_{(j\pm
1)1j}^{+}$ (see Eqs. (\ref{j}) and (\ref{ujp}) )appears complicated, but
specializing as in the singlet case, we find that the terms beyond the first
include higher order $\alpha $ terms from the various potential. \
Simplifying we find 
\begin{equation}
\frac{u_{(j-1)1j}^{-}}{r}=\frac{m}{E}\frac{u_{(j-1)1j}^{+}}{r}-\frac{\exp (3%
\mathcal{G})}{m^{2}\left( 2j+1\right) }\{[\frac{A_{mm}}{r^{2}}+\frac{F_{mm}}{%
r}\frac{d}{dr}]\frac{u_{(j-1)1j}^{+}}{r}\},
\end{equation}%
and%
\begin{equation}
\frac{u_{(j+1)1j}^{-}}{r}=\frac{m}{E}\frac{u_{(j+1)1j}^{+}}{r}-\frac{\exp (3%
\mathcal{G})}{m^{2}\left( 2j+1\right) }\{[\frac{A_{pp}}{r^{2}}+\frac{F_{pp}}{%
r}\frac{d}{dr}]\frac{u_{(j+1)1j}^{+}}{r}\}.
\end{equation}%
in which%
\begin{eqnarray}
A_{mm} &=&-2j(j+1)Q_{m}=-A_{pp},  \notag \\
F_{mm} &=&2j(j+1)(Q_{p}-Q_{m})=-F_{pp},
\end{eqnarray}%
where%
\begin{eqnarray}
Q_{p} &\equiv &\sqrt{\frac{1}{1-2A/w}}-1=(\sqrt{\frac{mr}{mr+\alpha }}%
-1)+O(\alpha ^{2}),  \notag \\
Q_{m} &\equiv &\sqrt{1-2A/w}-1=(\sqrt{\frac{mr+\alpha }{mr}}-1)+O(\alpha
^{2}).
\end{eqnarray}%
Thus%
\begin{eqnarray}
\frac{u_{(j-1)1j}^{-}}{r} &=&\sqrt{\frac{mr}{mr+\alpha }}\frac{%
u_{(j-1)1j}^{+}}{r}-\frac{2j(j+1)}{m^{2}\left( 2j+1\right) }\sqrt{\left( 
\frac{mr}{mr+\alpha }\right) ^{3}}  \notag \\
&&\times \{[-\frac{1}{r^{2}}(\sqrt{\frac{mr+\alpha }{mr}}-1)+\frac{1}{r}(%
\sqrt{\frac{mr}{mr+\alpha }}-\sqrt{\frac{mr+\alpha }{mr}})\frac{d}{dr}]\frac{%
u_{(j-1)1j}^{+}}{r}\},
\end{eqnarray}%
and%
\begin{eqnarray}
\frac{u_{(j+1)1j}^{-}}{r} &=&\sqrt{\frac{mr}{mr+\alpha }}\frac{%
u_{(j+1)1j}^{+}}{r}+\frac{2j(j+1)}{m^{2}\left( 2j+1\right) }\sqrt{\left( 
\frac{mr}{mr+\alpha }\right) ^{3}}  \notag \\
&&\times \{[-\frac{1}{r^{2}}(\sqrt{\frac{mr+\alpha }{mr}}-1)+\frac{1}{r}(%
\sqrt{\frac{mr}{mr+\alpha }}-\sqrt{\frac{mr+\alpha }{mr}})\frac{d}{dr}]\frac{%
u_{(j+1)1j}^{+}}{r}\}.
\end{eqnarray}

For the nonrelativistic wave functions we have%
\begin{equation*}
\frac{u_{(j\pm 1)1j}^{+}}{r}=R_{(j\pm 1)1j}(r)=r^{j\pm 1}\chi _{(j\pm
1)1j}(r).
\end{equation*}%
We also need the small component wave function%
\begin{eqnarray}
\frac{\upsilon _{j0j}}{r} &=&\frac{\exp (3\mathcal{G)}}{m}\{[\frac{%
(j-1)-2Q_{m}}{r}-(Q_{m}+1)\frac{d}{dr}]\sqrt{\frac{j}{2j+1}}\frac{%
u_{(j-1)1j}^{+}}{r}  \notag \\
&&+[\frac{(j+2)+2Q_{m}}{r}+(Q_{m}+1)\frac{d}{dr}]\sqrt{\frac{j+1}{2j+1}}%
\frac{u_{(j+1)1j}^{+}}{r}\},  \label{sml}
\end{eqnarray}

For $j=0=l-1$ we have%
\begin{eqnarray}
A_{0} &=&i\frac{2\pi e^{2}}{3}\int_{0}^{\infty }dr\exp (-mr)\big(%
(mr+1)(j_{0}(kr)+2j_{0}^{\prime }(kr)kr)\sqrt{\frac{mr}{mr+\alpha }}r\chi
_{110}(r)  \notag \\
&&-3j_{0}(kr)\sqrt{\left( \frac{mr}{mr+\alpha }\right) ^{3}}(Q_{m}+1)[2+r%
\frac{d}{dr}]r\chi _{110}(r)\big)  \notag \\
&=&i\frac{2\pi e^{2}}{3m}\int_{0}^{\infty }dx\exp (-x)\big(%
(x+1)(j_{0}(x)+2j_{0}^{\prime }(x)x)\sqrt{\frac{x}{x+\alpha }}r\chi _{110}(r)
\notag \\
&&-3j_{0}(x)\frac{x}{x+\alpha }[2+x\frac{d}{dx}]r\chi _{110}(r)\big){\large ,%
}
\end{eqnarray}%
where as before we let $x=mr=kr$. \ As before to lowest order we can replace 
$\frac{x}{x+\alpha }\rightarrow 1$ giving us%
\begin{equation}
A_{0}=i\frac{8\pi ^{2}\alpha }{3m^{2}}\chi _{110}(0)\int_{0}^{\infty }dx\exp
(-x)x\big((x+1)(j_{0}(x)+2j_{0}^{\prime }(x)x)-9j_{0}(x)\big){\large .}
\end{equation}%
(The factor of $9j_{0}(x)$ in the integrand would be absent if we had
ignored the small component $v~$of the wave function of Eq. (\ref{sml}).) \
Similarly%
\begin{equation}
B_{0}^{(2)}=i\frac{8\pi ^{2}\alpha }{3m^{2}}\chi _{110}(0)\int_{0}^{\infty
}dx\exp (-x)(x+1)x[2j_{2}(x)+j_{2}^{\prime }(x)x]
\end{equation}%
or since%
\begin{equation}
\frac{d}{dr}R_{110}(r)|_{r=0}=\frac{d}{dr}r\chi _{110}(r)|_{r=0}=\chi
_{110}(0)
\end{equation}%
we have%
\begin{eqnarray}
A_{0} &=&i\frac{8\pi ^{2}\alpha }{3m^{2}}\frac{d}{dr}R_{110}(r)|_{r=0}I_{1} 
\notag \\
B_{0}^{(2)} &=&i\frac{8\pi ^{2}\alpha }{3m^{2}}\frac{d}{dr}%
R_{110}(r)|_{r=0}J_{1}^{2}
\end{eqnarray}%
where%
\begin{eqnarray}
I_{1} &=&\int_{0}^{\infty }dx\exp (-x)x\big((x+1)(j_{0}(x)+2j_{0}^{\prime
}(x)x)-9j_{0}(x)\big)  \notag \\
J_{1}^{2} &=&\int_{0}^{\infty }dx\exp (-x)(x+1)x[2j_{2}(x)+j_{2}^{\prime
}(x)x]
\end{eqnarray}%
Integration by parts gives 
\begin{equation}
I_{1}=2K_{0,3}-3K_{0,2}-3K_{0,1}-9K_{0,1},
\end{equation}%
where%
\begin{equation}
K_{j,k}=\int_{0}^{\infty }dxj_{j}(x)\exp (-x)x^{j+k};~2j+\ k>-1,
\end{equation}%
and 
\begin{equation}
J_{1}^{(2)}=K_{2,1}.
\end{equation}%
Since the decay rate is 
\begin{equation}
\Gamma (^{3}P_{0}\rightarrow 2\gamma )=\frac{1}{(2\pi )^{4}}%
|A_{0}+2B_{0}^{(2)}|^{2}=\frac{4\alpha ^{2}}{9m^{4}}R_{110}^{^{\prime
}}(0)^{2}|I_{1}+2J_{1}^{(2)}|^{2},
\end{equation}%
and%
\begin{eqnarray*}
K_{0,3} &=&K_{0,2}=K_{0,1}=\frac{1}{2} \\
K_{2,1} &=&1
\end{eqnarray*}%
and so%
\begin{eqnarray*}
I_{1} &=&-\frac{13}{2} \\
J_{1}^{(2)} &=&1
\end{eqnarray*}%
and we obtain%
\begin{equation}
\Gamma (^{3}P_{0}\rightarrow 2\gamma )=\frac{9\alpha ^{2}}{m^{4}}\left\vert
R_{110}^{^{\prime }}(0)\right\vert ^{2},
\end{equation}%
Since%
\begin{equation*}
\left( R_{(n=2)110}^{^{\prime }}(0)\right) ^{2}=\frac{(m\alpha )^{5}}{%
(24)(32)}
\end{equation*}%
we obtain%
\begin{equation}
\Gamma (^{3}P_{0}\rightarrow 2\gamma )=\frac{3m\alpha ^{7}}{256}
\end{equation}%
which agrees with the results of \cite{alx} and \cite{beri}. We point out
that without the small component wave function (\ref{sml}) we would have
obtain zero for this rate.

Next we consider the case for $j=2,$ $l=j-1=1,~j^{\prime }=0,2,4.$ The
relevant amplitudes are (ignoring angular momentum coupling)%
\begin{equation}
A_{2}=-i\frac{2\pi e^{2}}{3}\int_{0}^{\infty }dr\exp (-mr)\big(%
(mr+1)(-j_{2}(kr)+j_{2}^{\prime }(kr)kr)\sqrt{\frac{2}{5}}\frac{%
u_{112}^{-}(r)}{r}-3j_{2}(kr)mr\frac{\upsilon _{202}^{+}(r)}{r}\big),
\end{equation}%
and%
\begin{eqnarray}
B_{2}^{(0)} &=&-i\frac{2\pi e^{2}}{3}\int_{0}^{\infty }dr\exp (-mr)(mr+1) 
\notag \\
&&\times \{[\frac{5}{2}j_{0}(kr)+\frac{1}{2}j_{0}^{\prime }(kr)kr]\sqrt{%
\frac{2}{5}}\frac{u_{112}^{-}(r)}{r}\},  \notag \\
B_{2}^{(2)} &=&+i\frac{2\pi e^{2}}{3}\int_{0}^{\infty }dr\exp (-mr)(mr+1) 
\notag \\
&&\times \{[\frac{5}{2}j_{2}(kr)+\frac{1}{2}j_{2}^{\prime }(kr)kr]\sqrt{%
\frac{2}{5}}\frac{u_{112}^{-}(r)}{r}\},  \notag \\
B_{2}^{(4)} &=&-i\frac{2\pi e^{2}}{3}\int_{0}^{\infty }dr\exp (-mr)(mr+1) 
\notag \\
&&\times \{[\frac{5}{2}j_{4}(kr)+\frac{1}{2}j_{4}^{\prime }(kr)kr]\sqrt{%
\frac{2}{5}}\frac{u_{112}^{-}(r)}{r}\},
\end{eqnarray}%
with the neglect of orbital mixing where 
\begin{eqnarray*}
\frac{u_{112}^{-}}{r} &=&\sqrt{\frac{mr}{mr+\alpha }}\frac{u_{112}^{+}}{r}-%
\frac{12}{5m^{2}}\sqrt{\left( \frac{mr}{mr+\alpha }\right) ^{3}} \\
&&\times \{[-\frac{1}{r^{2}}(\sqrt{\frac{mr+\alpha }{mr}}-1)+\frac{1}{r}(%
\sqrt{\frac{mr}{mr+\alpha }}-\sqrt{\frac{mr+\alpha }{mr}})\frac{d}{dr}]\frac{%
u_{112}^{+}}{r}\}, \\
\frac{\upsilon _{202}}{r} &=&\frac{1}{m}\sqrt{\left( \frac{mr}{mr+\alpha }%
\right) ^{3}}[\frac{3-2\sqrt{\frac{mr+\alpha }{mr}}}{r}-\sqrt{\frac{%
mr+\alpha }{mr}}\frac{d}{dr}]\sqrt{\frac{2}{5}}\frac{u_{112}^{+}}{r}
\end{eqnarray*}%
Using the above expressions for $\upsilon _{202}^{+}$ and $u_{112}^{-}(r)$
with%
\begin{equation}
\frac{u_{112}^{+}}{r}=R_{112}(r)=r\chi _{112}(r).
\end{equation}%
and approximations used earlier we find%
\begin{eqnarray*}
A_{2} &=&-i\frac{8\pi ^{2}\alpha }{3}\chi _{112}(0)\int_{0}^{\infty }dr\exp
(-mr)\big((mr+1)(-j_{2}(mr)+j_{2}^{\prime }(mr)mr) \\
&&\times \sqrt{\frac{2}{5}}r\{\sqrt{\frac{mr}{mr+\alpha }}-\frac{12}{5m^{2}}%
\sqrt{\left( \frac{mr}{mr+\alpha }\right) ^{3}}[-\frac{1}{r^{2}}(\sqrt{\frac{%
mr+\alpha }{mr}}-1) \\
&&+\frac{1}{r^{2}}(\sqrt{\frac{mr}{mr+\alpha }}-\sqrt{\frac{mr+\alpha }{mr}}%
)]\} \\
&&-3j_{2}(mr)mr\frac{1}{m}\sqrt{\left( \frac{mr}{mr+\alpha }\right) ^{3}}[%
\frac{3-2\sqrt{\frac{mr+\alpha }{mr}}}{r}-\frac{1}{r}\sqrt{\frac{mr+\alpha }{%
mr}}]\sqrt{\frac{2}{5}}\big),
\end{eqnarray*}%
In terms of dimensionless integration variables%
\begin{eqnarray}
A_{2} &=&-i\frac{8\pi ^{2}\alpha }{3}\frac{\chi _{112}(0)}{m^{2}}\sqrt{\frac{%
2}{5}}\mathcal{A(\alpha )}  \notag \\
\mathcal{A(\alpha )} &\mathcal{=}&\int_{0}^{\infty }dx\exp (-x)\big(%
(x+1)(-j_{2}(x)+j_{2}^{\prime }(x)x)  \notag \\
&&\times x\{\sqrt{\frac{x}{x+\alpha }}-\frac{12}{5}\sqrt{\left( \frac{x}{%
x+\alpha }\right) ^{3}}[-\frac{1}{x^{2}}(\sqrt{\frac{x+\alpha }{x}}-1) 
\notag \\
&&+\frac{1}{x^{2}}(\sqrt{\frac{x}{x+\alpha }}-\sqrt{\frac{x+\alpha }{x}})]\}
\notag \\
&&-9j_{2}(x)\sqrt{\left( \frac{x}{x+\alpha }\right) ^{3}}[1-\sqrt{\frac{%
x+\alpha }{x}}]\big),
\end{eqnarray}%
Expanding we find that%
\begin{eqnarray}
\mathcal{A(}0\mathcal{)} &\mathcal{=}&\int_{0}^{\infty }dx\exp
(-x)(x+1)(-j_{2}(x)+j_{2}^{\prime }(x)x)x  \notag \\
\mathcal{A}^{\prime }\mathcal{(}0\mathcal{)} &\mathcal{=}&\int_{0}^{\infty
}dx\exp (-x)\big((x+1)(-j_{2}(x)+j_{2}^{\prime }(x)x)  \notag \\
&&\times x\{-\frac{1}{2x}-\frac{12}{5}(1)[-\frac{1}{x^{2}}(\frac{1}{2x}) 
\notag \\
&&+\frac{1}{x^{2}}(\frac{1}{x})]\}-9j_{2}(x)(1)[-\frac{1}{2x}]\big)  \notag
\\
&=&\int_{0}^{\infty }dx\exp (-x)\big((x+1)(-j_{2}(x)+j_{2}^{\prime }(x)x) 
\notag \\
&&\times \{-\frac{1}{2}+\frac{6}{5x^{2}}\}-\frac{9j_{2}(x)}{2x}\big)
\end{eqnarray}%
Since $\mathcal{A}^{\prime }\mathcal{(}0\mathcal{)}$ is finite we can
neglect its contribution to the lowest order in $\alpha $. \ For$\mathcal{A(}%
0\mathcal{)}$ we use results below of%
\begin{equation}
\sqrt{\frac{2}{5}}\mathcal{A(}0\mathcal{)=}I_{1}=-\frac{1}{2}\sqrt{\frac{2}{5%
}}
\end{equation}%
and thus%
\begin{equation}
A_{2}=-i\frac{8\pi ^{2}\alpha }{3m^{2}}\frac{d}{dr}R_{112}(r)|_{r=0}\left( -%
\frac{1}{2}\sqrt{\frac{2}{5}}\right)
\end{equation}%
Similarly,%
\begin{eqnarray}
B_{2}^{(0)} &=&-i\frac{8\pi \alpha }{3m^{2}}\frac{d}{dr}%
R_{112}(0)J_{1}^{(0)}=-i\frac{8\pi \alpha }{3m^{2}}\frac{d}{dr}R_{112}(0)%
\frac{7}{4}\sqrt{\frac{2}{5}}  \notag \\
B_{2}^{(2)} &=&i\frac{8\pi \alpha }{3m^{2}}\frac{d}{dr}%
R_{112}(0)J_{1}^{(2)}=i\frac{8\pi \alpha }{3m^{2}}\frac{d}{dr}R_{112}(0)%
\frac{5}{4}\sqrt{\frac{2}{5}}  \notag \\
B_{2}^{(4)} &=&-i\frac{8\pi \alpha }{3m^{2}}\frac{d}{dr}%
R_{112}(0))J_{1}^{(4)}=-i\frac{8\pi \alpha }{3m^{2}}\frac{d}{dr}R_{112}(0))%
\frac{1}{2}\sqrt{\frac{2}{5}}
\end{eqnarray}%
where \ we have used%
\begin{eqnarray}
I_{1} &=&\sqrt{\frac{2}{5}}(-3K_{2,-1}-3K_{2,0}+K_{2,1})=-\frac{1}{2}\sqrt{%
\frac{2}{5}}  \notag \\
J_{1}^{(0)} &=&\sqrt{\frac{2}{5}}(\frac{3}{2}K_{0,1}+\frac{3}{2}K_{0,2}+%
\frac{1}{2}K_{0,3})=\frac{7}{4}\sqrt{\frac{2}{5}}  \notag \\
J_{1}^{(2)} &=&\sqrt{\frac{2}{5}}(\frac{3}{2}K_{2,-1}+\frac{3}{2}K_{2,0}+%
\frac{1}{2}K_{2,1})=\frac{5}{4}\sqrt{\frac{2}{5}}  \notag \\
J_{1}^{(4)} &=&\sqrt{\frac{2}{5}}(\frac{3}{2}K_{4,-3}+\frac{3}{2}K_{4,-2}+%
\frac{1}{2}K_{4,-1})=\frac{1}{2}\sqrt{\frac{2}{5}}.
\end{eqnarray}%
The rate given by Eq. (\ref{tpt}) is%
\begin{eqnarray}
\Gamma (^{3}P_{2} &\rightarrow &2\gamma )=\frac{1}{(2\pi )^{4}}[|A_{2}+\frac{%
2B_{2}^{(0)}}{5}+\frac{2B_{2}^{(2)}}{7}+\frac{36B_{2}^{(4)}}{35}|^{2}+24|%
\frac{B_{2}^{(0)}}{5}-\frac{2B_{2}^{(2)}}{7}+\frac{3B_{2}^{(4)}}{35}|^{2} 
\notag \\
&=&\frac{4\alpha ^{2}}{9m^{4}}\left( \frac{d}{dr}R_{112}(0)\right) ^{2}\frac{%
2}{5}\{|[\frac{1}{2}-\frac{7}{10}+\frac{5}{7}-\frac{18}{35}]|^{2}+24|[-\frac{%
7}{20}-\frac{5}{14}-\frac{3}{70}]|^{2}\}  \notag \\
&=&\frac{12}{5}\frac{\alpha ^{2}}{m^{4}}\left( \frac{d}{dr}R_{112}(0)\right)
^{2}  \notag \\
&=&\frac{m\alpha ^{7}}{320}
\end{eqnarray}%
in which we have assumed that%
\begin{eqnarray}
\frac{d}{dr}R_{112}(0) &=&\frac{d}{dr}R_{110}(0),  \notag \\
\left( R_{(n=2)110}^{^{\prime }}(0)\right) ^{2} &=&\frac{(m\alpha )^{5}}{%
(24)(32)}.
\end{eqnarray}%
we have therefore%
\begin{equation}
\frac{\Gamma (^{3}P_{0}\rightarrow 2\gamma )}{\Gamma (^{3}P_{2}\rightarrow
2\gamma )}=\allowbreak \frac{15}{4}.
\end{equation}%
Our amplitudes and rates for the $^{3}P_{0}$ and $^{3}P_{2}$ decays agree
with those first computed in \cite{alx}.$\allowbreak $

\subsubsection{Summation Evaluations}

Using Eq.(\ref{kjk}) and appropriate powers of the dummy variable $z$ we
obtain%
\begin{eqnarray}
K_{0,3} &=&\sum_{l=0}^{\infty }\frac{(-)^{l}(l)!(3+2l)!}{l!(2l+1)!}%
=\sum_{l=0}^{\infty }(-)^{l}(3+2l)(2+2l)z^{2l+1}  \notag \\
&=&\frac{d^{2}}{dz^{2}}z^{3}\sum_{l=0}^{\infty }(-)^{l}z^{2l}=\frac{d^{2}}{%
dz^{2}}\frac{z^{3}}{1+z^{2}}=\frac{1}{2}  \notag \\
K_{0,2} &=&\sum_{l=0}^{\infty }\frac{(-)^{l}(l)!(2+2l)!}{l!(2l+1)!}%
=\sum_{l=0}^{\infty }(-)^{l}(2+2l)  \notag \\
&=&2\frac{d}{dz}z\sum_{l=0}^{\infty }(-)^{l}z^{l}=2\frac{d}{dz}\frac{z}{1+z}=%
\frac{1}{2}  \notag \\
K_{0,1} &=&\sum_{l=0}^{\infty }\frac{(-)^{l}(l)!(1+2l)!}{l!(2l+1)!}%
=\sum_{l=0}^{\infty }(-)^{l}=\frac{1}{2}
\end{eqnarray}%
and \ thus 
\begin{eqnarray}
I_{1} &=&2K_{0,3}-3K_{0,2}-3K_{0,1}  \notag \\
&=&1-\frac{3}{2}-\frac{3}{2}=-2.
\end{eqnarray}%
Next is%
\begin{equation}
J_{1}^{(2)}=K_{2,1}
\end{equation}%
and%
\begin{eqnarray}
K_{2,1} &=&4\sum_{l=0}^{\infty }\frac{(-)^{l}(l+2)!(2l+5)!}{l!(2l+5)!}%
=4\sum_{l=0}^{\infty }(-)^{l}(l+2)(l+1)  \notag \\
&=&4\frac{d^{2}}{dz^{2}}\sum_{l=0}^{\infty }(-)^{l}z^{l+2}=4\frac{d^{2}}{%
dz^{2}}\frac{z^{2}}{(1+z)}=1.
\end{eqnarray}%
\ 

For $l=j-1$ we require the amplitudes%
\begin{eqnarray}
A_{j=l+1} &=&i\frac{8\pi ^{2}\alpha }{3m^{j}}(-i)^{j}\frac{d^{j-1}}{dr^{j-1}}%
R_{(j-1)1j}(0)I_{j-1}  \notag \\
B_{j=l+1}^{(j^{\prime })} &=&i\frac{8\pi \alpha }{3m^{j}}(-i)^{j^{\prime }}%
\frac{d^{j-1}}{dr^{j-1}}R_{(j-1)1j}(0)J_{j-1}^{(j^{\prime })},
\end{eqnarray}%
and the integrals%
\begin{eqnarray}
I_{j-1} &=&\frac{1}{(j-1)!}\int_{0}^{\infty }dx\exp
(-x)(x+1)x^{j-1}(-j_{j}(x)+\frac{2}{j}j_{j}^{\prime }(x)x)\sqrt{\frac{j}{2j+1%
}}  \notag \\
&=&\frac{1}{(j-1)!}\sqrt{\frac{j}{2j+1}}[-3K_{j,-1}-3K_{j,0}+\frac{2}{j}%
K_{j,1}],
\end{eqnarray}%
and%
\begin{eqnarray}
J_{j-1}^{(j^{\prime })} &=&\frac{1}{(j-1)!}\int_{0}^{\infty }dx\exp
(-x)(x+1)x^{j-1}((1+\frac{3}{j})j_{j^{\prime }}(kr)+\frac{1}{j}j_{j^{\prime
}}^{\prime }(x)x)\sqrt{\frac{j}{2j+1}}  \notag \\
&=&\frac{1}{(j-1)!}\sqrt{\frac{j}{2j+1}}(\frac{3}{j}K_{j^{\prime
},j-1-j^{\prime }}+\frac{3}{j}K_{j^{\prime },j-j^{\prime }}+\frac{1}{j}%
K_{j^{\prime },j+1-j^{\prime }}).
\end{eqnarray}

For $j=2$ and $j^{\prime }=0,2,4$ we have%
\begin{eqnarray}
I_{1} &=&\sqrt{\frac{2}{5}}(-3K_{2,-1}-3K_{2,0}+K_{2,1}),  \notag \\
J_{1}^{(0)} &=&\sqrt{\frac{2}{5}}(\frac{3}{2}K_{0,1}+\frac{3}{2}K_{0,2}+%
\frac{1}{2}K_{0,3}),  \notag \\
J_{1}^{(2)} &=&\sqrt{\frac{2}{5}}(\frac{3}{2}K_{2,-1}+\frac{3}{2}K_{2,0}+%
\frac{1}{2}K_{2,1}),  \notag \\
J_{1}^{(4)} &=&\sqrt{\frac{2}{5}}(\frac{3}{2}K_{4,-3}+\frac{3}{2}K_{4,-2}+%
\frac{1}{2}K_{4,-1}).
\end{eqnarray}%
Using once again Eq.(\ref{kjk}) and from above the result $K_{2,1}=1$ as
well as%
\begin{eqnarray}
K_{2,0} &=&4\sum_{l=0}^{\infty }\frac{(-)^{l}(l+2)!(4+2l)!}{l!(2l+5)!} 
\notag \\
&=&2\sum_{l=0}^{\infty }(-)^{l}lz^{l}+\sum_{l=0}^{\infty
}(-)^{l}z^{l}+3\sum_{l=0}^{\infty }(-)^{l}\frac{z^{2l+5}}{2l+5}  \notag \\
&=&(2z\frac{d}{dz}+1)\frac{1}{1+z}+3[\int_{0}^{z}dz^{\prime
}\sum_{l=0}^{\infty }(-)^{l}z^{\prime 2l}-z+\frac{z^{2}}{3}]  \notag \\
&=&(2z\frac{d}{dz}+1)\frac{1}{1+z}+3[\arctan z-z+\frac{z^{2}}{3}]  \notag \\
&=&-\frac{1}{2}+\frac{1}{2}+\frac{3}{4}\pi -3+1=-2+\frac{3}{4}\pi ,
\end{eqnarray}%
\begin{eqnarray}
K_{2,-1} &=&4\sum_{l=0}^{\infty }\frac{(-)^{l}(l+2)!(3+2l)!}{l!(2l+5)!} 
\notag \\
&=&2\sum_{l=0}^{\infty }\frac{(-)^{l}(l+1)}{(2l+5)}=\sum_{l=0}^{\infty
}(-)^{l}-3\sum_{l=0}^{\infty }\frac{(-)^{l}}{(2l+5)}  \notag \\
&=&\frac{1}{2}-\frac{3\pi }{4}+2=+\frac{5}{2}-\frac{3\pi }{4},
\end{eqnarray}%
and so%
\begin{eqnarray}
I_{1} &=&\sqrt{\frac{2}{5}}(-3K_{2,-1}-3K_{2,0}+\frac{2}{2}K_{2,1})  \notag
\\
&=&\sqrt{\frac{2}{5}}(\frac{9\pi }{4}-\frac{15}{2}+6-\frac{9\pi }{4}+1) 
\notag \\
&=&-\frac{1}{2}\sqrt{\frac{2}{5}}.
\end{eqnarray}%
For the next sum we need%
\begin{equation}
K_{0,1}=\sum_{l=0}^{\infty }\frac{(-)^{l}(1+2l)!}{(2l+1)!}=\frac{1}{2},
\end{equation}%
and%
\begin{eqnarray}
K_{0,2} &=&2\sum_{l=0}^{\infty }(-)^{l}(l+1)=2z\frac{d}{dz}%
\sum_{l=0}^{\infty }(-)^{l}z^{l}+1  \notag \\
&=&2z\frac{d}{dz}\frac{1}{1+z}+1  \notag \\
&=&-\frac{2z}{(1+z)^{2}}+1=\frac{1}{2},
\end{eqnarray}%
\begin{eqnarray}
K_{0,3} &=&\sum_{l=0}^{\infty }(-)^{l}(3+2l)2(l+1)  \notag \\
&=&2\sum_{l=0}^{\infty }(-)^{l}(2l^{2}+5l+3)  \notag \\
&=&4z\frac{d}{dz}z\frac{d}{dz}\frac{1}{1+z}+10z\frac{d}{dz}\frac{1}{1+z}+3 
\notag \\
&=&3-\frac{5}{2}=\frac{1}{2},
\end{eqnarray}%
and so%
\begin{eqnarray}
J_{1}^{(0)} &=&\sqrt{\frac{2}{5}}(\frac{3}{2}K_{0,1}+\frac{3}{2}K_{0,2}+%
\frac{1}{2}K_{0,3})  \notag \\
&=&\sqrt{\frac{2}{5}}(\frac{3}{4}+\frac{3}{4}+\frac{1}{4})=\frac{7}{4}\sqrt{%
\frac{2}{5}},
\end{eqnarray}%
\begin{eqnarray}
J_{1}^{(2)} &=&\sqrt{\frac{2}{5}}(\frac{3}{2}K_{2,-1}+\frac{3}{2}K_{2,0}+%
\frac{1}{2}K_{2,1})  \notag \\
&=&\sqrt{\frac{2}{5}}[\frac{3}{2}\left( \frac{5}{2}-\frac{3\pi }{4}\right) +%
\frac{3}{2}\left( -2+\frac{3}{4}\pi \right) +\frac{1}{2}]  \notag \\
&=&\sqrt{\frac{2}{5}}(\frac{5}{4}).
\end{eqnarray}%
And finally\ we compute 
\begin{eqnarray}
K_{4,-1} &=&16\sum_{l=0}^{\infty }\frac{(-)^{l}(l+4)!(7+2l)!}{l!(2l+9)!} 
\notag \\
&=&8\sum_{l=0}^{\infty }(-)^{l}[\frac{l^{2}}{2}+\frac{3l}{4}+\frac{17}{8}-%
\frac{105}{8\left( 2l+9\right) }]  \notag \\
&=&4z\frac{d}{dz}z\frac{d}{dz}\frac{1}{1+z}+6z\frac{d}{dz}\frac{1}{1+z}+%
\frac{17}{2}-105\sum_{l=4}^{\infty }(-)^{l}\frac{z^{2l+1}}{2l+1}  \notag \\
&=&7-105[\sum_{l=0}^{\infty }(-)^{l}\frac{z^{2l+1}}{2l+1}-z+\frac{z^{3}}{3}-%
\frac{z^{5}}{5}+\frac{z^{7}}{7}]  \notag \\
&=&83-\frac{105}{4}\pi
\end{eqnarray}%
and%
\begin{eqnarray}
K_{4,-2} &=&16\sum_{l=0}^{\infty }\frac{(-)^{l}(l+4)!(6+2l)!}{l!(2l+9)!} 
\notag \\
&=&4\sum_{l=0}^{\infty }\frac{(-)^{l}(l+3)(l+2)(l+1)}{(2l+7)}%
-4\sum_{l=0}^{\infty }\frac{(-)^{l}(l+3)(l+2)(l+1)}{(2l+9)}  \notag \\
&=&\frac{1}{2}K_{3,0}-\frac{1}{4}K_{4,-1}.
\end{eqnarray}%
Now, in addition to the result%
\begin{eqnarray*}
&&\frac{(l+3)(l+2)(l+1)}{(2l+9)} \\
&=&\frac{l^{2}}{2}+\frac{3l}{4}+\frac{17}{8}-\frac{105}{8(2l+9)},
\end{eqnarray*}%
we have%
\begin{eqnarray*}
&&\frac{(l+3)(l+2)(l+1)}{(2l+7)} \\
&=&\frac{l^{2}}{2}+\frac{5l}{4}+\frac{9}{8}-\frac{15}{8(2l+7)},
\end{eqnarray*}%
and so%
\begin{eqnarray}
&=&8\sum_{l=0}^{\infty }\frac{(-)^{l}(l+3)(l+2)(l+1)}{(2l+7)}  \notag \\
&=&4\sum_{l=0}^{\infty }(-)^{l}l^{2}z^{l}+10\sum_{l=0}^{\infty
}(-)^{l}lz^{l}+9\sum_{l=0}^{\infty }(-)^{l}z^{l}-15(-\sum_{l=3}^{\infty
}(-)^{l}\frac{z^{2l+1}}{2l+1})  \notag \\
&=&4z\frac{d}{dz}z\frac{d}{dz}\frac{1}{1+z}+10z\frac{d}{dz}\frac{1}{1+z}+%
\frac{9}{2}+15(\frac{\pi }{4}-z+\frac{z^{3}}{3}-\frac{z^{5}}{5})  \notag \\
&=&2+15(\frac{\pi }{4}-1+\frac{1}{3}-\frac{1}{5})=\frac{15\pi }{4}%
-11=K_{3,0},
\end{eqnarray}%
and so%
\begin{eqnarray}
K_{4,-2} &=&8\sum_{l=0}^{\infty }\frac{(-)^{l}(l+3)(l+2)(l+1)}{(2l+9)(2l+7)}
\notag \\
&=&\frac{15\pi }{8}-\frac{11}{2}-\frac{83}{2}+\frac{105}{8}\pi  \notag \\
&=&15\pi -47\allowbreak .
\end{eqnarray}%
The final sum to evaluate is 
\begin{eqnarray}
K_{4,-3} &=&16\sum_{l=0}^{\infty }\frac{(-)^{l}(l+4)!(5+2l)!}{l!(2l+9)!} 
\notag \\
&=&\sum_{l=0}^{\infty }(-)^{l}+\frac{15}{2}\sum_{l=0}^{\infty }\frac{(-)^{l}%
}{(2l+7)}-\frac{35}{2}\sum_{l=0}^{\infty }\frac{(-)^{l}}{(2l+9)}  \notag \\
&=&\frac{1}{2}-\frac{15}{2}[\frac{\pi }{4}-1+\frac{1}{3}-\frac{1}{5}]-\frac{%
35}{2}[\frac{\pi }{4}-1+\frac{1}{3}-\frac{1}{5}+\frac{1}{7}]  \notag \\
&=&\frac{59}{3}-\frac{25}{4}\pi ,
\end{eqnarray}%
and so 
\begin{eqnarray}
J_{1}^{(4)} &=&\sqrt{\frac{2}{5}}(\frac{3}{2}K_{4,-3}+\frac{3}{2}K_{4,-2}+%
\frac{K_{4,-1}}{2})  \notag \\
&=&\sqrt{\frac{2}{5}}[\frac{3}{2}(\frac{59}{3}-\frac{25}{4}\pi )+\frac{3}{2}%
(15\pi -47\allowbreak )+(\frac{83}{2}-\frac{105}{8}\pi )]  \notag \\
&=&\sqrt{\frac{2}{5}}\frac{1}{2}.
\end{eqnarray}%
Summarizing,%
\begin{equation*}
J_{1}^{(2)}=\sqrt{\frac{2}{5}}[\frac{5}{4}]
\end{equation*}%
\begin{equation*}
J_{1}^{(0)}=\frac{7}{4}\sqrt{\frac{2}{5}}
\end{equation*}%
\begin{equation*}
I_{1}=-\frac{1}{2}\sqrt{\frac{2}{5}}
\end{equation*}%
\begin{equation}
J_{1}^{(4)}=\sqrt{\frac{2}{5}}\frac{1}{2}.
\end{equation}


\begin{thebibliography}{99}
\bibitem{em} Eq. (\ref{em}) is for scalar and vector interactions acting
separately. \ When the two interaction are combined, \ mass and energy
potentials are $M_{i}^{2}=m_{i}^{2}+\exp (2\mathcal{G)(}2m_{w}S\mathcal{+}%
S^{2})$ and $E_{i}^{2}=\exp (2\mathcal{G)(\varepsilon }_{i}-A)^{2}~$with$%
~\exp (2\mathcal{G)=}1/(1-2A/w).$ This leads to Eq.\ref{b2}).
\end{thebibliography}
\end{document}